\documentclass[a4paper,12pt]{book}%
\usepackage[utf8]{inputenc}
\usepackage{amsmath,amssymb,mathrsfs}
\usepackage{graphicx}
\usepackage{slashed}
\usepackage{bbm}
\usepackage{fancyhdr}
\usepackage{a4wide}
\usepackage{dsfont}
\usepackage[usenames]{color}
\usepackage{lscape}
\usepackage{braket}
\usepackage{psfrag}
\usepackage[bf]{caption2}

\fancypagestyle{diploma}{
\fancyhf{}
\fancyhead[RO,LE]{\thepage}
\fancyhead[LO]{\rightmark}
\fancyhead[RE]{\leftmark}
}
\newcommand{\vtwo}[2]{\left(\begin{array}{c} #1 \\ #2 \end{array}\right)}
\newcommand{\vthree}[3]{\left(\begin{array}{c} #1 \\ #2\\ #3 \end{array}\right)}
\newcommand{\vfour}[4]{\left(\begin{array}{c} #1 \\ #2 \\ #3 \\ #4 \end{array}\right)}
\newcommand{\hatz}{\hat{\mathbf{z}}}
\newcommand{\hatx}{\hat{\mathbf{x}}}
\newcommand{\haty}{\hat{\mathbf{y}}}
\newcommand{\equa}[1]{(\ref{#1})}
\newcommand{\peta}{p_\eta}

\newcommand{\pplusbf}{ \mathbf{p_+}}
\newcommand{\kplusbf}{ \mathbf{k_+}}
\newcommand{\pminbf}{ \mathbf{p_-}}
\newcommand{\kminbf}{ \mathbf{k_-}}
\newcommand{\kbf}{ \mathbf{k}}
\newcommand{\pbf}{\mathbf{p}}
\newcommand{\Pbf}{\mathbf{P}}

\newcommand{\hpbf}{ \!\!\!\hat{\,\,\,\pplusbf}}
\newcommand{\nullbf}{ \mathbf{0}}
\newcommand{\epsbf}{{ \boldmath \mbox{$\mathbf{\epsilon}$}}}
\newcommand{\para}{\|}
\newcommand{\nn}{\nonumber}
\newcommand{\half}{ {\textstyle \frac{1}{2} }}

\newcommand{\fourth}{ {\textstyle \frac{1}{4} }}

\begin{document}

\begin{titlepage}
\begin{center}
\newcommand{\Rule}{\rule{\textwidth}{1mm}} 
\Rule
\vspace{5mm}

\Huge{Anomalous decays of pseudoscalar mesons}

\vspace{1mm}
\Rule
\vfill
\vspace{0.5cm}
\large
von\\
\vspace{1cm}
\LARGE
Thimo Petri\\
\vspace{1cm}
\Huge
Diplomarbeit in Physik\\
\vspace{1cm}
\Large
angefertigt am\\
\vspace{0.5cm}
Institut für Kernphysik\\
Forschungszentrum Jülich\\
\vspace{1cm}
\large
vorgelegt der\\ 
\vspace{0.5cm}
Mathematisch-Naturwissenschaftlichen Fakultät\\
der\\

Rheinischen Friedrich-Wilhelms-Universität Bonn\\
\vfill
im Juli/2010
\end{center}
\end{titlepage}

\newpage
\thispagestyle{empty}
\mbox{}

\newpage
\normalsize
\thispagestyle{empty}
\pagenumbering{roman}
\vspace*{14cm}
\hspace*{\fill}	\\[10mm]
Ich versichere, dass ich diese Arbeit selbständig verfasst und keine
anderen als die angegebenen Quellen und Hilfsmittel benutzt sowie die
Zitate kenntlich gemacht habe.
\vspace{3cm}
\begin{tabbing}
xxxxxxxxxxxxxxxxxx \=			\kill
Referent:	\> PD Dr. Andreas Wirzba \\
Koreferent: 	\> Prof. Dr. Ulf-G. Meißner  \\
\end{tabbing}

\newpage
\thispagestyle{empty}
\mbox{}
\tableofcontents
           
\pagestyle{diploma}
\clearpage
\pagenumbering{arabic}


\chapter{Introduction}
In this work we will discuss decays of the pseudoscalar mesons
($\pi^0$, $\eta$ and $\eta'$) which are governed by the chiral anomaly. The
chiral anomaly is the non conservation of the axial vector current under
quantization when gauge fields are present.\medskip

The anomalous decays of the $\pi^0$ were first 
measured in the 1960s (\cite{Samios:1961zz, Budagov:1960zz}and \cite{Joseph:1960zz})
and have been updated in the recent years \cite{Beddall:2008zza}. Moreover, many
calculations were performed in this sector, see e.g. \cite{Landsberg:1986fd, Kroll:1955zu, Miyazaki:1974qi, Barker:2002ib} and \cite{Lih:2009np}. Some of these decays are of special interest because their study permits
a deep insight into aspects of modern physics. Through the decays $\pi^0 \to e^+e^-\gamma$ and
$\pi^0 \to e^+e^-e^+e^-$ the single or double off-shell behavior, respectively,
of the pion-2-photon vertex can be analyzed. The decay $\pi^0\to e^+ e^-$ was
first measured by \cite{Deshpande:1993zn, McFarland:1993wv}. Even most recent calculations (\cite{Dorokhov:2007bd, Dorokhov:2008cd, Dorokhov:2009xs}) are still three standard deviations away from the modern
measurements (\cite{Abouzaid:2008cd}). This may open a window for the
search of new physics.
\smallskip

By recent measurements of
so-called $\eta$-factories, e.g. WASA@CELSIUS, WASA@COSY
(\cite{Berlowski:2008zz}), CB@MAMI, KLOE@DA$\Phi$NE (\cite{Ambrosino:2008cp}) or the
CMD-2 Collaboration (\cite{Akhmetshin:2000bw}), the decays of the $\eta$ meson
have become an important subject of modern hadron physics. Analogous to the
$\pi^0$-decays, the theoretical calculations of the decays $\eta\to l^+l^-\gamma$, $\eta\to
l^+l^-l^+l^-$ and $\eta\to l^+l^-$ can now be tested by modern
measurements. While all the above mentioned decays proceed via the
triangle anomaly, a study of the box anomaly is possible as well. This can be done by analyzing the decays $\eta\to \pi^+\pi^-\gamma$
and $\eta\to \pi^+\pi^- l^+ l^-$, see e.g. \cite{Picciotto:1991ae,
  Geng:2000fs} and \cite{Picciotto:1993aa, Nissler:2007zz}, respectively. Particularly these decays admit a study of CP-violations beyond the standard
model. The decay $\eta \to
\pi^+\pi^- \gamma$ allows a test of a formally unchecked CP-violating formalism. In the decay $\eta \to
\pi^+\pi^-l^+l^-$  CP-violation can be observed via asymmetry measurements of
the $\pi^+\pi^-$- with respect to the $l^+l^-$ decay planes. Theoretical
discussions can be found in \cite{Gao:2002gq}, while measurements were done
recently by KLOE@DA$\Phi$NE (\cite{Ambrosino:2008cp}) and are in progress by
WASA@COSY. \smallskip

In the $\eta'$ sector experimental data are very scarce and only few
theoretical calculations were done. New measurements are in progress, but it will take some time until
precise findings can be expected. Still, theoretical predictions can be made. In
addition to the analogous decays of the $\eta$-sector, also the
decay $\eta'\to \pi^+\pi^-\pi^+\pi^-$ is
kinematically allowed. Since the aim of this work is the study of the
anomalous $\eta$-decays and analogous decays in the $\pi^0$- and
$\eta'$-sector, $\eta'$-decays as the above mentioned $\eta' \to \pi^+ \pi^-
\pi^+ \pi^-$ process will not be discussed here. \medskip

Extrapolated to the chiral point, all anomalous decays are solely determined by the
Wess-Zumino-Witten Lagrangian (\cite{Wess:1971yu, Witten:1983tw}). As we
will see for decays into off-shell photon(s), we have to take the momentum dependency of the
form factors into account. To describe that we will apply vector meson
dominance (VMD)
models. Over the years the so called hidden gauge model (\cite{Bando:1984pw, Fujiwara:1984mp,Meissner:1986ka}) has reached good
agreement with the experimental data. Modern refinements were introduced by the
authors of \cite{Benayoun:2007cu, Benayoun:2009im} and \cite{Benayoun:2009fz} in the last years. We will calculate the decay rates and
branching ratios for all mentioned decays using the hidden gauge model and its
modern update in order to discuss the ability of these models to describe the experimental
data.

\bigskip

This work is organized as follows. In the second Chapter we will give a short overview
over pseudoscalar mesons and their decay channels. We will discuss symmetries
in general in order to analyze the axial anomaly in particular. Thereafter we derive
the Wess-Zumino-Witten Lagrangian according to \cite{Wess:1971yu, Witten:1983tw} and present the contact
terms which determine the chiral triangle and box anomaly at the chiral point. The
concept of vector mesons as dynamical gauge bosons of hidden local symmetries
will be discussed to explain the solutions of the anomaly equation in the presence of
vector mesons. Thereby we can present the mentioned vector meson
dominance models (the hidden gauge model and the modified vector meson dominance model \cite{Benayoun:2007cu, Benayoun:2009im} and \cite{Benayoun:2009fz})
which will be used in the following chapters. \smallskip

The third Chapter contains the derivations of the decay rates and
branching ratios of all mentioned anomalous decays. We will differentiate between the
decays proceeding via the triangle anomaly and the ones progressing via the
box anomaly, because each sector is separately governed by closely related form factors. We also present extensive
discussions of the vector meson form factors for each decay. In the case of the
box-anomaly decays, we will emphasize the contributions of the CP-violating
form factors in addition to the leading-order contribution. \smallskip

The calculations are presented in detail to be comprehensible for future
students. Up to Chapter 3, this work can be seen as a handbook for
anomalous decays, which is actually the reason for the detailed presentation of the
calculations. \smallskip

In the fourth Chapter we will discuss the results calculated for the decay
rates and branching ratios with the various vector meson dominance models. We will
show the dependence of the decay rates on the invariant masses of the outgoing
particles and compare our results for the branching ratios to other
theoretical works and experimental data. \smallskip

In the last Chapter we give a conclusion of our work and a brief outlook on
further studies.

\chapter{Theoretical background}
\section{Pseudoscalar mesons}
In the quark model the different hadrons are classified according to their quark
content. Because these particles are color-neutral states, hadrons
have to be
constructed from a quark and an antiquark or three valence quarks (or antiquarks). The hadrons
constructed by two valence quarks , a quark and an anti-quark with color and
'anti-color', respectively, are called mesons. The hadrons with three quarks with suitable
colors are called baryons. These valence
quarks give rise to the quantum numbers of the hadrons via their
flavor and via their symmetry $J^{PC}$. Here $J=L+S$ is the total angular
momentum containing orbital angular momentum $L$ and spin $S$, while $P=(-1)^{L+1}$ and $C= (-1)^{L+S}$ stands for parity and
charge conjugation. Baryons are constructed from three quarks, respectively, three
antiquarks. Thus they are fermions. Mesons contain a quark-antiquark
pair and thus are bosons. In the following
work we are only interested in light mesons built  by up, down or strange
quarks, which are subject to an approximate U(3) flavor symmetry. The resulting nine states
can be decomposed into a singlet and an octet state. Written in group notation
this means:
\begin{equation}
3\times {\bar 3} = 8 + 1
\end{equation}
The different mesons can be classified into types according to their spin
configurations. 
\begin{table}[!htp]
\begin{center}
\renewcommand{\arraystretch}{1.0}
\begin{tabular}{c||c|c|c|c||c}
Type & $S$ & $L$ & $P$ & $J$ & $J^P$ \\ \hline \hline
Pseudoscalar meson & 0 & 0 & - & 0 & $0^-$\\ \hline 
Axial vector meson & 0 & 1 & + & 1 & $1^+$\\ \hline 
Vector meson & 1 & 0 & - & 1 & $1^-$\\ \hline 
Scalar meson & 1 & 1 & + & 0 & $0^+$\\ \hline 
Tensor meson & 1 & 1 & + & 2 & $2^+$ \\ 
\multicolumn{6}{c}{\dots}
\end{tabular}
{\footnotesize  \caption{ Types of mesons\label{tab:typemeson}}}
\end{center}
  
\end{table}

\begin{figure}
  \begin{minipage}[b]{8.5 cm}
 \includegraphics[width=8.5cm]{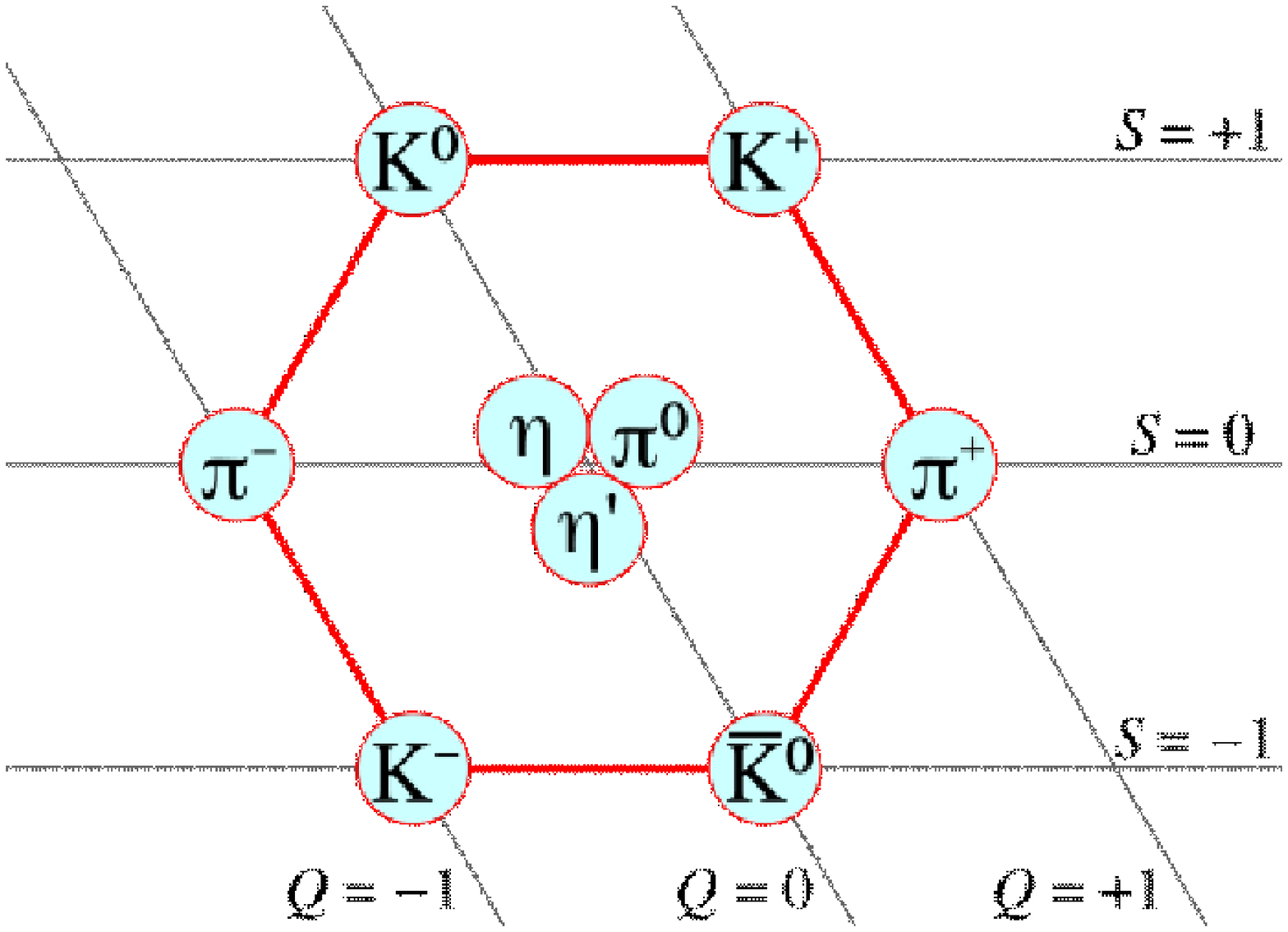}
  \caption{nonet of pseudoscalar mesons}
  \label{fig:pseudoscalar}
  \end{minipage}\hspace{0.2cm}
  \begin{minipage}[b]{8.5 cm}
 \includegraphics[width=8.5cm]{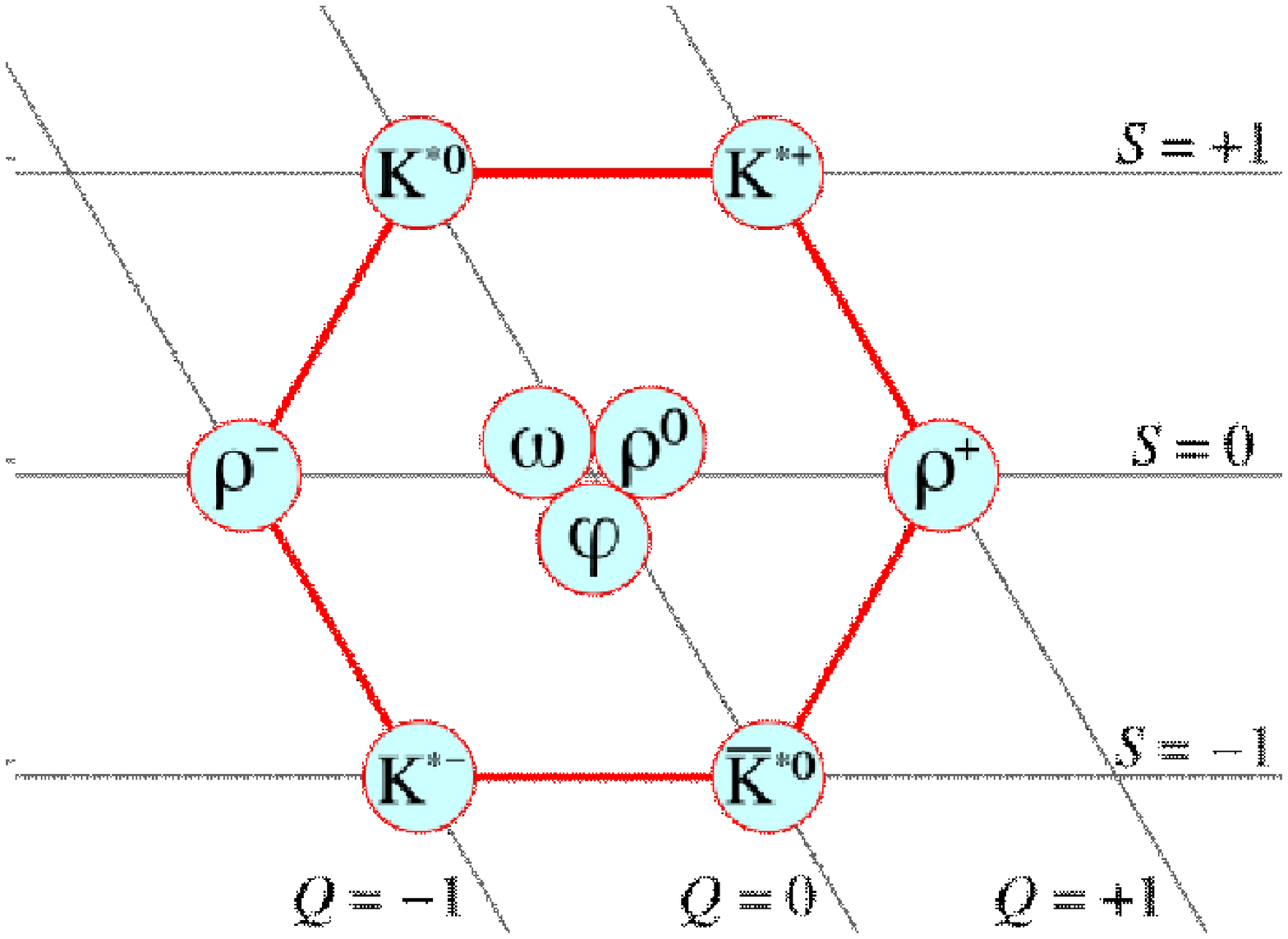}
  \caption{nonet of vector mesons}
  \label{fig:vector}
  \end{minipage}
\end{figure}
The different types are shown in Table \ref{tab:typemeson}. The nonet of the pseudoscalar mesons ($J^P=0^-$) and the nonet
of the vector mesons are shown in Figure \ref{fig:pseudoscalar} and Figure
\ref{fig:vector}. Here the charge increases towards the right and the
strangeness increases towards the upward direction. Note, that $\eta$ and $\eta'$ are not the exact octet and singlet
states, respectively. These are denoted by $\eta_0$ and $\eta_8$. The physical, measured particles
are mixings between the $\eta_0$ and $\eta_8$ states with an $\eta$-$\eta'$-mixing
angle $\theta_{mix} \approx - 20^\circ$ \cite{Holstein:2001bt}. These states can be constructed from the flavor states
according to
\begin{equation}
\begin{pmatrix} \eta \\ \eta'  \end{pmatrix} =  \begin{pmatrix}  -\sin
  \theta_{{\rm mix}} & \cos \theta_{{\rm mix}}  \\ \cos \theta_{{\rm mix}} & \sin
  \theta_{{\rm mix}}   \end{pmatrix} \cdot \begin{pmatrix} \eta_0 \\ \eta_8  \end{pmatrix}.
\end{equation}

Since we want to study the decay of the pseudoscalar mesons $\pi^0, \eta$ and
$\eta'$, we first give some general informations of this particles and then
show their different decay modes.

\bigskip

The $\pi^0$ is the lightest meson with a mass of $m_{\pi^0}= (134.9766 \pm 0.0006)
{\rm MeV}$ \cite{Yost:1988ke}. The quark content is:
\begin{equation}
 \pi^0 : \frac{1}{\sqrt{2}}\left(u \bar u  - d \bar d \right).
\end{equation}
Like all mesons, the pion is unstable, however w.r.t. electroweak decays. The decay modes are given in Table
\ref{tab:pi0}.
\begin{table}[!htp]
\begin{center}
\renewcommand{\arraystretch}{1.08}
{\small
\begin{tabular}{l|c}												
Mode	& Branching ratio \\ \hline \hline	
$\pi^0 \to 2\gamma$	 &   $ (98.823 \pm 0.034)  \cdot 10^{-2}$ \\	
$\pi^0 \to e^+ e^-\gamma$  &  $  (1.174 \pm 0.035)  \cdot 10^{-2}$ \\
$\pi^0 \to \gamma $ positronium   &  $ (1.82 \pm 0.29)  \cdot 10^{-9}$\\
$\pi^0 \to  e^+ e^+ e^- e^-	$  &  $( 3.34 \pm 0.16)  \cdot 10^{-5}$\\
$\pi^0 \to  e^+ e^-$  &  $ (6.46 \pm 0.33)  \cdot  10^{-8}$\\
$\pi^0 \to 4\gamma$	&  $<2 \cdot 10^{-8}$\\
$\pi^0 \to \nu \bar \nu$  &  $<2.7 \cdot 10^{-7}$\\
$\pi^0 \to \nu_e \bar \nu_e$  &  $<1.7 \cdot 10^{-6}$\\
$\pi^0 \to \nu_{\mu} \bar \nu_{\mu}$  &  $<1.6 \cdot 10^{-6}$\\
$\pi^0 \to \nu_\tau \bar \nu_\tau $  &  $<2.1 \cdot 10^{-6}$\\
$\pi^0 \to \gamma \nu \bar \nu$	 &  $<6 \cdot 10{-4}$
\end{tabular}
  \caption{branching ratios of the $\pi^0$ decays \cite{Amsler:2008zzb}\label{tab:pi0}}}
\end{center}
  
\end{table}\\

The $\eta$ and $\eta'$ - in comparison to the pion - have also strange quark content:
\begin{eqnarray}
 \eta_8 &:& \frac{1}{\sqrt{6}}\left(u \bar u  + d \bar d  - 2 s \bar s \right),\\
 \eta_0 &:& \sqrt{\frac{2}{3}}\left(u \bar u  + d \bar d  +  s \bar s \right).
\end{eqnarray}
The masses of the mesons are $m_\eta = 547.853 \pm 0.024 \, {\rm MeV}$ and
$m_{\eta'}=(957.78 \pm 0.06) \, {\rm MeV} $. The total decay widths were measured to
$\Gamma_\eta = (1.30 \pm 0.07) \, {\rm keV} $ for the $\eta$-meson and $\Gamma_{\eta'} =
(0.204 \pm 0.015) \, {\rm MeV} $ for the $\eta'$. The decay modes and branching ratios are
given in Table \ref{tab:eta} and Table \ref{tab:eta'}.
\begin{table}[!htp]
\begin{center}
\renewcommand{\arraystretch}{1.08}
{\small
\begin{tabular}{l|c}												
Mode	& Branching ratio \\ \hline \hline	
$\eta \to 2\gamma$	 &   $ (39.30 \pm 0.20)  \cdot 10^{-2}$ \\
$\eta \to 3 \pi^0$  &  $ (32.56 \pm 0.23)  \cdot 10^{-2}$ \\
$\eta \to \pi^0 2\gamma$	 &  $(2.7 \pm 0.5)  \cdot 10^{-4}$\\
$\eta \to \pi^0 \pi^0 \gamma \gamma$  &  $<1.2 \cdot 10^{-3}$\\
$\eta \to 4 \gamma$  &  $<2.8 \cdot 10^{-4}$\\
$\eta \to$ invisible  &  $<6 \cdot 10^{-4}$\\
$\eta \to \pi^+ \pi^- \pi^0$  &  $ (22.73 \pm 0.28 ) \cdot 10^{-2}$\\
$\eta \to \pi^+\pi^- \gamma$ & $ (4.60 \pm 0.16)  \cdot 10^{-2}$\\
$\eta \to e^+ e^- \gamma$  &  $ (7.0 \pm 0.7)  \cdot 10^{-3}$\\
$\eta \to \mu^+\mu^- \gamma$  &  $ (3.1 \pm 0.4) \cdot 10^{-4}$\\
$\eta \to e^+ e^-$  &  $<2.7 \cdot 10^{-5}$ \\
$\eta \to \mu^+\mu^-$  &   $(5.8 \pm 0.8) \cdot 10^{-6}$\\
$\eta \to e^+ e^- e^+ e^-$  &  $<6.9 \cdot 10^{-5}$\\
$\eta \to \pi^+ \pi^- e^+ e^- $  &  $ (4.2- 1.3+ 1.5 ) \cdot 10^{-4}$\\
$\eta \to e^+ e^- \mu^+ \mu^- $  &  $<1.6 \cdot 10^{-4}$\\
$\eta \to \mu^+\mu^-\mu^+\mu^-$  &  $<3.6 \cdot 10^{-4}$\\
$\eta \to \mu^+\mu^-\pi^+ \pi^-$  &  $<3.6 \cdot 10^{-4}$\\
$\eta \to \pi^+ \pi^- 2\gamma$  &  $<2.0 \cdot 10^{-3}$\\
$\eta \to \pi^+ \pi^-\pi^0 \gamma$  &  $<5 \cdot 10^{-4}$ \\
$\eta \to \pi^0 \mu^+ \mu^- \gamma $  &  $<3 \cdot 10^{-6}$
\end{tabular}
  \caption{branching ratios of the $\eta$ decays \cite{Amsler:2008zzb}\label{tab:eta}}}
\end{center}
  
\end{table}

\newpage

\begin{table}[!htp]
\begin{center}
\renewcommand{\arraystretch}{1.08}
{\small
\begin{tabular}{l|c}												
Mode	& Branching ratio \\ \hline \hline
$\eta' \to \pi^+ \pi^- \eta$  &  $ (44.6 \pm 1.4) \cdot 10^{-2}$\\
$\eta' \to \rho_0 \gamma$ (including non-resonant $\pi^+ \pi^- \gamma$)  &  $(29.4 \pm
0.9) \cdot 10^{-2}$ \\
$\eta' \to \pi^0 \pi^0 \eta$  &  $ (20.7 \pm 1.2) \cdot 10^{-2}$ \\
$\eta' \to \omega \gamma$&  $ (3.02 \pm 0.31) \cdot 10^{-2}$ \\
$\eta' \to \gamma \gamma$  &  $(2.10 \pm 0.12) \cdot 10^{-2}$ \\
$\eta' \to 3 \pi^0$  &  $(1.61 \pm 0.23 ) \cdot 10^{-3}$\\
$\eta' \to \mu^+ \mu^-\gamma$  &  $( 1.03 \pm 0.26 ) \cdot 10^{-4}$\\
$\eta' \to \pi^+ \pi^- \mu^+ \mu^-$  &  $<2.3 \cdot 10^{-4}$\\
$\eta' \to \pi^+ \pi^- \pi^0$  &  $( 3.7- 1.0+ 1.1 ) \cdot 10^{-3}$\\
$\eta' \to \pi^0 \rho ^0$  &  $<4 \cdot 10^{-2}$\\
$\eta' \to 2(\pi^+ \pi^-)$  &  $<2.5 \cdot 10^{-4}$ \\
$\eta' \to \pi^+ \pi^- 2 \pi^0$  &  $<2.6 \cdot 10^{-3}$ \\
$\eta' \to 2(\pi^+\pi^-)$ neutrals  &  $<1 \cdot 10^{-2}$\\
$\eta' \to 2(\pi^+ \pi^-) \pi^0$  &  $<1.9 \cdot 10^{-3}$ \\
$\eta' \to 2(\pi^+\pi^-) 2\pi^0$  &  $<1 \cdot 10^{-2}$ \\
$\eta' \to 3(\pi^+\pi^-)$  &  $<5 \cdot 10^{-4}$\\
$\eta' \to \pi^+ \pi ^- e^+ e^-$  &  $( 2.5- 1.0+ 1.3 ) \cdot 10^{-3}$ \\
$\eta' \to e^+ e^- \gamma$  &  $<9 \cdot 10^{-4}$ \\
$\eta' \to \pi^0 \gamma \gamma$  &  $<8 \cdot 10^{-4}$ \\
$\eta' \to 4 \pi^0$  &  $<5 \cdot 10^{-4}$ \\
$\eta' \to e^+ e^- $  &  $<2.1 \cdot 10^{-7}$ \\
$\eta' \to invisible$  &  $<9 \cdot 10^{-4}$ 
\end{tabular}
  \caption{branching ratios of the $\eta'$ decays \cite{Amsler:2008zzb}\label{tab:eta'}}}
\end{center}
  
\end{table}

\section{Symmetries and anomalies}
Transformations which do not change the physics of a system are symmetry
transformations. In classical physics this means that the
action and thereby the equation of motion are unchanged. In a quantum mechanical
formulation, e.g. in a path integral formalism, a symmetry is given if the
Lagrangian and the path integral measure are invariant under the respective
transformation. The relationship between symmetries and conversation laws is
expressed via the Noether theorem which says that for every continuous
transformation that leaves the action invariant there exists a time
independent classical charge $ Q $ and a corresponding conserved current
$\partial_\mu J^\mu = 0$.\\
There exist many different kinds of symmetries, which are all
realized by nature. We give a short overview and examples:\\
\begin{itemize}
\item exact symmetry: examples for exact symmetries are the electromagnetic
  gauge $U(1)$ or the $SU(3)$ color symmetry of QCD;
\item anomalous symmetry: If a classical symmetry is broken in quantum physics it is
  called anomalous. It is not a true
  symmetry. An example is the axial $U(1)$ symmetry, which will be discussed below;
\item explicitly broken symmetry: the symmetry is explicitly broken by a term in the
  Lagrangian that does not preserve the invariance. If this term is very small it
  is an approximate symmetry. The isospin symmetry is a common example;
%
\item hidden symmetry: if the Lagrangian is invariant, but not the ground
  state the symmetry is called hidden. Examples are the broken
  $SU(2)_L$ invariance by Higgs fields (spontaneous symmetry breaking) or the
  $SU(2)_L \times SU(2)_R$ chiral symmetry in the strong interactions
  (dynamical symmetry breaking).
%
\end{itemize} 
We are interested in the chiral $U_A(1)$ axial anomaly. The concept of anomalies was introduced by Adler, Bell and Jackiw (\cite{Adler:1969gk,Bell:1969ts}) and also by Fujikawa \cite{Fujikawa:1979ay} via path integral formalism. We will give a short
overview of the calculations given in Chapter 19 of
\cite{Peskin:1995ev}.\\
In the massless Dirac
Lagrangian the left- and right- handed
fermions are decoupled and the Lagrangian is therefore invariant under the transformation of the fields\footnote[1]{We use the standard notation of the $\gamma$-matrices according to \cite{Bjorken:1979dk}. The parameter $\theta$ is real valued and $\varepsilon^{\mu\nu\alpha\beta}$ is the total antisymmetric tensor in 3+1 dimensions}:
\begin{equation}
\Psi \rightarrow \Psi' =
e^{- i \theta \gamma_5}\Psi
\label{trafoqed}
\end{equation}
The corresponding axial current
\begin{equation}
j_{5\mu} = \bar \Psi \gamma_\mu \gamma_5 \Psi   
\end{equation}
is classically conserved,
\begin{equation}
\partial^\mu j_{5\mu} = 0.
\label{classj}
\end{equation}

\begin{figure}[t]
\begin{center}
 \includegraphics[width=15cm]{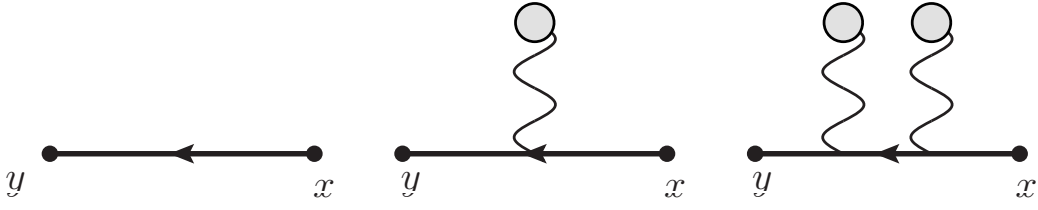}
  \caption{higher order radiative corrections of $\Psi(y) \bar\Psi (x)$}
\label{fig:radcor}
\end{center}
\end{figure}

This does not hold quantum mechanically when gauge fields are present. The axial vector
current is built from two fermion fields. Because
the product of two local operators can induce singularities, we separate their
locations $x$ and $y$, and take the limit $(y-x) \to 0$ in the
end. This is visualized in Figure \ref{fig:radcor}.\\
The lowest order contribution (without background gauge fields) results in zero, because we have to take the trace over three $\gamma$-matrices. The next
order contribution instead gives a nonvanishing result. Therefore the
divergence of the current has the following form,
\begin{equation}
\partial^\mu j_{5\mu} = - \frac{e^2}{16 \pi^2} \varepsilon^{\mu\nu\alpha\beta}
F_{\mu\nu}F_{\alpha\beta},
\label{ABJ}
\end{equation}
which is known as Adler-Bell-Jackiw anomaly \cite{Adler:1969gk, Bell:1969ts}. $F_{\mu\nu}$ is the electromagnetic field strength tensor, $F_{\mu\nu}=\partial_\mu A_\nu - \partial _\nu A_\mu$.

Another approach uses the path integral method. The result is that the
conservation of the axial current
clashes with the gauge invariance of the fields. The transformation of the
functional measure of the path integral gives an additional contribution via
the Jacobian $J$
\begin{equation}
D \Psi' D \bar \Psi ' = J^{-2} \cdot D \Psi D \bar \Psi
\end{equation}
In this case the divergence of the axial vector
current reads as
\begin{equation}
\partial^\mu j_{5\mu} = (-1)^{n+1} \frac{2e^n}{n!(4 \pi)^2} \varepsilon^{\mu_1
\mu_2 \dots \mu_n} F_{\mu_1\mu_2} \dots F_{\mu_{2n-1}\mu_{2n}}
\end{equation}
where $n=d/2$ with d the number of space time dimensions. Taking $d=4$ reproduces exactly the Adler-Bell-Jackiw anomaly given
in \equa{ABJ}.
The discussion in QCD is quite similar. The transformation has
the following form
\begin{equation}
{\rm Q} = \begin{pmatrix} u \\ d \\ s  \end{pmatrix} \rightarrow {\rm Q}' =
e^{-i  T^a \Theta_a \gamma_5}{\rm Q}
\label{trafoqcd}
\end{equation}
where $T^a=\half \lambda^a$ with the Gell-Mann Matrix $\lambda^a$. The corresponding axial current reads:
\begin{equation}
j^{\mu5a} = \bar {\rm Q} \gamma^\mu \gamma^5 T^a {\rm Q}
\end{equation}
The calculations done for the QED sector are valid here as well, either in the
point splitting formalism via
radiative corrections or in the path integral formalizm due to the jacobian. This yields
the following result
\begin{equation}
\partial_\mu j^{5\mu a} = - \frac{g^2}{16 \pi^2} \varepsilon^{\mu\nu\alpha\beta}
G_{\mu\nu}^c G_{\alpha\beta}^d \cdot {\rm Tr} [ T^a t^c t^d ]
\end{equation}
with the gluon field strength tensor $G_{\mu\nu}^c$ and the color matrices $t^c$. Because of the flavor trace this term
actually vanishes. We see that the axial flavor currents have no anomaly in
QCD. But there is an anomaly in the electromagnetic sector which is
given by
\begin{equation}
\partial_\mu j^{5\mu a} = - \frac{e^2}{16 \pi^2} \varepsilon^{\mu\nu\alpha\beta}
F_{\mu\nu}F_{\alpha\beta} \cdot {\rm Tr} [T^a Q^2].
\end{equation}
Here $Q$ is the matrix of the electric charges of the quarks,
\begin{equation}
Q = \begin{pmatrix} \frac{2}{3} & 0 &0 \\ 0 & -\frac{1}{3} &0 \\0 & 0 & -\frac{1}{3} \end{pmatrix}
\end{equation}
Since 
\begin{equation}
\lambda^3 = \begin{pmatrix} 1 & 0 &0 \\ 0 & -1 &0 \\0 & 0 &
  0 \end{pmatrix}, \hspace{1cm} \partial_\mu j^{5\mu 3} = - \frac{e^2}{32 \pi^2} \varepsilon^{\mu\nu\alpha\beta}
F_{\mu\nu}F_{\alpha\beta}
\end{equation}
Note, that $ j^{5 3}_\mu$ annihilates a $\pi^0$ and we therefore get the
anomaly contribution to the decay $\pi^0 \to \gamma \gamma$.

\section{Wess-Zumino-Witten Lagrangian (WZW)}
We want to present briefly the effective Wess-Zumino-Witten Lagrangian, which
summarizes and determines the effects of anomalies in current algebra. The
discussion will follow the presentation of \cite{Wess:1971yu} and \cite{Witten:1983tw}.\\
The QCD Lagrangian is given by:
\begin{equation}
 \mathcal {L}_{QCD} = -\half {\ Tr} [G_{\mu\nu} G^{\mu\nu}] + \bar q \left( i\gamma_\mu D^\mu - m \right)q
\end{equation}
with 
\begin{eqnarray}
 G_{\mu\nu} &=& \partial_\mu G_\nu - \partial_\nu G_\mu -i g \left[ G_\mu , G_\nu \right]\nn \\
D_\mu q &=& \left(\partial_\mu - i g G_\mu \right)q
\end{eqnarray}
where $G_\mu = G_\mu^a\lambda^a/2$ is the vector field of the gluons, $G_{\mu\nu}$ is the field strength tensor. Because of the inherent nonlinearity and
the large effective coupling constant in the non-perturbative regime, it is difficult to derive predictions about hadrons directly from the QCD Lagrangian. The most
successful way to deal with it is lattice QCD, although it needs very large
computational power and the theoretical insight is limited. In the low-energy
sector chiral perturbation theory ($\chi$PT) as low-energy effective field theory of
QCD has been successfully applied. Low energy $\chi$PT exploits the global $SU(3)_L
\times SU(3)_R $ symmetry of the QCD Lagrangian in the limit of vanishing
quark masses, which was mentioned in the previous Section. The lowest order effective chiral action is given by:
\begin{equation}
 \mathcal {S}_{{\rm eff}} = \frac{f_\pi^2}{4} \int {\rm d}^4 x {\rm Tr}
          [\left(D_\mu U\right) \left(D^\mu U^\dagger\right)]
\label{Seff}
\end{equation}
with the chiral unitary matrix
\begin{equation}
 U = {\rm exp} \left( \frac{i\sqrt{2}}{f_\pi}  P \right)
\label{U}
\end{equation}
and $f_\pi=92.4 {\rm MeV}$ is the physical pion decay constant and $P$ are the pseudoscalar fields
\begin{equation}
P = \begin{pmatrix}
  \frac{1}{\sqrt{2}}\pi^0 + \frac{1}{\sqrt{6}}\eta_8 +\frac{1}{\sqrt{3}}\eta_0
  & \pi^+ & K^+  \\ \pi^- & -\frac{1}{\sqrt{2}}\pi^0 +
  \frac{1}{\sqrt{6}}\eta_8 +\frac{1}{\sqrt{3}}\eta_0 & K^0 \\ K^- & \bar K^0 &
  -\sqrt{\frac{2}{3}}\eta_8 + \frac{1}{\sqrt{3}}\eta_0
\label{Phi}
    \end{pmatrix}.
\end{equation}
This Lagrangian does not only  contain all symmetries of the QCD Lagrangian, but possesses
extra symmetries, which are not present in the QCD Lagrangian. According to
\cite{Witten:1983tw} these symmetries are naive parity
conjugation and a $U \leftrightarrow U^\dagger$ symmetry which counts the number of the
Goldstone bosons modulo two. These separately are not symmetries of the QCD Lagrangian,
but form a symmetry (total parity) if they are combined.\\
The equation of motion derived from \equa{Seff} in the case without external fields is given by:
\begin{equation}
\partial^\mu \left(  f_\pi^2 U^\dagger \partial_\mu U \right) = 0.
\end{equation}

As shown in \cite{Witten:1983tw}, the equation of motion which violates naive
parity can be constructed by adding a symmetry violating extra term with the smallest possible number
of derivatives. This is given by
\begin{equation}
\partial_\mu \left(  f_\pi^2 U^\dagger \partial^\mu U \right) + \lambda \epsilon^{\mu\nu\alpha\beta} U^\dagger\left( \partial_\mu U \right)
U^\dagger\left( \partial_\nu U \right) U^\dagger\left( \partial_\alpha U
\right) U^\dagger\left( \partial_\beta U \right) = 0.
\end{equation}
Here $\lambda$ is a constant. Note, that typically there appears a four-dimensional antisymmetric tensor due to the violation of the naive parity conjugation.\\
 The next step is to construct a higher order Lagrangian which leads to this
 equation of motion. The derivation is quite complicated and can be found
 in \cite{Witten:1983tw}.\\
Because it is not possible to construct a closed form expression in four dimensions which breaks the symmetries the Wess Zumino action is constructed in five dimensions \cite{Witten:1983tw}:
\begin{equation}
 \Gamma_{WZ}(U)=-\frac{iN_C}{240\pi^2} \int\limits_{M^5} d^5 x \,\, \epsilon^{ijklm} Tr\left[\left((\partial_i U)U^\dagger\right)\left((\partial_j U)U^\dagger\right)\left((\partial_k U)U^\dagger\right)\left((\partial_l U)U^\dagger\right)\left((\partial_m U)U^\dagger\right)\right]
\end{equation}
This integral over a five-dimensional Manifold $M^5$ can be expressed by Stokes' theorem as an integral over the boundary of $M^5$, which is ordinary Minkowski space $M^4$.\\
This action is invariant under global charge rotations $U\rightarrow U + i \epsilon [Q,U]$, where $\epsilon$ is a constant and $Q$ the electric charge matrix of quarks. By turning this into a local symmetry $U\rightarrow U + i \epsilon (x) [Q,U]$ also the Wess-Zumino action changes to:
\begin{eqnarray}
 \tilde \Gamma (U,A_\mu) &=& \Gamma (U) - \frac{e}{48\pi^2} \varepsilon^{\mu\nu\alpha\beta} \int {\rm d}^4 x A_\mu {\rm Tr}\biggl[  Q \left( \partial_\nu U U^\dagger \right) \left( \partial_\alpha U U^\dagger \right)\left( \partial_\beta U U^\dagger \right) \nn \\
&& \mbox{} \qquad \qquad \qquad \qquad \qquad \qquad \quad +  Q \left( U^\dagger \partial_\nu U  \right) \left( U^\dagger \partial_\alpha U  \right)\left(U^\dagger \partial_\beta U  \right) \biggr] \nn \\
&& \mbox{} \qquad + \frac{ie^2}{24\pi^2}\int {\rm d}^4 x \varepsilon^{\mu\nu\alpha\beta} \left( \partial_\mu A_\nu \right) A_\alpha \nn \\
&& \mbox{} \qquad \qquad\times \left[ Q^2 \left( \partial_\beta U \right)
   U^\dagger + Q^2 U^\dagger \left( \partial_\beta U \right)  +  Q U Q
   U^\dagger \left( \partial_\beta \right) U^\dagger \right]
\label{WZWA}
\end{eqnarray}
and the effective Lagrangian becomes:
\begin{equation}
 \mathcal {L} = \frac{f_\pi^2}{4}\int {\rm d}^4 x {\rm Tr} [\left(D_\mu U\right) \left(D^\mu U^\dagger\right)] + n \tilde \Gamma
\end{equation}
which can be found in many publications, {\it e.g}
\cite{Kaymakcalan:1983qq,Gomm:1984at,Kaymakcalan:1984bz,Jain:1988se} and also
in \cite{Holstein:2001bt}.\\
Note that after expanding $U$ and integrating by parts one can find 
\begin{equation}
 A = \frac{n e^2}{96 \pi^2 f_\pi^2} \pi^0 \varepsilon^{\mu\nu\alpha\beta} F_{\mu\nu} F_{\alpha\beta},
\end{equation}
with $n=N_c$ the number of colors, which is exactly the part that describes the decay $\pi^0 \to
\gamma\gamma$. The author of ref \cite{Witten:1983tw} also found that the Noether coupling
$-eA_\mu J^\mu$ describes a $\gamma \pi^+ \pi^0 \pi^-$-vertex:
\begin{equation}
B=-\frac{1}{12}\frac{n}{\pi^2
  f_\pi^3}\epsilon^{\mu\nu\alpha\beta}A_\mu\partial_\nu \pi^+ \partial_\alpha
\pi^- \partial_\beta \pi^0.
\end{equation}
This vertex describes the coupling of a photon to three pseudoscalar mesons
and therefore the decays $\eta/\eta' \to \pi^+\pi^-\gamma$. \\
In summary the
Wess-Zumino-Witten Lagrangian already determines the triangle anomaly sector
via $A$ and the box anomaly sector via $B$.

\section{Vector meson dominance (VMD)}
Using the WZW Lagrangian the decay of a pseudoscalar particle into two photons is described very well, as we will see in the next Chapter. Because the theoretical predictions for other anomalous decays differ from the experimental data an extended model is needed. Vector meson dominance models have reached good agreement in describing the experimental data of the triangle anomaly sector and are also valid in the box anomaly sector.\\ 
Original work for the so-called hidden gauge model has been done by \cite{Bando:1984pw, Bando:1985rf}. Overviews can be
found in \cite{Meissner:1987ge} and \cite{Harada:2003jx}. We will present this subject according to \cite{Fujiwara:1984mp}
 and discuss the so-called 'total vector meson dominance', the 'hidden gauge model' and modern
 improvements done by  \cite{Benayoun:2007cu, Benayoun:2009im} and \cite{Benayoun:2009fz} afterwards.\\
\subsubsection{Vector mesons as dynamical gauge bosons of hidden local symmetries and the low-energy theorem}
The nonlinear sigma model based on the manifold $G/H = U(3)_L \times
U(3)_R/U(3)_V$ is a low energy effective theory of massless 3-flavored
QCD. The global symmetry $G_{global}=U(3)_L \times
U(3)_R$ is spontaneously broken down to the diagonal subgroup
$H_{local}=U(3)_V$. The dynamical gauge bosons of the hidden local symmetry
$H_{local}$ can be modelled as the vector mesons ($\rho, \omega \hspace{2mm} {\rm
  and} \hspace{2mm} \phi$). In this framework the KSFR relation and the
universally of $\rho$-coupling can be shown (see \cite{Kawarabayashi:1966kd, Riazuddin:1966sw}).\\
Following \cite{Fujiwara:1984mp} we introduce the variables $\xi_L(x)$ and $\xi_R(x)$ as
\begin{equation}
U(x) = \xi_L(x)^\dagger \xi_R(x)
\end{equation}
and non-abelian gauge fields $V_\mu(x) = V_\mu^a(x) T^a$.
The global group $U(3)_L \times U(3)_R$ is gauged with external fields
$A_{L \mu}$ and $A_{R \mu}$ such that the corresponding covariant derivatives
are given as
\begin{eqnarray}
D_\mu\xi_L &=& (\partial_\mu - ig V_\mu)\xi_L + i \xi_LA_{L\mu} , \nn \\
D_\mu\xi_R &=& (\partial_\mu - ig V_\mu)\xi_R + i \xi_RA_{R\mu}.
\end{eqnarray}
In the following we simplify
\begin{equation}
A_{L \mu} = A_{R \mu} = e B_\mu Q , \hspace{0.7cm} Q = \begin{pmatrix}
  \frac{2}{3} & 0 & 0  \\ 0 & -\frac{1}{3} & 0 \\ 0 & 0 &
  -\frac{1}{3}  \end{pmatrix},
\label{quarkmass}
\end{equation} 
where $B_\mu$ is the electromagnetic field.\\
Using this, the Lagrangian reads
\begin{eqnarray}
\mathcal {L} = \mathcal {L}_A + a \mathcal {L}_V + \mathcal
         {L}_{{\rm gauge} \hspace{1mm} {\rm fields}}
\end{eqnarray}
with an arbitrary parameter $a$. In this term $\mathcal {L}_{{\rm gauge} \hspace{1mm} {\rm fields}}$ stands for the kinetic terms of the gauge fields and
\begin{eqnarray}
\mathcal {L}_A = -\frac{f_\pi^2}{4} {\rm Tr}\left( D_\mu \xi_L\cdot
\xi_L^\dagger -  D_\mu \xi_R\cdot
\xi_R^\dagger \right)^2 ,\nn \\
\mathcal {L}_V = -\frac{f_\pi^2}{4} {\rm Tr}\left( D_\mu \xi_L\cdot
\xi_L^\dagger +  D_\mu \xi_R\cdot
\xi_R^\dagger \right)^2 .\nn
\end{eqnarray}
By fixing the hidden local $U(3)$ gauge to $\xi_R = \xi_L^\dagger =
e^{i\pi/f_\pi}$, the Lagrangians, when expanded, yield the following
relations:
\begin{eqnarray}
{\rm vector-photon \hspace{2mm} mixing} \hspace{1cm} g_V &=& a g f_\pi^2, \\ 
{\rm vector-\pi-\pi \hspace{2mm} coulping} \hspace{1cm} g_{V\pi\pi} &=& \half a g,  \\
{\rm vector-meson \hspace{2mm} mass} \hspace{1cm} m_V^2 &=& a g^2 f_\pi^2 .
\label{KSFR}
\end{eqnarray}
When the parameter $a$ is eliminated from the first two relations the
low-energy theorem of hidden local symmetry follows,
\begin{equation}
\frac{g_V}{g_{V\pi\pi}}=2f_\pi^2,
\end{equation}
which is known as KSFR relation (\cite{Fujiwara:1984mp}). By setting the parameter $a=2$
we get another KSFR relation namely $m_V^2=2g_{V\pi\pi} f_\pi^2$
and the universality of the vector meson coupling $g_{V\pi\pi} = g$. This was postulated by Sakurai in 1960 \cite{Sakurai:1960ju}.

\subsubsection{Solutions to the anomaly equation in the presence of vector mesons}
In the presence of vector mesons the anomaly equation has the form \cite{Fujiwara:1984mp}
\begin{equation}
\delta \Gamma (\xi_L, \xi_R, V, A_L, A_R) = -10 C \cdot i \int_{M^4} {\rm Tr}
  \left[ \epsilon_L \left( (dA_L)^2 -\frac{i}{2} dA_L^3 \right)  -
    (L \leftrightarrow R) \right],
\end{equation}
with a constants $ C = - i \frac{N_c}{240 \pi^2}$. Here the gauge variation $\delta$ contains also a hidden local symmetry
transformation, such that $\delta = \delta_L(\epsilon_L) + \delta_V(v) + \delta_R(\epsilon_R) $.
The anomaly equation is an inhomogeneous linear differential equation. So the
solution will be a special solution of the inhomogeneous equation, for which
we take the Wess-Zumino action \equa{WZWA}, and general solutions of the
homogenous equation. As shown in \cite{Fujiwara:1984mp, Meissner:1987ge} and
\cite{Benayoun:2009im} six invariants that conserve total parity but violate intrinsic
parity can be found. According to \cite{Bando:1987br, Furui:1986ep} and
\cite{Jain:1987sz} two of these invariants are charge conjugation odd and can
therefore be omitted. So the interesting part of the action is given by the
Wess-Zumino action and the remaining four invariants,
\begin{equation}
\Gamma = \Gamma_{WZ} + \sum_{i=1}^4 \int_{M^4}c_i \mathcal {L}_i.
\end{equation}
These additional terms do not contribute to anomalous processes like
$\pi^0 \to \gamma \gamma$ in the chiral limit, but can contribute away from the chiral limit via
vector mesons decaying into photons, since they have the same quantum numbers. So the anomalous Lagrangian including all couplings
is given by \cite{Benayoun:2009im}:
\begin{equation}
\mathcal {L}_{anomalous} = \mathcal {L}_{VVP} + \mathcal {L}_{AVP} + \mathcal {L}_{AAP} + \mathcal {L}_{VPPP} +\mathcal {L}_{APPP}.
\end{equation}
Here $V$, $P$ and $A$ denote the vector meson, pseudoscalar  and
electromagnetic fields. The different Lagrangian pieces are given by
\begin{eqnarray}
\mathcal {L}_{VVP} &=& -\frac{N_c g^2}{32\pi^2f_\pi} c_3
\varepsilon^{\mu\nu\alpha\beta} {\rm Tr} [\partial_\mu V_\nu \partial_\alpha
  V_\beta P] ,\nn \\
\mathcal {L}_{AVP} &=& -\frac{N_c ge}{32\pi^2f_\pi} (c_3-c_4) 
\varepsilon^{\mu\nu\alpha\beta} \partial_\mu
A_\nu  {\rm Tr} [\{ \partial_\alpha
  V_\beta, Q \} P] ,\nn \\
\mathcal {L}_{AAP} &=& -\frac{N_c e^2}{8\pi^2f_\pi} (1-c_4)
\varepsilon^{\mu\nu\alpha\beta} \partial_\mu A_\nu \partial_\alpha
  A_\beta {\rm Tr} [Q^2 P] ,\nn \\
\mathcal {L}_{VPPP} &=& - i \frac{N_c g}{64\pi^2f_\pi^3} (c_1-c_2-c_3)
\varepsilon^{\mu\nu\alpha\beta} {\rm Tr} [V_\mu \partial_\nu P \partial_\alpha
  P \partial_\beta P] ,\nn \\
\mathcal {L}_{APPP} &=& - i \frac{N_c e}{24\pi^2f_\pi^3} [1-\frac{3}{4}(c_1-c_2+c_4)]
\varepsilon^{\mu\nu\alpha\beta} A_\mu {\rm Tr} [Q\partial_\nu P
  \partial_\alpha P \partial_\beta P ] .
\label{lvmd}
\end{eqnarray}
Here $g$ is the universal vector coupling constant, $Q$ is the quark charge
matrix given in \equa{quarkmass} and $N_C$ is the number of colors. The
pseudoscalar fields
are  defined as
\begin{equation}
P = P_8 + P_0 = \sqrt{2} \begin{pmatrix}
  \frac{1}{\sqrt{2}}\pi^0 + \frac{1}{\sqrt{6}}\eta_8 +\frac{1}{\sqrt{3}}\eta_0
  & \pi^+ & K^+  \\ \pi^- & -\frac{1}{\sqrt{2}}\pi^0 +
  \frac{1}{\sqrt{6}}\eta_8 +\frac{1}{\sqrt{3}}\eta_0 & K^0 \\ K^- & \bar K^0 &
  -\sqrt{\frac{2}{3}}\eta_8 + \frac{1}{\sqrt{3}}\eta_0
    \end{pmatrix}\\
\end{equation}
and the vector meson fields read
\begin{equation}
V_\mu =  \sqrt{2} \begin{pmatrix}
  \frac{\rho^I + \omega^I}{\sqrt{2}} & \rho^+ & K^{\star+}  \\ \rho^- &
  -\frac{\rho^I + \omega^I}{\sqrt{2}} & K^{\star 0} \\ K^{\star -} & \bar
  K^{\star 0} & \phi^I
    \end{pmatrix}_\mu .\\
\end{equation}
Note that the prefactors of \cite{Benayoun:2009im} based on the normalizations of \cite{Benayoun:2007cu} differ from ours.\\
The original terms of the WZW Lagrangian are of course the ones in \equa{lvmd} ${\cal L}_{AAP}$ and ${\cal L}_{APPP}$
without vector
meson couplings. The first three Lagrangians contribute to the triangle
anomaly, while the last two contribute to the box anomaly sector.

\subsubsection{The various vector meson dominance models}
As one can see, the Lagrangians only depend on \textbf{three} different parameter
combinations, namely $c_1-c_2$, $c_3$ and $c_4$. These parameters have to be
fixed by comparison with data. For different choices of these parameter sets different vector meson dominance models result.\\
Before we present the different models, we can give some general conditions on
the parameters by discussing briefly the triangle sector. In Ref
\cite{Benayoun:2009im} the amplitude for the decay
$\pi^0 \to \gamma \gamma$ is derived from the Lagrangians given in \equa{lvmd}:
\begin{equation}
A(\pi^0 \to \gamma \gamma) = -i \frac{\alpha}{\pi
  f_\pi}[\underbrace{1-c_4}_{AAP \rm  \,\, term} + \underbrace{c_3-c_4}_{AVP \rm
   \,\,term} + \underbrace{c_3}_{VVP \rm  \,\,term}]  = -i \frac{\alpha}{\pi f_\pi}[1+2(c_3-c_4)].
\end{equation}
To make sure that the vector meson dominance factor is normalized to one at $q^2=0$ and
to recover the original WZW term we
set $c_3=c_4$. In this case the second Lagrangian will always vanish. We now
present the different VMD models.\\
The first one is the '\textbf{full vector meson dominance}':
\begin{equation}
c_3=c_4 =1 \hspace{1cm} {\rm and} \hspace{1cm} c_1-c_2 = \frac{1}{3}.
\end{equation}
Inserting these values of the parameters in \equa{lvmd}, one can easily see that
the second, third and last Lagrangian, which are the ones involving photon
fields, vanish and so photons can only couple to pseudoscalar mesons via a
vector meson. Although this model is successfully describing data in the triangle
sector we will not deal with it in our work for the following reasons: (i) In the
box anomaly sector at the chiral point, full VMD does not
produce the WZW interaction. As pointed out in
\cite{Holstein:2001bt}, the full VMD model gives a 50\% larger value
at zero four momentum than the anomaly. (ii) The authors of
\cite{Fujiwara:1984mp} and \cite{Harada:2003jx} calculated that the decay rate
of the decay $\omega\to \pi^0\pi^+\pi^-$ is $2/3$ times smaller than
the respective experimental data. \\
The following model, which we present, has the same predictions as the full VMD
model in the triangle sector, because the values of $c_3$ and $c_4$ are the
same, but it also agrees reasonably well with the accepted data referring to the
box anomaly as was shown in \cite{Holstein:2001bt}. The parameters are in this case:
\begin{equation}
c_3=c_4 =1 = c_1-c_2 = 1.
\end{equation}
Here only the first and last Lagrangian in \equa{lvmd} give contributions,
while the others vanish. We will refer to this model as the \textbf{hidden gauge model}
(see e.g. \cite{Fujiwara:1984mp}).\\
The next model, \textbf{modified VMD}, which is of interest, is a further extention of the hidden gauge
model. The authors of Refs. \cite{Benayoun:2007cu, Benayoun:2009im} and \cite{Benayoun:2009fz}
fitted the c-parameters of the different Lagrangians and the coupling constant
$g$ to the data of the the processes
\begin{eqnarray}
e^+ e^- &\to& \pi^+ \pi^- ,\nn \\
e^+ e^- &\to& (\pi^0/\eta)\gamma ,\nn \\
e^+ e^- &\to& \pi^+ \pi^- \pi^0 \nn 
\end{eqnarray}
below a CM energy of 1.05 GeV. Note that the fitted processes do not include $\eta'$, so it will be of special
interest and a nice proof of the extended model to compare the results for
the different models in the $\eta'$ sector.\\
Ref. \cite{Benayoun:2009im} presented various fits, where different data sets have been taken into
account. Two of these fits were indicated as the ones that represent the data
the best. These fits are given in table 3 of \cite{Benayoun:2009im} and were denoted by the conditions (i) Global Fit with ND+CMD and (ii) Global Fit
with ND+CMD++CMD2. The corresponding values for $g, c_3$ and $c_1 -
c_2$ are listed in Table \ref{tab:cfit}.\\
\begin{table}[!hbt]
\begin{center}
\renewcommand{\arraystretch}{1.3}
\begin{tabular}{c|c|c}
 & (i) & (ii)  \\ \hline \hline
$g$ & $5.566\pm0.010$  & $5.568\pm0.011$ \\ \hline 
$a$ & $2.364\pm0.011$  & $2.365\pm0.011$ \\ \hline 
$c_3$ & $0.927\pm0.010$ & $0.930\pm0.011$ \\ \hline 
$c_1 - c_2$ & $1.168\pm0.069$ & $1.210\pm0.043$
\end{tabular}
  \caption{Values of the fitted parameters given by the 'best data sets' in \cite{Benayoun:2009im}}
\end{center}
  \label{tab:cfit}
\end{table}

Because the fits indeed give very similar values in our calculations we
will refer to them as one model and state the discrepancies as an error range.\\
The vector meson mass is related to the coupling constant $g$ and the parameter $a$ via
\equa{KSFR}. Therefore there appear different vector meson masses in \cite{Benayoun:2009im} which  vary between
$m_V = 760 \,{\rm MeV}$ and  $m_V = 791 \,{\rm MeV}$ in some of the fits. This will
contribute to the error of the results in the end.

\newpage

\section{Definitions}

\subsection{The decay momenta}
In the next Chapter we will discuss the kinematics of the different decays of
the pseudoscalar mesons. Therefore we try to make common definition for the
four-momenta. This will work for every decay mode except the one into four
leptons, which we will relabel in the corresponding Chapter. The general
definitions are discussed here. Some further relations as well as a
comparison to the essential kinematics can be found in Appendix A.\\
The decay momenta of a pseudoscalar meson $P$ into two, three or four
particles $p_i$ are defined as follows:
\begin{eqnarray}
P(P) &\to& p_1(p) + p_2 (k) ,\nn \\
P(P) &\to& p_1^+(p_+) + p_2^-(p_-) + p_3(k), \nn \\
P(P) &\to& p_1^+(p_+) + p_2^-(p_-) + p_3^+(k_+) + p_4^-(k_-) , \nn
\end{eqnarray}
such that the following relations for the four-momenta are valid in any frame:
\begin{eqnarray}
P &=& p + k ,\nn \\
P &=& \underbrace{p_+ + p_-}_{\equiv p} + k ,\nn \\
P &=& \underbrace{p_+ + p_-}_{\equiv p} + \underbrace{k_+ + k_-}_{\equiv k} .\nn
\end{eqnarray}
Furthermore, we use the following notations: a four-momentum is denoted by an italic letter, {\it e.g.}
$q$, whereas the corresponding three-momentum is signalled by a 
bold-faced letter $\mathbf{q}$, {\it i.e.}
\begin{equation}
   q\equiv  \vtwo{q_0}{\mathbf{q}} \equiv \vfour{q^0}{q^x}{q^y}{q^z} \equiv 
    \vthree{q^0}{\mathbf{q}_\perp}{\mathbf{q}_{||}} \,.
\end{equation}
Moreover, 
\begin{eqnarray}
  \mathbf{q} &\equiv& q^x \,\hatx + q^y \,\haty  + q^z \,\hatz\,,
\\
   \mathbf{q_{\|}} &\equiv & q^z \,\hatz   \label{qpara}\,,
\\ \mathbf{q_\perp} &\equiv& \mathbf{q} - \mathbf{q_{\|}} = \mathbf{q} -\hatz
   \left(\mathbf{q}\cdot  \hatz \right) = q^x \,\hatx + q^y \,\haty\,,
  \label{qperp}
\end{eqnarray} 
where $\mathbf{q_{\|}}$is the momentum in the $z$-direction.

\subsection{The identification of the momenta}

By comparing with the text below the cross section formula (38.19) 
of Ref.\,\cite{Amsler:2008zzb}, the following identifications can easily
be made:
\begin{eqnarray}
  \pbf_+^\star &\equiv& \left(|\pbf^\star_1|,\Omega_1^\star\right)\!, 
  \,\mbox{see below Eq.\,(38.19) of Ref\,\cite{Amsler:2008zzb}},  \nn \\
 |\pbf_+^\star| = | {\pbf_+^\star}_\perp+{\pbf_+^\star}_\para| 
  &\equiv&   |{\bf p}^\star_1| ,\ \mbox{see Eq.\,(38.20a) of Ref\,\cite{Amsler:2008zzb},} \nn \\
  &=& \sqrt {\fourth s_{pp}-m_p^2}
  = \half \sqrt{  s_{pp} } 
     \underbrace{ \sqrt{ 1 -\frac{4 m_{p}^2}{s_{pp}} } }_{\equiv \beta_p}, \label{past}
 \end{eqnarray}
 {\it i.e.}, $\pbf_+^\star$ is the three-momentum of particle 1 in the rest
 frame of particle 1 and 2.
Note that in analogy to Eq.\,\equa{past} 
\begin{equation}
|\kbf_-^\diamond| =\sqrt{\fourth s_{kk} - m_k^2}=\half \sqrt{s_{kk}}\,\sqrt{1 -\frac{4 m_k^2}{s_{kk}}} \equiv \half \sqrt{s_{kk}}\ \beta_k\,.
\end{equation}

 Furthermore,
 \begin{eqnarray}
    \widetilde\kbf &\equiv& \left(|\pbf_3|,\Omega_3\right)\!, \,\mbox{see below Eq.\,(38.19) of Ref\,\cite{Amsler:2008zzb},}\\
  |{\widetilde \kbf}| =|{\widetilde \kbf_\para}| &\equiv& |{\bf p}_3|,     \ \mbox{see Eq.\,(38.20b) 
  of Ref\,\cite{Amsler:2008zzb}} \nn \\
   &\equiv& \frac{\lambda^{1/2}(m_P^2, s_{pp}, s_{kk}) }{2 m_P}   
  = \frac{ \lambda^{1/2}(s_{pp}, m_P^2, s_{kk}) }{ 2 m_P } 
    = \frac{ \lambda^{1/2}(s_{pp}, s_{kk},m_P^2) }{ 2 m_P } 
\end{eqnarray}
where
\begin{equation}
\lambda(a,b,c) \equiv a^2 +b^2 +c^2 -2 (ab +bc + ca) \,.
\label{deflambda}
\end{equation}
Thus, $\widetilde \kbf$ is the momentum of particle 3 
in the rest frame of the decaying particle. 
Note that Eq.\,\equa{deflambda} allows to relate moduli of momenta in different frames:
\begin{equation}
  |\widetilde\kbf|  =  \frac{\lambda^{1/2}(m_P^2, s_{p p}, s_{kk}) }{2 m_P}   
    =\frac{\sqrt{s_{pp}}}{m_P}\,
    \frac{ \lambda^{1/2}(s_{pp},m_P^2, s_{kk}) }
    {2  \sqrt{s_{pp}} } 
    \equiv
    \frac{\sqrt{s_{pp}}}{m_P}\, |\kbf^\star|\,.
\end{equation}
The variables $s$, $\beta$ and $\lambda$ are invariant in all frames. We will
give our results in terms of these standard variables.

\subsection{The identification of the angles}

Let us define the angle $\widetilde\theta_{p^+}$ as the
angle between the three-momentum  $\widetilde\pbf_+$ and the direction 
$\hat{\kbf}=\hat{\kbf_\para}=\hatz$ in the $P$ rest frame:
\begin{equation}
  \widetilde\theta_{p^+} \equiv \arccos\left(  \widetilde\pbf_+ \cdot \hat{\kbf} 
/|\widetilde\pbf_+| \right)\equiv   \angle\left(\widetilde\pbf_+,\hat{\kbf}\right) \,.
\end{equation}
Furthermore, we define the   angle  $\theta^\star_{p^+}$  as the
angle between the three-momentum $\pbf_+^\star$ and the direction 
$\hat{\kbf}=\hat{\kbf_\para}=\hatz$ in the $p_+ p_-$ rest frame:
\begin{equation}
  \theta_{p^+}^\star \equiv \arccos\left(  \pbf_+^\star \cdot \hat{\kbf} 
/|\pbf_+^\star| \right)\equiv   \angle\left(\pbf_+^\star,\hat{\kbf}\right)
\equiv \theta_p \,.
\end{equation}
$\theta_{p^+}^\star$ is actullay the angle we will use in our discussions, so
we will refer to $\theta_{p^+}^\star$ as $\theta_p$ to make things easier. In
analogy we can define $\theta_{k^-}^\diamond$ as the angle between the $k^-$ three-momentum and the $P$ three-momentum in the $k^-k^+$
 rest frame.  Note that  the $P$ three-momentum in the $k^-k^+$
 rest frame points into the negative $z$-direction
\begin{equation}
  \theta_{k^-}^\diamond \equiv \arccos\left( - \kbf_-^\diamond \cdot \hat{\kbf} 
/|\kbf_-^\diamond| \right)\equiv   \angle\left(\kbf_-^\diamond,-\hat{\kbf}\right) \equiv \theta_k \,.
\end{equation}
We also define $\phi$, the azimuthal 
angle between the plane formed by $p_+ p_-$ in the $P$ rest frame  and the
corresponding plane formed by $k^-k^+$.

\chapter{Anomalous decays}
In the following Chapter we will discuss the decays of the neutral pseudoscalar mesons
$P\in \{\pi^0,\eta,\eta'\}$ that are induced by  the chiral anomaly. We differentiate
between the ones which are governed by the triangle anomaly and the ones
resulting from the box anomaly, because the structure of the pertinent
form factors will be quite similar in the respective
cases.\\

\begin{center}
\begin{figure}[hbt]
  \begin{minipage}[b]{6 cm}
   \includegraphics[height=3cm]{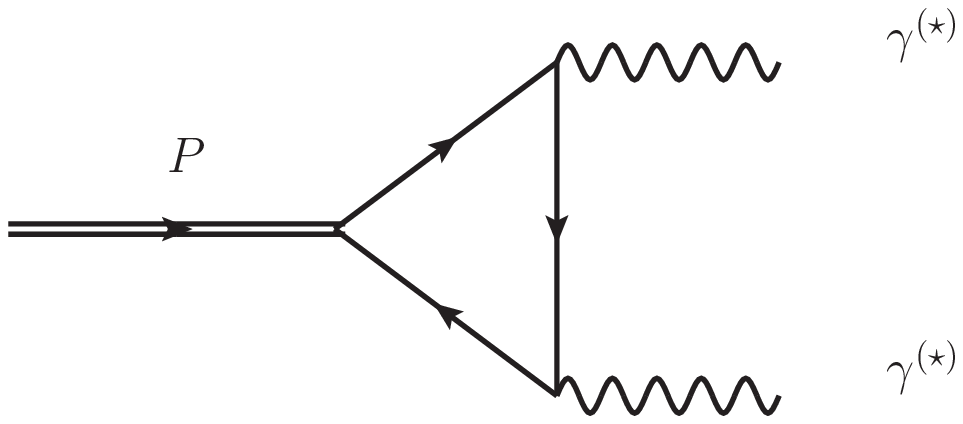}  
\caption{triangle anomaly}
  \end{minipage}
\hspace{1cm}
  \begin{minipage}[b]{6 cm}
    \includegraphics[height=3cm]{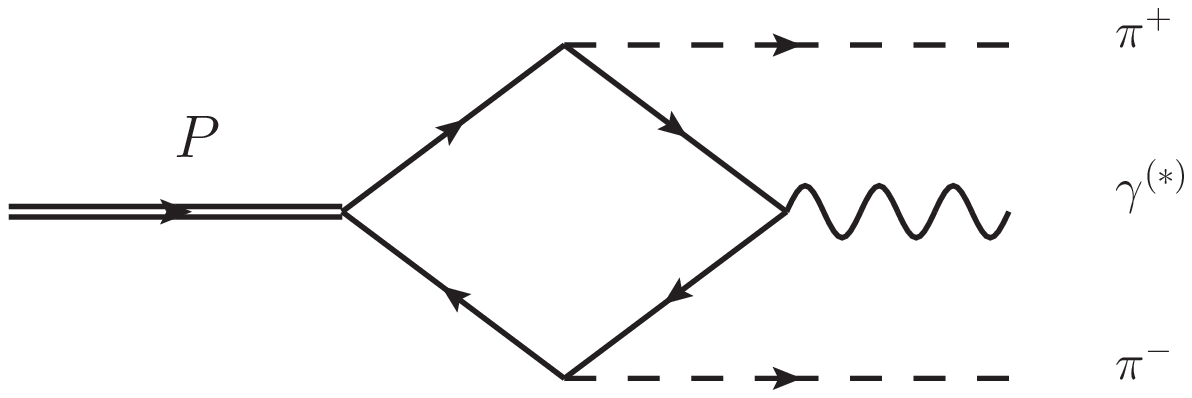} 
\caption{box anomaly}
  \end{minipage}
\end{figure}
\end{center}

The leading decays induced by the triangle anomaly are discussed next.
We add here the qualifier 'leading' in order to discriminate these decays from
those which involve subleading sequential decays as, e.g., Bremsstrahlung
corrections etc. The discussed decays are\\
\begin{eqnarray}
 && P \to \gamma \gamma, \nn \\
 && P \to l^+ l^- \gamma, \nn \\
 && P \to l^+ l^- l^+l^-, \nn \\
 && P \to l^+ l^-,\nn 
\end{eqnarray}\\
where $l^+l^-$ are lepton-antilepton pairs. Obviously only electrons and muons are
involved, because the tauons are too heavy. In the
case of the box anomaly the $\pi^0$ decays are dynamically forbidden. The
leading decays induced by the box anomaly are:\\
\begin{eqnarray}
 && P \to \pi^+ \pi^- \gamma, \nn \\
 && P \to \pi^+ \pi^- l^+l^- .\nn
\end{eqnarray}\\
Note that we are not dealing with the decay $\eta'\to \pi^+ \pi^- \pi^+ \pi^-$, although it would be very interesting because it was never measured.\\
We will present the squared matrix elements and the decay rates of all decays
listed above. The decays induced by the triangle anomaly will be related to the
decay into two photons, while the decays induced by the box anomaly are related to
the $P \to \pi^+ \pi^- \gamma$ decay. We will also give explicit expressions for the
form factors and the different vector meson dominance models which currently
are used in the description of the respective decays.
\section{$P \to \gamma \gamma$}
\begin{figure}[hbt]
  \begin{minipage}[b]{5 cm}
 \includegraphics[width=5cm]{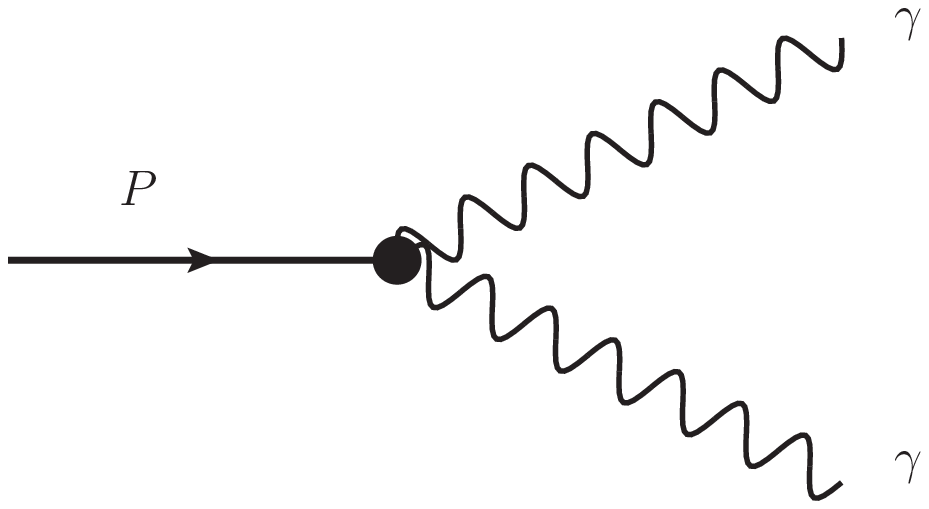}
  \caption{direct contribution $\mathcal{L}_{PAA}$}
  \label{fig:ggdirekt}
  \end{minipage}\hspace{0.5cm}
  \begin{minipage}[b]{5 cm}
 \includegraphics[width=5cm]{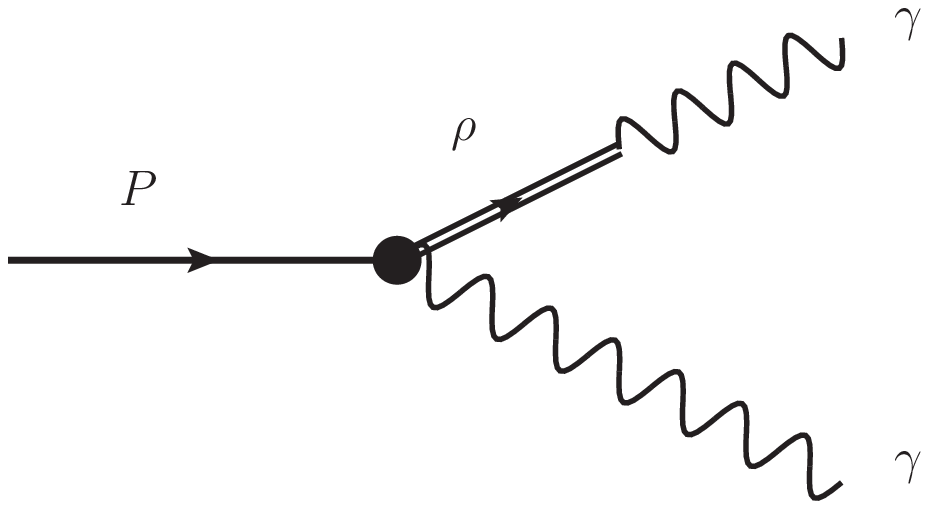}
  \caption{mixed term $\mathcal{L}_{PVA}$}
  \label{fig:ggmix}
  \end{minipage}\hspace{0.5cm}
  \begin{minipage}[b]{5 cm}
 \includegraphics[width=5cm]{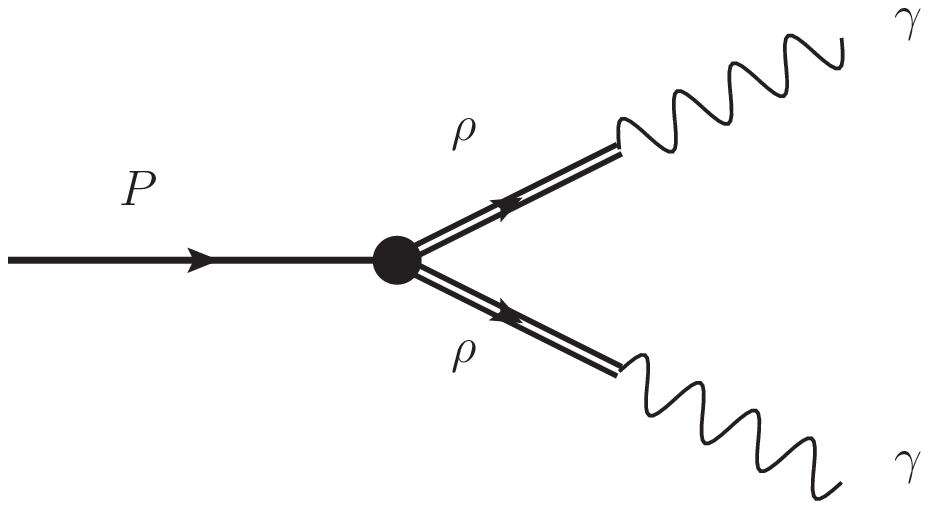}
  \caption{full VMD term $\mathcal{L}_{PVV}$}
  \label{fig:ggvoll}
  \end{minipage}
\end{figure}

We start with the decay $P \to \gamma \gamma$. This is probably the most famous
anomalous decay, because historically it was the first process wherein
anomalies were discovered. Expressed in terms of the respective momenta it reads: $P_P \to \gamma(\epsilon,p) \gamma(\epsilon,k)$, where $\epsilon$ and $\epsilon'$ are the polarizations of the photons. The four momentum of the decaying meson is $P= p+k$. As required by Lorentz invariance, parity conservation and gauge invariance, the amplitude has the general structure:
\begin{equation}
 {\cal A}(P_P \to \gamma(\epsilon_1,p) \gamma(\epsilon_2,k))= {\cal M}_P(p^2=0,k^2=0) \varepsilon_{\mu\nu\rho\sigma}\epsilon_1^\mu p^\nu \epsilon_2^\rho k^\sigma
\end{equation}
The form factor ${\cal M}_P(p^2=0,k^2=0)$ holds the information of the decaying
particle and since the decay products are on-shell photons which are massless it is given by a constant
\begin{equation}
 {\cal M}_P=\begin{cases}
         {\displaystyle\frac{\alpha}{\pi f_{\pi}}} & \mbox{if $P=\pi^0$};\\
        {\displaystyle\frac{\alpha}{\pi f_\pi} \frac{1}{\sqrt{3}} }\left( \frac{f_\pi}{f_8} \cos\theta_{mix} -2\sqrt{2} \frac{f_\pi}{f_0} \sin\theta_{mix} \right)& \mbox{if $P=\eta$};\\
        {\displaystyle\frac{\alpha}{\pi f_\pi} \frac{1}{\sqrt{3}}} \left( \frac{f_\pi}{f_8} \sin\theta_{mix} +2\sqrt{2} \frac{f_\pi}{f_0} \cos\theta_{mix} \right)& \mbox{if $P=\eta'$}.
 \label{MP}
\end{cases}
\end{equation}
Here $\alpha=e^2/4\pi \approx 1/137$ is the electromagnetic
fine structure constant, $f_\pi \approx 92.4 \,{\rm MeV}$ is the physical value
of the pion-decay constant and $f_0 \approx 1.04 f_\pi$ and $f_8 \approx 1.3
f_\pi$ are the singlet and octet Pseudo-Goldstone meson decay constants (see \cite{Holstein:2001bt}). One
can nicely see the mixing between $\eta_0$ and $\eta_8$ to the pseudoscalar
mesons $\eta$ and $\eta'$ via the mixing angle $\theta_{mix} \approx -
20^\circ$ (\cite{Holstein:2001bt}).\\
Note that the vector meson dominance (VMD) factor is set to unity. In all triangle
anomaly cases the three different terms $\mathcal{L}_{PAA}$, $\mathcal{L}_{PVV}$
and $\mathcal{L}_{PVA}$ contribute, {\it i.e.} the coupling of the pseudoscalar meson directly to the two on-shell photons,
the full vector meson coupling and the mixed form, where the pseudoscalar
meson couples directly to a photon and to a vector meson, which decays into a
photon. This is shown in Figures \ref{fig:ggdirekt}, \ref{fig:ggmix} and \ref{fig:ggvoll}.

Therefore the amplitude of the  $\eta\to \gamma\gamma$-decay is proportional to\\
\begin{equation}
{\cal M}_P \to {\cal M}_P \times \left(\underbrace{1-c_4}_{\rm direct \,\, term} + \underbrace{c_3-c_4}_{\rm mixed \,\,term} + \underbrace{c_3}_{\rm full\,\, VMD \,\,term}\right)  ={\cal M}_P \times \left(1+2(c_3-c_4)\right).
\end{equation}
As already discussed the coefficient $c_3$ is equal to $c_4$ in order to
recover the usual WZW term and to make sure that the VMD term is normalized
to 1 in the decay to two on-shell photons. So the contribution of the mixed term vanishes.
\subsection{Squared matrix element}
The squared matrix element of the decay $P_P \to \gamma(\epsilon_1,p) \gamma(\epsilon_2,k)$ is given by
\begin{equation}
 |{\cal A}(P_P \to \gamma(\epsilon,p) \gamma(\epsilon,k))|^2 = |{\cal M}|_P^2
 \varepsilon_{\mu\nu\rho\sigma}\varepsilon_{\mu'\nu'\rho'\sigma'} \epsilon_1^\mu
 p^\nu \epsilon_2^\rho k^\sigma \epsilon_1^{\mu'} p^{\nu'} \epsilon_2^{\rho'}
 k^{\sigma'}.
\label{App}
\end{equation}
If we assume that the polarizations of the photons remain unobserved, the photon polarization vectors can be summed over. We use the following relation:
\begin{eqnarray}
 {\cal O}_{\mu \mu'}^{\rm photon} &\equiv& \!\!\sum_{\lambda=1}^2 \epsilon_\mu (\pbf,\lambda) \epsilon_{\mu'} (\pbf,\lambda) \nn \\
  &=&   - g_{\mu \mu'}  + \frac{\eta_\mu p_{\mu'} \eta_{\mu}p_{\mu'}}{\eta \cdot p}  -\eta^2
  \frac{p_\mu p_{\mu'} }{(\eta \cdot p)^2} .
\label{Ogamma}
\end{eqnarray}
In this formula $\lambda$ labels the two transversal polarizations of the
photons, while $\pbf$ is its three momentum. Since this expression will be
contracted with the four-dimensional antisymmetric tensor the second and third
term cancel out, because they contain either an additional $p$ or an
additional $k$. Thus we actually have
the simplified form:
\begin{equation}
 \sum_{polarizations} \epsilon_\mu \epsilon_{\mu'} = - g_{\mu \mu'}
\label{Ogammas}
\end{equation}
This expression can be found in standard quantum field theory books,
e.g. equation (5.75) of \cite{Peskin:1995ev}. Inserting \equa{Ogammas} into
\equa{App}, one fiends:
\begin{equation}
  |{\cal A}(P \to \gamma \gamma)|^2 = {\cal M}_P^2 \varepsilon_{\mu\nu\rho\sigma}\varepsilon^{\mu\nu}_{\,\,\,\,\rho'\sigma'} p^\rho p^{\rho'} k^\sigma k^{\sigma'}.
\end{equation}
Using $p^2= k^2=0$, because the photons are on-shell, and the following well-known
identity, (e.g. equation (A.30) of \cite{Peskin:1995ev})
\begin{equation}
 \varepsilon_{\mu\nu\rho\sigma}\varepsilon^{\mu\nu}_{\,\,\,\,\rho'\sigma'} = -2 (g_{\rho\rho'}g_{\sigma\sigma'}-g_{\rho\sigma'}g_{\rho'\sigma}),
\end{equation}
we can derive the final expression of the squared amplitude of the decay $P \to \gamma \gamma$ as:\\
\begin{eqnarray}
 |{\cal A}(P \to \gamma \gamma)|^2 &=& {\cal M}_P^2 \,\,2(p\cdot k)^2 = {\cal M}_P^2 \half \,\,(p+k)^4\nn \\
&=&\half {\cal M}_P^2m_P^4.
\label{a^2gg}
\end{eqnarray}
\subsection{Decay rate}
We can now discuss the decay rate of $P \to \gamma \gamma$. We use the general
form for the decay into two particles (e.g. (38.17) of \cite{Amsler:2008zzb})
expressed in terms of the squared matrix element.
\begin{eqnarray}
{\rm d} \Gamma_{P\to\gamma\gamma} = \frac{1}{2m_P} \frac{1}{16\pi^2} S |{\cal A}(P \to \gamma \gamma)|^2 \frac{|\kbf^\star|}{E_{\rm CM}} {\rm d} \Omega.
\end{eqnarray}
Here $S=1/2$ is the symmetry factor which appears because of the Bose symmetry of the two
outgoing photons. Inserting the relations $|\kbf^\star|=
E_\gamma^\star$, because the photon is massless, $ E_{\rm CM} =  m_P$, and the
squared matrix element \equa{a^2gg}, we find the  expression of the decay rate:\\
\begin{equation}
 \Gamma_{P \to \gamma \gamma} = \frac{1}{64\pi}{\cal M}_P^2 m_P^3.
\label{gammapp}
\end{equation}
For the decay $\pi^0\to\gamma\gamma$
this expression simplifies to
\begin{equation}
 \Gamma_{\pi^0\to \gamma\gamma}=\frac{\alpha^2 m_\pi^3}{64\pi^3 f_\pi^2},
\end{equation}
where we inserted ${\cal M}_P=\alpha/(\pi f_\pi)$ from \equa{MP}. This is the same as given in \cite{Barker:2002ib}, namely
\begin{equation}
 \Gamma_{\pi^0\to \gamma\gamma}=\frac{m_P{\widetilde g}^2_P}{16 \pi}
\end{equation}
with
\begin{equation}
 {\widetilde g}_P=\frac{\alpha m_P}{2\pi f_\pi} \times \frac{{\cal M}_P}{{\cal M}_{\pi^0}},
\end{equation}
where we just have to use the different form factors given in \equa{MP} for the respective decay. This leads to the results:
\begin{table}[!h]
\begin{center}
\renewcommand{\arraystretch}{1.3}
\begin{tabular}{c|c|c|c}
 & $\Gamma_{\pi^0\to\gamma\gamma}$ & $\Gamma_{\eta\to\gamma\gamma}$ & $\Gamma_{\eta'\to\gamma\gamma}$ \\ \hline\hline
 theoretical values & 7.73 eV & 0.471 keV& 4.841 keV  \\ \hline
 experimental data \cite{Amsler:2008zzb} & $7.7\pm0.6$ eV & $0.511\pm0.003$ keV&   $4.284\pm0.245$ \\ 
\end{tabular}
  \caption{Decay rates with the input data values of
    \cite{Holstein:2001bt}(see Eq. \ref{MP} and the values below) and experimental data \cite{Amsler:2008zzb} of the decay $P\to \gamma\gamma$\label{tab:pigg}}
\end{center}
For the decay $\pi^0 \to \gamma\gamma$ the theoretical value represents the
experimental data very nicely. It is consistent with most of the results
calculated via ChPT, {\it e.g} $\Gamma_{\pi^0\to\gamma\gamma}=7.78 {\rm
 \, eV}$ \cite{Bijnens:1988kx} and $\Gamma_{\pi^0\to\gamma\gamma}=7.74 {\rm
  \,eV}$ \cite{Ametller:2001nq}. Other ChPT calculations lead to higher values: $\Gamma_{\pi^0\to\gamma\gamma}=8.14 {\rm
  \,eV}$ \cite{PhysRevD.66.076014} and $\Gamma_{\pi^0\to\gamma\gamma}=8.07 \frac{f_{\pi^+}^2}{f_{\pi^0}^2}{\rm
  \,eV}$ \cite{PhysRevD.51.4939}. Recent calculations gain the result
\cite{Ioffe:2007eg}: $\Gamma_{\pi^0\to\gamma\gamma}=7.93 {\rm
  \, eV} \pm 1.5 \%$. It is planned by the PrimEx experiment at JLab to reduce the error in
the experimental value down to 1-2\% \cite{PrimEx:2006}.\\
For the decays $\eta / \eta' \to \gamma \gamma$
this is not the case. The reason is the values of
\cite{Holstein:2001bt}. By using other values we could be closer to the
experimental data. In the following we will give the values for the branching
ratios with respect to the decay $P \to \gamma\gamma$ and therefore the terms where
these values occur would vanish anyway.
\end{table}

\section{$P \to l^+ l^- \gamma$}
\begin{center}
\begin{figure}[hbt]
  \begin{minipage}[b]{7.5 cm}
 \includegraphics[width=7.5cm]{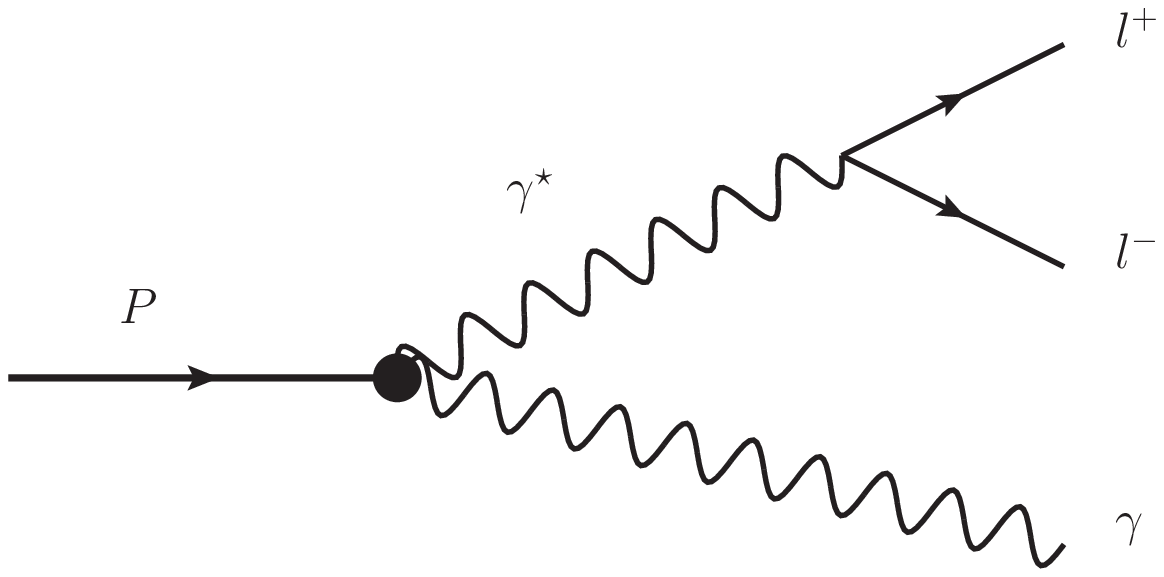}
  \caption{direct contribution $\mathcal{L}_{PAA}$}
  \label{fig:llgdirekt}
  \end{minipage}\hspace{0.5cm}
  \begin{minipage}[b]{7.5 cm}
 \includegraphics[width=7.5cm]{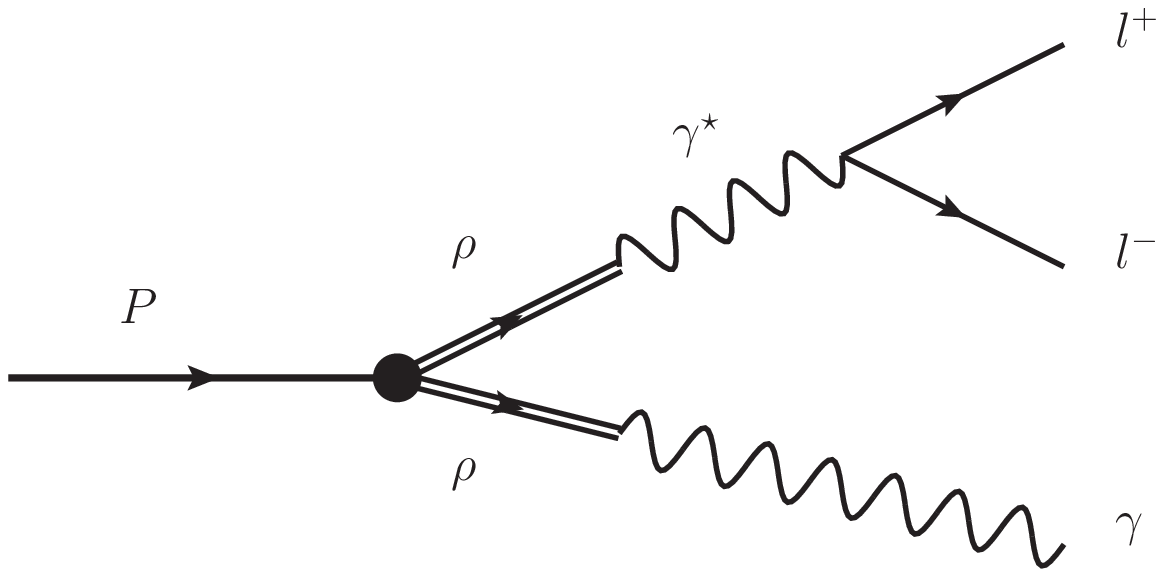}
  \caption{full VMD term $\mathcal{L}_{PVV}$}
  \label{fig:llgvoll}
  \end{minipage}
\end{figure}
\end{center}

The next decay that we want to discuss is the one of a pseudoscalar meson $P$ into a photon $\gamma$ and a lepton-antilepton
pair $l^+ l^-$, the so-called single off-shell decay or
Dalitz decay. It is related to the decay into two photons, but in this case
one of the photons is off-shell ($\gamma^\star$) and decays into the lepton
pair. Hence the form factors will be very similar, except that there is the
invariant mass dependence in this case. That is why we will present the final
result for the decay rate in terms of the double on-shell decay.\\
The leptons can be either electrons or muons, but this does not have an effect
on the kinematical formulae we present. We define the four-momenta for the process $P_P
\to \gamma^\star(p) \gamma(k) \to l^+(p_+) l^-(p_-) \gamma(k)$ so that $P = p
+ k = p_+ + p_- +k$ holds.
Since $k^2=0$, the only occuring standard variables are
\begin{eqnarray*}
 s_{pp}&=&(p_++p_-)^2 ,  \\
  \beta_{p} &=&  \sqrt{ 1 -\frac{4   m_{l}^2}{s_{pp}} } , \\
\lambda(m_P^2,p^2,k^2)  &=& \lambda(m_P^2,p^2,0) = (m_P^2-p^2)^2.
\end{eqnarray*}
There is only one relevant angle appearing, namely $\theta_p$. This can be
taken as the
angle between the outgoing lepton, $l^+(p_+)$, and the pseudoscalar $P_P$ in
the $l^+l^-$ rest frame, namely $\Theta_p$.\\
The amplitude for the decay $P_P \to \gamma^\star(p) \gamma(k) \to l^+(p_+)
l^-(p_-) \gamma(k)$ is given by the following expression:
\begin{equation}
{\cal A}(P\to l^+l^- \gamma) = 2 {\cal M}_P(p^2,k^2=0) \varepsilon_{\mu\nu\rho\sigma} \frac{1}{p^2} j^\mu(p_+,s_+;p_-,s_-) p^\nu \epsilon^\rho k^\sigma.
\end{equation}
Compared to the amplitude of the decay into two photons we see that the polarization of the off-shell photon turned into the current $j^\mu$ of the lepton pair, which is given by
\begin{equation}
 j^\mu = e \bar u(k_-,s_-) \gamma^\mu v(k_+,s_-)
\end{equation}
where $s_\pm$ are the helicities of the outgoing leptons $l^\pm$. The factor
two is a symmetry factor, because each of the photons can go off-shell and
would therefore turn into the current. The
polarization of the outgoing photon is $\epsilon$. The form factor ${\cal M}_P(p^2,k^2=0)$ can be written as follows:
\begin{equation}
 {\cal M}_P(p^2,k^2=0) = {\cal M}_P \times VMD(p^2).
\end{equation}
Here the factor ${\cal M}_P$ is the one given in the decay to two photons \equa{MP}
and still includes the information of the decaying particle. The factor $VMD(p^2)$ is the
vector meson dominance factor. In this case the diagrams
contributing are given by the direct
$\mathcal{L}_{PAA}$ term and the full VMD term $\mathcal{L}_{PVV}$, see Figure \ref{fig:llgdirekt} and Figure \ref{fig:llgvoll}. As mentioned the mixed term cancels because of the factor $c_3-c_4=0$.\\
Thus the vector meson dominance factor becomes
\begin{equation}
VMD(p^2)=1-c_3+c_3\frac{1}{1-\frac{p^2}{m_V^2}-i\frac{\Gamma(p^2)}{m_V^2}}.
\end{equation}
Note, that this factor becomes unity in the on-shell case, $p^2=0$, such that the
normalization holds.

In this equations $\Gamma$ is the width of the vector meson. It is
necessary to consider this width in the
$\eta'$ decays. Otherwise there would be singularities because the $\eta'$ mass is larger than the vector meson mass. The width is given by
\begin{equation}
\Gamma(s)= g_{m_V}\left( \frac{s}{m_V^2}\right) \left(
  \frac{1-\frac{4m^2}{s}}{1-\frac{4m^2}{m_V^2}}\right)^{3/2} \Theta \left( s-4 m_\pi^2\right)
\label{width}
\end{equation}
with $g_{m_V}= 149.1\,{\rm MeV}$, see e.g. \cite{Ericson:1988gk}. Here we can switch between the different vector meson dominance models by
inserting different values of $c_3$. In the hidden gauge case ($c_3=1$)
the direct term cancels and there is no direct coupling between a
pseudoscalar meson and a photon. In the case of the modified model
($c_3=0.927$ or $c_3=0.930$, respectively) this term
will give a small additional contribution.
\subsection{Squared matrix element}
In order to calculate the squared amplitude we will use the following
projection tensor, which is derived in \equa{projtensor}:
\begin{eqnarray}
&& \!\!\!\! {\cal O}_{\mu \mu'}(p_-,p_+) \equiv 
  \!\!\sum_{s_-=-1/2}^{1/2}  \sum_{s_+=-1/2}^{1/2} 
\!\!  j_\mu(p_-,s_-;p_+,s_+) \, j^\dagger _{\mu'}(p_-,s_-;p_+,s_+)  \nn \\
  &=& e^2 p^2 \times 2  \left[  -\left( g_{\mu \mu'}  - \frac{p_\mu p_{\mu'}}{p^2} \right) -
  \frac{(p^+ \!-\! p^-)_\mu (p^+ \!-\! p^-)_{\mu'}}{p^2} \right] .
\label{projtensor2}
\end{eqnarray}
The operator actually becomes ${\cal O}_{\nu \nu'}^{\rm photon}$ given in
\equa{Ogamma} if the photon goes on-shell.
Because of the total antisymmetric tensor in the amplitude the second term of
\equa{projtensor2} cancels. By the same reason only the first
term of \equa{Ogamma} gave a contribution. The squared amplitude then reads:
\begin{eqnarray}
 &&|{\cal A}(P\to l^+l^- \gamma)|^2 = \frac{4e^2 {\cal M}_P^2 \,\, |VMD(p^2)|^2}{(p^2)^2} \varepsilon_{\mu\nu\rho\sigma}\varepsilon_{\mu'\nu'\rho'\sigma'} {\cal O}^{\mu \mu'}(p_-,p_+){\cal O}^{\nu \nu'}_{\rm photon} p^\rho k^\sigma p^{\rho'} k^{\sigma'}\nn\\
&&= \frac{4e^2 {\cal M}_P^2 |VMD(p^2)|^2}{(p^2)^2} \varepsilon_{\mu\nu\rho\sigma}\varepsilon_{\mu'\nu'\rho'\sigma'}\left[  - g^{\mu \mu'}  -
  \frac{(p^+ \!-\! p^-)^\mu (p^+ \!-\! p^-)^{\mu'}}{p^2} \right] (-g^{\nu \nu'}) p^\rho k^\sigma p^{\rho'} k^{\sigma'} .\nn\\
\end{eqnarray}
The structure gets simpler when the following momentum relations are inserted
and the antisymmetric structure of the epsilon tensor is used again:
\begin{eqnarray}
 k = P - p \hspace{5mm};\hspace{5mm} p_-= p -p_+ \,.
\end{eqnarray}
The squared amplitude is then:
\begin{eqnarray}
 &&|{\cal A}(P\to l^+l^- \gamma)|^2 \nn \\
&&= \frac{4e^2 {\cal M}_P^2 |VMD(p^2)|^2}{(p^2)^2} \left[\varepsilon_{\mu\nu\rho\sigma}\varepsilon_{\mu\nu\rho'\sigma'} p^\rho p^{\rho'} P^\sigma P^{\sigma'} + \frac{4}{p^2} \varepsilon_{\mu\nu\rho\sigma}\varepsilon_{\mu\nu'\rho'\sigma'} p_+^\nu p_+^{\nu'} p^\rho p^{\rho'} P^\sigma P^{\sigma'} \right] .\nn\\
\end{eqnarray}
We can now switch to the rest frame of the pseudoscalar meson where the
relation $P^\mu=m_P\delta^{\mu0}$ holds. Note the sign change due to
$g^{ii'}=-\delta^{ii'}$. Thus the squared amplitude reads now:
\begin{eqnarray}
|{\cal A}(P\to l^+l^- \gamma)|^2 &=& \frac{4e^2 {\cal M}_P^2 |VMD(p^2)|^2}{(p^2)^2} m_P^2 \left[\varepsilon^{ijk}\varepsilon^{ijk'} p^k p^{k'} - \frac{4}{p^2} \varepsilon^{ijk}\varepsilon^{ij'k'} p_+^j p^k p_+^{j'} p^{k'} \right] \nn \\
&=& \frac{2e^2 {\cal M}_P^2 |VMD(p^2)|^2}{(p^2)^2} m_P^2 \left[ 2|\pbf|^2-\frac{4}{p^2}|\pbf_+|^2 |\pbf|^2 \sin^2\theta_p \right].
\end{eqnarray}
The squared matrix element of the decay $P\to l^+ l^- \gamma$ can now be given
in terms of the standard variables as\\
\begin{eqnarray}
 |{\cal A}(P\to l^+l^- \gamma)|^2 = \frac{4e^2 {\cal M}_P^2 |VMD(s_{pp})|^2}{s_{pp}} m_P^2 (m_P^2- s_{pp})^2 \left[2-\beta_k^2\sin^2\theta_k \right].
\end{eqnarray}
\subsection{Decay rate}
In order to calculate the decay rate we have to deal with the phase space
first. The phase space for a three-body decay can be found,
e.g., in (38.19) of \cite{Amsler:2008zzb}:
\begin{equation}
{\rm d}\Phi_{3}(P_P;p_+,p_-,k)=\frac{1}{(2\pi)^9}\frac{1}{8m_P}|\pbf_+^*| |\widetilde \kbf|
{\rm d}m_{p_+p_-} {\rm d}\Omega_{p_+}^\ast {\rm d} \widetilde \Omega_k.
\end{equation}
In our case this leads to the following form of the decay rate:
\begin{equation}
 {\rm d}\Gamma = \frac{1}{(2\pi)^5}\frac{1}{32m_P} |{\cal A} |^2 \sqrt{s_{pp}} \beta_k E_\gamma \,\,{\rm d}\sqrt{s_{pp}} \,\,{\rm d} \cos \theta_k\,\, {\rm d}\cos \theta_p \,\,{\rm d}\phi \,\,{\rm d}\phi_\gamma .
\end{equation}
After integration over the angles we find the following result 
\begin{eqnarray}
 {\rm d}\Gamma = \frac{{\cal M}_P^2m_P^3}{64\pi}\frac{(4\pi\alpha)
   |VMD(s_{pp})|^2}{12 \pi^2 m_P^6 s_{pp}} \left(m_P^2-s_{pp}\right)^3  \beta \left[3-\beta^2\right]{\rm d} s_{pp},
\end{eqnarray}
where we inserted the relations
\begin{eqnarray}
 s_{pp}&=&m_P^2(1-\frac{2E_\gamma}{m_P}),\nn\\
\sqrt{s_{pp}}{\rm d} \sqrt{s_{pp}}&=&\half {\rm d} s_{pp} = - m_P {\rm d} E_\gamma.
\end{eqnarray}
Note that the first factor is exactly $\Gamma_{P\to\gamma\gamma}=\frac{{\cal
    M}_P^2m_P^3}{64\pi}$. The final expression for the decay rate of the decay
$P\to l^+ l^- \gamma$ can be expressed in terms of the decay $P\to \gamma \gamma$ as\\
\begin{eqnarray}
 {\rm d}\Gamma(P\to l^+l^-\gamma) = \Gamma_{P\to\gamma\gamma}\frac{\alpha
   |VMD(s_{pp})|^2}{3 \pi m_P^6 s_{pp}} \left(m_P^2-s_{pp}\right)^3 \beta \left[3-\beta^2\right]{\rm d} s_{pp}.
\end{eqnarray}

\section{$P \to l^+ l^- l^+ l^-$}
We will now discuss the double off-shell decay of an pseudoscalar particle $P$
in two lepton pairs. The decay is again related to the $P\to \gamma \gamma$ decay, but in this case both of the photons go off-shell $(\gamma^\star\gamma^\star)$. The general structure of the form factors will be very similar, but there will be a dependence on the invariant masses of both lepton pairs.\\
The leptons can be either muons or electrons, so that we are
dealing with the following three decays:
\begin{eqnarray*}
&&  (i)\qquad P\to \mu^+\mu^- e^+ e^-  , \\
&&  (ii)\qquad P\to \mu^+\mu^-\mu^+\mu^-  , \\
&&  (iii)\qquad P\to  e^+ e^- e^+ e^-.   
\end{eqnarray*}
In the case that the pseudoscalar meson is a $\pi^0$, only the decay into two
electron pairs $(iii)$ is possible, of course.\\
We define the momenta of the decay $P(P)\to l_1^+(p_1)l_1^-(p_2) l_2^+(p_3) l_2^-(p_4)$ as follows:
\begin{eqnarray*}
&& P = p_1 + p_2 + p_3 + p_4 \,, \\
&& p_{ij}=p_i+p_j  .
 \end{eqnarray*}
The relevant variables can now be written as:
\begin{eqnarray*}
&& s_{ij}=(p_i+p_j)^2 ,  \\
&&  \beta_{ij}=  \sqrt{ 1 -\frac{4   m_{ij}^2}{s_{ij}} }  ,
\end{eqnarray*}
with $m_{ij}=m_l$ is the lepton mass of the outgoing leptons $i$ and $j$.
We define the angle $\theta_{ij}$ as the angle between $p_i$ and $p_P$ in
the $l(p_i)l(p_j)$ rest frame, the angle $\phi$ as the
angles between the planes formed by $l(p_1)l(p_2)$ and  $l(p_3)l(p_4)$ and $\widetilde\phi$ as the
angles between the planes formed by $l(p_1)l(p_4)$ and  $l(p_2)l(p_3)$.

\subsection{Amplitudes}
\begin{figure}[!hbt]
  \begin{minipage}[b]{8cm}
\begin{center}
 \includegraphics[width=7cm]{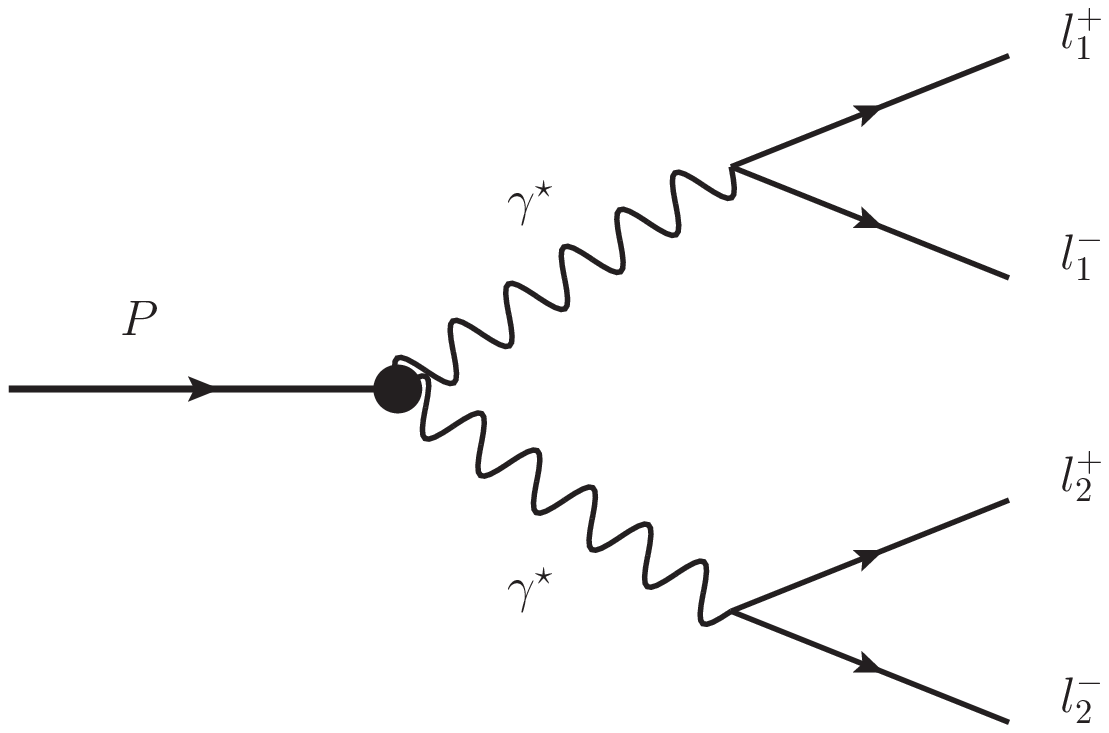}
  \caption{Double Dalitz diagram for ${\cal A}_1$}
  \label{fig:l11l22}
\end{center}
  \end{minipage}\hspace{0.5cm}
  \begin{minipage}[b]{8 cm}
\begin{center}
 \includegraphics[width=7cm]{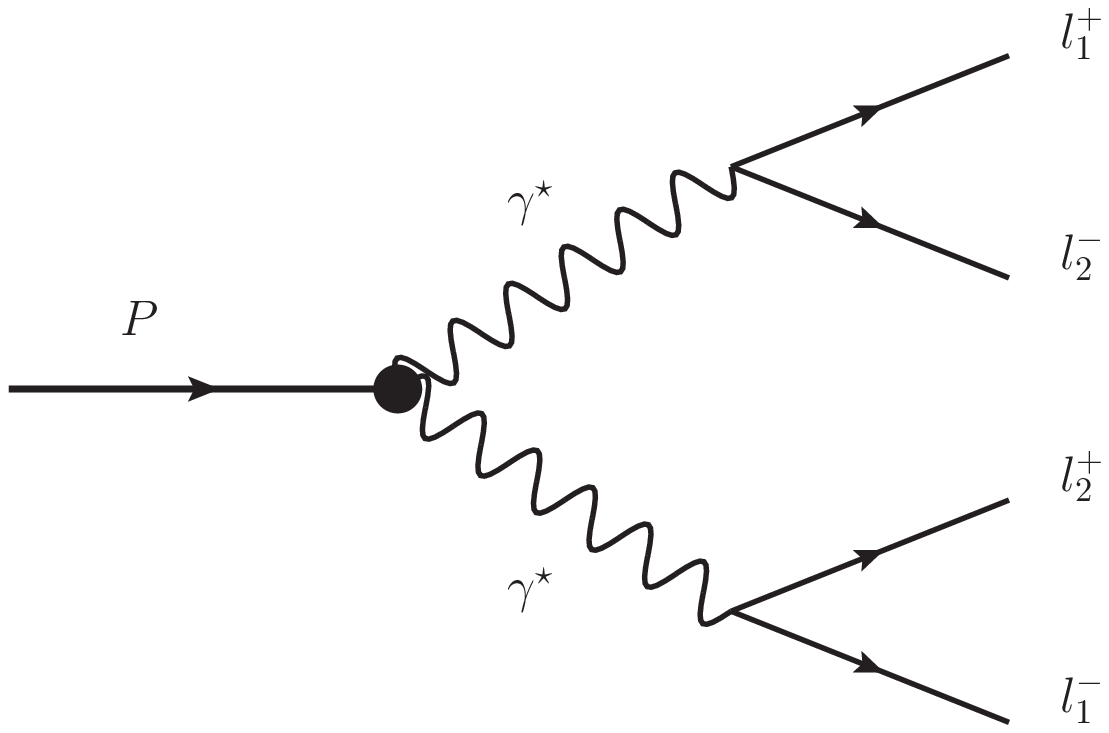}
  \caption{Double Dalitz diagram for ${\cal A}_2$}
  \label{fig:l12l21}
\end{center}
  \end{minipage}
\end{figure}

Calculating the decay of a pseudoscalar meson $P$ into two lepton pairs we have to differentiate
between the decay into two different lepton pairs $(i)$ and the decay into two
identical lepton pairs $(ii, iii)$. The matrix element squared $ |{\cal  A}_1|^2$ of the decay $(i)$ can be calculated relatively straight
forward by using the invariant decay amplitude (see Figure \ref{fig:l11l22}):
\begin{equation}
{\cal A}_1(P\to\mu^+\mu^-e^+e^-) = \frac{|{\cal M}|}{s_{12}^2 s_{34}^2} 
\varepsilon_{\mu\nu\rho\sigma}(p_1 + p_2)^\nu (p_3+ p_4)^\sigma
\overline{u}(p_2)\gamma^\mu v(p_1) \cdot \overline{u}(p_4)\gamma^\rho v(p_3).
\end{equation}

For the decays into two identical particle pairs $(ii,iii)$ we have to take another
amplitude ${\cal A}_2$ into account (see Figure \ref{fig:l12l21}):
\begin{equation}
{\cal A}_2(P \to l^+l^-l^+l^-) = - \frac{|{\cal M}|}{s_{14}^2 s_{23}^2} 
\varepsilon_{\mu\nu\rho\sigma}(p_1 + p_4)^\nu (p_2 + p_3)^\sigma
\overline{u}(p_4)\gamma^\mu v(p_1) \cdot \overline{u}(p_2)\gamma^\rho v(p_3) .
\end{equation}
In this case the a total amplitude is ${\cal A} = {\cal A}_1 + {\cal
  A}_2$. In order to give the squared matrix element $ |{\cal  A}|^2$ it is necessary to calculate  the direct term $ |{\cal  A}_1|^2$, but 
also the crossed term $ |{\cal  A}_2^2|$ and the interference term $2 Re( {\cal
  A}_1  {\cal  A}_2^\star)$. The crossed term will look the same as the direct
term when the variables are exchanged. To determine the decay rate we will
have to integrate over the variables. Therefore the direct and crossed term
will give the same contribution to the decay rate.
\subsection{The reaction $P \to \mu^+ \mu^-e^+ e^- $}
The invariant decay amplitude mentioned above can also be written in the
following form:
\begin{equation}
{\cal A}_1(P\to\mu^+\mu^-e^+e^-) = \frac{|{\cal M}|}{p_{12}^2 p_{34}^2} 
\varepsilon_{\mu\nu\alpha\beta}j^{\mu}_{(ee)}(p_3,p_4) p_{34}^{\nu}
j^{\alpha}_{(\mu\mu)}(p_1,p_2) p_{12}^{\beta}\, .
\end{equation}
Here $j^{\mu}_{(ee)}$ and $j^{\alpha}_{(\mu\mu)}$ are currents of the respective lepton pairs. In order to calculate the squared amplitude we need the projection
tensor, which we used in \equa{projtensor2}, see also \equa{projtensor}. Here
it reads
\begin{equation}
 {\cal O}_{\mu \mu'}(p_1,p_2)= e^2 p_{12}^2 \times 2  \left[ - \left( g_{\mu \mu'}  - \frac{p_{12\mu} p_{12\mu'}}{p_{12}^2} \right) -
  \frac{(p_1 \!-\! p_2)_\mu (p_1 \!-\! p_2)_{\mu'}}{p_{12}^2} \right].
\end{equation}
The squared matrix amplitude then reads:
\begin{eqnarray}
 \overline{|{\cal A}_1|^2}(P\to \mu^+\mu^-e^+e^-) =  \frac{ |{\cal M}|^2}{s_{12}^2 s_{34}^2} \varepsilon_{\mu \nu\alpha\beta}\varepsilon_{\mu'\nu'\alpha'\beta'} {\cal
  O}^{\mu\mu'}(p_1,p_2) p_{12}^{\nu} p_{12}^{\nu'} {\cal O}^{\alpha\alpha'}(p_3,p_4) p_{34}^{\beta}.
p_{34}^{\beta'}  
\end{eqnarray}
The following steps are as usual. We insert the relations between the momenta
to make use of the antisymmetric tensor:
\begin{eqnarray*}
p_2 &=& P-p_1   -p_{34}\,, \\
p_3 &=& P-p_4  - p_{12}\,, \\
p_{34}   &=& P - p_{12}.
\end{eqnarray*}
We can now switch to the rest frame of the pseudoscalar meson, where we use
$P^\mu=m_P\delta^{\mu0}$ and $g^{ii'}=-\delta^{ii'}$. The squared
amplitude can now be expressed in standard variables as follows:
\begin{eqnarray}
&& \overline{|{\cal A}_1|^2}(P\to \mu^+\mu^-e^+e^-) \nn  \\
&& = \frac{ e^4 |{\cal M}|^2} { s_{12} s_{34}} \lambda(m_{P}^2,s_{12},s_{34}) \biggl[2  -\beta_{12}^2\sin^2\theta_{12} -\beta_{34}^2 \sin^2\theta_{34} +\beta_{12}^2 \beta_{34}^2  \sin^2\theta_{12}
\sin^2\theta_{34} \sin^2\phi\biggr]\nn \\
&& = \frac{ e^4 |{\cal M}|^2}{ s_{12} s_{34}}\lambda(m_{P}^2,s_{12},s_{34})
\biggl[ \biggl( 1 + (1-\beta_{12}^2\sin^2\theta_{12}) (1-\beta_{34}^2 \sin^2\theta_{34}) \biggr)
\sin^2\phi \nn  \\
&&\qquad\qquad\qquad\qquad\qquad\mbox{}+ \biggl( 2 - \beta_{12}^2\sin^2\theta_{12} -
\beta_{34}^2 \sin^2\theta_{34} \biggr) \cos^2\phi \biggr].
\label{A1}
\end{eqnarray}
The result is symmetric under the exchange of the momenta $p_1$, $p_2$ and
$p_3$, $p_4$.\\
Note that the last line agrees with Eq.(16) of \cite{Barker:2002ib}, expressed in standard variables. It also agrees with the result given in Appendix
B of \cite{Miyazaki:1974qi}.
\subsection{The reaction $P \to l^+ l^- l^+ l^-$}
\subsubsection{Direct and Mixed Term}
We will now calculate the squared matrix element of the decay into two
identical lepton pairs $(ii,iii)$. The part of  $ |{\cal  A}_1|^2$ is of course
the one given in \equa{A1}. We can get $ |{\cal  A}_2|^2$ by replacing $p_2$ and
$p_4$:
\begin{eqnarray}
&& \overline{|{\cal A}_2|^2}(P\to l^+ l^- l^+ l^-) \nn  \\
&& = \frac{ e^4 |{\cal M}|^2} { s_{14} s_{23}} \lambda(m_{P}^2,s_{14},s_{23}) \biggl[2  -\beta_{14}^2\sin^2\theta_{14} -\beta_{23}^2 \sin^2\theta_{23} +\beta_{14}^2 \beta_{23}^2  \sin^2\theta_{14}
\sin^2\theta_{23} \sin^2\widetilde\phi\biggr] \nn \\
&& = \frac{ e^4 |{\cal M}|^2}{ s_{14} s_{23}}\lambda(m_{P}^2,s_{14},s_{23})
\biggl[ \biggl( 1 + (1-\beta_{14}^2\sin^2\theta_{14}) (1-\beta_{23}^2 \sin^2\theta_{23}) \biggr)
\sin^2\widetilde\phi \nn  \\
&&\qquad\qquad\qquad\qquad\qquad\mbox{}+ \biggl( 2 - \beta_{14}^2\sin^2\theta_{14} -
\beta_{23}^2 \sin^2\theta_{23} \biggr) \cos^2\widetilde\phi \biggr].
\end{eqnarray}
Note that this formula is given in different variables than used in all other
decays. To compare this terms it is necessary to express the new variables in
terms of the ones used before. The translation formulae are given in Appendix B.

\subsubsection{Interference Term}
We will now present the term  $2 {\rm Re}( {\cal A}_1  {\cal  A}_2^\star)$. The
calculation is much more extensive than the calculations presented before, so
we will give less steps than before.\\
We have to deal with the traces at first. In the calculation there are traces over products of four, six and eight
$\gamma$-matrices. We therefore use the formula
\begin{equation*}
Tr(\slashed {a}_1 \slashed{a}_2 ...\slashed{a}_{2n}) = a_1\cdot a_2
Tr(\slashed {a}_3 ...\slashed{a}_{2n}) - a_1\cdot a_3
Tr(\slashed {a}_2 ...\slashed{a}_{2n}) + ... + a_1\cdot a_{2n}
Tr(\slashed {a}_2 ...\slashed{a}_{2n-1}),
\end{equation*}
which can be looked up in standard books on quantum field theory e.g. \cite{Itzykson:1980rh}.\\
This leads to very long calculations in which errors, especially in signs, can
easily occur. Hence the following result was rechecked with the algebraic
program FORM (see \cite{Vermaseren:2000nd}).
\begin{eqnarray}
&& 2 Re( {\cal A}_1  {\cal  A}_2^\star) = -\frac{e^4 |{\cal M}|^2 \varepsilon_{\mu
    \nu\rho\sigma}\varepsilon_{\mu'\nu'\rho'\sigma'} g^{\rho \rho'} }{p_{12}^2
p_{34}^2 p_{14}^2 p_{23}^2} (p_1 + p_2)^\nu (p_3 + p_4)^\sigma (p_1 + p_4)^{\nu'}
(p_2 + p_3)^{\sigma'} \nn  \\
&& \qquad\qquad\qquad \times \Biggr[ (m^2 + p_1\cdot p_2) \left(p_3^\mu
  p_4^{\mu'}- p_3^{\mu'} p_4^\mu\right) + (m^2 + p_3\cdot p_4) \left(p_1^\mu
  p_2^{\mu'} - p_1^{\mu'} p_2^\mu \right) \nn  \\
&& \qquad\qquad\qquad\quad - (m^2 - p_2\cdot p_4) \left(p_1^\mu
  p_3^{\mu'} + p_1^{\mu'} p_3^\mu\right) + (m^2 - p_1\cdot p_3) \left(p_2^\mu
  p_4^{\mu'} - p_2^{\mu'} p_4^\mu\right) \nn  \\
&& \qquad\qquad\qquad\quad - (m^2 + p_2\cdot p_3) \left(p_1^\mu
  p_4^{\mu'} - p_1^{\mu'} p_4^\mu\right) + (m^2 + p_1\cdot p_4) \left(p_2^\mu
  p_3^{\mu'} - p_2^{\mu'} p_3^\mu\right)\Biggl]. \nn \\
\end{eqnarray}
This equation is invariant under the permutation of $p_2$ and $p_4$. As one
can easily check, the first and second term will turn into the sixth and
fifth term, while the third and fourth term stay invariant.\\
Note that this is not the formula given in Appendix B of
\cite{Miyazaki:1974qi}, which has an typological error. This was recognized by
the authors of Ref. \cite{Barker:2002ib}, too.\\
Next we have to deal with the four-dimensional $\varepsilon$-tensors. Here we have to contract the
$\varepsilon$-tensors explicitly. We will again use FORM to control the
calculation. Inserting the  relations $p_2 = p_{12}-p_1$ and $ p_3 =
p_{34}-p_4$ and identifying $p^2=m^2$ we find:
\begin{eqnarray}
&& \overline{|{\cal A}_{mix}|^2} = -\frac{e^4 |{\cal M}|^2  }{p_{12}^2
p_{34}^2 p_{14}^2 p_{23}^2}\nn \\
&&\times 16 m^4 \Biggl\{ (p_{12}\cdot p_{34})^2 - p_{12}^2p_{34}^2\Biggr\} \nn \\
&& + 4 m^2 \Biggl\{ 8 p_1 \cdot p_4 \left[ (p_{12} \cdot p_{34})^2 -
  p_{12}^2p_{34}^2 \right] - 2 p_{12}^2 p_{34}^2 p_{12} \cdot p_{34}\nn \\
&& \mbox{} \qquad - 2p_{12} \cdot p_4 p_1 \cdot p_{34} \left[ p_{12}^2 +
  p_{34}^2 + 4 p_{12} \cdot p_{34} \right] + 2 (p_{12} \cdot p_4)^2 p_{34}^2 +
2 (p_1 \cdot p_{34})^2 p_{12}^2 \nn \\
&& \mbox{} \qquad + p_{12} \cdot p_4 \left[ (p_{12}^2-p_{34}^2) p_{12} \cdot
  p_{34} + 4 p_{12}^2 p_{34}^2 \right] + p_{34} \cdot p_1 \left[
  (p_{34}^2-p_{12}^2) p_{12} \cdot p_{34} + 4 p_{12}^2 p_{34}^2 \right]
\Biggr\} \nn \\
&& + 2 \Biggl\{ - 8 p_1 \cdot p_4 p_{12} \cdot p_4 p_1 \cdot p_{34} p_{12}
\cdot p_{34} - 2 p_{12} \cdot p_4 p_1 \cdot p_{34} \bigl[ p_{12}^2 p_{34}^2- 2
(p_{12} \cdot p_4)^2 - 2 (p_1 \cdot p_{34})^2 \bigr] \nn \\
&& \mbox{} \qquad +4 p_1 \cdot p_4 \bigl[ 2 p_{12}^2 p_{34}^2 \left( p_{12}
  \cdot p_4 + p_1 \cdot p_{34} \right) - p_{12} \cdot p_{34} \left( (p_{12}
  \cdot p_4)^2 + (p_1 \cdot p_{34})^2 \right) - p_{12}^2 p_{34}^2 (p_{12}
\cdot p_{34}) \bigr] \nn \\
&& \mbox{} \qquad + 8 (p_1 \cdot p_4)^2 \bigl[ (p_{12} \cdot p_{34})^2 - p_{12}^2 p_{34}^2 \bigr] - p_{12}^2 p_{34}^2 \bigl[ (p_{12} \cdot p_4)^2 + (p_1 \cdot p_{34})^2 \bigr] +\frac{1}{2} p_{12}^4 p_{34}^4 \Biggr\}.
\end{eqnarray}
All terms of this formula are still valid for any Lorentz frame. So we
can make use of the invariants given in
Appendix A.2. This leads to another long calculation, but a check by any
algebraic program (e.g. Maple) shows that this formula, written in standard variables, is given by
\begin{eqnarray}
&& \overline{|{\cal A}_{mix}|^2} =  -\frac{e^4 |{\cal M}|^2  }{p_{12}^2
p_{34}^2 p_{14}^2 p_{23}^2}  \frac{\lambda
  (m_P^2,p_{12}^2,p_{34}^2)}{16} \nn \\
&& \mbox{} \qquad \times \Bigl\{ 4 s_{12} \beta_{12}^2 s_{34} \beta{34}^2
\sin^2\theta_{12} \sin^2\theta_{34} \cos^2\phi + 64 m^4 \nn \\
&& \mbox{} \qquad - \sqrt{s_{12}} \beta_{12} \sqrt{s_{34}} \beta_{34}
\sin\theta_{12} \sin\theta_{34} \cos\phi \biggl[ 32 m^2 -
(m^2-p_{12}^2-p_{34}^2)\bigl( \beta_{12} \cos\theta_{12} - \beta_{34}
\cos\theta_{34} \bigr)^2 \biggr] \nn \\
&& \mbox{} \qquad - 2p_{12}^2 p_{34}^2 \bigl(  1 - \beta_{12} \beta_{34}
\cos\theta_{12} \cos\theta_{34} \bigr)  \bigl(  2 - \beta_{12}^2
\cos\theta_{12} -\beta_{34}^2 \cos\theta_{34}  \bigr) \nn \\
&& \mbox{} \qquad + 8 m^2 \bigl(  \beta_{12} \cos\theta_{12} - \beta_{34} \cos\theta_{34} \bigr)  
 \bigl(  p_{12}^2 \beta_{12} \cos\theta_{12} - p_{34}^2 \beta_{34} \cos\theta_{34} \bigr)  \Bigr\}.
\end{eqnarray}
Note that this formula is exactly the one given in \cite{Barker:2002ib}. Only
the definitions of the angles differ, as mentioned in the Section referring to
the kinematics (see Appendix A).\\
The $s_{23}$ and $s_{14}$ terms in the denominator can be expressed again via
the relations given in \equa{llllid1} and \equa{llllid2}.

\subsection{Decay rate}

The decay rate can now  be stated in terms of the four-body phase space which
is given in \equa{4dr} in a general form:
\begin{equation}
{\rm d}\Gamma=|A|^2 \,\frac{ \beta_{12} \,\beta_{34}\, \lambda^{1/2}(s_{12},m_P^2,s_{34})}{m_P^3\cdot 2^{15}\cdot \pi^6} \,{\rm d}s_{12}\,  {\rm d}s_{34}\, {\rm d}\!\cos\theta_{12} \,{\rm d}\varphi\, 
 {\rm d}\!\cos\theta_{34}.
\end{equation}
Inserting the squared matrix element \equa{A1} and integrating over the angles, we get the final expression for 
the decay rate of the
process $P\to \mu^+\mu^- e^+ e^-$:
\begin{eqnarray}
{\rm d}\Gamma_1& = & \frac{e^4\,|{\cal M}(s_{12},s_{34})|^2 \beta_{12}\beta_{34}
  \lambda^{\frac{3}{2}}(m_P^2,s_{12},s_{34})}{m_P^3\cdot2^{10}\cdot\pi^{5} \cdot s_{12}
  \cdot s_{34}} \left[1 -
\frac{1}{3}\left(\beta_{12}^2 + \beta_{34}^2\right) +\frac{1}{9}\beta_{12}^2\beta_{34}^2 \right]{\rm
  d}s_{12}\,{\rm d}s_{34}.\nn \\
\end{eqnarray}
In all decays induced by the triangle anomaly we will later give the branching
ratios with respect to the decay rate of $P \to \gamma \gamma$. This
simplifies the calculations. Moreover, they will also be more precise, because many terms
of the form factors are the same. Thus we neither have to take care of the
various values of the $\eta_0$ /$\eta_8$ mixing angle nor the coupling constants $f_0$
and $f_8$. In this case the branching ratio simplifies to:
\begin{eqnarray}
\dfrac{{\rm d}\Gamma_1}{{\rm d}\Gamma_{\gamma \gamma}}& = & \frac{2 \alpha^2 \beta_{12}\beta_{34}
  \lambda^{\frac{3}{2}}(m_P^2,s_{12},s_{34})}{ 9 \pi^2 \cdot s_{12}
  \cdot s_{34}} \nn \\
&& \mbox{} \times \left[1 -
\frac{1}{3}\left(\beta_{12}^2 + \beta_{34}^2\right) +\frac{1}{9}\beta_{12}^2\beta_{34}^2 \right]{\rm
  d}s_{12}\,{\rm d}s_{34}\times |VMD_1(s_{12},s_{34})|^2. \nn \\
\end{eqnarray}
Here the factor $VMD_1$ represents the vector meson dominance model that we used.\\
The interference term in this case becomes:
\begin{eqnarray}
&& \dfrac{\Gamma_{12}}{\Gamma_{\gamma \gamma}} = \frac{ \alpha^2
}{4\pi^3}\frac{\lambda (m_P^2,p_{12}^2,p_{34}^2)}{s_{12}s_{34}s_{14} s_{23}}
\times {\rm Re} (VMD_{12}(s_{12},s_{34},s_{14},s_{23})) \nn \\
&& \mbox{} \qquad\Bigl\{ 4 s_{12} \beta_{12}^2 s_{34} \beta_{34}^2
\sin^2\theta_{12} \sin^2\theta_{34} \cos^2\phi + 64 m^4 \nn \\
&& \mbox{} \qquad - \sqrt{s_{12}} \beta_{12} \sqrt{s_{34}} \beta_{34}
\sin\theta_{12} \sin\theta_{34} \cos\phi \biggl[ 32 m^2 -
(m^2-p_{12}^2-p_{34}^2)\bigl( \beta_{12} \cos\theta_{12} - \beta_{34}
\cos\theta_{34} \bigr)^2 \biggr] \nn \\
&& \mbox{} \qquad - 2p_{12}^2 p_{34}^2 \bigl(  1 - \beta_{12} \beta_{34}
\cos\theta_{12} \cos\theta_{34} \bigr)  \bigl(  2 - \beta_{12}^2
\cos\theta_{12} -\beta_{34}^2 \cos\theta_{34}  \bigr) \nn \\
&& \mbox{} \qquad + 8 m^2 \bigl(  \beta_{12} \cos\theta_{12} - \beta_{34} \cos\theta_{34} \bigr)  
 \bigl(  p_{12}^2 \beta_{12} \cos\theta_{12} - p_{34}^2 \beta_{34}
 \cos\theta_{34} \bigr)  \Bigr\} 
\label{interference}
\end{eqnarray}
In this case the vector meson dominance factor depends on all four variables. 
\subsection{Vector meson dominance factor}
\begin{figure}[hbt]
  \begin{minipage}[b]{8 cm}
\begin{center}
 \includegraphics[width=6.5cm]{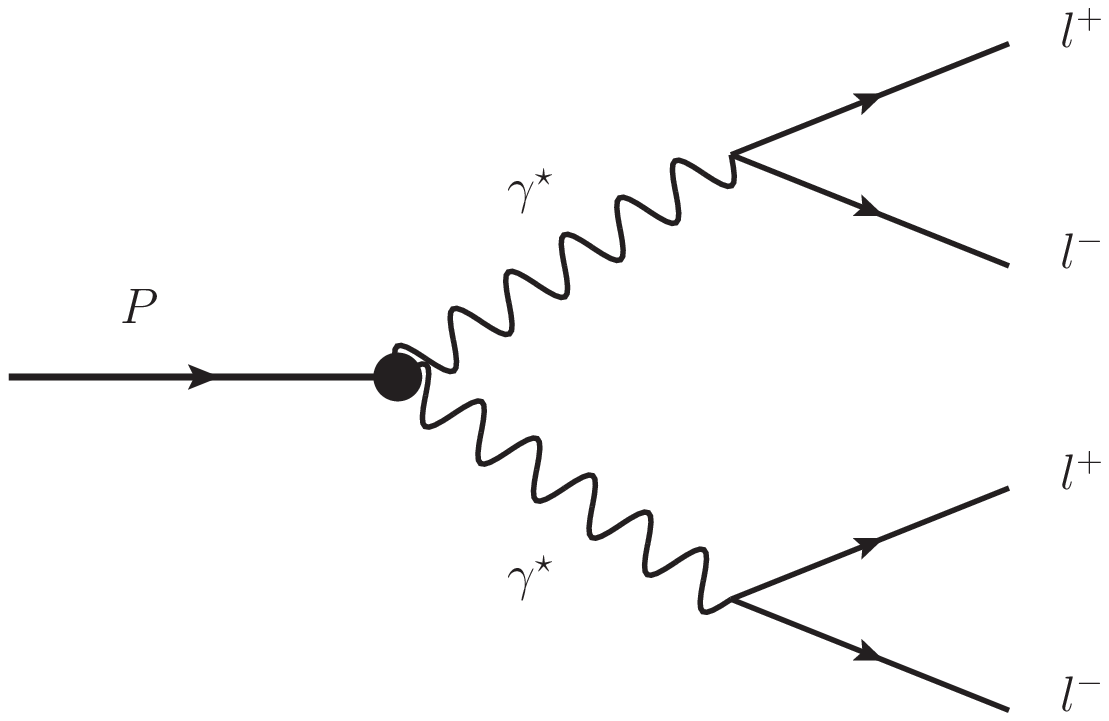}
  \caption{direct contribution $\mathcal{L}_{PAA}$}
  \label{fig:lllldirekt}
\end{center}
  \end{minipage}\hspace{0.5cm}
  \begin{minipage}[b]{8 cm}
\begin{center}
 \includegraphics[width=6.5cm]{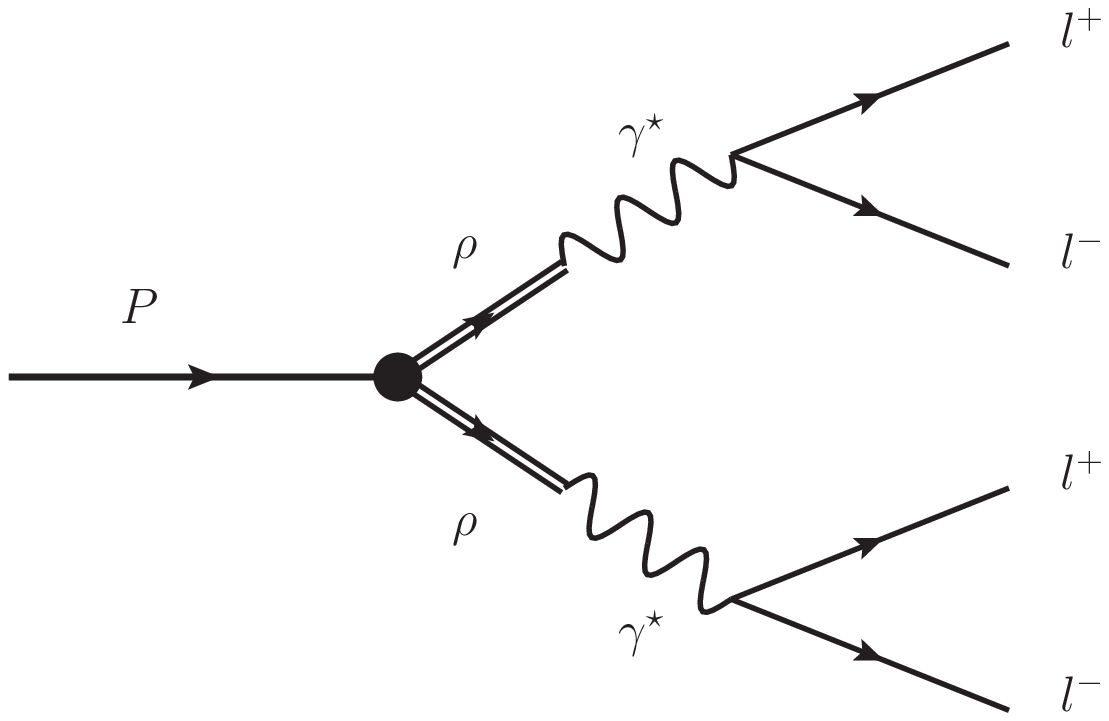}
  \caption{full VMD term $\mathcal{L}_{PVV}$}
  \label{fig:llllvoll}
\end{center}
  \end{minipage}
\end{figure}

In the final step we will show the terms of the used vector meson dominance
model. We have again contributions from the
terms $\mathcal{L}_{PAA}$ and $\mathcal{L}_{PVV}$, while the
$\mathcal{L}_{PVA}$-term vanishes, because of the prefactor $c_3 - c_4$. This
is shown in Figure \ref{fig:lllldirekt} and Figure \ref{fig:llllvoll}.

Thus the factor for the direct and crossed term can be constructed as:
\begin{eqnarray}
VMD_1(s_{12}, s_{34}) = 1-c_3 +c_3 \frac{m_V^2}{m_V^2-s_{12}-i m_V
  \Gamma(s_{12})} \,\, \frac{m_V^2}{m_V^2-s_{34}-i m_V \Gamma(s_{34})}
\end{eqnarray}
The normalization to unity can be tested by setting the invariant masses to
zero.\\ 
In this equations, $\Gamma$ is again the width of the vector meson as given in \equa{width}.\\
The corresponding factor for the interference term looks somewhat different. We build it
up via the expressions $VMD(s_{12}, s_{34})$ and $VMD^\star( s_{14}, s_{23})$.
 Thus we have to calculate the real part of the term
\begin{eqnarray}
VMD_{12}(s_{12}, s_{34}, s_{14}, s_{23}) &=& 1-c_3 + c_3
\frac{m_V^2}{m_V^2-s_{12}-i m_V \Gamma(s_{12})} \frac{m_V^2}{m_V^2-s_{34}-i
  m_V \Gamma(s_{34})}\nonumber \nn \\
&&\mbox{} \qquad   \times \frac{m_V^2}{m_V^2-s_{23}+i m_V
  \Gamma(s_{23})}\frac{m_V^2}{m_V^2-s_{14}+i m_V \Gamma(s_{14})} .\nn \\
\end{eqnarray}
Note the dependence on the four variables $s_{12}, s_{34}, s_{14}$ and
$s_{23}$. The variables $s_{14}$ and $s_{23}$ can be expressed by the usual
variables as reported in (\ref{llllid1}) and (\ref{llllid2}).\\
In the discussion of the results we will compare the values of two different
vector meson dominance models, the 'hidden gauge' and the 'modified' one, with the values calculated without a VMD-factor. The different models can be specified by setting $c_3=1$ (hidden gauge) and $c_3=0.927$, respectively, $c_3=0.930$ (modified VMD model). The two values for $c_3$ of the modified model will give approximately the same results, so we include them via the errors.

\pagebreak
\section{$P \to l^+ l^-$}
The final missing decay that proceeds via the triangle anomaly is the one of $P(q)\to l^+(p') l^-(p)$. A
lot of work was done on this subject already by several groups
e.g. \cite{Young:1967zz, Pratap:1972tb, Martin:1970ai, Bergstrom:1982zq,Bergstrom:1983ay} or
\cite{Silagadze:2006rt}. We will mainly
follow the work of \cite{Dorokhov:2007bd, Dorokhov:2008uk}. There the dependence on many
different form factors was discussed, so we can study how the new optimized
vector meson dominance model acts in comparison. In \cite{Dorokhov:2009xs}
additional corrections were handled and the results were completed.
\begin{figure}[hbt!]
\begin{center}
 \includegraphics[width=9cm]{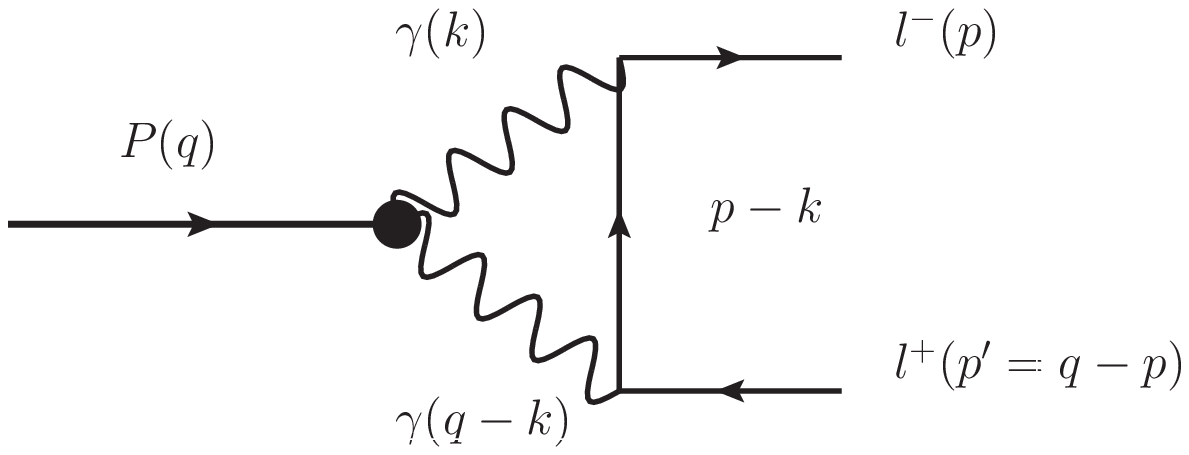}
  \caption{The amplitude of $P \to l^+ l^-$}
\end{center}
\end{figure}\\
We define the momenta of the decay $P(q) \to l^+(p') l^-(p)$ such that the following relation holds:
\begin{equation}
q=p+p'.
\end{equation}
We use the variable $\beta_l(q^2)=\sqrt{1-\frac{4m_l^2}{q^2}}$ and of
course the momenta and masses of the respective leptons to describe the decays.\\ 
According to \cite{Silagadze:2006rt} the Feynman amplitude is given by:
\begin{eqnarray}
{\cal M} (P \to l^+ l^-) &=& \frac{e^4}{f_\pi} \bar u(p,s_-)  A v(p',s_+)
\nn \\
&&= \frac{e^4}{f_\pi} Tr \left[ v(p',s_+) \bar u(p,s_-)   A \right]
\label{All}
\end{eqnarray}
where
\begin{equation}
  A = \int \frac{d^4 k}{(2 \pi)^4}\frac{\gamma_\mu \left( \slashed p -\slashed k
  +m_l \right) \gamma_\nu \varepsilon^{\mu\nu\sigma\tau} k_\sigma
q_\tau}{\left[ (p-k)^2 -m_l^2  \right] k^2 (q-k)^2 } \times VMD(k^2,(q-k)^2).
\end{equation}
Here the form factor $VMD(k^2,(q-k)^2)$ contains the vector meson dominance input and will
be discussed later in detail.\\
Using the informations given in Appendix C we can write the branching ratio of the decay $p \to
l^+l^-$ in terms of the reduced, dimensionless amplitude ${\cal A}$:\\
\begin{equation}
\frac{\Gamma_{P \to l^+l^-}}{\Gamma_{\gamma\gamma}}=2\left( \frac{\alpha
    m_l}{\pi m_P}  \right)^2 \beta_l(q^2) |{\cal A} (m_P^2)|^2
\end{equation}
with
\begin{equation}
{\cal A} (q^2)= \frac{2i}{\pi^2 m_p^2}\int d^4 k
\frac{ \left( q^2 k^2 - (k \cdot q)^2  \right)  }{\left[ (p-k)^2 -m_l^2  + i
    \varepsilon \right] (k^2+i \varepsilon) \left((q-k)^2 + i \varepsilon \right) } \times VMD(k^2,(q-k)^2)
\label{integralR}
\end{equation}\\
as given in \cite{Bergstrom:1982zq} and \cite{Dorokhov:2007bd}. Note that
\cite{Silagadze:2006rt} differs from other calculation,
e.g. \cite{Bergstrom:1982zq} by a factor 2 in the form factor.\\
To solve this integral we follow the
dispersion approach applied in many publications before, e.g. \cite{Young:1967zz, Pratap:1972tb, Bergstrom:1982zq, Dorokhov:2007bd}. Therefore, we will first calculate the
imaginary part of the amplitude ${\cal A}$ using the Cutkosky rules
\cite{Cutkosky:1960sp, Mandelstam:1959bc} (for further calculations see
Appendix C):
\begin{equation}
{\rm Im} {\cal A} (q^2) = \frac{\pi}{2\beta_l(q^2)}{\rm ln}\frac{1-\beta_l(q^2)}{1+\beta_l(q^2)}.
\end{equation}
This is the contribution of two on-shell photons in the intermediate
state. Therefore, the form factor is trivial and model independent, {\it i.e.} $VMD(0,0)=1$. We can give a lower limit for the
reduced branching ratio by using only the imaginary part $(|{\cal A}|^2 \ge
{\rm Im} {\cal A}^2)$. These unitary bounds are given in Chapter 4.\\
In order to calculate the branching ratio we have to deal also with the
real part of ${\cal A}$. According to \cite{Bergstrom:1982zq} the once-subtracted dispersion
relation is given by:
\begin{equation}
{\rm Re} {\cal A} (q^2) = {\cal A} (q^2=0) + \frac{q^2}{\pi}\int_0^\infty ds \frac{{\rm Im} {\cal A} (s)}{s(s-q^2)}.
\end{equation}
The integral can be calculated using analytical programs. We are only interested in
the integral for $q^2\ge 4
m_l^2$ \footnote[1]{The expressions for $q^2\le 0 $ and $0 \le q^2 \le 4
m_l^2$ are given in Ref. \cite{Dorokhov:2007bd}.}, where $m_l$ is again the lepton mass, which is given by the following expression \cite{Dorokhov:2007bd}:
\begin{equation}
{\rm Re} {\cal A} (q^2) = {\cal A} (q^2=0) + \frac{1}{\beta_l(q^2)}\left[ \frac{1}{4}
  {\rm ln}^2\left(\frac{1-\beta_l(q^2)}{1+\beta_l(q^2)}\right) +\frac{\pi^2}{12} + Li_2\left(-\frac{1-\beta_l(q^2)}{1+\beta_l(q^2)}\right)  \right]
\end{equation}
with the dilogarithm function $Li_2(z)=-\int_0^z (dt/t)\,\, \ln(1-t)$. To the
leading order in $\left(m_l/m_P\right)^2$ this is given by
\begin{equation}
{\rm Re} {\cal A} (m_P^2) = {\cal A} (q^2=0) + {\rm ln}^2\left( \frac{m_l}{m_P} \right) + \frac{\pi^2}{12}.
\end{equation}
Thus all the nontrivial dynamics of the process are contained in the
subtraction constant $A(q^2=0)$. The derivation of this constant is summarized in
Appendix C according to the procedure of Ref. \cite{Dorokhov:2007bd}. To the first order in $m_l/m_V$, where $m_V$ is the vector meson mass, which appears in the calculations given in Appendix C, the expansion is given as
follows\footnote[2]{Higher order corrections can be found in Ref. \cite{Dorokhov:2009xs}.}:
\begin{equation}
{\cal A} (q^2=0)=3{\rm ln} \left(\frac{m_l}{\mu}\right) + \chi_p(\mu)
\label{subconstA}
\end{equation}
where $\mu$ is a scale parameter, which can be set to $m_V$. The constant
$\chi(\mu)$ depends on the VMD factor in symmetrized Euclidean kinematics:
\begin{eqnarray}
\chi_p(\mu)&=&-\frac{5}{4}+\frac{3}{2}\int_0^\infty dt \,\, {\rm ln}\left(\frac{t}{\mu^2}\right)
\frac{\partial VMD(-t,-t)}{\partial t} \nn \\
&=&-\frac{5}{4}-\frac{3}{2} \left[ \int_0^{\mu^2} dt \frac{VMD(-t,-t)-1}{t} + \int_{\mu^2}^\infty dt \frac{VMD(-t,-t)}{t} \right].
\end{eqnarray}
The VMD factor now depends only on one invariant mass $t$ which simplifies the
calculations. Note that ${\cal A} (q^2=0)$ is independent on the parameter $\mu$ after
integration of $t$. Also note that $VMD$ is expressed in Euclidean space-time and we therefore
do not need a width, because there are no singularities. The form factor can be written as
follows:
\begin{equation}
VMD(-t,-t) = 1-c_3 +c_3 \frac{m_V^2}{m_V^2+t} \frac{m_V^2}{m_V^2+t},
\end{equation}
where $c_3$ is again the c-parameter that differentiates between the various VMD models.\\
In the case of $\eta'$ decays there is an additional contribution to the
imaginary part(\cite{Dorokhov:2009xs}). Because the mass of the $\eta'$ is greater than the ones of the
$\omega$ and $\rho$ mesons, one vector meson can be on-shell. This contribution
can be calculated again with the help of the Cutkosky rules. The additional contribution is given as:
\begin{equation}
\Delta {\rm Im} {\cal A}=-\frac{\pi}{\beta}\left( 1-\frac{1}{z}\right)^2 {\rm
  ln} \left(\frac{1-\beta}{1+\beta}\right) \Theta(z-1)
\end{equation}
where $z=\left(m_p/m_V\right)^2$. Note that in comparison to \cite{Dorokhov:2009xs} the numerator and
the denominator are exchanged.

\section{$P \to \pi^+ \pi^- \gamma$ and $P \to \pi^+ \pi^- l^+ l^-$}
In the next Section we will deal with the decays $P\to \pi^+ \pi^-\gamma$, $P
\to \pi^+ \pi^- e^+ e^-$ and $P
\to \pi^+ \pi^- \mu^+ \mu^-$. The three decays are very similar in their basic
structure, since the photon in the $P \to \pi^+ \pi^- \gamma$ decay can be
replaced by an off-shell one that decays into a lepton pair. Because of
kinematic reasons only the $\eta$  and
$\eta'$ decays are possible, but not the $\pi^0$ decay. All of the decays
governed by box anomaly proceed as shown
in Figure \ref{fig:box}.
\begin{figure}[hbt]
\begin{center}
 \includegraphics[width=8.5cm]{box.eps}
  \caption{box anomaly}
  \label{fig:box}
\end{center}
\end{figure}

\begin{figure}[hbt]
\begin{center}
 \includegraphics[width=16.5cm]{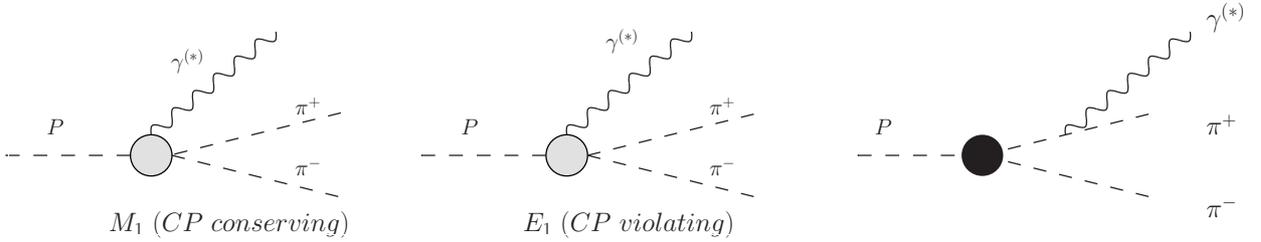}
  \caption{CP conserving magnetic $(M_1)$, CP violating electric $(E_1)$ and Bremsstrahlung contribution}
  \label{fig:box_M1}
\end{center}
\end{figure}

These reactions are quite interesting, because there is the possibility to
measure CP violation as in the equivalent kaonic
decays \cite{Bijnens:1994me}. To model this we have to take the usual magnetic
vertex with a form factor $M$ into account and add another vertex, which is constructed as an CP violating electric dipole operator. The corresponding form factor $E$ was introduced by \cite{Geng:2002ua} for the $P\to
\pi^+ \pi^- \gamma$ decay and modified by \cite{Gao:2002gq} for the $P \to \pi^+ \pi^- l^+ l^-
$ decay. The form factors will be presented after we have handled the
kinematics of the respective decays. The structure of the decay amplitude is visualized
in Figure \ref{fig:box_M1}. Note that the contribution of the
Bremsstrahlung term is negligibly small ($BR(\eta\to \pi \pi)<3.5\cdot
10^{-14}$), see \cite{Gorchtein:2008pe}, as otherwise the electric dipole moment of the
neutron would have been measured with a value bigger than the experimental
upper bound tabulated by PDG (\cite{Amsler:2008zzb}).

In the squared amplitude
there appear terms proportional to $|M|^2$ and $|E|^2$ which are both CP
conserving. The relevant terms for a CP violating contribution are
proportional to Re($ME^\star$). As we will see it is not easy to construct these terms
and even harder to measure them. \\
 We will use the definitions of Ref.\,\cite{Geng:2002ua} for the four-momenta of the decay $P(P) \to$ \\
$\pi^+(p_+) \, \pi^-(p_-) \,\gamma(k)$,
  such that the following 
 relation is valid in any frame:
 \begin{equation}
      P = \underbrace{p_+ + p_-}_{\equiv p} + k  \,,
    \label{pgeng}
 \end{equation}
 whereas the four-momenta for the decay $P(P) \to \pi^+(p_+) \, \pi^-(p_-) \,l^+ (k_+) \,l^-(k_-)$ are defined according to Ref.\,\cite{Gao:2002gq}
as
 \begin{equation}
      P = \underbrace{p_+ + p_-}_{\equiv p} + \underbrace{k_+ + k_-}_{\equiv k} \,.     
    \label{pmometa}
 \end{equation}
In order to make the kinematics more obvious we will rename the standard
variables $s_{pp}$, $\beta_p$ and $\theta_p$ as $s_{\pi\pi}$, $\beta_\pi$ and
$\theta_\pi$ labelled by the respective particles and not by the momenta. Of course
$s_{kk}$ turns into $s_{ee}$ and so on for the $P \to \pi^+ \pi^- l^+ l^-$
case.
\subsection{The reaction $P \to \pi^+ \pi^- \gamma$}
We start out with the decay $P \to \pi^+ \pi^- \gamma$ and therefore with the invariant decay amplitude given in Eq.\,(2) of
Ref.\,\cite{Geng:2002ua}:
\begin{equation}
{\cal A}(P\to \pi^+ \pi^- \gamma)= \frac{i}{m_P^3} 
\left ( M_G \varepsilon_{\mu\nu\alpha\beta} \epsilon^\mu k^\nu p_+^\alpha p_- ^\beta
 + E_G \left[ ( \epsilon \cdot p_+) (k \cdot p_-) - (\epsilon\cdot p_-) (k \cdot p_+)
 \right]\right).
 \label{AmpGeng}
 \end{equation}
Here we can nicely see the usual magnetic form factor $M$ attached to the total
antisymmetric tensor, which is typical for anomalous decays, and the new
contribution of the CP
violating electric form factor $E$. Note that the form factors of Ref.\,\cite{Geng:2002ua} and Ref.\,\cite{Gao:2002gq} differ in structure and
normalization. Therefore, the subindex $G$ is added here.\\
The following steps are straight forward. We first insert $p_- = P - p_+
-k$, so that we can make use of the antisymmetric $\varepsilon$-tensor. The
amplitude then reads:
\begin{eqnarray}
&& {\cal A}(P \to \pi^+ \pi^- \gamma)=  \frac{i}{m_P^3} 
\biggl( M_G\varepsilon_{\mu\nu\alpha\beta} \epsilon^\mu k^\nu p_+^\alpha P ^\beta + 
  E_G \left[ ( \epsilon \cdot p_+) (k \cdot P ) 
   - (\epsilon\cdot P) (k \cdot p_+)\right]\biggr).\nn \\
 \end{eqnarray}
 In the $P$ rest frame, where $P^\mu = m_P \delta^{\mu 0}$, the amplitude is then
 given by
 \begin{eqnarray}
{\cal A}(P\to \pi^+ \pi^- \gamma)&=&
 \frac{i}{m_P^2} 
\biggl( -M_G\varepsilon^{ijk}  \epsilon^i \widetilde{k}^j \widetilde{p}_+^k  
+ E_G\left[    (  -{\bf\epsilon} \cdot{\widetilde{\bf p}_+}) (\widetilde E_\gamma ) 
   \right]\biggr)\nn \\
 &=&
 \frac{i \widetilde E_\gamma}{m_P^2} 
\biggl( M_G\,\hat{\kbf} \cdot ( \epsbf\wedge \widetilde\pbf_+)  
- E_G \,\epsbf \cdot \widetilde \pbf_+ \biggr) \,.
\end{eqnarray}
The squared amplitude reads
 \begin{eqnarray}
|{\cal A}|^2(P \to \pi^+ \pi^- \gamma)&=&
  \frac{\widetilde E_\gamma^2}{m_P^4} 
\biggl( |M_G|^2\,|\hat{\kbf} \cdot ( \epsbf\wedge {\pbf_+}_{\!\perp})|^2 + |E_G|^2|\epsbf\cdot {\pbf_+}_{\!\perp}|^2\nn \\
&& +E_G^* M_G \left[ \hat\kbf \cdot ( {\pbf_+}_{\!\perp}\wedge \epsbf) \right] 
(\epsbf \cdot {\pbf_+}_{\!\perp} )^*
+  M_G^* E_G \left[ \hat\kbf \cdot ({ \pbf_+}_{\!\perp}\wedge \epsbf) \right]^* 
(\epsbf \cdot {\pbf_+ }_{\!\perp})\biggr).\nn
\end{eqnarray}   
In Ref.\,\cite{Geng:2002ua} the following polarization vectors are used:
\begin{eqnarray}
\epsbf_1 &=& \frac{\left( \pbf_+ \wedge \bf k \right ) \wedge \kbf}
                                 {|{\left( \pbf _+\wedge \bf k \right ) \wedge \kbf}|}
                      =\frac{   \hat\kbf (\pbf_+ \cdot \hat\kbf) - \pbf _+ }
                                { | \hat\kbf (\pbf _+\cdot \hat\kbf) - \pbf_+ | }
                    = -{ \hpbf }_{\!\! \perp}                                                                          \,, \nn\\
 \epsbf_2 &=&   \frac{ \pbf_+\wedge \bf k }
                                 {|  \pbf_+\wedge \bf k |} 
                          = \frac {  {\pbf_+}_{\!\perp}\wedge \hat\kbf}
                          { | {\pbf_+}_{\!\perp}\wedge \hat\kbf |} 
                          = \hat\kbf \wedge (- { \hpbf }_{\!\! \perp}) 
                        \,. \nn
\end{eqnarray}
Thus the unpolarized squared decay amplitude reads
 \begin{eqnarray}
\sum_{{\rm pol}=1}^2 \!\!|{\cal A}|^2(P\to \pi^+ \pi^- \gamma)\!\!&=&\!\!
  \frac{\widetilde{E_\gamma}^2 \overbrace{|\pbf_+^\star|^2\sin^2\theta_\pi}^{|\pplusbf_{\!\!\perp}^2|}}{m_P^4} 
\biggl( \underbrace{|M_G(s_{\pi\pi})|^2}_{\epsbf_2\,\mbox{part}}+
\underbrace{|E_G|^2}_{\epsbf_1\,\mbox{part}}\! \biggr)\!.
\label{AmpGaSq}
\end{eqnarray}
The contributions of the mixed term vanished  when both photon polarizations
had been summed. This means that the CP
violation in this decay cannot be found, if the polarizations of the
photons are not measured explicitly.\\
The squared amplitude \equa{AmpGaSq}, expressed in standard variables $s_{\pi\pi}$ and $\theta_{\pi}$, reads 
\begin{eqnarray}
\sum_{{\rm pol}=1}^2 \!\!|{\cal A}_{P \to \pi^+ \pi^- \gamma}|^2(s_{\pi\pi},\theta_\pi)
\!\!&=&\!\!
  \frac{\lambda(m_P^2,s_{\pi\pi},0) s_{\pi\pi} \beta_\pi^2  \sin^2\theta_\pi}
  {16\, m_P^6} 
\biggl( |M_G|^2 + |E_G|^2 \biggr) .
\end{eqnarray}
In terms of $\theta_\pi^\star$, $\widetilde E_\gamma$ and 
\[
 \beta_\pi=\beta_\pi(\widetilde E_\gamma) 
= \sqrt{1-\frac{4 m_\pi^2}{s_{\pi\pi}(\widetilde E_\gamma)} }
= \sqrt{1-\frac{4 m_\pi^2}{m_P^2- 2 m_P \widetilde E_\gamma}} 
\]
(see Eq.\,\equa{Egamma}) it is given by
\begin{eqnarray}
\sum_{{\rm pol}=1}^2 \!\!|{\cal A}_{P \to \pi^+ \pi^- \gamma}|^2(\widetilde E_\gamma,\theta_\pi)
\!\!&=&\!\!
  \frac{{\widetilde E_\gamma}^2 \left(1- 2 \widetilde E_\gamma/m_\eta\right)
   \beta_\pi^2  \sin^2\theta_\pi}
  {4\, m_P^2} 
\biggl( |M_G|^2 + |E_G|^2 \biggr). 
\end{eqnarray}

\subsection{Decay rate}
The decay rate for the three-body decay $P\to \pi^+\pi^-\gamma$ is given by
relation (38.19) of Ref.\,\cite{Amsler:2008zzb}:
\begin{eqnarray}
{\rm d}\Gamma &=& \frac{1}{ (2\pi)^5 } \,\frac{1}{ 16 m_P^2}  \, |{\cal A}|^2 \,|\pplusbf^\star | 
\underbrace{|\widetilde \kbf |}_{\widetilde E_\gamma} \, {\rm d} 
\underbrace{m_{\pi \pi}}_{\sqrt{s_{\pi\pi}}}\  
{\rm d}\!\cos\theta_\pi\  {\rm d} \phi_\pi^\star\  {\rm d}\!\cos\widetilde\theta_\gamma\  {\rm d}
\widetilde \phi_\gamma \nn \\
&=& \frac{1}{2^{12} \pi^3} \,\left( 1-\frac{s_{\pi\pi}}{m_P^2} \right)^3\, \frac{s_{\pi\pi}^{\,\,3/2}}
{m_P^3}\, \beta_\pi^3 \sin^2\!\theta_\pi
\left(|M_G(s_{\pi\pi)} |^2 + |E_G|^2\right) {\rm d}\sqrt{s_{\pi\pi}} \,
{\rm d}\!\cos\theta_\pi\ \nn \\
&=& \frac{1}{512 \pi^3} \,\frac{\widetilde E_\gamma^3}{m_P^3}\,  
\beta_\pi^3 \left( 1 - \frac{2 \widetilde E_\gamma}{m_P} \right)
 \sin^2\!\theta_\pi
\left(|M_G(s_{\pi\pi)} |^2 + |E_G|^2\right) {\rm d}\widetilde E_\gamma  \,
{\rm d}\!\cos\theta_\pi\ \nn \\
\label{resdGa}
\end{eqnarray}  
where the definition \equa{past} was inserted for $|\pbf_+^\star|$  and the
relations 
\begin{eqnarray}
  s_{\pi\pi} &\equiv & \widetilde p^2 = m_P^2 - 2m_P  \widetilde E_\gamma
= m_P^2\left(1 - \frac{2\widetilde E_\gamma}{m_P} \right),  \label{Egamma} \\
 \sqrt{s_{\pi\pi}}\, {\rm d}\sqrt{s_{\pi\pi}} &=&\half {\rm d} s_{\pi\pi} =
  - m_P \,{\rm d}\widetilde E_\gamma \label{dspipi}
\end{eqnarray}
 were used in the last line.  Note that the minus sign in Eq.\,\equa{dspipi} cancels against
 a minus sign resulting under a switch of the upper and lower integration limits of
 the $d \widetilde E_\gamma$ integration, since $\widetilde E_\gamma^{\rm max}=
 \widetilde E_{\gamma}(s_{\pi\pi}^{\rm min})$ and vice versa, see
 Eq.\,\equa{Egamma}. Also note that equation \equa{resdGa} is exactly the result (6) of Ref.\,\cite{Geng:2002ua}.

\subsection{The form factors for  $P\to\pi^+\pi^-\gamma$}

\begin{center}
\begin{figure}[hbt]
  \begin{minipage}[b]{5 cm}
 \includegraphics[width=5cm]{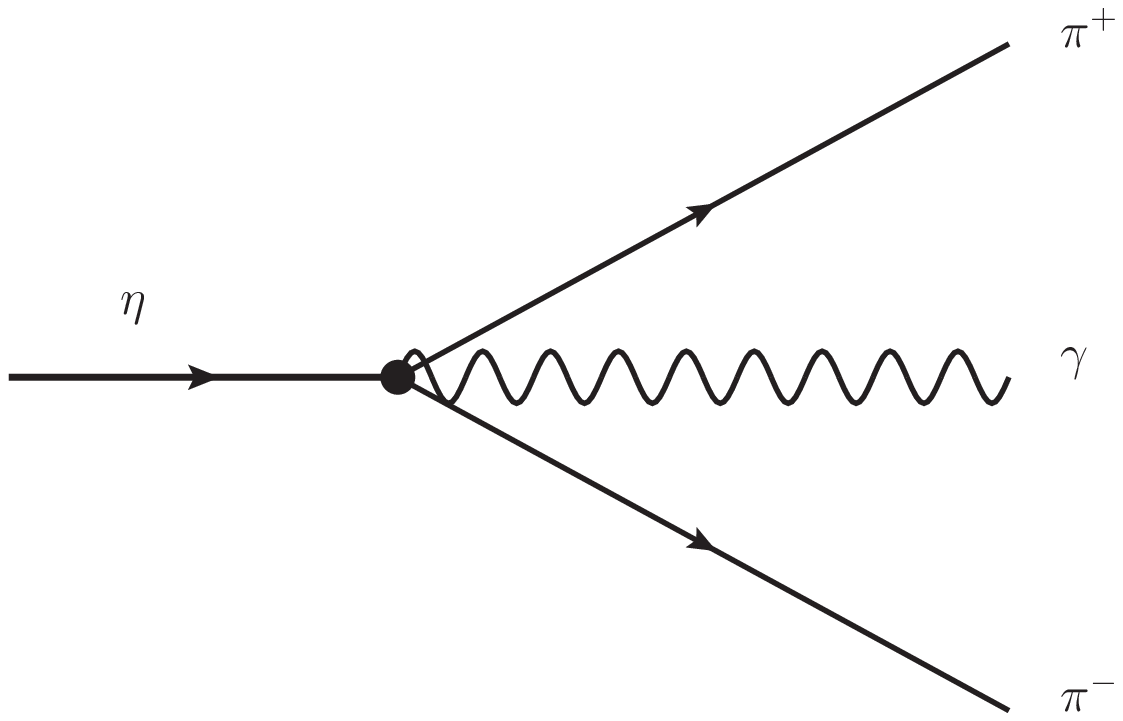}
  \caption{direct contribution $\mathcal{L}_{PPPA}$}
  \label{fig:VMDppg1}
  \end{minipage}\hspace{0.5cm}
  \begin{minipage}[b]{5 cm}
 \includegraphics[width=5cm]{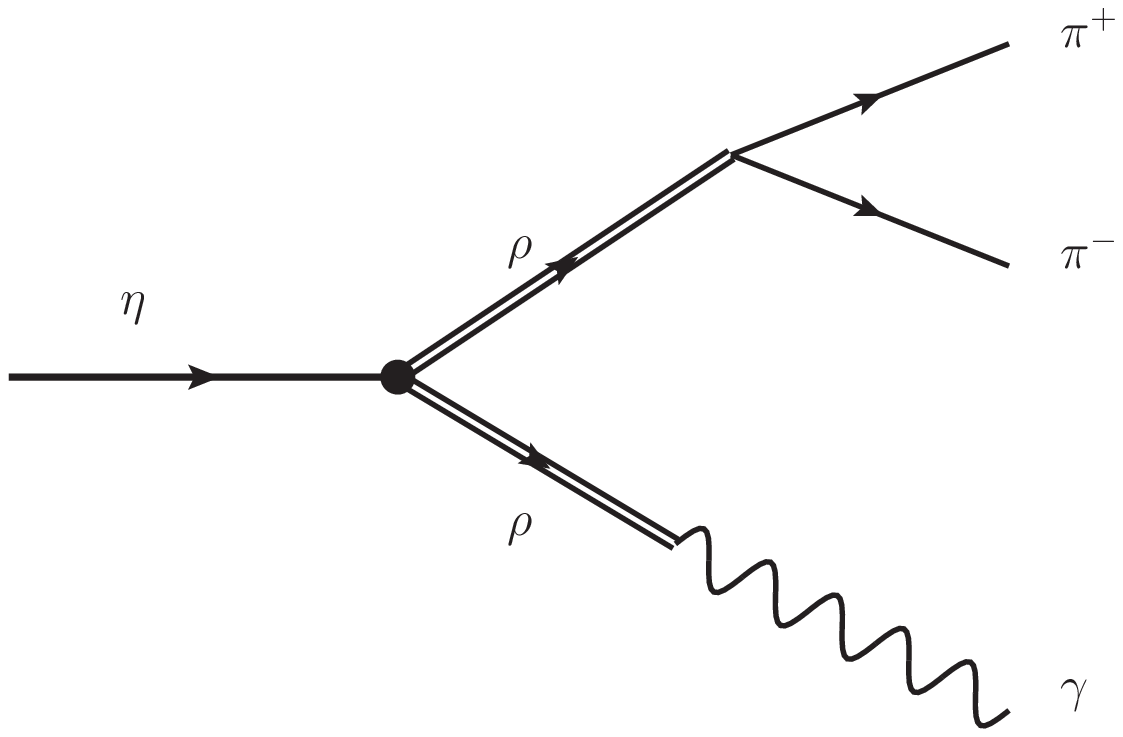}
  \caption{direct contribution $\mathcal{L}_{PVV}$}
  \label{fig:VMDppg2}
  \end{minipage}\hspace{0.5cm}
  \begin{minipage}[b]{5 cm}
 \includegraphics[width=5cm]{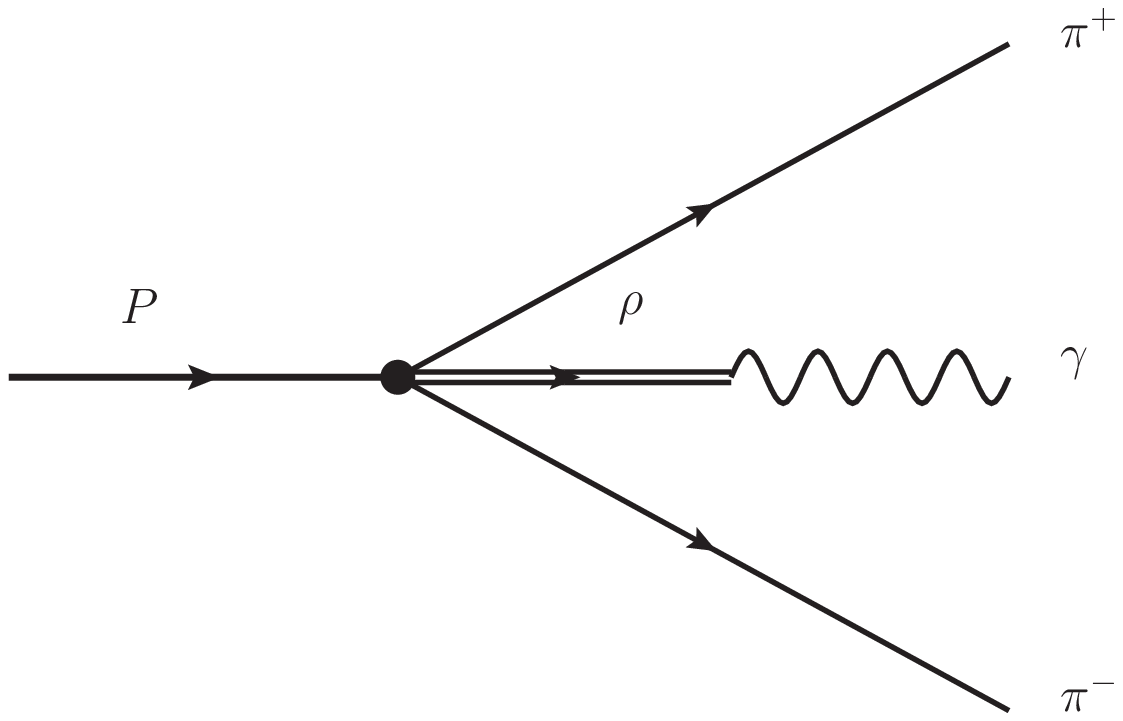}
  \caption{full VMD term $\mathcal{L}_{PPPV}$}
  \label{fig:VMDppg3}
  \end{minipage}
\end{figure}
\end{center}

We have marked the form factors  of Ref.\,\cite{Geng:2002ua} 
for the decay $P\to\pi^+\pi^- \gamma$ (with a real photon)
by a subindex $G$, {\it i.e.} $M_G$ and $E_G$. They differ in normalization and structure
from the corresponding form factors $M$ and $E_\pm$ of Ref.\,\cite{Gao:2002gq} for the decay
 $P\to\pi^+\pi^- e^+ e^-$. Let us first focus on the magnetic form factor $M_G$.
 Comparing the corresponding amplitudes \equa{AmpGeng} with \equa{AmpGao} ({\it i.e.}
 Eq.\,(2) of Ref.\,\cite{Geng:2002ua} with Eq.\,(1) of Ref.\,\cite{Gao:2002gq})
we can read off the following relation between $M_G$ and $M$:
\begin{eqnarray}
M_G(s_{\pi\pi}) &=& m_P^3 M(s_{\pi\pi}, k^2=0).
\end{eqnarray} 
Here $M(s_{\pi\pi}, k^2=0)$ contains the information of the decaying particle
and the vector meson dominance input:
\begin{eqnarray}
M(s_{\pi\pi}, k^2=0)= {\cal M} \times VMD(s_{\pi\pi})
\end{eqnarray}
with 
\begin{equation}
{\cal M}=\begin{cases}
         {\displaystyle \frac{e}{8\pi^2 f_{\pi}^3}} & \mbox{if $P=\pi^0$};\\
       {\displaystyle \frac{e}{8\pi^2 f_{\pi}^3}\frac{1}{\sqrt{3}}} \left( \frac{f_\pi}{f_8} \cos\theta_{mix} -2\sqrt{2} \frac{f_\pi}{f_0} \sin\theta_{mix} \right)& \mbox{if $P=\eta$};\\
       {\displaystyle \frac{e}{8\pi^2 f_{\pi}^3} \frac{1}{\sqrt{3}}} \left( \frac{f_\pi}{f_8} \sin\theta_{mix} +2\sqrt{2} \frac{f_\pi}{f_0} \cos\theta_{mix} \right)& \mbox{if $P=\eta'$}.
\end{cases}
\end{equation}
This means that $M_G(s_{\pi\pi})$ is dimensionless, whereas $M(s_{\pi\pi},0)$ has the dimension $[{\rm mass}]^{-3}$.\\
The vector meson dominance factor contains the  contributions from the
terms $\mathcal{L}_{PPPA}$, $\mathcal{L}_{PVV}$  and also a small contribution
from the $\mathcal{L}_{PPPV}$ , which actually vanishes for the hidden gauge case. This
is shown in Figures \ref{fig:VMDppg1}, \ref{fig:VMDppg2} and \ref{fig:VMDppg3}.

Therefore the vector meson dominance factor is given by
\begin{eqnarray}
VMD_1(s_{\pi\pi}) = 1-\frac{3}{2}c_3  + \frac{3}{2}c_3 \frac{m_V^2}{m_V^2-s_{\pi\pi}-i m_V \Gamma(s_{\pi\pi})} .
\end{eqnarray}
 Note that  neither form factor is normalized to one for the special cases  
 $s_{\pi\pi} \to 4 m_\pi^2 $ or $s_{\pi\pi}\to 0$. The vector meson dominance
 factor instead is normalized to one.\\ 
 The case of the electric form factor $E_G$ is more complicated since it differs also in
 structure (and not only in normalization) from $E_\pm$ from \cite{Gao:2002gq}.
 \begin{itemize}
 \item[0.] To leading order the form factor $E_G$ could simply be put to zero.
 \item[1.] A model with a intermediate CP-violating decay $P\to \pi^+\pi^-$
   and a subsequent Bremsstrahlung corresponds to the following structure of
   the electric form factor $E_G$ (\cite{Geng:2002ua}):
 \begin{eqnarray}
  E_G(s_{\pi\pi},\theta_{\pi}) &=& \frac{e\, m_\eta^3 \,g_{\eta \pi\pi} } { (p_+\cdot k)(p_-\cdot k)}
                = \frac{16 \,e\,m_\eta^3 \,g_{\eta \pi\pi} } { (m_\eta^2-s_{\pi\pi})^2\,\left(1 
                        -\beta_\pi^2\,\cos^2\theta_{\pi}\right)}\nn \\
                  &=& \frac{4 \,e\,m_\eta \,g_{\eta \pi\pi} } { {\widetilde E_\gamma}^2\,\left(1 
                        -\beta_\pi^2\,\cos^2\theta_{\pi}\right)}
                        =\frac{4 \,e\,m_\eta \,g_{\eta \pi\pi} } { {\widetilde E_\gamma}^2\,\left\{1
                        -\left(1-\frac{4m_\pi^2}{ m_\eta^2-2m_\eta\widetilde E_\gamma} \right)\!\cos^2\!\theta_\pi\right\}} \nn \\
  &\equiv& E_G(\widetilde E_\gamma,\theta_\pi) \,. 
 \end{eqnarray}
This model is very unlikely to lead to measurable results, since theory and phenomenology predict that the
 $g_{\eta\pi^+\pi^-}$  coupling constant is very tiny: $|g_{\eta\pi^+\pi^-}|$ is estimated to be 
less than $2.6 \times 10^{-16}$ in the Standard Model (via
the CKM phase), less than $2 \times 10^{-10}$ under the presence of a {\em strong $\theta$ term}
in QCD, and less than $5 \times 10^{-11}$ in spontaneous CP violating
models (with more than one Higgs particle), see Refs.\,\cite{Geng:2002ua,Gao:2002gq} and references therein.  
 \item[2.] In Ref.\,\cite{Geng:2002ua} a form factor was constructed in terms
   of a 4-quark-operator $\mathcal O$. This
operator is assumed to be unconventional,
which means that it is not constrained by known physics. Especially, it should not contribute
directly to the decay $P\to \pi^+ \pi^-$ and the well studied $K^0$ decays. It
also should be a
flavor-conserving CP violating four-fermion operator with explicit $s \bar s$
quark content, such that constraints from the empirical bounds on the electric
dipole moment of the neutron are excluded as well. The following short-range operator
does have these features:
\begin{equation}
{\mathcal O} = \frac{1}{m_P^3} G \bar s i \sigma_{\mu\nu}\gamma_5(p-k)^\nu s
\hspace{2mm} \bar u \gamma^\mu u \, .
\end{equation}
Here $G$ is a free, dimensionless, 'natural' model-coefficient. The latter
means that it
is of order  $G \leq {\cal
  O}(1)$. It parameterizes the strength of the operator.
Geng et al. (\cite{Geng:2002ua}) calculated the contribution under the
assumption that the production of the photon results from the strangeness
containing part, while the $u$- and $d$-quark part is responsible for the
dipion-production from the vacuum. This means:
\begin{equation}
\bra{P}{\mathcal O}\ket{\pi^+ \pi^- \gamma} \sim \frac{1}{m_P^3} G \bra{P}
\bar s i \sigma_{\mu\nu}\gamma_5(p-k)^\nu s \ket{\gamma}  \bra{0}\bar u
\gamma^\mu u \ket{\pi^+ \pi^-}
\end{equation}
which is visualized in Figure \ref{fig:FFEa}.\\
\begin{figure}[hbt]
\begin{center}
 \includegraphics[width=10cm]{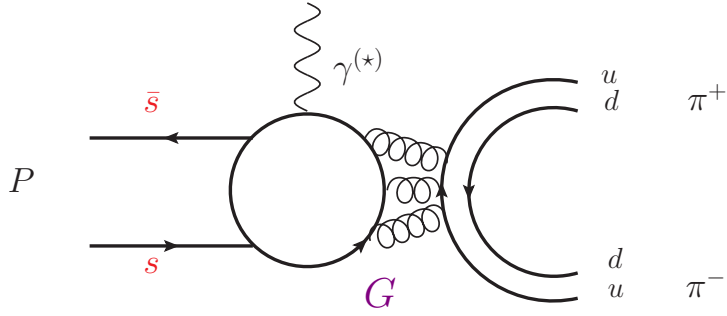}
  \caption{Structure-dependent contribution of the unconventional operator of
    Ref. \cite{Geng:2002ua} to $P\to \pi^+ \pi^- \gamma$ .}
  \label{fig:FFEa}
\end{center}
\end{figure}

The authors of Ref \cite{Geng:2002ua} introduced for the $P\to \gamma$
transition the form factor $F(s)$:
\begin{equation}
\bra{P} \bar s i \sigma_{\mu\nu}\gamma_5(p-k)^\nu s
\ket{\gamma}=ie[\epsilon_\mu(k\cdot p)-(\epsilon \cdot p) k_\mu]\frac{F(s)}{m_P^3}.
\end{equation}
It can be calculated in the framework of quark models (see
\cite{Geng:2000fs, Geng:1997ws} and \cite{Poblaguev:1990tv}). In Ref.
\cite{Geng:2002ua} it was suggested to use $F(s_{\pi\pi})\sim F(0) \approx 0.19$.\\
Assuming this value we find
 \begin{equation}
 E_G(s_{\pi\pi}) \ \sim\  2 e F(s_{\pi\pi}) G \ = \ 
  2 e F\left(m_\eta^2-2m_\eta \widetilde E_\gamma \right) G  
 \ \equiv\
 E_G(\widetilde E_\gamma)\, .
 \end{equation}  
\end{itemize}
In our calculations we will use the form factor given in item 2.

 \subsection{The reaction $P\to \pi^+ \pi^- l^+ l^-$ }
We will now concentrate on the decay $P\to \pi^+ \pi^- l^+ l^-$. We start out with the invariant decay amplitude given in  Eq.\,(1) of Ref.\,\cite{Gao:2002gq}:
 \begin{equation}
    {\cal A}(P \to \pi^+\! \pi^-\! l^+\! l^-) 
    \!=\! \frac{1}{k^2} \underbrace {e \,\bar u(k_-,s_-)\gamma^\mu v(k_+,s_+)}_{ \equiv j ^\mu(k_-,k_+)}
\! \left( M \varepsilon_{\mu \nu\alpha\beta} k^\nu p_+^\alpha p_-^\beta + E_+ p_+^\mu + E_- p_-^\mu \right ).
 \label{AmpGao}
 \end{equation}
Note that the polarizations appearing in the process $P\to \pi^+ \pi^- \gamma$ are
now replaced by a current into a lepton pair. Furthermore, we will use the following `projection' tensor constructed from the bilinear combination
of the current
$j_\mu(k_-,s_-; k_+,s_+)$ 
under a summation over the final spins $s_-$ and $s_+$ of the outgoing $l^- l^+$
pair, which can be found in Appendix A5: 
\begin{eqnarray}
 \!\!\!\! {\cal O}_{\mu \mu'}(k_-,k_+)  &=& e^2 k^2 \times 2  \left[  -\left( g_{\mu \mu'}  - \frac{k_\mu k_{\mu'}}{k^2} \right) -
  \frac{(k^+ \!-\! k^-)_\mu (k^+ \!-\! k^-)_{\mu'}}{k^2} \right]  .
\end{eqnarray}
If the Lorentz-indices $\mu,\mu'$ are space-like  (as it is   the case for  the on-shell photon, $\epsilon^\mu=(0,\vec \epsilon_\perp)$),
then  $ {\cal O}_{\mu \mu'}(k_-,k_+) $ gives a positive contribution.

Note that $(k^+ - k^-)_\mu k^\mu =0$ for on-shell $l^+$ and $l^-$.
Thus the spin-summed squared amplitude reads
\begin{eqnarray}
&& \overline{|{\cal A}|^2}(P \to \pi^+\pi^-l^+l^-) = \frac{1}{(k^2)^2} {\cal O}^{\mu\mu'}
 \left(  M \varepsilon_{\mu \nu\alpha\beta} k^\nu p_+^\alpha p_-^\beta + E_+ p_+^\mu + E_- p_-^\mu \right ) \nn \\
 && \qquad \qquad\qquad \qquad \mbox{}\times \left(  M \varepsilon_{\mu' \nu'\alpha'\beta'} k^{\nu'} p_+^{\alpha'} 
 p_-^{\beta'} + E_+ p_+^{\mu'} + E_- p_-^{\mu'} \right ) \nn
 \\
&& = \frac{2 e^2}{(k^2)^2} \biggl \{ |M|^2 \Bigl[ 
\varepsilon_{\mu \nu\alpha\beta} \,\varepsilon_{\mu' \nu' \alpha'\beta'} (k^2) 
(-g^{\mu\mu'})k^\nu p_+^\alpha p_-^\beta k^{\nu'} p_+^{\alpha'}p_-^{\beta'} \nn\\
&& \qquad\qquad\qquad\mbox{}
\mbox{}-\varepsilon_{\mu \nu\alpha\beta} \,\varepsilon_{\mu' \nu' \alpha'\beta'} (k_+^\mu \!-\!k_-^\mu) k^\nu p_+^\alpha p_-^\beta   (k_+^{\mu'} \!-\!k_-^{\mu'}) k^{\nu'} p_+^{\alpha'}p_-^{\beta'}\Bigr] \nn 
\\
&& \quad \mbox{}- \{ M E^*_+ + E_+ M^*\} 
\varepsilon_{\mu \nu\alpha\beta}   (k_+^\mu  \! - \! k_-^\mu )
k^\nu p_+^\alpha p_-^\beta  \,
 (k^+_{\mu'}  \! - \! k^-_{\mu'} ) p_+^{\mu'} \nn
 \\
 && \quad\mbox{}- \{ M E^*_- + E_- M^*\} 
\varepsilon_{\mu \nu\alpha\beta}   (k_+^\mu  \! - \! k_-^\mu )
k^\nu p_+^\alpha p_-^\beta  \,
 (k^+_{\mu'}  \! - \! k^-_{\mu'} ) p_-^{\mu'} \nn\\
&&\quad\mbox{}- |E_+|^2 [ k^2p_+^2 - (k\cdot p_+)^2 +((k_+-k_-)\cdot p_+)^2 ] \nn\\
&&\quad\mbox{}- |E_-|^2 [ k^2p_-^2 - (k\cdot p_-)^2 +((k_+-k_-)\cdot p_-)^2 ] \nn\\
&&\quad\mbox{}- (E_+^\star E_- + E_+ E_-^\star) [ k^2(p_+\cdot p_-) - (k\cdot p_+)(k\cdot p_-) \nn\\
&& \qquad\qquad\qquad\qquad\qquad\mbox{}+(k_+-k_-)\cdot p_+(k_+-k_-)\cdot p_- ]\biggr\}.
  \label{Asquared}
\end{eqnarray}
Inserting $p_- = P - p_+ - k$ into the antisymmetric products in \equa{Asquared} and making use of the total antisymmetric tensor, we get
\begin{eqnarray}
&& \overline{|{\cal A}|^2}(P \to \pi^+\pi^-l^+l^-) 
 =\frac{2 e^2}{(k^2)^2} \biggl \{ |M|^2 \Bigl[ 
\varepsilon_{\mu \nu\alpha\beta} \,\varepsilon_{\mu \nu' \alpha'\beta'} (k^2) 
(-g^{\mu\mu'}) k^\nu p_+^\alpha P^\beta k^{\nu'} p_+^{\alpha'}P^{\beta'} \nn\\
&& \qquad\qquad\qquad\mbox{}-\varepsilon_{\mu \nu\alpha\beta} \,\varepsilon_{\mu' \nu' \alpha'\beta'} (k_+^\mu \!-\!k_-^\mu) k^\nu p_+^\alpha P^\beta   (k_+^{\mu'} \!-\!k_-^{\mu'}) k^{\nu'} p_+^{\alpha'}P^{\beta'}\Bigr] \nn 
\\
&& \quad \mbox{}- {\rm Re}\{ M E^*_+\} 
\varepsilon_{\mu \nu\alpha\beta}   (k_+^\mu  \! - \! k_-^\mu )
k^\nu p_+^\alpha P^\beta  \,
 (k^+_{\mu'}  \! - \! k^-_{\mu'} ) (p_+^{\mu'} - p_-^{\mu'} +P^{\mu'} ) \nn
 \\
 && \quad \mbox{}+ {\rm Re}\{ M E^*_-\} 
\varepsilon_{\mu \nu\alpha\beta}   (k_+^\mu  \! - \! k_-^\mu )
k^\nu p_+^\alpha P^\beta  \,
 (k^+_{\mu'}  \! - \! k^-_{\mu'} ) (p_+^{\mu'} - p_-^{\mu'} -P^{\mu'})\nn\\
&&\quad\mbox{}- |E_+|^2 [ k^2p_+^2 - (k\cdot p_+)^2 +((k_+-k_-)\cdot p_+)^2 ] \nn\\
&&\quad\mbox{}- |E_-|^2 [ k^2p_-^2 - (k\cdot p_-)^2 +((k_+-k_-)\cdot p_-)^2 ] \nn\\
&&\quad\mbox{}- (E_+^\star E_- + E_+ E_-^\star) [ k^2(p_+\cdot p_-) - (k\cdot p_+)(k\cdot p_-) \nn\\
&& \qquad\qquad\qquad\qquad\qquad\mbox{}+(k_+-k_-)\cdot p_+(k_+-k_-)\cdot p_- ]\biggr\}.
\end{eqnarray}
In the rest frame of the $P$ meson, where $P^\mu  \equiv (\widetilde P)^\mu  = m_P \delta^{\mu 0}$, the
spin-summed squared amplitude reads
\begin{eqnarray}
&& \overline{|{\cal A}|^2}(P \to \pi^+\pi^-l^+l^-) 
 =\frac{2 e^2}{(k^2)^2} \biggl \{ |M|^2 m_P^2 \Bigl[ (k^2) 
\varepsilon^{ijk} \,\varepsilon^{i j' k'}  \widetilde k^j \widetilde p_+^k  \widetilde k^{j'} 
\widetilde p_+^{k'} \nn\\
&& \qquad\qquad\qquad\mbox{}-\varepsilon^{i j k} \,\varepsilon^{i' j' k'}
 (\widetilde k_+^i \!-\! \widetilde k_-^i) \widetilde k^j \widetilde p_+^k 
  (\widetilde k_+^{i'} \!-\! \widetilde k_-^{i'}) \widetilde k^{j'} \widetilde p_+^{k'} \Bigr] \nn 
\\
&& \quad \mbox{}+ {\rm Re}\{ M (E^*_+   +   E^*_-)\} m_P
\varepsilon^{i j k }   (\widetilde k_+^i  \! - \!  \widetilde k_-^i )
\widetilde k^j  \widetilde p_+^k     \,
 (\widetilde k^+_{\mu'} \! - \! \widetilde k^-_{\mu'} ) P^{\mu'}  \nn\\
&& \quad \mbox{}+ {\rm Re}\{ M (E^*_+   -   E^*_-)\} m_P
\varepsilon^{i j k }   (\widetilde k_+^i  \! - \!  \widetilde k_-^i )
\widetilde k^j  \widetilde p_+^k     \,
 (\widetilde k^+_{\mu'}  \! - \! \widetilde k^-_{\mu'} )( \widetilde
 p_+^{\mu'}-\widetilde p_-^{\mu'}) \nn\\
&&\quad\mbox{}- |E_+|^2 [ k^2p_+^2 - (k\cdot p_+)^2 +((k_+-k_-)\cdot p_+)^2 ] \nn\\
&&\quad\mbox{}- |E_-|^2 [ k^2p_-^2 - (k\cdot p_-)^2 +((k_+-k_-)\cdot p_-)^2 ] \nn\\
&&\quad\mbox{}- (E_+^\star E_- + E_+ E_-^\star) [ k^2(p_+\cdot p_-) - (k\cdot p_+)(k\cdot p_-) \nn\\
&& \qquad\qquad\qquad\qquad\qquad\mbox{}+(k_+-k_-)\cdot p_+(k_+-k_-)\cdot p_-
]\biggr\}\, ,
\end{eqnarray}
where we used $-g^{ii'} = \delta^{ii'}$ in the first term.

\bigskip
Since the scalar product over four-vectors is invariant in any frame, we can drop the 
tilde in the expressions
\begin{eqnarray*}
(\widetilde k^+_{\mu'}  \! - \! \widetilde k^-_{\mu'} )( \widetilde p_+^{\mu'}-\widetilde p_-^{\mu'}) 
&\equiv& (\widetilde k_+  \! - \! \widetilde k_- )\cdot ( \widetilde p_+-\widetilde p_-) 
= (k_+  \! - \! k_- )\cdot (  p_+-p_-)\,,\\
(\widetilde k^+_{\mu'}  \! - \! \widetilde k^-_{\mu'} ) \widetilde P^{\mu'} 
&\equiv& (\widetilde k_+  \! - \! \widetilde k_- )\cdot  \widetilde P 
= (k_+  \! - \! k_- )\cdot P\,.
\end{eqnarray*}
Formulated in terms of 3-vectors, the spin-summed squared amplitude reads then
\begin{eqnarray}
&& \overline{|{\cal A}|^2}(P \to \pi^+\pi^-l^+l^-)  \nn \\
&&= \frac{2 e^2}{(k^2)^2} \biggl \{ |M|^2 m_P^2 |\widetilde \kbf |^2 
\left[ ( k^2) 
 ( \pbf^+_\perp )^2
-\left( ( \kbf^+_\perp  -  \kbf^-_\perp) \cdot \hatz \wedge \pbf^+_\perp \right)^2 
    \right] \nn 
\\
&& \quad \mbox{}+ {\rm Re}\{ M (E^*_+   -   E^*_-)\} m_P |\widetilde \kbf |\,
 (\kbf^+_\perp \! - \! \kbf^-_\perp )\cdot
\hatz \wedge \pbf^+_\perp     \,
  ( k_+  \! - \! k_-) \cdot (p_+ \!-\! p_-) 
   \nn 
\\
&& \quad \mbox{}+ {\rm Re}\{ M (E^*_+   +   E^*_-)\} m_P^2 |\widetilde \kbf |\,
 (\kbf^+_\perp \! - \! \kbf^-_\perp )\cdot
\hatz \wedge \pbf^+_\perp     \,
  ( k_+ - k_-)\cdot p_\eta\nn\\
&&\quad\mbox{}- |E_+|^2 [ k^2p_+^2 - (k\cdot p_+)^2 +((k_+-k_-)\cdot p_+)^2 ] \nn\\
&&\quad\mbox{}- |E_-|^2 [ k^2p_-^2 - (k\cdot p_-)^2 +((k_+-k_-)\cdot p_-)^2 ] \nn\\
&&\quad\mbox{}- (E_+^\star E_- + E_+ E_-^\star) [ k^2(p_+\cdot p_-) - (k\cdot p_+)(k\cdot p_-) \nn\\
&& \qquad\qquad\qquad\qquad\qquad\mbox{}+(k_+-k_-)\cdot p_+(k_+-k_-)\cdot p_- ]\biggr\}. \nn
\end{eqnarray}
After (i)  the relations\,\equa{pess} and \equa{kess} are applied, 
(ii) the  scalar products 
$(k^+-k^-)\cdot(p^+-p^-)$ and
$(k^+ - k^-)\cdot P$ 
are simplified with the help of  \equa{ppppmkpmkm} and  \equa{pmmpmkpmkm},
respectively,
and (iii)  the relations\,\equa{kpp},  \equa{kpm} and  \equa{kmkppm} are inserted in the
electric terms,
the spin-summed squared amplitude reads
\begin{eqnarray}
&& \overline{|{\cal A}|^2}(P \to \pi^+\pi^-l^+l^-)  \nn \\
&& 
= \frac{2 e^2}{(k^2)^2} \Biggl \{ |M(s_{\pi\pi})|^2\, m_P^2\, |\widetilde \kbf |^2 \,
|\pbf_+^\star |^2
\sin^2\theta_\pi
\biggl[ ( k^2)  - 4 |\kbf_-^\diamond|^2  \sin^2\theta_l \sin^2\varphi
    \biggr] \nn 
\\
&& \mbox{}+ {\rm Re}\{ M (E^*_+  \! -\!   E^*_-)\} m_P \beta_{\pi} \beta_l |\widetilde \kbf |
|\pbf_+^\star| |\kbf_-^\diamond| \,\sin\theta_\pi\sin\theta_l\nn
\\
&& \quad \mbox{} \biggl ( - ( m_P^2\!-\!s_{\pi\pi}\! -\! s_{ll} ) 
\cos\theta_{\pi} \cos\theta_{l}  \sin\varphi 
+2\sqrt{s_{\pi\pi}}\sqrt{s_{ll}}\sin\theta_{\pi}\sin\theta_l \sin\varphi\cos\varphi\biggr) \nn
\\
&& \mbox{}+ {\rm Re}\{ M (E^*_+   +   E^*_-)\} m_P |\widetilde \kbf |
|\pbf_+^\star| |\kbf_-^\diamond|\beta_l\lambda^{1/2}(s_{ll},m_P^2,s_{\pi\pi} )
\sin\theta_\pi\sin\theta_l\cos\theta_l\sin\varphi \nn\\
&&\mbox{}+|E_+|^2 \biggl[ \frac{1}{16} \biggl[\lambda^{1/2}(s_{\pi\pi}, m_P^2,k^2) - 
\left(m_P^2-s_{\pi\pi}-k^2\right) \beta_\pi\cos\theta_\pi\biggr]^2 \cdot 
\left( 1 - \beta_l^2 \cos^2\theta_l \right) \nn \\
&&\quad\mbox{}+\fourth s_{\pi\pi}s_{ll}\beta^2_\pi \sin^2\theta_\pi 
  \left(1- \beta^2_l\sin^2\theta_l \cos^2 \varphi  \right) \nn \\
&&\quad\mbox{}-\fourth \biggl[\lambda^{1/2}(s_{\pi\pi}, m_P^2,k^2) - 
(m_P^2-s_{\pi\pi}-k^2) \beta_\pi \cos\theta_\pi \biggr] \sqrt{s_{\pi\pi}}\beta_\pi \sqrt{s_{ll}}\beta_l^2 \sin\theta_\pi  \sin\theta_l \cos\theta_l \cos\varphi \biggr]\nn \\
&&\mbox{}+ |E_-|^2 \biggl[ \frac{1}{16} \biggl[\lambda^{1/2}(s_{\pi\pi}, m_P^2,k^2) + 
\left(m_P^2-s_{\pi\pi}-k^2\right) \beta_\pi\cos\theta_\pi\biggr]^2 \cdot 
\left(1-\beta_l^2 \cos^2\theta_l \right) \nn \\
&&\quad\mbox{}+\fourth s_{\pi\pi}s_{ll}\beta^2_\pi \sin^2\theta_\pi
 \left(1- \beta^2_l\sin^2\theta_l \cos^2 \varphi \right) \nn \\
&&\quad\mbox{}+\fourth \biggl[\lambda^{1/2}(s_{\pi\pi}, m_P^2,k^2) + 
(m_P^2-s_{\pi\pi}-k^2) \beta_\pi \cos\theta_\pi \biggr] \sqrt{s_{\pi\pi}}\beta_\pi \sqrt{s_{ll}}\beta_l^2 \sin\theta_\pi  \sin\theta_l \cos\theta_l \cos\varphi \biggr]\nn \\
&&\mbox{}+ (E_+^\star E_- + E_+ E_-^\star) \biggl[ \frac{1}{16} \lambda(s_{\pi\pi}, m_\eta^2,k^2) \biggl(1-\beta_\pi^2 \cos^2\theta_\pi \biggr)\left( 1- \beta_l^2 \cos^2\theta_l \right)\nn \\
&&\quad\mbox{}-\fourth s_{\pi\pi}s_{ll}\beta^2_\pi \biggl( 1 - \beta^2_l \sin^2\theta_\pi \sin^2\theta_l \cos^2 \varphi  - \beta_l^2 \cos^2\theta_\pi \cos^2\theta_l\biggr)\nn \\
&&\quad\mbox{}-\fourth \sqrt{s_{\pi\pi}} \sqrt{s_{ll}} \beta_\pi^2 
\beta_l^2(m_P^2-s_{\pi\pi}-k^2)\sin\theta_\pi \cos\theta_\pi  \sin\theta_l \cos\theta_l
\cos\varphi \biggr]\Biggr\}. 
\label{Ampeefin}
\end{eqnarray}
We also used
\[
\cos\varphi_{p_+} (-\sin \varphi_{k_-}) - \sin \varphi_{p_+} (-\cos \varphi_{k_-} )
= \sin \left( \varphi_{p_+} - \varphi_{k_-}\right )  \equiv +\sin\varphi
\]
and
\begin{eqnarray}
 \hatz \cdot ( \pbf^+_\perp \wedge \kbf^-_\perp) 
 &=& (p^+_\perp)^x (k^-_\perp)^y - (p^+_\perp)^y (k^-_\perp)^x \nn \\
 &=& \underbrace{|\pbf_+| \sin\theta_\pi}_{|\pbf^+_\perp|} 
   \,\underbrace{|\kbf_-^\diamond|\sin\theta_l}_{|\kbf^-_\perp|}
   \, \underbrace{\left(   \cos \varphi_{\pi^+}
 \sin\varphi_{l^-} - \sin\varphi_{\pi^+} \cos\varphi_{l^-} \right)}_{\sin\left( \varphi_{l^-} -\varphi_{\pi^+} \right) = 
- \sin\varphi}  \nn \\
 & =&- |\pbf_+ | \sin\theta_\pi \, |\kbf_-^\diamond|\sin\theta_l\, \sin\varphi\,,
\end{eqnarray}
where $\varphi$ is the angle between 
$\pbf^+_\perp=-\pbf^-_\perp$  and
$\kbf^+_\perp= - \kbf^-_\perp$ 
in any frame (especially
also in the $\eta$ rest frame). Since $\pbf^+_\para$,  $\pbf^-_\para$,  $\kbf^-_\para$ and $\kbf^+_\para$ are parallel or antiparallel to each other and to the positive $z$-direction,
$\varphi$ is also the angle between the $l^+ l^-$ plane and the $\pi^+ \pi^-$ plane (see the text
below Eq.\,(1) of Ref.\cite{Gao:2002gq}). \\
Expressed in terms of the standard variables the squared matrix element
for the decay $P\to \pi^+ \pi^- l^+ l^-$ reads:
\begin{eqnarray}
&& \overline{|{\cal A}_{P \to \pi^+\pi^-l^+l^-} |}^2 (s_{\pi\pi},s_{ll},\theta_{\pi},\theta_{l},\varphi)
\nn \\
&& =  \frac{e^2 }{8 \,(k^2)^2}   \Biggl \{ 
 |M(s_{\pi\pi},s_{ll})|^2 \, 
\lambda(m_P^2,s_{\pi\pi},s_{ll}) \left[  1 - \beta_l^2  \sin^2\theta_{l} \sin^2\varphi
    \right] s_{\pi\pi}\,\beta_\pi^2 \sin^2\theta_{\pi}\nn 
\\
&& \mbox{}+ 4{{\rm Re}\!\left\{\! M(s_{\pi\pi},s_{ll}) (E^*_+   -   E^*_-)\! \right\}} \,
\lambda^{1/2}(m_P^2,s_{\pi\pi},s_{ll})
\beta_l^2\,\beta_{\pi}^2 \sqrt{s_{\pi\pi}}\sqrt{s_{ll}}\nn\\
&& \quad\ \mbox{}\times \left(-\half
(m_P^2\!-\!s_{\pi\pi}\!-\!s_{ll})\sin\theta_{\pi}\cos\theta_{\pi} \cos\theta_{l} \sin\theta_{l} \sin\varphi +
  \sqrt{s_{\pi\pi} s_{ll}}\sin^2\theta_{\pi}\sin^2\theta_l \sin\varphi\cos\varphi\right) \nn
\\
&& \mbox{}+2{\rm Re}\{ M(s_{\pi\pi},s_{ll}) (E^*_+   +   E^*_-)\}
\sqrt{s_{\pi\pi} s_{ll}}\beta_l^2 \beta_{\pi}\lambda(m_P^2,s_{\pi\pi},s_{ll})\sin\theta_\pi\sin\theta_l\cos\theta_l\sin\varphi\nn\\
&&\mbox{}+\sum_{\pm}  |E_\pm|^2 
\biggl\{  \left [  \lambda^{1/2}(s_{\pi\pi}, m_P^2,k^2) \mp 
\left(m_P^2-s_{\pi\pi}-k^2\right) \beta_\pi\cos\theta_\pi \right]^2 \cdot
 \left(1-\beta_l^2 \cos^2\theta_l \right) \nn \\
&&\quad\mbox{}+4 s_{\pi\pi}s_{ll}\beta^2_\pi \sin^2\theta_\pi 
 \left(1- \beta^2_l\sin^2\theta_l \cos^2 \varphi \right) \nn \\
&&\quad\mbox{}\mp 4 \left [\lambda^{1/2}(s_{\pi\pi}, m_P\right] \sqrt{s_{\pi\pi}}\beta_\pi \sqrt{s_{ll}}\beta_l^2 \sin\theta_\pi  \sin\theta_l \cos\theta_l \cos\varphi \biggr\}\nn \\
&&\mbox{}+ 2{\rm Re}[E_+^\ast E_- ] \biggl [ 
 \lambda(s_{\pi\pi}, m_P^2,k^2) \left(1-\beta_\pi^2 \cos^2\theta_\pi \right)\left( 1- \beta_l^2 \cos^2\theta_l \right)\nn \\
&&\quad\mbox{}-4 s_{\pi\pi}s_{ll}\beta^2_\pi \left( 1 - \beta^2_l \sin^2\theta_\pi \sin^2\theta_l \cos^2 \varphi  - \beta_l^2 \cos^2\theta_\pi \cos^2\theta_l\right)\nn \\
&&\quad\mbox{}-4 \sqrt{s_{\pi\pi}} \sqrt{s_{ll}} \beta_\pi^2\, \beta_l^2
\left(m_P^2-s_{\pi\pi}-k^2\right)\sin\theta_\pi \cos\theta_\pi  \sin\theta_l \cos\theta_l 
\cos\varphi\biggr]\Biggr\}. \nn\\
\label{Decay_beta}
\end{eqnarray}

\subsection{Decay rate}

We will now insert the expression of the squared matrix element \equa{Decay_beta}
into the formula of the decay rate 
\equa{4dr}.
\subsubsection*{Magnetic term}
We will start with the magnetic term. As we will see in the next Chapter, it is
the leading contribution. The final expression (with $k^2\equiv s_{ll}$) is:
\begin{eqnarray}
{\rm d}\Gamma(\mbox{\small $|M|^2$})
&=& \frac{2 e^2}{(k^2)^2}|M|^2\frac{1}{m_P^2\sqrt{s_{ll}}}\frac{1}{2^{12}\pi^{6}}
   m_P^2 |\kbf_-^\diamond |\,|\pbf_+^\ast|^3|\widetilde \kbf|^3  
   k^2 \left[  1 - \beta_l^2  \sin^2\theta_{l} \sin^2\varphi
    \right] \sin^2\!\theta_{\pi} \, (s_{\pi\pi})^{-\frac{1}{2}}\nn \\
&& \mbox{}\qquad \qquad \qquad \qquad \qquad \qquad \qquad \qquad  \qquad
\qquad  \quad \times {\rm d}s_{\pi\pi} 
\,{\rm d}\!\cos\theta_l\, {\rm d}\!\cos\theta_{\pi} \,{\rm d}\varphi\, {\rm d}k^2 \nn \\
&=& \frac{e^2}{k^2}|M|^2\frac{1}{m_P^3 2^{16}\pi^{5}}\beta_l
\left(1-\frac{1}{3} \beta_l^2\right)  s_{\pi\pi}  \beta_{\pi}^3 \lambda^{\frac{3}{2}}(m_P^2,s_{\pi\pi},k^2) \sin^2\!\theta_{\pi} \,{\rm d}s_{\pi\pi}
{\rm  d}\!\cos\theta_{\pi}\,{\rm d}k^2\,\nn \\
&=&\frac{e^2}{s_{ll}}|M|^2\frac{1}{m_P^3 \, 3^2\cdot 2^{13}\,\pi^{5}} s_{\pi\pi}  \beta_{\pi}^3 \lambda^{\frac{3}{2}}(m_P^2,s_{\pi\pi},s_{ll})  \beta_l \frac{ 3-\beta_l^2}{2}\,{\rm d}s_{\pi\pi} \, {\rm d}s_{ll} .
\label{GaeeMsq}
\end{eqnarray}
This expression exactly agrees
with Eq.\,(4) of Ref.\,\cite{Picciotto:1993aa}.

\subsubsection{Mixed term}

With the same definitions as before we can calculate the decay rate of the mixed term,
weighted by ${\rm sign}(\sin\varphi\cos\varphi)$. Note that only the second
part of the mixed term ($\propto {\rm Re}[M(E^*_+ -E^*_-)]$) will give a contribution. The other two terms will be equal to zero for the following reasons: (i) they vanish if we integrate over the angle $\theta_{\pi}$ (and also over  $\theta_l$ in case of  the last term), because of the 
$\cos\theta_\pi \sin\theta_\pi$ dependence; (ii) the main reason -- valid also for the
case of  generalized form factors -- is  the integration over the angle
$\varphi$: with the weight ${\rm sign}(\sin\varphi\cos\varphi)$ this integration reduces all of the
above
mentioned  terms to zero, unless there is the structure $\sin\varphi\cos\varphi$.\\
Note that this is actually  the CP-violating term because
the remaining part is proportional to ${\rm Re}[M(E^*_+ -E^*_-)]$, which is in
fact CP-violating. The latter term would vanish, as in the $P \to \pi^+
\pi^- \gamma$-decay, if the norm
$|\sin\varphi\cos\varphi|$ were not taken. This term can be measured by
analyzing the forward-backward asymmetry of the angle $\phi$. If $\phi$ is not
zero in average, there does exist a term proportional to ${\rm
  Re}[M(E^*_+ -E^*_-)]$ and so CP-violation can be measured. 

\bigskip
The  decay rate for the mixed term is given as
\begin{eqnarray}
&&{\rm d}\Gamma( \mbox{\small ${\rm Re}((E_+-E_-)M^*){\rm sign}(\sin\varphi\cos\varphi)$} )\nn\\
&&
=\frac{2^4 e^2}{s_{ll}^2}\frac{m_P}{m_P^2\sqrt{s_{ll}}}\frac{1}{2^{12}\pi^{6}}{\rm Re}[M(E^*_+ -E^*_-)]|\pbf_+^\ast|^3 \, |\widetilde \kbf|^2\, |\kbf_-^\diamond|^3 \nn  \\
&&\quad\mbox{}\times \sin^2\!\theta_{\pi}\, \sin^2\!\theta_{l}\, |\sin\varphi \cos\varphi|\, (s_{\pi\pi})^{-\frac{1}{2}}
\,{\rm d}s_{\pi\pi} \,{\rm d}\!\cos\theta_l\, {\rm d}\!\cos\theta_{\pi}\, {\rm d}\varphi\, 
{\rm d}s_{ll}\, \nn  \\
&&=\frac{e^2\, {\rm Re}[M(E^*_+ \!-\!E^*_-)]} 
{3\cdot 2^{13}\,\pi^{6} \,m_P^3 \, s_{ll}} \lambda(m_P^2,s_{\pi\pi},s_{ll}) s_{\pi\pi}\beta_{\pi}^3 \,\sin^2 \theta_{\pi}\,\beta_l^3 
{\rm d}s_{\pi\pi}\,  {\rm d}\!\cos\theta_{\pi} \,{\rm d}s_{ll} \nn \\
&&\approx \frac{e^2\, {\rm Re}[M(E^*_+ \!-\!E^*_-)]} 
{3\cdot 2^{13}\,\pi^{6} \,m_P^3 } \frac{ \lambda(m_P^2,s_{\pi\pi},s_{ll})}{s_{ll}} \, s_{\pi\pi}\beta_{\pi}^3 \,\sin^2\! \theta_{\pi}  \,
{\rm d}s_{\pi\pi}\,  {\rm d}\!\cos\theta_{\pi}\,{\rm d}s_{ll}\, .
\label{GaeeMix}
\end{eqnarray}
In the last step the approximation $\beta_l \approx 1$ 
was inserted. This can be justified for the case that the leptons are electrons.\\
The CP-violating observable $A_{\rm CP}$ which can be measured by the experiments is given
by the mixed term normalized to the total decay width (which is given to
leading order by the integral of Eq.\,\equa{GaeeMsq}):
\begin{eqnarray}
A_{\rm CP} &=& \frac{\int_0^{2\pi} {\displaystyle \frac{{\rm d} \Gamma(\eta\to \pi^+\pi^- e^+
    e^- )}{{\rm d \phi}}} {\rm d} \phi {\rm sign}(\sin\phi \cos\phi)}
{\int_0^{2\pi} {\displaystyle \frac{{\rm d} \Gamma(\eta\to \pi^+\pi^- e^+
    e^- )}{{\rm d \phi}}} {\rm d} \phi}\nn \\
&=& \frac{e^2\, } 
{3\cdot 2^{13}\,\pi^{6} \,m_P^3 \,\Gamma(\eta\to \pi^+\pi^-e^+e^-)} \nn
\\
&&\times\int \frac{\lambda(m_P^2,s_{\pi\pi},s_{ll})}{ s_{ll}} s_{\pi\pi}\beta_{\pi}^3 \,\sin^2 \theta_{\pi}\,\beta_l^3 {\rm Re}[M(E^*_+ \!-\!E^*_-)]
{\rm d}s_{\pi\pi}\,  {\rm d}\!\cos\theta_{\pi} \,{\rm d}s_{ll} \, .\nn \\
\label{ACP}
\end{eqnarray}

Note that our result for $A_{\rm CP}$ agrees  with Eq.\,(3) of Ref.\,\cite{Gao:2002gq}
when the mentioned approximation $\beta_l \approx 1$ is inserted.
\subsubsection{Electric terms}

The decay rates of the electric terms can be calculated in the same way as the ones of the magnetic and mixed terms before. 

In the following, we will  separately construct the decay rates of 
the squared electric terms which are proportional to
 $|E_\pm|^2$
 and the decay rate of the mixed electric terms  proportional to $(E_+^\ast E_- + E_+E_-^\ast) = 2 {\rm Re}[E_+^\ast E_-]$: 

\begin{eqnarray}
&&{\rm d}\Gamma( |E_\pm|^2 )\nn\\
&&=\frac{e^2
|E_\pm|^2 }{8 s_{ll}^2}
\Biggl\{  \biggl[\lambda^{1/2}(s_{\pi\pi}, m_P^2,k^2) \mp 
\left(m_P^2-s_{\pi\pi}-s_{ll}\right) \beta_\pi\cos\theta_\pi\biggr]^2 \cdot (1 -\beta_l^2 \cos^2\theta_l) \nn \\
&&\quad\quad\quad\quad\quad\mbox{}+4 s_{\pi\pi}s_{ll}\beta^2_\pi \sin^2\theta_\pi \biggl(1 -  \beta_l^2 \sin^2\theta_l \cos^2 \varphi  \biggr)+O( \sin\theta_e \cos\theta_l
\cos\varphi) \Biggr\}\nn \\
&&\qquad\quad\quad\mbox{}\times\frac{1}{m_P^3 \cdot 2^{15} \cdot \pi^6}
\beta_{\pi}\beta_l \lambda^{1/2}(s_{\pi\pi}, m_P^2,s_{ll})\,{\rm d}s_{\pi\pi} \,{\rm d}\!\cos\theta_l\, {\rm d}\!\cos\theta_{\pi}\, {\rm d}\varphi\, 
{\rm d}s_{ll}\, \nn  \\
&&
= \frac{e^2 |E_\pm|^2 }{s_{ll}^2 \cdot m_P^3 \cdot3 \cdot  2^{15}
\cdot \pi^5 } \cdot  \frac{ 3 - \beta_l^2 }{2}
 \nn \\
&&\qquad \mbox{}\times
\left\{  \left [\lambda^{1/2}(s_{\pi\pi}, m_P^2,s_{ll}) \mp 
\left(m_P^2\!-\!s_{\pi\pi}\!-\!s_{ll}\right) \beta_\pi\cos\theta_\pi\right]^2+4 s_{\pi\pi}s_{ll}\beta^2_\pi \sin^2\theta_\pi \right\} \nn  \\
&&\quad\quad\mbox{}\times\beta_{\pi}\beta_l
\lambda^{1/2}(s_{\pi\pi}, m_P^2,s_{ll})\,{\rm d}s_{\pi\pi} \, {\rm d}\!\cos\theta_{\pi}\, 
{\rm d}s_{ll}\,. 
\end{eqnarray}
The mixed term has a slightly different form compared to the previous expression and reads:
\begin{eqnarray}
&&{\rm d}\Gamma( (E_+^\star E_- + E_+E_-^\star ))\nn\\
&&=\frac{e^2 (E_+^\star E_- + E_+E_-^\star) }{8 s_{ll}^2}
\Biggl\{  \lambda(s_{\pi\pi},
m_P^2,s_{ll})  \biggl(1-\beta_\pi^2 \cos^2\theta_\pi \biggr) \cdot 
 (1 - \beta_l^2 \cos^2\theta_l) \nn \\
&&\quad\mbox{}
-4 s_{\pi\pi}s_{ll}\beta^2_\pi 
 \biggl(1- \beta_l^2 \cos^2 \theta_\pi \cos^2\theta_l - \beta_l^2 \sin^2\theta_\pi  \sin^2\theta_l \cos^2\varphi \biggr)+O( \sin\theta_l \cos\theta_l \cos\varphi) \Biggr\}\nn \\
&&\quad\quad\mbox{}\times \frac{1}{m_P^3 \cdot 2^{15} \cdot \pi^6}\beta_{\pi} \beta_l
\lambda^{1/2}(s_{\pi\pi}, m_P^2,s_{ll})\,{\rm d}s_{\pi\pi} \,{\rm d}\!\cos\theta_l\, {\rm d}\!\cos\theta_{\pi}\, {\rm d}\varphi\, 
{\rm d}s_{ll}\, \nn  \\
&&
=\frac{e^2 {\rm Re}[E_+^\ast E_-] } {(k^2)^2 \cdot m_P^3 \cdot 3 \cdot
  2^{14} \cdot \pi^5} \,\frac{3 -\beta_l^2}{2} \,
 \biggl\{  \lambda(s_{\pi\pi}, m_P^2,s_{ll})
\biggl(1\!-\!\beta_\pi^2\cos^2\theta_\pi\biggr)-  4 s_{\pi\pi}s_{ll}\beta^2_\pi
\biggr\} \nn \\
&& \qquad\mbox{}\times \beta_{\pi}\beta_l
 \lambda^{1/2}(s_{\pi\pi}, m_P^2,s_{ll})\,{\rm d}s_{\pi\pi} \, {\rm d}\!\cos\theta_{\pi}\, 
{\rm d}s_{ll}\,. \nn 
\end{eqnarray}

\subsection{The form factors for the decay  $\eta\to\pi^+\pi^-l^+l^-$}

Here we discuss the magnetic and electric form factors $M$ and $E$ for the
decay  $P\to\pi^+\pi^- l^+ l^-$ as defined in Ref.\,\cite{Gao:2002gq} and as used in the
last Section. The form factors are modeled according to the one of the $P \to
\pi^+ \pi^- \gamma$-decay, but as mentioned before, they differ in the powers of
masses.\\
We start again with the magnetic form factor $M(s_{\pi\pi},
s_{ll})$. This is again:
\begin{eqnarray}
M(s_{\pi\pi}, s_{ll})= {\cal M} \times VMD(s_{\pi\pi},s_{ll})
\end{eqnarray}
where $VMD$ is the pertinent vector meson dominance factor and ${\cal M}$ is
given by:
\begin{equation}
{\cal M}=\begin{cases}
         {\displaystyle \frac{e}{8\pi^2 f_{\pi}^3}} & \mbox{if $P=\pi^0$};\\
       {\displaystyle \frac{e}{8\pi^2 f_{\pi}^3}\frac{1}{\sqrt{3}}} \left( \frac{f_\pi}{f_8} \cos\theta_{mix} -2\sqrt{2} \frac{f_\pi}{f_0} \sin\theta_{mix} \right)& \mbox{if $P=\eta$};\\
       {\displaystyle \frac{e}{8\pi^2 f_{\pi}^3} \frac{1}{\sqrt{3}}} \left( \frac{f_\pi}{f_8} \sin\theta_{mix} +2\sqrt{2} \frac{f_\pi}{f_0} \cos\theta_{mix} \right)& \mbox{if $P=\eta'$}
\end{cases}
\end{equation}
with the pion decay constant $f_\pi \approx 92.4\,{\rm MeV}$, 
the octet pseudoscalar decay constant $f_8\approx 1.3 f_\pi$, the singlet pseudoscalar decay constant 
$f_0 \approx 1.04 f_\pi$ and the $\eta$--$\eta'$ mixing angle
$\theta_{\rm mix}\approx -20^\circ$, see Ref.\, \cite{Holstein:2001bt}.\\
The VMD form factor again contains contributions from the
terms $\mathcal{L}_{PPPA}$, $\mathcal{L}_{PVV}$  and $\mathcal{L}_{PPPV}$ as
it was already shown in figure \ref{fig:VMDppg1}, figure \ref{fig:VMDppg2} and figure \ref{fig:VMDppg3}.
Especially  $VMD(s_{\pi\pi},s_{ll})$ differs from  $VMD(s_{\pi\pi})$ of
Ref.\,\cite{Gao:2002gq} by an additional form factor related to this part of the off-shell photon (decaying
into the $l+ l-$ pair) that according to Ref.\,\cite{Fujiwara:1984mp} 
does not directly couple 
to the $P\pi^+\pi^-$ complex, see Ref.\,\cite{Picciotto:1993aa}: 
\begin{equation}
  F_{\gamma^*}(s_{ll}) = \frac{m_V^2}{m_V^2-s_{ll}}\, .
\end{equation}
Therefore the vector meson dominance factor has the following form:
\begin{eqnarray}
VMD_1(s_{\pi\pi}, s_{ll}) &=& 1-\frac{3}{4}(c_1-c_2+c_3) + \frac{3}{4}(c_1-c_2-c_3)
 \frac{m_V^2}{m_V^2-s_{ll}-i m_V \Gamma(s_{ll})}\nn \\
&& + \frac{3}{2} c_3 \frac{m_V^2}{m_V^2-s_{ll}-i m_V \Gamma(s_{ll})} \, \, \frac{m_V^2}{m_V^2-s_{\pi\pi}-i m_V \Gamma(s_{\pi\pi})}.
\end{eqnarray}
By adjusting the values of the $c_i$-parameters, we can switch between the various VMD
models. Note that for the hidden gauge case the $\mathcal{L}_{PPPV}$ term vanishes.

The electric form factors $E_{\pm}$ which contribute via the combination ${\rm Re}[M (E_+^* - E_-^*)]$ to the
squared CP-breaking amplitude are model-dependent. 
\begin{itemize}
\item[0)]
To leading order,  $E_\pm$ 
can be put
to zero. 
\end{itemize}
In Refs.\,\cite{Geng:2002ua,Gao:2002gq} two models for $E_\pm$ can be found:
\begin{itemize}
\item[1)] The first model consists of induced Bremsstrahlung of an $\pi^+\pi^-$ 
intermediate state which violates  CP symmetry (see Eqs.(7)--(14) of Ref.\,\cite{Gao:2002gq}):
\begin{eqnarray}
   E_{\pm}(s_{\pi\pi},s_{ll},\theta_\pi) 
   &=& \pm\frac {2 \,e\, g_{P \pi^+ \pi^-}} {k^2+ 2 k\cdot p_\pm } \, \nn  \\
                  &=& \pm\frac {4 \,e\, g_{P \pi^+ \pi^-}} {m_P^2  \!+\!q^2\! -\! s_{\pi\pi} 
                  \mp \beta_\pi \lambda^{1/2}(m_P^2,s_{\pi\pi},q^2)\cos\theta_\pi} .
\end{eqnarray}
As argued in the previous Section, this model will not play a role in our discussions.
\item[2)]
The second model of Refs.\,\cite{Geng:2002ua,Gao:2002gq} describes the CP violating 
$P\to\pi^+\pi^- l^+ l^-$ decay to a short-range $E_1$ operator, which was
constructed for the $P\to \pi^+ \pi^-\gamma$-decay. According to Eqs.(15--17)
of Ref.\,\cite{Gao:2002gq} the electric form factors $E_\pm$ have the following parameterization:
\begin{eqnarray}
  E_{\pm}(s_{\pi\pi},s_{ll},\theta_\pi) 
   &=& \pm \frac {e\, F(s_{\pi\pi} )\, G}{m_P^3}  \left(k^2 + 2 k \cdot p_{\mp}\right) \nn \\
                     &=& \frac {e\, F(s_{\pi\pi} )\, G}{2 m_P^3} 
                      \left( \pm\left(m_P^2  +s_{ll} -s_{\pi\pi}\right) + \ \beta_\pi \lambda^{1/2}(m_P^2,s_{\pi\pi},s_{ll}) \cos\theta_\pi \right)\nn\\
\end{eqnarray}
 with the form factor 
 $F(s_{\pi\pi})\sim F(0) \approx 0.19$ and  $G$ a  free model coefficient of order
  $G \leq {\cal O}(1)$.

Note that in this case 
\begin{eqnarray}
  E_+(s_{\pi\pi}, s_{ll}, \theta_{\pi}) -  E_-(s_{\pi\pi}, s_{ll}, \theta_{\pi})
 &=& \frac{e F(s_{\pi\pi}) G}{ m_P^3} \left( m_P^2 \!+\!s_{ll}\!-\!s_{\pi\pi}\right) \nn \\
 = E_+(s_{\pi\pi}, s_{ll}) -  E_-(s_{\pi\pi}, s_{ll}) &&
\end{eqnarray}
which is in fact independent of $\theta_{\pi}$,
while
\begin{eqnarray}
&& |E_\pm (s_{\pi\pi},s_{ll},\theta_{\pi^+})|^2 \nn\\
&& = \frac{e^2 F(s_{\pi\pi})^2 G^2}{4 m_P^6 s_{\pi\pi}^2}\biggl\{
 \left(m_P^2\!+\!s_{ll}\!-\!s_{\pi\pi} \right)^2 s_{\pi\pi}^2
 + \beta_\pi \lambda (m_{P}^2,s_{\pi\pi},s_{ll}) \cos^2\theta_{\pi}\nn \\
 && \mbox \qquad\qquad\qquad\quad\mp 2 s_{\pi\pi}\left(m_{P}^2\!+\!s_{ll}\!-\!s_{\pi\pi}\right) \beta_\pi \lambda^{1/2}(m_P^2,s_{\pi\pi},s_{ll} )\cos\theta_{\pi} \biggr\} \nn\\
 &&\equiv |E (s_{\pi\pi},s_{ll},\cos^2\theta_{\pi})|^2 \pm \Delta E^2(s_{\pi\pi},s_{ll}) \cos\theta_{\pi}
\end{eqnarray}
contains even and odd powers in $\cos\theta_{\pi}$.

\end{itemize}
In general, the difference between $E_+(s_{\pi\pi},s_{ll},\theta_{\pi})$ and 
$E_-(s_{\pi\pi},s_{ll},\theta_{\pi})$ is a function of only even powers of $\cos\theta_{\pi}$, {\it i.e.}
\begin{equation}
E_+(s_{\pi\pi},s_{ll},\theta_{\pi})- 
E_-(s_{\pi\pi},s_{ll},\theta_{\pi})
\equiv
\Delta E_{\pm}(s_{\pi\pi},s_{ll},\cos^2\theta_{\pi})\,, 
\end{equation}
whereas the squared moduli of $E_+$ and $E_-$ have the following dependencies:
\begin{eqnarray}
 |E_+(s_{\pi\pi},s_{ll},\theta_{\pi})|^2 &\equiv& |E (s_{\pi\pi},s_{ll},\cos^2\theta_{\pi})|^2 -
  \Delta E^2(s_{\pi\pi},s_{ll},\cos^2\theta_{\pi}) \cos\theta_{\pi} \,,\nn \\
   |E_-(s_{\pi\pi},s_{ll},\theta_{\pi})|^2 &\equiv& |E (s_{\pi\pi},s_{ll},\cos^2\theta_{\pi})|^2 +
  \Delta E^2(s_{\pi\pi},s_{ll},\cos^2\theta_{\pi}) \cos\theta_{\pi}\,. \nn \\
\end{eqnarray}

\chapter{Results}
In this Chapter we will state the results for the branching ratios of pseudoscalar
mesons $P = \pi^0, \eta, \eta' $ for all discussed modes. The calculations
have to be done numerically. In practice, we applied the Gaussian routine of
the CERNLIB library \cite{CERNLIB} to perform the integration over the squared invariant masses of the
outgoing particles. The decays into three particles, namely $P\to l^+ l^-
\gamma$ and $P \to \pi^+ \pi^- \gamma$, can be handled easily, because one has
to integrate over just one variable. In comparison, the decays into four
particles, $P \to l^+l^-l^+l^-$ and $P \to \pi^+ \pi^- l^+ l^-$,
are more complicated, since one has to integrate over two variables which
depend on each other.\\
We compare the results for the different vector meson dominance models (the hidden
gauge and the new modified one) with the  experimental data
and other already published theoretical calculations.\\
The listed errors can be traced back to the different vector meson masses
which occur in the VMD factor. As pointed out in the second Chapter, see Eq. \equa{KSFR}, the coupling constant $g$ is
related to the vector meson mass. So according to the different fits given in
\cite{Benayoun:2009im} various vector meson masses are used in our
calculations to take this into account. For the hidden gauge model the vector meson mass varies from
$m_V=760\,{\rm MeV}$ to $m_V=782\,{\rm MeV}$. For the modified model we use a vector
meson masses from $m_V=760\,{\rm MeV}$ to $m_V=791\, {\rm MeV}$, because the relevant fits generate very high coupling constants.\\
As for the modified model,
there exist also a small contribution to the total uncertainty by the
ambiguity of the two fits of Ref \cite{Benayoun:2009im}. This contribution will be in range
of a couple of per cent of the total errors, so that we
will not notice them in the most cases. Because we normalize the decays $P\to
l^+l^-\gamma$ and $P\to l^+l^-l^+l^-$ to the two-photon decay, there appear no
errors from the mixing angle of $\eta_0$ and $\eta_8$ and the coupling
constants $f_0$ and $f_8$. They are indeed present when the absolute decay rates for
the $P\to \pi^+ \pi^- \gamma$- and $\pi^+\pi^-l^+l^-$-decays are given.

\section{$P \to l^+ l^- \gamma$}
We start with the discussion of the decay of a pseudoscalar meson into
a lepton-antilepton pair and a photon. We show the dependence of the branching ratio expressed relative to the decay into two photons on the invariant masses of the leptons and give the branching ratios expressed relative to the decay into two photons:\\
\begin{equation}
{\rm BR}^{\rm rel} \left( P\to l^+l^-\gamma^{(\star)} \right)=\frac{{\rm BR}(P\to l^+l^-\gamma^{(\star)})}{{\rm BR}(P\to\gamma\gamma)} .
 \label{BRrel}
\end{equation}

\subsection*{$\pi^0\to e^+ e^- \gamma$}
The dependence for the relative branching ratio of $\pi^0\to e^+e^-\gamma$ on the invariant mass of the electrons, $\sqrt{s_{ee}}$, is shown in Figure
\ref{fig:pieeg}:

\begin{center}
\begin{figure}[!hbt]
\begin{center}
    \psfrag{x}[tc]{\tiny $\sqrt{s_{ee}}$ \hspace{0.2cm} [MeV]}
    \psfrag{y}[c]{\tiny $\partial {\rm BR}^{\rm rel} (\pi^0\to e^+e^-\gamma)  /\partial{\sqrt{s_{ee}/m_\pi}}$}
    \includegraphics[width=8cm]{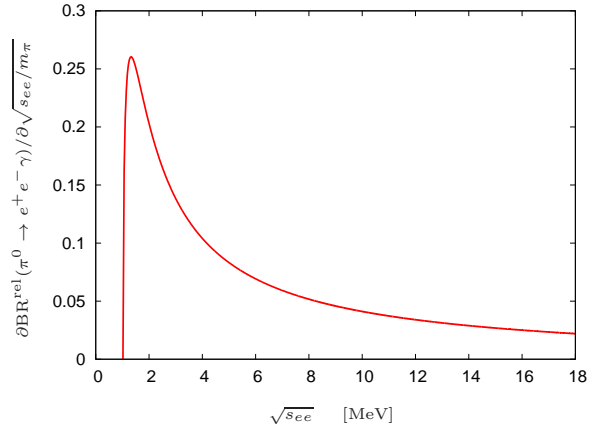}
\caption{Dependency of the differential relative branching ratio (normalized to BR($\pi^0\to\gamma\gamma$), see Eq. \equa{BRrel}) of the decay $\pi^0\to e^+e^-\gamma$ on
the change of the invariant mass of the electron pairs.\label{fig:pieeg}}

\end{center}
\end{figure}
\end{center}

The strong peak for small masses is typical for branching ratios plotted
against the invariant mass of electrons, $\sqrt{s_{ee}}$, as we will see in
following sections. The errors are smaller than the width of the line\footnote[1]{Here and in the following calculations based on two on/off-shell photons the differential branching ratios are calculated as ratios to the total 2-photon decay rate, such that the coefficients of the $f_0$, $f_8$, and the mixing angle $\theta_{mix}$ drop out. The plotted values are therefore dimensionless.}. Also
the different models, the one without VMD, the hidden gauge, and the modified VMD
model all give the same graph.\\
In Table \ref{tab:pillg} the results for the branching ratios of the decay $\pi^0 \to e^+ e^- \gamma$
are listed for the cases (i) without a VMD
factor, (ii) with the hidden gauge model, and (iii) with the modified hidden gauge model. The errors are smaller than the number of decimal places, so we do not show them.\\
\begin{table}[!hbt]
\begin{center}
\renewcommand{\arraystretch}{1.3}
\begin{tabular}{l|c|c|c}
& without VMD & hidden gauge & modified VMD\\ \hline\hline
${\rm BR}^{\rm rel}$($\pi^0\to e^+e^-\gamma$)($10^{-2}$) &$1.185$&$1.188$&$1.187$
\end{tabular}
  \caption{Relative branching ratios (normalized to BR($\pi^0\to\gamma\gamma$), see Eq. \equa{BRrel}) of the decay $\pi^0\to e^+e^-\gamma$ calculated
    without VMD, with the hidden gauge model, and the modified VMD model. \label{tab:pillg}}
\end{center}
\end{table}\\
There is a small difference between the values with and without a VMD
factor, but the difference between the two VMD models is hardly noticeable.\\
In Table \ref{tab:pillga} we list the corresponding  values calculated by other groups and compare
them to the experimental data.\\
\begin{table}[!hbt]
\begin{center}
\renewcommand{\arraystretch}{1.3}
\begin{tabular}{l|c|c|c|c}
& \cite{Kroll:1955zu} & \cite{Miyazaki:1974qi}  & \cite{Lih:2009np}&  exp. data \cite{Amsler:2008zzb}\\ \hline\hline
${\rm BR}^{\rm rel}$($\pi^0\to e^+e^-\gamma$)($10^{-2}$) &$1.18$ &$1.18$ &$1.18$&$1.188\pm0.034$
\end{tabular}
  \caption{Published theoretical values and experimental data for the relative branching ratios
    of the decay $\pi^0\to e^+e^-\gamma$. \label{tab:pillga}}
\end{center}
 \end{table} 

A comparison of Table \ref{tab:pillg} and \ref{tab:pillga} confirms that our
results agree with the published ones and are consistent with the current data
\cite{Amsler:2008zzb}.

\subsection*{$\eta \to l^+ l^- \gamma$}

\begin{figure}[!hbt]
  \begin{minipage}[b]{8 cm}
    \psfrag{x}[tc]{\tiny $\sqrt{s_{ee}}$ \hspace{0.2cm} [MeV]}
    \psfrag{y}[c]{\tiny $\partial {\rm BR}^{\rm rel} (\eta \to e^+e^- \gamma)  /\partial{\sqrt{s_{ee}/m_{\eta}}}$}
    \includegraphics[width=8 cm]{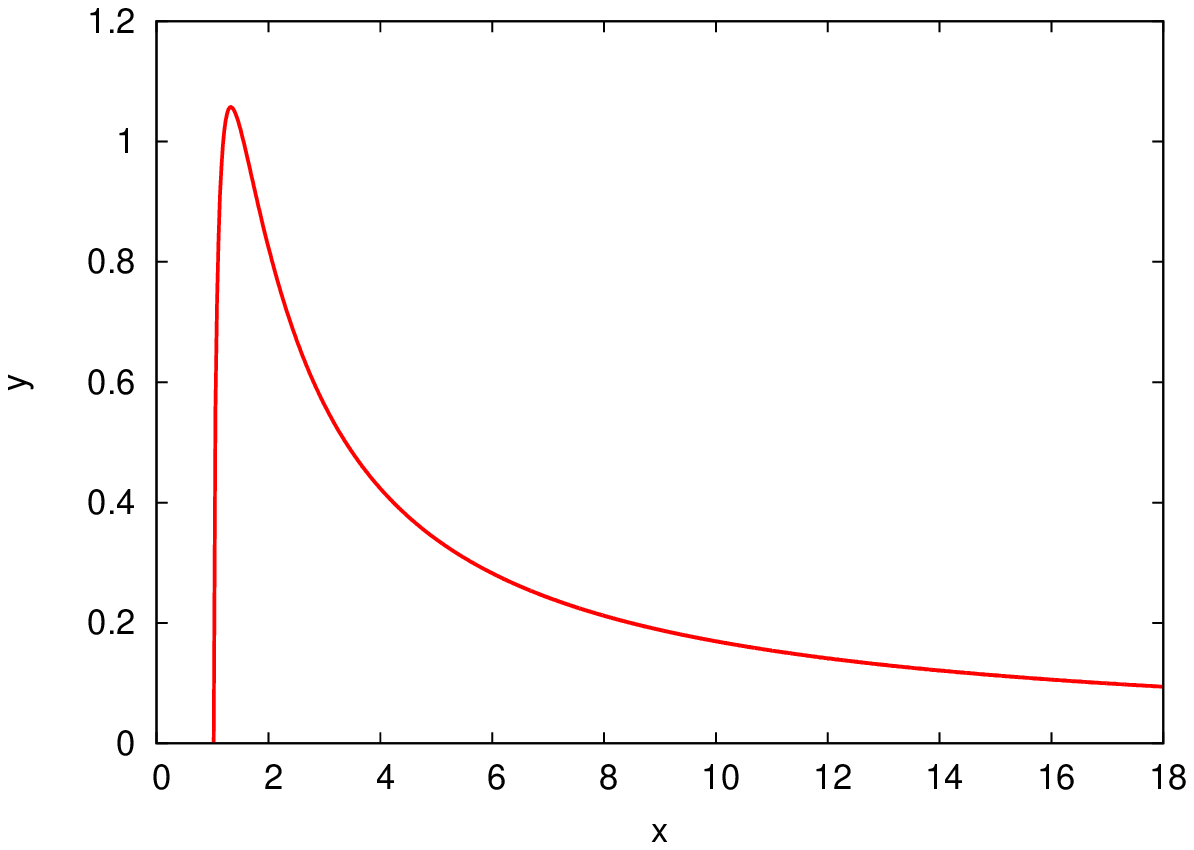}
  \end{minipage}
  \begin{minipage}[b]{8 cm}
    \psfrag{x}[tc]{\tiny $\sqrt{s_{\mu\mu}}$\hspace{0.2cm} [MeV]}
    \psfrag{y}[c]{\tiny $\partial {\rm BR}^{\rm rel} (\eta\to \mu^+\mu^- \gamma)  /\partial{\sqrt{s_{\mu\mu}/m_{\eta}}}$}
    \psfrag{without VMD}[c]{\,\,\,\,\,\,\,\,\tiny without VMD}
    \psfrag{hidden gauge}[c]{\,\,\,\,\,\,\,\,\,\,\,\tiny hidden gauge}
    \psfrag{modified VMD}[c]{\,\,\,\,\,\,\,\,\,\,\,\,\,\tiny modified VMD}
    \includegraphics[width=8 cm]{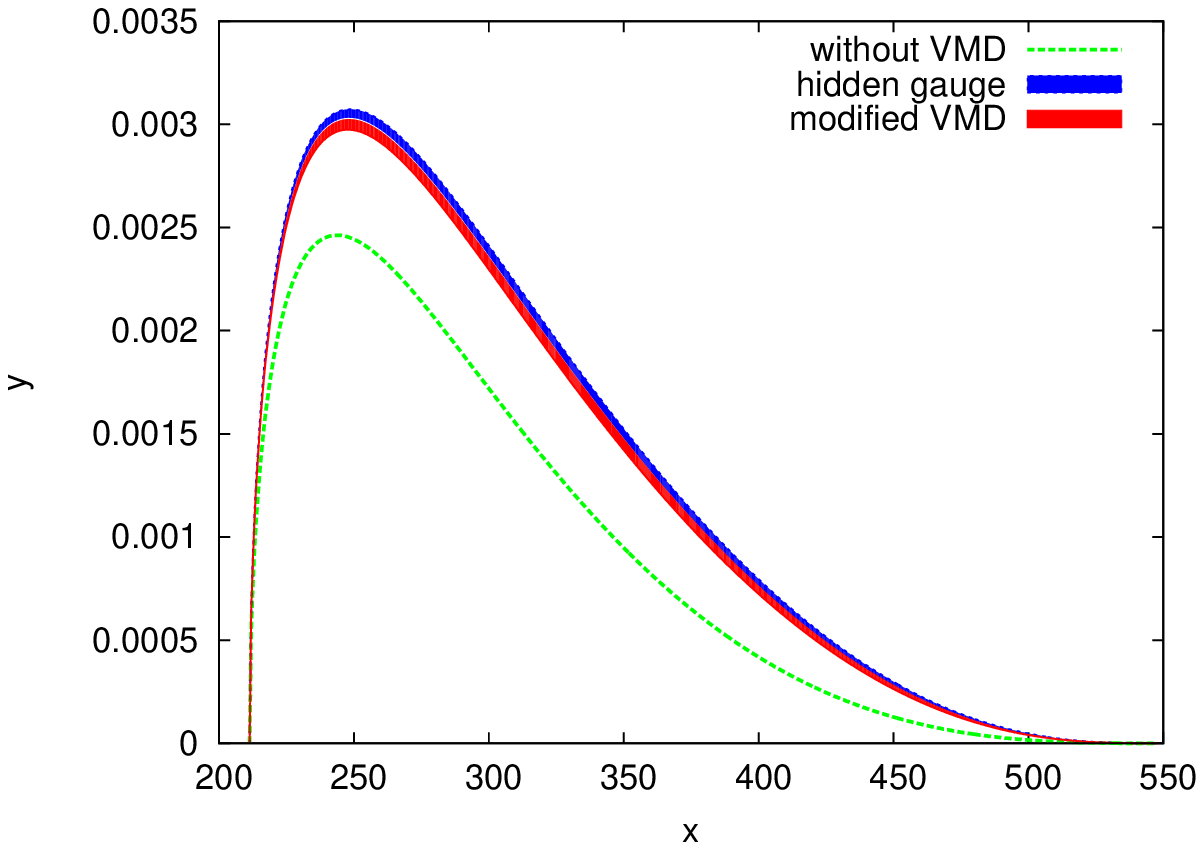}
  \end{minipage}

\caption{Dependence of the differential relative branching ratio (normalized to BR($\eta\to\gamma\gamma$), see Eq. \equa{BRrel}) of the decays $\eta\to e^+e^- \gamma$ and $\eta\to \mu^+\mu^-\gamma$
  on
the change of the invariant mass of the electron-pairs and the muon-pairs, respectively.\label{fig:etallg}}
\end{figure}

The next results which we present are the ones of the decays $\eta \to e^+ e^-
\gamma$ and $\eta \to \mu^+ \mu^- \gamma$. The dependences on the invariant
masses of the dileptons are shown in Figure \ref{fig:etallg}:\\
\\
The curve plotted against $\sqrt{s_{ee}}$ is almost the same as in the
analogous $\pi^0$-decay, except for the amplitude. The graph plotted against
$\sqrt{s_{\mu\mu}}$ is wider and also has a smaller amplitude. The errors are
again smaller than the width of the lines, but the contribution of
the VMD factor is visible now.\\
The branching ratios for these decays are listed in Table \ref{tab:etallg}.\\

\begin{table}[!hbt]
\begin{center}
\renewcommand{\arraystretch}{1.3}
\begin{tabular}{l|c|c|c}
& without VMD & hidden gauge & modified VMD  \\ \hline\hline
${\rm BR}^{\rm rel}$($\eta\to e^+e^-\gamma$)($10^{-2}$)
&$1.619$&$1.666\pm0.002$&$1.662\pm 0.002$ \\ \hline
${\rm BR}^{\rm rel}$($\eta\to \mu^+\mu^-\gamma$)($10^{-4}$) &$5.51$&$7.75\pm0.09$&$7.54\pm 0.11$
\end{tabular}
  \caption{Relative branching ratios (normalized to BR($\eta\to\gamma\gamma$), see Eq. \equa{BRrel}) of the decays $\eta\to e^+e^-\gamma$ and $\eta
    \to \mu^+ \mu^- \gamma$ calculated
    without VMD, with the hidden gauge model, and the modified VMD model. \label{tab:etallg}}
\end{center} 
\end{table}

Qualitatively, the results of the decay $\eta\to e^+e^-\gamma$ are similar to
the ones in the $\pi^0$ sector. The differences between the values
calculated without
VMD and the with VMD are still small, while the two VMD models generate
almost the same results.\\
For the decay $\eta\to \mu^+\mu^-\gamma$, however, there is a clear discrepancy between
the predictions with and without VMD. The two VMD models are still in the same range, although the
difference has increased to 0.3\% now.\\
Other theoretical values and experimental data are listed in Table
\ref{tab:etallga}.\\

\begin{table}[!hbt]
\begin{center}
\renewcommand{\arraystretch}{1.3}
\begin{tabular}{l|c|c|c}
& \cite{Miyazaki:1974qi}  & \cite{Lih:2009np}&  exp. data \cite{Amsler:2008zzb} \\ \hline\hline
${\rm BR}^{\rm rel}$($\eta\to e^+e^-\gamma$)($10^{-2}$)
&$1.62$&$1.77$&$1.78\pm0.19$ \\ \hline
${\rm BR}^{\rm rel}$($\eta\to \mu^+\mu^-\gamma$)($10^{-4}$) &$5.54$&$7.48$&$7.9\pm1.1$
\end{tabular}
  \caption{Theoretical values and experimental data for the relative branching ratios of the decays $\eta\to e^+e^-\gamma$ and $\eta
    \to \mu^+ \mu^- \gamma$. \label{tab:etallga}}
\end{center}  
\end{table}

The values given in Ref. \cite{Miyazaki:1974qi} agree with our calculation without
a VMD factor. These values fit the data for the decay $\eta\to
e^+e^-\gamma$, but have no overlap with the experimental data of the decay
$\eta\to \mu^+\mu^-\gamma$. For the  case of the $\eta\to
e^+e^-\gamma$-decay, the values of \cite{Lih:2009np} are larger than ours,
while both calculations are consistent for the $\eta\to
\mu^+\mu^-\gamma$-decay. Note that our VMD values represent the data very
well, compare Table \ref{tab:etallg} and \ref{tab:etallga}.\\
The authors of \cite{Bijnens:1999jp} calculated also the corresponding
branching ratio in the framework of VMD, {\it i.e}. with the hidden gauge model and
without a VMD factor, respectively. The corresponding values are totally analogous to ours.\\

\subsection*{$\eta'\to l^+ l^- \gamma$}

\begin{figure}[!hbt]
  \begin{minipage}[b]{8 cm}
    \psfrag{x}[tc]{\tiny $\sqrt{s_{ee}}$ \hspace{0.2cm} [MeV]}
    \psfrag{y}[c]{\tiny $\partial {\rm BR}^{\rm rel} (\eta' \to e^+e^- \gamma)  /\partial{\sqrt{s_{ee}/m_{\eta'}}}$}
    \includegraphics[width=8 cm]{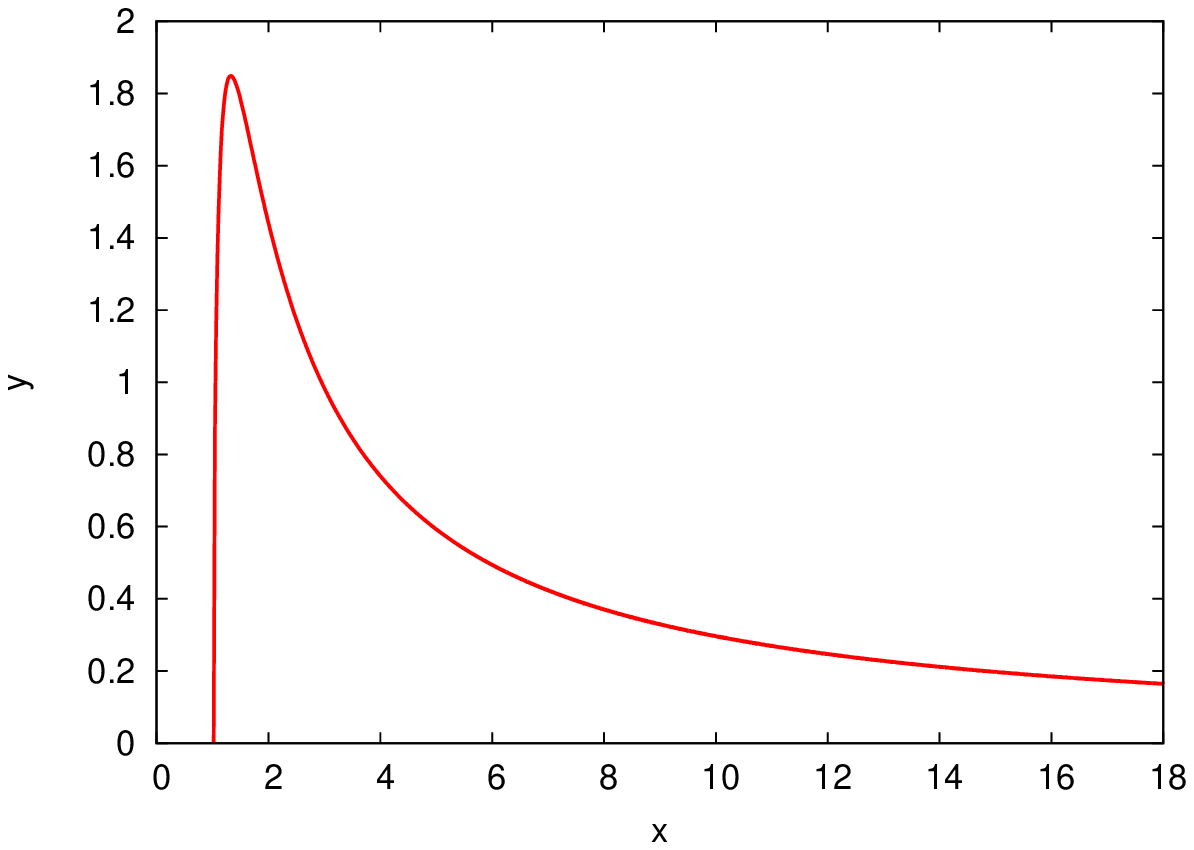}
  \end{minipage}
  \begin{minipage}[b]{8 cm}
    \psfrag{x}[tc]{\tiny $\sqrt{s_{ee}}$ \hspace{0.2cm} [MeV]}
    \psfrag{y}[c]{\tiny $\partial {\rm BR}^{\rm rel} (\eta' \to e^+e^- \gamma)  /\partial{\sqrt{s_{ee}/m_{\eta'}}}$}
    \psfrag{hidden gauge}[c]{\,\,\,\,\,\,\,\,\,\,\,\tiny hidden gauge}
    \psfrag{modified VMD}[c]{\,\,\,\,\,\,\,\,\,\,\,\,\,\tiny modified VMD}
    \includegraphics[width=8 cm]{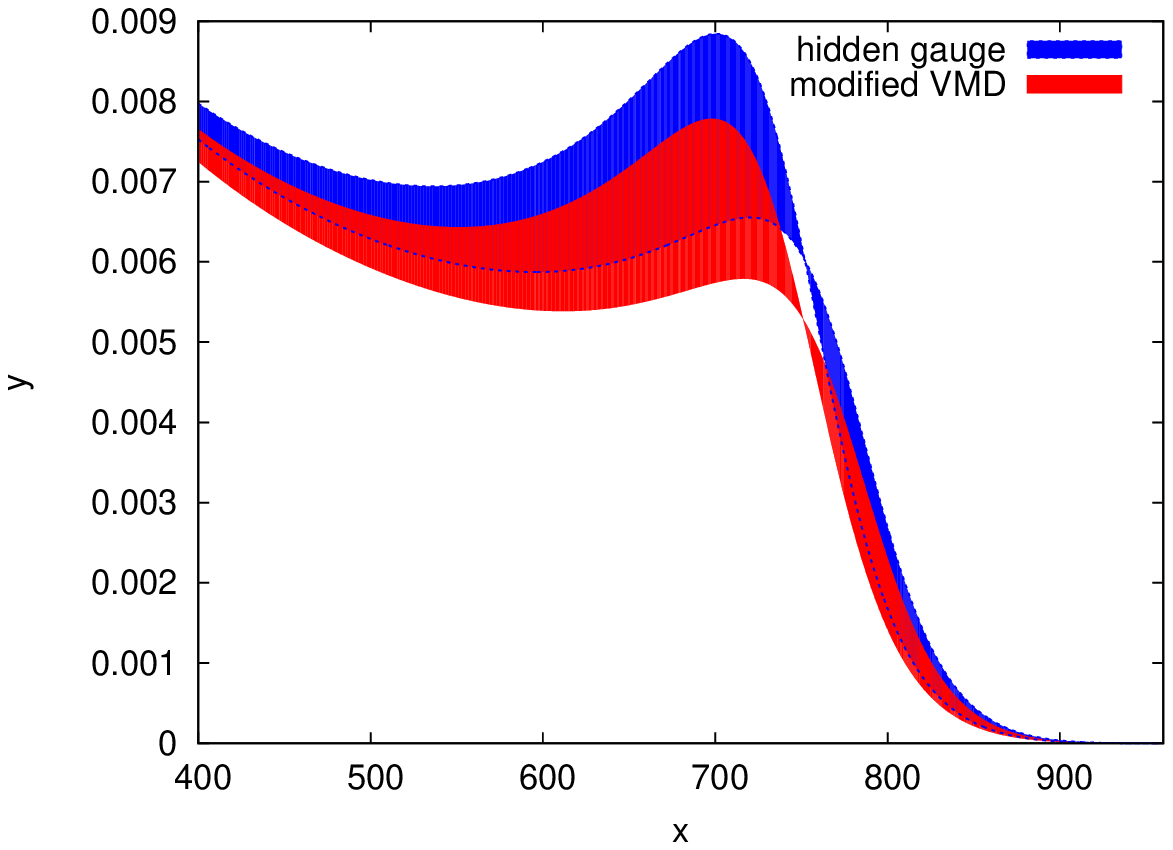}
  \end{minipage}

\caption{Dependency of the differential relative branching ratio (normalized to BR($\eta'\to\gamma\gamma$), see Eq. \equa{BRrel}) of the decay $\eta'\to e^+e^- \gamma$ on
the change of the invariant mass of the electron-pair in different energy regions.\label{fig:etapeeg}}
\end{figure}

Finally, we discuss the results for the $\eta'$ decays. The dependence of the
differential branching ratio of the decay $\eta' \to e^+ e^- \gamma$ looks
basically the same as in the $\eta$ sector. As shown in Figure \ref{fig:etapeeg} there is again a very strong peak for small energies. But there is also a contribution
from the width, a bend in the region of the vector meson mass. To visualize
that we plotted also the relevant sector (see Figure \ref{fig:etapeeg}).

\begin{center}
\begin{figure}[!hbt]
\begin{center}
\psfrag{x}[tc]{\tiny $\sqrt{s_{\mu\mu}}$\hspace{0.2cm} [MeV]}
    \psfrag{y}[c]{\tiny $\partial {\rm BR}^{\rm rel} (\eta'\to \mu^+\mu^- \gamma)  /\partial{\sqrt{s_{\mu\mu}/m_{eta'}}}$}
    \psfrag{hidden gauge}[c]{\,\,\,\,\,\,\,\,\,\,\,\tiny hidden gauge}
    \psfrag{modified VMD}[c]{\,\,\,\,\,\,\,\,\,\,\,\,\,\tiny modified VMD}
    \includegraphics[width=8 cm]{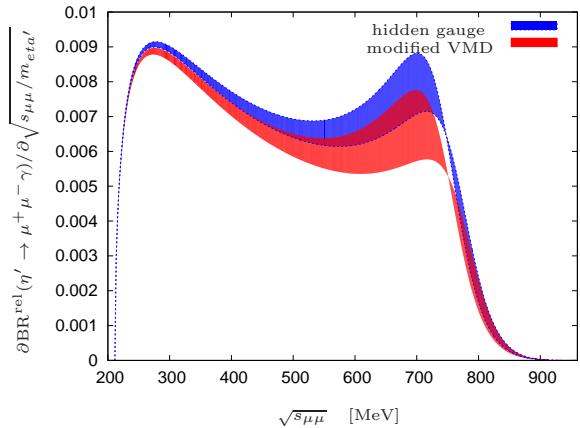}
\caption{Dependency of the differential relative branching ratio (normalized to BR($\eta'\to\gamma\gamma$), see Eq. \equa{BRrel}) of the decay $\eta' \to \mu^+ \mu^- \gamma$ on
the change of the invariant mass of the muon pair.\label{fig:etapmumug}}

\end{center}
\end{figure}
\end{center}

In the decay $\eta' \to \mu^+ \mu^- \gamma$ the contribution of the width is
very large. It is shown also in Figure \ref{fig:etapmumug}.

\newpage

The branching ratios are listed in
Table \ref{tab:eta'llg}.
\begin{table}[!hbt]
\begin{center}
\renewcommand{\arraystretch}{1.3}
\begin{tabular}{l|c|c|c|c}
& without VMD & hidden gauge & modified VMD &  exp. data \cite{Amsler:2008zzb} \\ \hline\hline
${\rm BR}^{\rm rel}$($\eta'\to e^+e^-\gamma$)($10^{-2}$)
&$1.79$&$2.10\pm0.02$&$2.06\pm0.02$&$<4.5$ \\ \hline
${\rm BR}^{\rm rel}$($\eta'\to \mu^+\mu^-\gamma$)($10^{-3}$) &$1.72$&$4.45\pm0.15$&$4.11\pm0.18$&$4.9\pm1.2$
\end{tabular}
  \caption{Relative branching ratios (normalized to BR($\eta'\to\gamma\gamma$), see Eq. \equa{BRrel}) of the decays $\eta'\to e^+e^-\gamma$ and
    $\eta'\to \mu^+\mu^-\gamma$ calculated
    without VMD, with the hidden gauge model, the modified VMD model, and
    the experimental data. \label{tab:eta'llg}}
\end{center}
  
\end{table}

The trend of the $\eta$ decay channels is repeated  here. The values of the
various VMD models hardly differ, while the values calculated without VMD are clearly
smaller.\\
All results are compatible with the upper experimental limit of the $\eta'\to
e^+e^-\gamma$ decay. For the case of the decay $\eta'\to \mu^+\mu^-\gamma$, the value
calculated without a VMD factor is out of the experimental range. Both VMD
results are smaller than the experimental data, and only the
hidden gauge model has a small overlap. This may be significant, but notice
that the
experimental accuracy in the $\eta'$ sector is also not very high.\\

\subsection*{Summary}

In general, the theoretical values represent the data very well. The
contribution of the VMD factor is very small for the decays $P \to
e^+ e^- \gamma$ and increases with the mass of the decaying particle. One can
see that for the decays $\eta / \eta' \to \mu^+ \mu^- \gamma$ a vector meson
dominance model is needed for an accurate description of the data, but no
preference of any model can be given.

\section{$P\to l^+l^-l^+l^-$}
We will now present the results for the rates of the decays of a
pseudoscalar meson, $P = \pi^0, \eta, \eta' $, into four leptons.\\
 The direct term and the crossed term give the same result under integration,
 so we just need to calculate one of them. The interference term is more complicated to
 calculate, because there appear functions of the three respective angles,
 namely $\theta_{12}$, $\theta_{34}$ and $\phi$ see Eq. \equa{interference},
   \equa{llllid1}, \equa{llllid2} and \equa{idangles}, in the denominator. Thus the
 angle integrations are no longer trivial. So instead of a
 two-dimensional integration the  task now is to perform a Gaussian integration over five variables.\\

\subsection*{$\pi^0 \to e^+e^-e^+e^-$}

\begin{center}
\begin{figure}[!hbt]
\begin{center}
    \psfrag{x}[tc]{\tiny $\sqrt{s_{ee}}$ \hspace{0.2cm} [MeV]}
    \psfrag{y}[c]{\tiny $\partial {\rm BR}^{\rm rel} (\pi^0\to e^+e^-e^+e^-)  /\partial{\sqrt{s_{ee}/m_{\pi}}}$}
    \psfrag{direct and crossed term}[c]{\,\,\,\,\,\,\,\,\,\,\,\,\,\,\,\,\,\,\,\,\,\,\,\,\,\,\tiny direct and crossed term}
    \psfrag{-100 * interference term}[c]{\,\,\,\,\,\,\,\,\,\,\,\,\,\,\,\,\,\,\,\,\,\,\,\tiny $-100 \,\,\times$ interference term}
    \includegraphics[width=8cm]{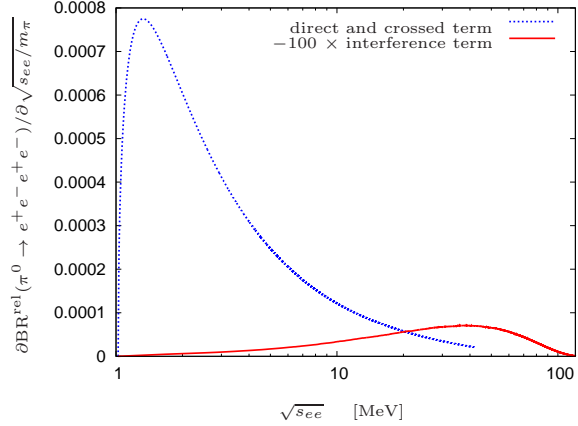}
\caption{Dependency of the relative branching ratio (normalized to BR($\pi^0\to\gamma\gamma$), see Eq. \equa{BRrel}) of the decay $\pi^0\to e^+e^-e^+e^-$ on
the change of the invariant mass of the electrons.\label{fig:pieeee}}

\end{center}
\end{figure}
\end{center}

We start with the four lepton decay of the $\pi^0$. Here only the decay into
$e^+e^-e^+e^-$ is possible. The dependence of the branching ratio on the
invariant mass of the electron ($\sqrt{s_{ee}}$) is shown in Figure
\ref{fig:pieeee}. Note that the x-axis is of logarithmic scale and the
interference term is scaled up by a factor of $-100$.

For the direct (crossed) term we can see a peak at very small
$\sqrt{s_{ee}}$. The interference term in contrast is negative and wider. The different VMD
models give almost the same curves for the direct (crossed) and the
interference term. So we just plotted one curve, respectively. The errors
are smaller than the width of the lines.\\
The calculated values of the branching ratios are given in Table
\ref{tab:pillll}. The errors are again very small so that we do not show them.\\

\begin{table}[!hbt]
\begin{center}
\renewcommand{\arraystretch}{1.3}
\begin{tabular}{l|c|c|c}
 $\pi^0\to e^+e^-e^+e^- (10^{-5})$& without VMD & hidden gauge & modified VMD  \\ \hline\hline
 ${\rm BR}^{\rm rel}_{1+2}$ & $\:\;\;3.456$ & $\:\;\;3.469$ & $\:\;\;3.468$   \\ \hline
${\rm BR}^{\rm rel}_{12}$ & $-0.036$ & $-0.037$ & $-0.036$   \\ \hline
${\rm BR}^{\rm rel}_{total}$& $\:\;\;3.420$ & $\:\;\;3.432$ & $\:\;\;3.431$ 
\end{tabular}
  \caption{Relative branching ratios (normalized to BR($\pi^0\to\gamma\gamma$), see Eq. \equa{BRrel}) of the decay $\pi^0\to e^+e^-e^+e^-$ calculated
    without VMD, with the hidden gauge model, and with the modified VMD model
    for the direct and crossed term  ${\rm BR}^{\rm rel}_{1+2}$, the interference term
    ${\rm BR}^{\rm rel}_{12}$, and the total one ${\rm BR}^{\rm rel}_{total}$ = ${\rm BR}^{\rm rel}_{1+2}$ + ${\rm BR}^{\rm rel}_{12}$.\label{tab:pillll}}
\end{center}
\end{table}

As one can see the contribution of the interference term (${\rm BR}^{\rm rel}_{12}$) is of the order of one per cent
compared with the leading direct and crossed term (${\rm BR}^{\rm rel}_{1+2}$). Note that the
interference term has a negative
sign. The difference between the VMD models is very small for this
decay. We display four decimal places just to see a difference in the values
of the interference term. Moreover the errors are small. This is
compatible with the small
contribution of the VMD.\\
We can compare our results with other theoretical calculations and the
experimental data, see Table \ref{tab:pilllla}.\\
\begin{table}[!hbt]
\begin{center}
\renewcommand{\arraystretch}{1.3}
\begin{tabular}{l|c|c|c|c|c}
$\pi^0\to e^+e^-e^+e^- (10^{-5})$ & \cite{Kroll:1955zu} & \cite{Miyazaki:1974qi} &\cite{Barker:2002ib} &\cite{Lih:2009np}& Data \cite{Amsler:2008zzb}\\ \hline\hline
${\rm BR}^{\rm rel}_{1+2}$  &  & $\:\;\;3.46$ & $\:\;\;3.456$ & & \\ \cline{1-1}\cline{3-4}
${\rm BR}^{\rm rel}_{12}$  &  & $-0.18$ & $-0.036$ & & \\ \cline{1-1}\cline{3-4}
${\rm BR}^{\rm rel}_{total}$& 3.47 & $\:\;\;3.28$ & $\:\;\;3.42$ &3.29& $\:\;\;3.38\pm0.16$ 
\end{tabular}
  \caption{Theoretical calculations and experimental data for the relative branching
    ratios of the decay $\pi^0\to e^+e^-e^+e^-$. \label{tab:pilllla}}
\end{center}

\end{table}\\
For the case of the total rate we agree with all values of the recent references, namely
\cite{Kroll:1955zu}, \cite{Miyazaki:1974qi},\cite{Barker:2002ib} and
\cite{Lih:2009np}. Note that the comparison has to be done with the first column of Table \ref{tab:pillll}, because
none of them used a vector meson dominance model. With respect to the interference
term, we totally agree with the values given in \cite{Barker:2002ib}. In
agreement with this reference we also differ by a factor of 5 compared with the result of Ref. \cite{Miyazaki:1974qi}. All of
the predictions for the total branching ratio are consistent with the experimental data.\\

\subsection*{$\eta \to l^+ l^- l^+ l^-$}

\begin{figure}[!hbt]
  \begin{minipage}[b]{8 cm}
    \psfrag{x}[tc]{\tiny $\sqrt{s_{ee}}$ \hspace{0.2cm} [MeV]}
    \psfrag{y}[c]{\tiny $\partial {\rm BR}^{\rm rel} (\eta\to e^+e^-e^+e^-)  /\partial{\sqrt{s_{ee}/m_{\eta}}}$}
    \includegraphics[width=8 cm]{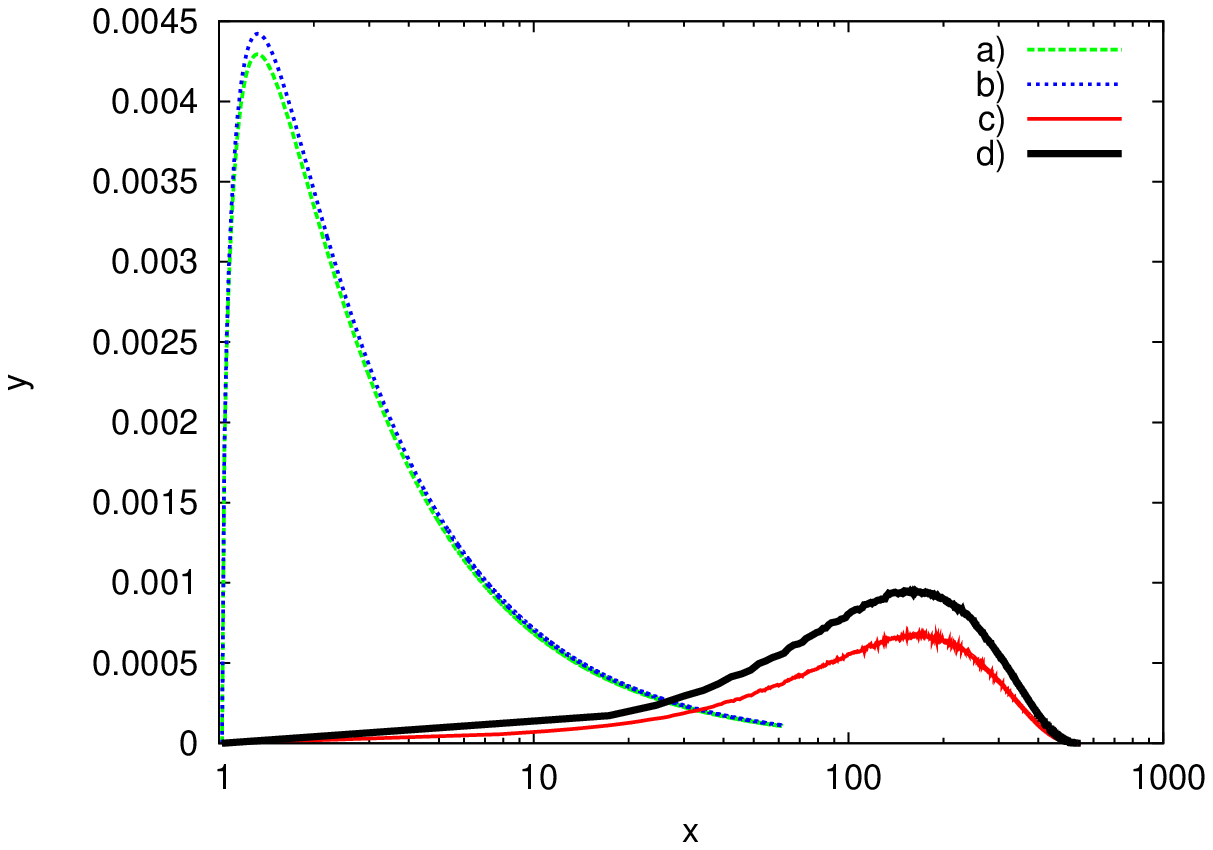}
  \end{minipage}
  \begin{minipage}[b]{8 cm}
    \psfrag{x}[tc]{\tiny $\sqrt{s_{\mu\mu}}$\hspace{0.2cm} [MeV]}
    \psfrag{y}[c]{\tiny $\partial {\rm BR}^{\rm rel} (\eta\to \mu^+\mu^-\mu^+\mu^-)  /\partial{\sqrt{s_{\mu\mu}/m_{\eta}}}$}
    \includegraphics[width=8 cm]{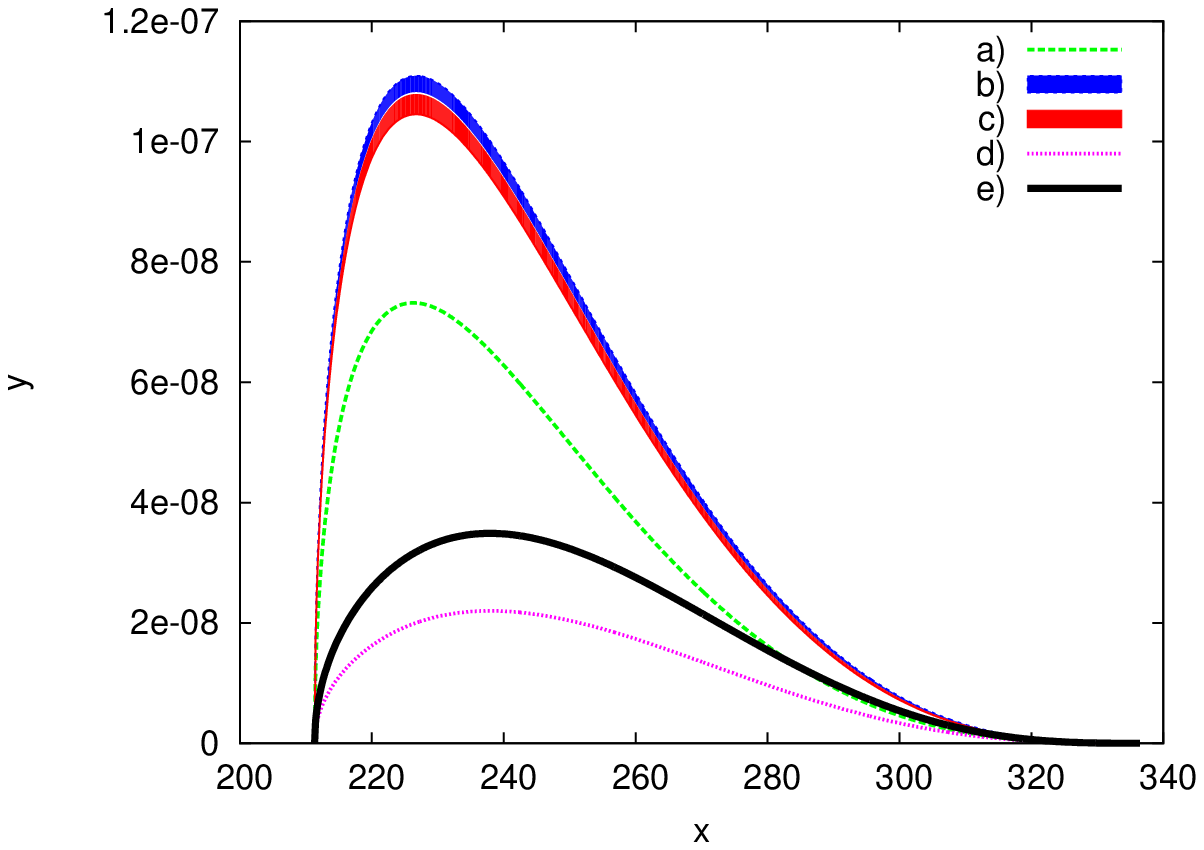}
  \end{minipage}

\caption{Dependency of the differential relative branching ratio (normalized to BR($\eta\to\gamma\gamma$), see Eq. \equa{BRrel}) of the decays $\eta\to e^+e^-e^+e^-$ and $\eta\to \mu^+\mu^-\mu^+\mu^-$
  on
the change of the invariant mass of the electrons and the muons,
respectively. In the left graph a) is the curve for the direct and crossed term calculated
without a VMD factor, b) the one calculated with a VMD factor, c) is the
curve for the interference term calculated without VMD factor, and d) the
respective one calculated with a VMD factor. The values for the interference
terms were scaled up by a factor $-1000$. The fluctuations in this curves are due to the imprecision of the Gaussian integrations. In the right graph a), b) and c) are
the curves for the direct and crossed terms calculated without a VMD factor, the hidden gauge
model, and the modified model, respectively. d) and e)
are the curves of the interference term calculated without a VMD factor and
with a VMD factor. The curves of the interference term were scaled up by
a factor $-5$.\label{fig:etal1l2}}

\end{figure}

\begin{figure}[!hbt]
  \begin{minipage}[b]{8 cm}
    \psfrag{x}[tc]{\tiny$\sqrt{s_{ee}}$ \hspace{0.2cm} [MeV]}
    \psfrag{y}[c]{\tiny $\partial {\rm BR}^{\rm rel} (\eta\to \mu^+\mu^-e^+e^-)  /\partial{\sqrt{s_{ee}/m_{\eta}}}$}
    \psfrag{without VMD}[c]{\,\,\,\,\,\,\,\,\tiny without VMD}
    \psfrag{hidden gauge}[c]{\,\,\,\,\,\,\,\,\,\,\,\tiny hidden gauge}
    \psfrag{modified VMD}[c]{\,\,\,\,\,\,\,\,\,\,\,\,\,\tiny modified VMD}
    \includegraphics[width=8 cm]{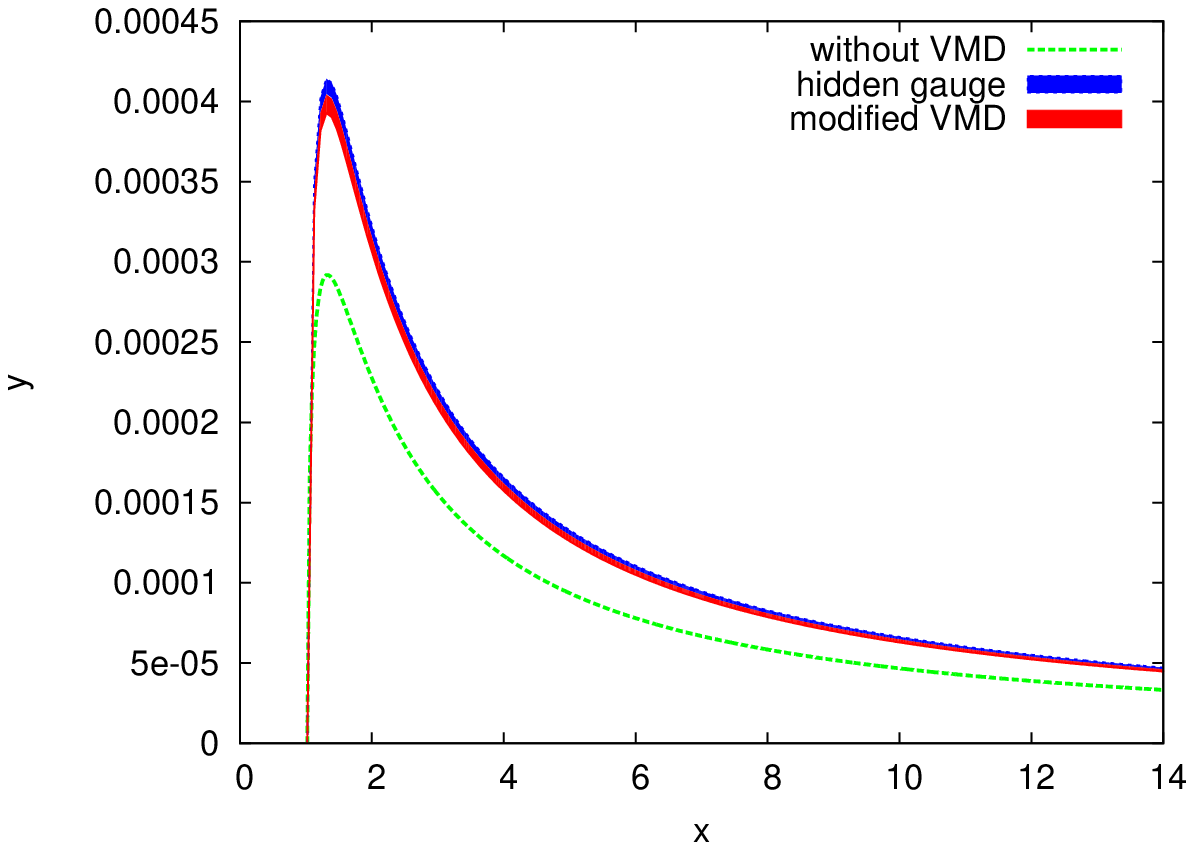}
  \end{minipage}
  \begin{minipage}[b]{8 cm}
    \psfrag{x}[tc]{\tiny$\sqrt{s_{\mu\mu}}$\hspace{0.2cm} [MeV]}
    \psfrag{y}[c]{\tiny $\partial {\rm BR}^{\rm rel} (\eta\to \mu^+\mu^-e^+e^-)  /\partial{\sqrt{s_{\mu\mu}/m_{\eta}}}$}
    \psfrag{without VMD}[c]{\,\,\,\,\,\,\,\,\tiny without VMD}
    \psfrag{hidden gauge}[c]{\,\,\,\,\,\,\,\,\,\,\,\tiny hidden gauge}
    \psfrag{modified VMD}[c]{\,\,\,\,\,\,\,\,\,\,\,\,\,\tiny modified VMD}
    \includegraphics[width=8 cm]{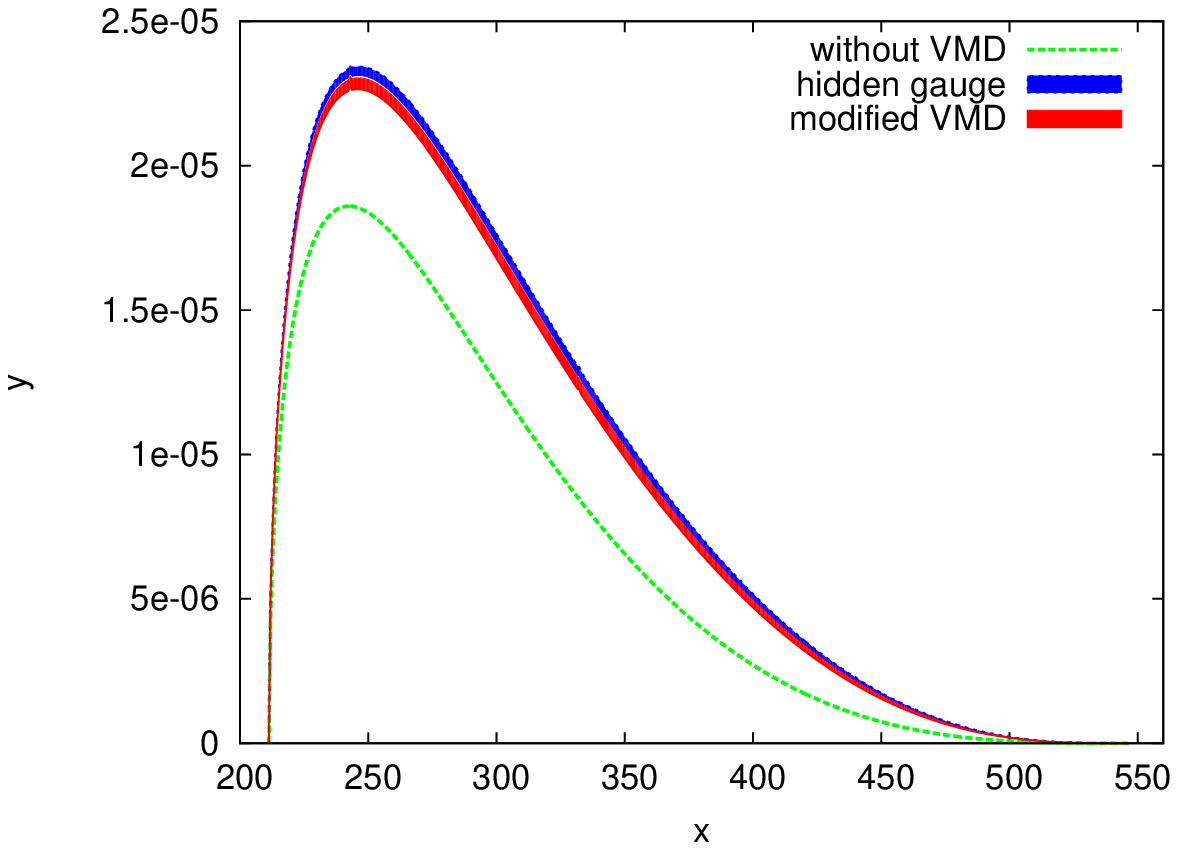}
  \end{minipage}

\caption{Dependency of the differential relative branching ratio (normalized to BR($\eta\to\gamma\gamma$), see Eq. \equa{BRrel}) of the decay $\eta\to \mu^+\mu^-e^+e^-$ on
the change of the invariant mass of the electrons and the muons, respectively.\label{fig:etallll}}

\end{figure}

In the case of the $\eta$, all three decays into two $e^+e^-$-pairs, two
$\mu^+\mu^-$-pairs and an $e^+e^-$- plus a $\mu^+\mu^-$-pair are possible. The
dependence of the differential branching ratio on the invariant masses can be seen in
Figures \ref{fig:etallll} and \ref{fig:etal1l2}.

The decay into two $e^+e^-$-pairs is similar to the analogous decay of the
$\pi^0$. The errors are again smaller than the width of the line, but we can
see the difference between the used models. The curve without VMD is smaller
than the ones calculated with VMD models. For the direct and crossed term we
can also see a small difference between the hidden gauge and the modified VMD
model.\\
The graph for the decay into two $\mu^+\mu^-$-pairs is very different. We
do not need a logarithmic scale to present the values appropriately because
the direct and crossed terms are much wider. Also the amplitude is
smaller. The difference between the used models is obvious, even for the
interference term. For the first time we can see the errors in the direct and
crossed terms.

The differential ratio for the decay $\eta \to
\mu^+\mu^-e^+e^-$ plotted versus $\sqrt{s_{ee}}$ has basically the same
structure than the decay into two $e^+e^-$-pairs. Here only the amplitude is
smaller. On the other hand, the differential ratio plotted versus $\sqrt{s_{\mu\mu}}$ is similar to the decay into two $\mu^+\mu^-$-pairs, with
bigger and wider amplitude.\\
Now we present our results of the branching ratios for these three
decays, calculated without VMD, with the hidden gauge, and
with the modified model, and compare them with other theoretical values
as well as the experimental data (Table \ref{tab:etallll}).\\
\begin{table}[!hbt]
\begin{center}
\renewcommand{\arraystretch}{1.3}
\begin{tabular}{l|c|c|c|c|c|c}
 & without VMD & hidden gauge & modified VMD & \cite{Miyazaki:1974qi}& \cite{Lih:2009np}  & Data \cite{Amsler:2008zzb}\\ \hline\hline
\multicolumn{5}{l}{$\eta\to e^+e^-e^+e^-(10^{-5})$}\\\hline
${\rm BR}^{\rm rel}_{1+2}$ & $\:\;\;6.497$ & $\:\;\;6.848\pm0.012$ & $\:\;\;6.817\pm0.015$ & $\:\;\;6.50$& \\ \cline{1-5}
${\rm BR}^{\rm rel}_{12}$ & $-0.034$ & $-0.048\pm0.001$ & $-0.047\pm0.001$&$-0.36$ &  \\ \cline{1-5}
${\rm BR}^{\rm rel}_{total}$& $\:\;\;6.463$ & $\:\;\;6.800\pm0.013$ & $\:\;\;6.770\pm0.016$ &$\:\;\;6.14$& 6.26&$<17.5$ \\ \hline\hline
\multicolumn{5}{l}{$\eta\to \mu^+\mu^-e^+e^-(10^{-6})$}\\\hline
${\rm BR}^{\rm rel}$ & $\:\;\;3.996$ & $\:\;\;5.616\pm0.063$ & $\:\;\;5.465\pm0.079$ &$\:\;\;1.99$& 1.48 &$<40.5$\\ \hline\hline
\multicolumn{5}{l}{$\eta\to \mu^+\mu^-\mu^+\mu^-(10^{-9})$}\\\hline
${\rm BR}^{\rm rel}_{1+2}$ & $\:\;\;6.560$ & $\;10.031\pm0.129$ & $\:\;\;9.708\pm0.162$ &$\:\;\;6.73$&  \\ \cline{1-5}
${\rm BR}^{\rm rel}_{12}$ & $-0.049$ & $-0.078\pm0.001$ & $-0.074\pm0.001$ &$-0.50$&  \\ \cline{1-5}
${\rm BR}^{\rm rel}_{total}$& $\:\;\;6.511$ & $\:\;\;9.953\pm0.053$ & $\:\;\;9.634\pm0.163$ &$\:\;\;6.23$&4.27& $<9.1\cdot 10^{-4}$ 
\end{tabular}
  \caption{Relative branching ratios (normalized to BR($\eta\to\gamma\gamma$), see Eq. \equa{BRrel}) of the decay $\eta\to l^+l^-l^+l^-$ with
    different VMD models, other theoretical values \cite{Miyazaki:1974qi}, \cite{Lih:2009np} and
    experimental data. \cite{Amsler:2008zzb}\label{tab:etallll}}
\end{center}

\end{table}\\
The relative branching ratio of the $\eta\to \mu^+\mu^-\mu^+\mu^-$-decay is much smaller
than the branching ratios of the other 4-lepton decays. This is caused by the very small
phase space in this reaction. The interference term is again of the order of
one per
cent.\\
In Ref. \cite{Bijnens:1999jp} the decays
$\eta\to e^+e^-e^+e^-$ and $\eta\to \mu^+\mu^-e^+e^-$ were
calculated with VMD models and without. These values
totally agree with ours.\\
Our values of the decays $\eta\to e^+e^-e^+e^-$ and $\eta\to
\mu^+\mu^-\mu^+\mu^-$ calculated without a VMD form factor approximately agree with the results given in
\cite{Miyazaki:1974qi}. The difference may be caused by the improved input data
that we
used. We also agree with \cite{Lih:2009np} for the $\eta\to e^+e^-e^+e^-$-decay
while their values for $\eta\to
\mu^+\mu^-\mu^+\mu^-$ are much smaller. We
disagree again for the values of the interference term. Our interference term
is here about 10 times smaller than the values given in
\cite{Miyazaki:1974qi}. Ref. \cite{Lih:2009np} did not calculate an
interference term. We also disagree in the total branching ratio of the
decay $\eta\to \mu^+\mu^-e^+e^-$, but the factor 2 of \cite{Miyazaki:1974qi} may be due to a
typographical error in their calculations. This was already pointed out by the
authors of \cite{Barker:2002ib}, when they compared the respective results of the decay
$K_L \to \mu^+\mu^-e^+e^-$. It is clearly visible that Ref. \cite{Lih:2009np} gives even
smaller values than \cite{Miyazaki:1974qi}.\\
The vector meson dominance factor now gives a big contribution. Especially for
the $\eta\to \mu^+\mu^-e^+e^-$- and $\eta\to \mu^+\mu^-\mu^+\mu^-$-decays this
factor is essential. While the results of the two VMD models are still very close to
each other, the one without VMD is distinctly
smaller. Unfortunately the existing experimental data only give upper bounds
which all of the models can meet. More precise measurements are needed to
falsify the VMD model.\\

\subsection*{$\eta' \to l^+ l^- l^+ l^-$}

\begin{center}
\begin{figure}[!hbt]
\begin{center}
    \psfrag{x}[tc]{\tiny $\sqrt{s_{\mu\mu}}$ \hspace{0.2cm} [MeV]}
    \psfrag{y}[c]{\tiny $\partial {\rm BR}^{\rm rel} (\eta' \to \mu^+\mu^-e^+e^-)  /\partial{\sqrt{s_{\mu\mu}/m_{\eta'}}}$}
    \psfrag{hidden gauge}[c]{\,\,\,\,\,\,\,\,\,\,\,\tiny hidden gauge}
    \psfrag{modified VMD}[c]{\,\,\,\,\,\,\,\,\,\,\,\,\,\tiny modified VMD}
    \includegraphics[width=8cm]{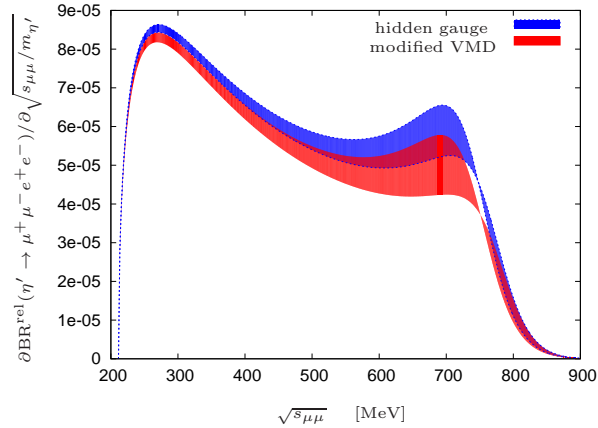}
\caption{Dependency of the differential relative branching ratio (normalized to BR($\eta'\to\gamma\gamma$), see Eq. \equa{BRrel}) of the decay $\eta' \to \mu^+\mu^-e^+e^-$ on
the change of the invariant mass of the muon pair.\label{fig:etapmumuee}}

\end{center}
\end{figure}
\end{center}

Also for the $\eta'$ case all three decay modes ($e^+ e^- e^+
e^-$, $\mu^+ \mu^- \mu^+ \mu^-$ and $\mu^+ \mu^- e^+ e^-$) are possible. The
dependence of the branching ratios on the invariant mass is almost the same as
in the analogous $\eta$-decays. In general, the amplitudes in the
$\eta'$-decays are greater. Also the differences between the VMD models are
bigger. This holds also for the errors. Differences can be seen in the curve of the
decay $\eta' \to \mu^+ \mu^- e^+e^-$, where the width gives a very strong
contribution to the branching ratio in the region of the vector meson mass when
plotted against the invariant mass of the muons $\sqrt{s_{\mu\mu}}$. This
can be seen in Figure \ref{fig:etapmumuee}.

\begin{figure}[!hbt]
  \begin{minipage}[b]{8 cm}
    \psfrag{x}[tc]{\tiny$\sqrt{s_{ee}}$ \hspace{0.2cm} [MeV]}
    \psfrag{y}[c]{\tiny $\partial {\rm BR}^{\rm rel} (\eta' \to e^+e^-e^+e^-)  /\partial{\sqrt{s_{ee}/m_{\eta'}}}$}
    \psfrag{hidden gauge}[c]{\,\,\,\,\,\,\,\,\,\,\,\tiny hidden gauge}
    \psfrag{modified VMD}[c]{\,\,\,\,\,\,\,\,\,\,\,\,\,\tiny modified VMD}
    \includegraphics[width=8 cm]{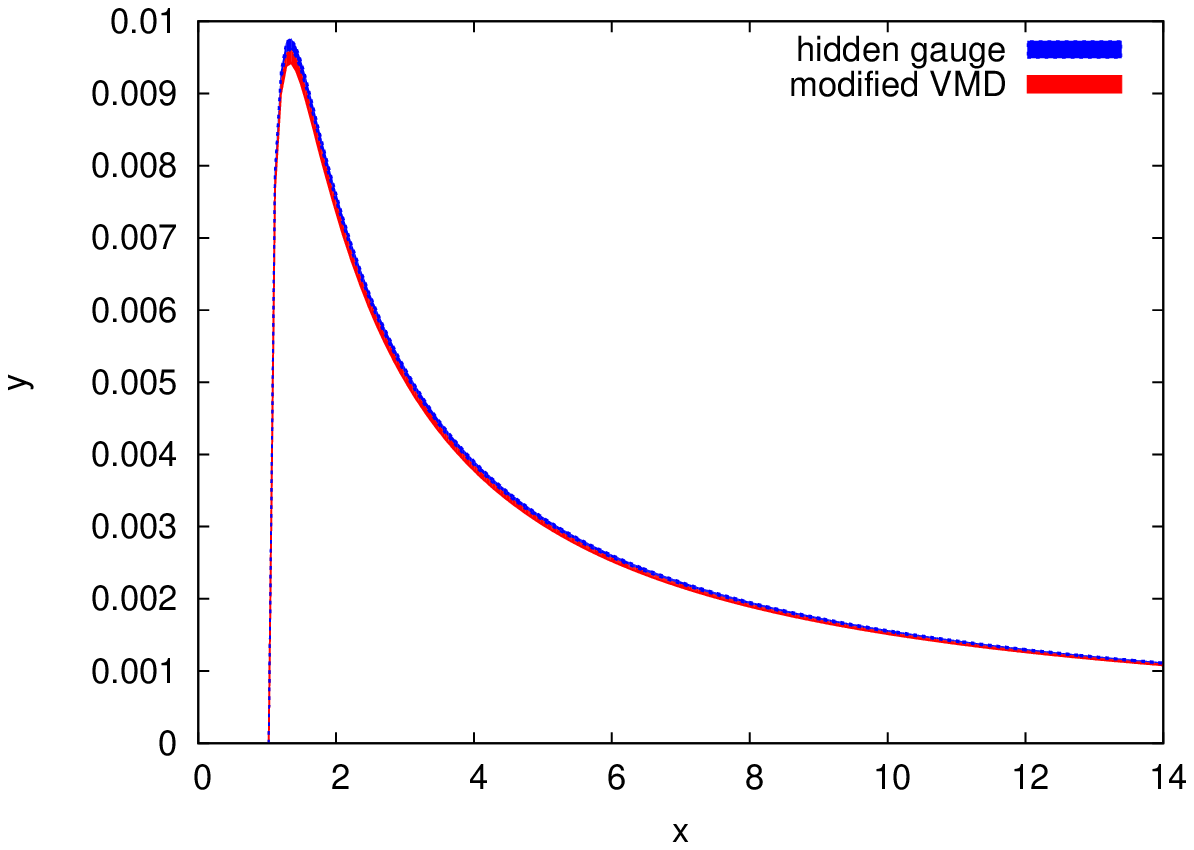}
  \end{minipage}
  \begin{minipage}[b]{8 cm}
    \psfrag{x}[tc]{\tiny$\sqrt{s_{ee}}$\hspace{0.2cm} [MeV]}
    \psfrag{y}[c]{\tiny $\partial {\rm BR}^{\rm rel} (\eta'\to e^+e^-e^+e^-)  /\partial{\sqrt{s_{ee}/m_{\eta'}}}$}
    \psfrag{hidden gauge}[c]{\,\,\,\,\,\,\,\,\,\,\,\tiny hidden gauge}
    \psfrag{modified VMD}[c]{\,\,\,\,\,\,\,\,\,\,\,\,\,\tiny modified VMD}
    \includegraphics[width=8 cm]{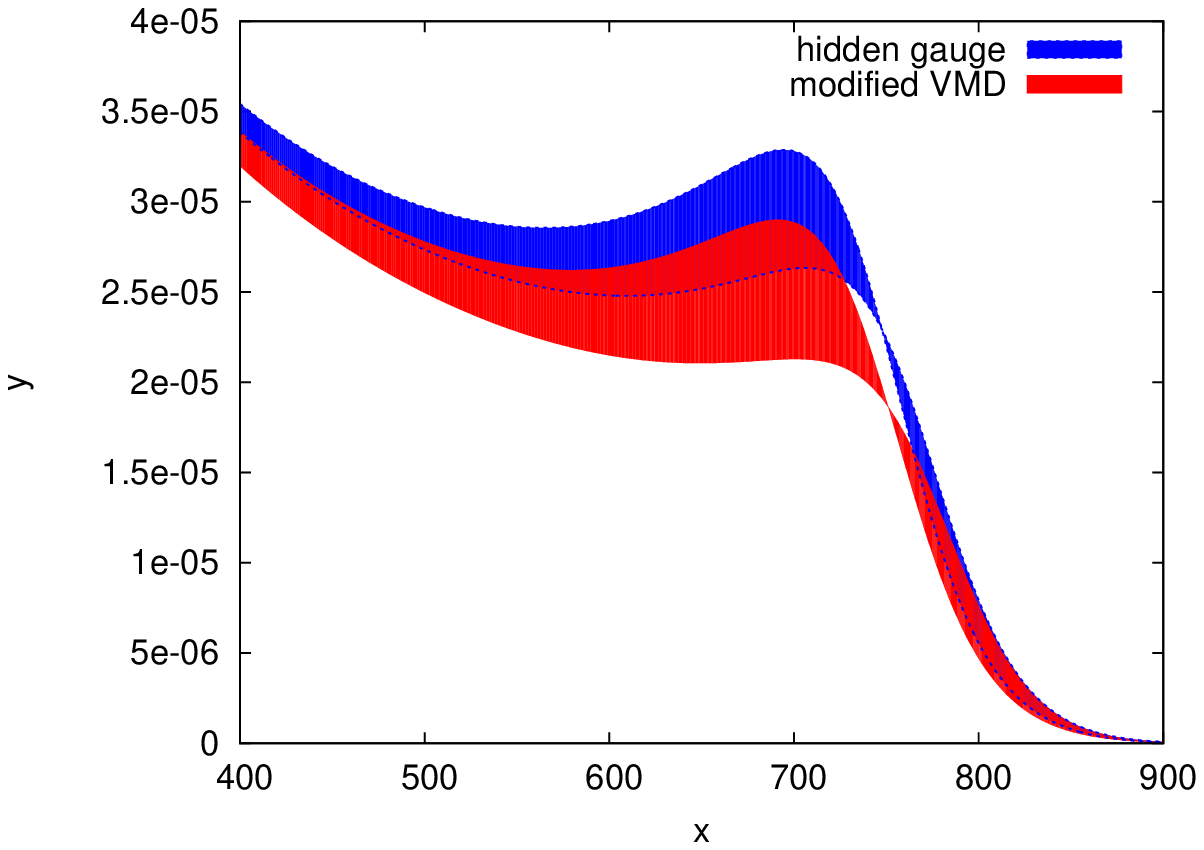}
  \end{minipage}

\caption{Dependency of the differential relative branching ratio (normalized to BR($\eta'\to\gamma\gamma$), see Eq. \equa{BRrel}) of the decay $\eta'\to e^+ e^-e^+e^-$ on
the change of the invariant mass of the electrons.\label{fig:etapeeee}}
\end{figure}

Plotted against the invariant mass
of the electrons one can not see the contribution of the width. Because of the
high mass of the muons the
energy of the electron pair does not get in the region of the vector meson mass.\\
In the curve of the decay $\eta'
\to e^+e^-e^+e^-$ the contribution of the width gives a small bend in the region of the
vector meson mass but this is hardly noticeable because the values in this area
are very small. Therefore we plotted the respective range in Figure \ref{fig:etapeeee}.

For the decay $\eta' \to \mu^+ \mu^-\mu^+\mu^-$ again the width does not play
a role, because the energy of the muons is lower than the vector meson
mass. Therefore the dependence of the differential branching ratio on the invariant mass is similar to the one in the decay $\eta\to \mu^+\mu^-\mu^+\mu^-$.

The branching ratios corresponding to the $\eta'$ decays are given in Table \ref{tab:eta'llll}.\\
\begin{table}[!hbt]
\begin{center}
\renewcommand{\arraystretch}{1.3}
\begin{tabular}{l|c|c|c}
 & without VMD & hidden gauge & modified VMD  \\ \hline\hline
\multicolumn{4}{l}{$\eta'\to e^+e^-e^+e^-(10^{-4})$}\\\hline
${\rm BR}^{\rm rel}_{1+2}$ & $\:\;\;0.7972$ & $\:\;\;1.0445\pm0.0136$ & $\:\;\;1.0148\pm0.0170$   \\ \hline
${\rm BR}^{\rm rel}_{1+2}$ & $-0.0033$ & $-0.0110\pm0.0004$ & $-0.0104\pm0.0003$   \\ \hline
${\rm BR}^{\rm rel}_{total}$& $\:\;\;0.7939$ & $\:\;\;1.0335\pm0.0140$ & $\:\;\;1.0044\pm0.0173$ \\ \hline\hline
\multicolumn{4}{l}{$\eta'\to \mu^+\mu^-e^+e^-(10^{-5})$}\\\hline
${\rm BR}^{\rm rel}$ & $\:\;\;1.462$ & $\:\;\;3.739\pm0.132$ & $\:\;\;3.458\pm0.160$ \\ \hline\hline
\multicolumn{4}{l}{$\eta'\to \mu^+\mu^-\mu^+\mu^-(10^{-6})$}\\\hline
${\rm BR}^{\rm rel}_{1+2}$ & $\:\;\;0.443$ & $\:\;\;1.205\pm0.047$ & $\:\;\;1.119\pm0.056$   \\ \hline
${\rm BR}^{\rm rel}_{1+2}$ & $-0.050$ & $-0.182\pm0.003$ & $-0.172\pm0.011$   \\ \hline
${\rm BR}^{\rm rel}_{total}$& $\:\;\;0.393$ & $\:\;\;1.023\pm0.050$ & $\:\;\;0.947\pm0.067$ 
\end{tabular}
  \caption{Relative branching ratios (normalized to BR($\eta'\to\gamma\gamma$), see Eq. \equa{BRrel}) of the decay $\eta'\to l^+l^-l^+l^-$ with
    different VMD models.\label{tab:eta'llll}}
\end{center}
  
\end{table}\\
The contribution of the interference term is again of the same range. Because of the larger $\eta'$ mass and thereby a bigger phase space, the
difference between the $\eta'\to \mu^+\mu^-\mu^+\mu^-$ decay mode and the
other modes is smaller than for the $\eta$-decays. We have again a huge effect of the
vector meson dominance factor for the $\eta'\to e^+e^-\mu^+\mu^-$ case, and
even more so for the $\eta'\to \mu^+\mu^-\mu^+\mu^-$-decay. However, the theoretical and experimental status
is much worse than for the $\pi^0$ and $\eta$ cases, since no data exist.\\

\subsection*{Summary}

In summary, the effect of VMD again increases with the value of
the mass of the decaying particle and the lepton mass. It is very small for the case
of the decay into four electrons and increases roughly by a factor of
$1.5$ or $2.5$, respectively, for the cases of the decay into four muons. The effect
of the modified VMD model is noticeable, but not very large. A test
whether a VMD model is needed or not and especially which version of the
VMD model should be applied will be at least very difficult for these decay modes due
to the lack of precise data.

\section{$P \to l^+ l^-$}
The last of the decays via the triangle anomaly that we want to discuss is the one
of a pseudoscalar meson into a lepton pair $P\to l^+ l^-$. As mentioned in the
Chapter 3.2 a lot of work was done there before.\\

\medskip

As discussed in Chapter 3 we can give a value for the subtraction constant
${\cal A}(0)$, see \equa{subconstA}, which contains all of the nontrivial dynamics. The results calculated with the
hidden gauge model and the modified VMD model are compared to theoretical
values and experimental data in Table \ref{tab:A(0)}. \\

\begin{table}[!hbt]
\begin{center}
\renewcommand{\arraystretch}{1.3}
\begin{tabular}{l|c|c|c|c}
  & hidden gauge & modified VMD  & \cite{Dorokhov:2007bd}& exp. data \cite{Abouzaid:2006kk} \\ \hline\hline
${\cal A}_e(q^2=0)$&$-21.42\pm0.04$&$-21.56\pm0.06$&$-21.9\pm0.3$&$-18.6\pm0.9$\\ \hline
${\cal A}_\mu(q^2=0)$&$-5.42\pm0.02$& $-5.57\pm0.06$
\end{tabular}
  \caption{Subtraction constant ${\cal A}(q^2=0)$ for decays into electrons and
    muons (denoted by the index $e$ and $\mu$). \label{tab:A(0)}}
\end{center}
\end{table}

In \cite{Dorokhov:2007bd} the authors presented the values of ${\cal A}(0)$ and the resulting
branching ratios calculated for various phenomenological models. All these
results basically fall into the same region (the predictions from the quark models
are even higher in absolute magnitude, varying between $-22$ and $-24.5$), higher in magnitude than the
experimental data. Our result matches very nicely the theoretical predictions,
but does not overlap with the experimental data.\\

\subsection*{$\pi^0 \to e^+ e^-$}

This situation persists for the branching ratios of the decay $\pi^0\to e^+ e^-$. We present the
unitary bound as discussed in Chapter 3 and also the branching ratios with
respect to the total decay rate and the one into two photons in Table \ref{tab:piee}.

\begin{table}[!hbt]
\begin{center}
\renewcommand{\arraystretch}{1.3}
\begin{tabular}{lc|c|c|c}
 && unitary bound & hidden gauge & modified VMD \\ \hline\hline
$\Gamma_{\pi^0 \to e^+ e^-}/ \Gamma_{total}$
&{\footnotesize$\left(10^{-8}\right)$}& $\ge 4.69$ &
$6.38\pm0.01$&$6.33\pm0.02$ \\ \hline
$\Gamma_{\pi^0 \to e^+ e^-}/ \Gamma_{\pi^0\to\gamma\gamma}$  &{\footnotesize$\left(10^{-8}\right)$}&
$\ge 4.75$ & $6.45\pm0.01$&$6.41\pm0.02$ 
\end{tabular}
  \caption{Unitary bound and branching ratio for the decay $\pi^0 \to e^+
    e^-$ calculated with different VMD models.\label{tab:piee}}
\end{center}
\end{table}

\begin{center}
\begin{figure}[!hbt]
\begin{center}
    \psfrag{x}[tc]{ \large Year}
    \psfrag{y}[c]{ \large $BR(\pi^0\to e^+ e^-)\times 10^{-8}$}
    \psfrag{a}[tc]{\,\,\,\,\,\,\,\,\,\,\,\,\,\,\,\,\,\,\,\small Babu et al.}
    \psfrag{b}[tc]{\small \,\,\,\,\,\,\,\,\,\,\,\,Bergstrom}
    \psfrag{c}[tc]{\small \,\,\,\,\,\,Savage et al.}
    \psfrag{d}[tc]{\small \,\,\,\,\,\,\,\,\,\,\,\,\,\,\,\,\,\,\,\,\,\,\,Ametller et al.}
    \psfrag{e}[tc]{ unitary bound}
    \psfrag{f}[tc]{ CLEO bound}
    \psfrag{g}[tc]{\small \,\,\,\,\,\,\,\,\,\,\,\,\,\,\,Gomez Dumm et al.}
    \psfrag{h}[tc]{\small \,\,\,\,\,\,\,Knecht et al.}
    \psfrag{i}[tc]{\small \,\,\,\,\,\,\,KTeV}
    \psfrag{j}[tc]{\small \!\!\!\!\!\!\!\!Dorokhov et al.}
    \includegraphics[width=16cm]{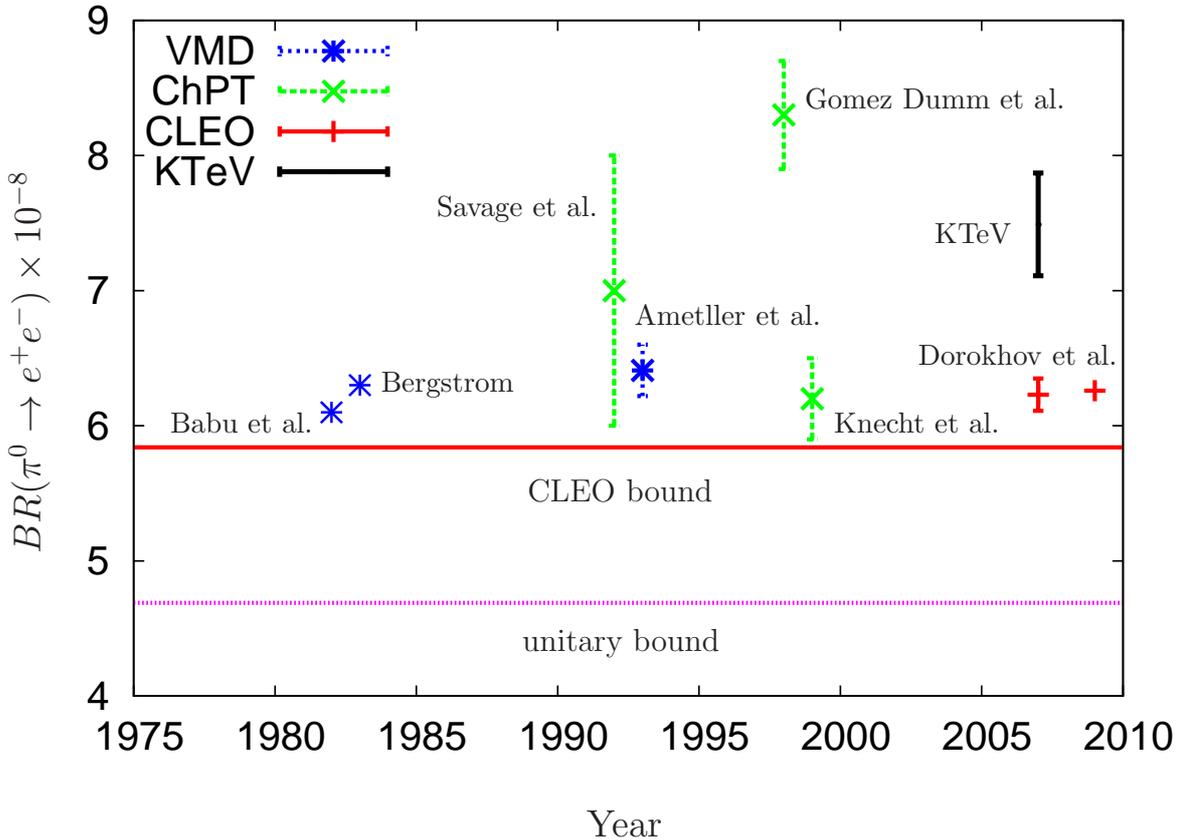}
\caption{Theoretical values calculated with VMD models \cite{Babu:1982yz,Bergstrom:1983ay,Ametller:1993we}, with ChPT \cite{Savage:1992ac,GomezDumm:1998gw,Knecht:1999gb}, modern calculations by \cite{Dorokhov:2007bd,Dorokhov:2009xs} (CLEO), and recent experimental data \cite{Abouzaid:2006kk} of the KTeV Collaboration  of the branching ratio  $BR(\pi^0\to e^+ e^-)$ according to \cite{Dorokhov:2008uk}.\label{fig:theopiee}}

\end{center}
\end{figure}
\end{center}

For the decay $\pi^0 \to e^+ e^-$ an overview of theoretical predictions
calculated via different models and a comparison with the latest KTeV result
\cite{Abouzaid:2006kk} can be found in \cite{Dorokhov:2008uk}, see Figure \ref{fig:theopiee}.

Most of these theoretical values can not describe the data
(\cite{Abouzaid:2006kk}) and this is the same for our calculations. The
values calculated via VMD models (\cite{Babu:1982yz,Bergstrom:1983ay,Ametller:1993we}) are basically consistent with each other and
also with the values of modern calculations \cite{Dorokhov:2007bd,Dorokhov:2009xs}. The
overview contains calculations of ChPT
\cite{Savage:1992ac,GomezDumm:1998gw,Knecht:1999gb}. The calculated values of
\cite{Savage:1992ac} have very large uncertainties and are therefore the only
values, which can describe the data. The values given in
\cite{GomezDumm:1998gw} and \cite{Knecht:1999gb} have smaller uncertainties and
an overlap with the values of \cite{Savage:1992ac}. Though these values do not
overlap with each other. The authors of \cite{GomezDumm:1998gw} provided a
consistent description of the decays $\pi^0 \to e^+ e^-$, $\eta \to \mu^+
\mu^-$, and $K_L\to \mu^+\mu^-$. According to \cite{Knecht:1999gb} these
calculations did not take contributions of $1/N_C$ suppressed counterterms into account. In contrast the calculations of \cite{Knecht:1999gb} contain
these contributions. Note that the values of \cite{Knecht:1999gb} are
consistent with the values of VMD models and of modern calculations.\\

\medskip

We will compare our results to the recent calculations given in \cite{Dorokhov:2007bd}
and \cite{Dorokhov:2009xs} and the existing experimental data explicitly, and also show
briefly the values of the various ChPT calculations.\\
The theoretical calculations of \cite{Dorokhov:2007bd} and \cite{Dorokhov:2009xs} and experimental data are given in Table \ref{tab:pieeA}.

\begin{table}[!hbt]
\begin{center}
\renewcommand{\arraystretch}{1.3}
\begin{tabular}{lc|c|c|c}
 && \cite{Dorokhov:2007bd} &\cite{Dorokhov:2009xs}& exp. data \cite{Abouzaid:2006kk}\\ \hline\hline
$\Gamma_{\pi^0 \to e^+ e^-}/ \Gamma_{total}$
&{\footnotesize$\left(10^{-8}\right)$}&  $6.23\pm0.09$&$6.26$&$7.49\pm0.38$
\end{tabular}
  \caption{Theoretical values and experimental data of the branching ratios of the decay $\pi^0\to e^+
    e^-$. \label{tab:pieeA}}
\end{center}
\end{table}

Our calculated values agree with these theoretical
predictions. But although our value is a little bit higher, the theoretical
predictions are still approximately three standard deviations lower than the
experimental data.\\
The branching ratio calculated by \cite{Savage:1992ac} is the following:
\begin{equation}
 \Gamma_{\pi^0 \to e^+ e^-}/ \Gamma_{total}= (7\pm1)\,\times \,10^{-8} .
\end{equation}
As mentioned this value overlaps with the experimental data, but has a very large uncertainty. The braching ratio of \cite{GomezDumm:1998gw} is
\begin{equation}
 \Gamma_{\pi^0 \to e^+ e^-}/ \Gamma_{total}= (8.3\pm0.4)\,\times \,10^{-8} .
\end{equation}
This value is higher than all other calculated values and also higher than the
data. The values of \cite{Knecht:1999gb} are given with respect to the decay
into two photons and are consistent with our values:
\begin{equation}
 \Gamma_{\pi^0 \to e^+ e^-}/ \Gamma_{\pi^0\to\gamma\gamma}= (6.2\pm0.3)\,\times \,10^{-8} .
\end{equation}

\subsection*{$\eta \to l^+ l^-$}

We will now discuss the situation in the $\eta$-sector. In addition to the decay
$\eta\to e^+e^-$ also the decay $\eta\to \mu^+\mu^-$ is possible. Our
calculated values are given in Table \ref{tab:etall}.

\begin{table}[!hbt]
\begin{center}
\renewcommand{\arraystretch}{1.3}
\begin{tabular}{lc|c|c|c}
 && unitary bound & hidden gauge & modified VMD  \\ \hline\hline
$\Gamma_{\eta\to e^+ e^-} / \Gamma_{total}$
&{\footnotesize$\left(10^{-9}\right)$}& $\ge 1.78$ & $4.68\pm0.01$& $4.65\pm0.01$ \\ \hline
$\Gamma_{\eta\to e^+ e^-} / \Gamma_{\eta\to\gamma\gamma}$  &{\footnotesize$\left(10^{-9}\right)$}&
$\ge 4.51$ & $11.89\pm0.02$& $11.84\pm0.03$ \\ \hline \hline
$\Gamma_{\eta\to \mu^+ \mu^-} / \Gamma_{total}$
&{\footnotesize$\left(10^{-6}\right)$}& $\ge 4.36$ & $4.87\pm0.02$& $4.96\pm0.06$ \\ \hline
$\Gamma_{\eta\to \mu^+ \mu^-} / \Gamma_{\eta\to\gamma\gamma}$  &{\footnotesize$\left(10^{-6}\right)$}&
$\ge 11.04$ & $12.40\pm0.02$& $12.62\pm0.95$
\end{tabular}
  \caption{Unitary bound and branching ratios for the decays $\eta \to e^+e^-$
    and $\eta \to \mu^+\mu^-$ calculated with different VMD models. \label{tab:etall}}
\end{center}
\end{table}

We can compare our values to the theoretical ones and the experimental
data, see Table \ref{tab:etallA}.

\begin{table}[!hbt]
\begin{center}
\renewcommand{\arraystretch}{1.3}
\begin{tabular}{lc|c|c|c}
 && \cite{Dorokhov:2007bd} &\cite{Dorokhov:2009xs}& exp. data
 \cite{Amsler:2008zzb, Abegg:1994wx, Berlowski:2008zz}\\ \hline\hline
$\Gamma_{\eta \to e^+ e^-}/ \Gamma_{total}$
&{\footnotesize$\left(10^{-9}\right)$}&  $4.60\pm0.09$&$5.24$&$\le 2.7\times 10^4$ \\
\hline
$\Gamma_{\eta \to \mu^+ \mu^-}/ \Gamma_{total}$
&{\footnotesize$\left(10^{-6}\right)$}&  $5.12\pm0.276$&$4.64$&$5.8\pm0.8$
\end{tabular}
  \caption{Theoretical values and experimental data of the branching ratios of the decays $\eta\to e^+
    e^-$ and $\eta \to \mu^+\mu^-$. \label{tab:etallA}}
\end{center}
\end{table}

Our results fall between the two different theoretical calculations given
in \cite{Dorokhov:2007bd} and \cite{Dorokhov:2009xs}. It is interesting to see
that the approach given in \cite{Dorokhov:2009xs}, which is a little bit closer to the
experimental data of the decay $\pi^0 \to e^+ e^-$ gives now worse predictions
for the decay $\eta \to \mu^+ \mu^-$ if we compare it to the experimental
data. As in \cite{Dorokhov:2007bd}, we also reach a small
overlap with the experimental data. For the decay $\eta \to e^+ e^-$ all
values meet the experimental bound.\\
The authors of \cite{Savage:1992ac} gave predictions for the decay $\eta \to e^+ e^-$:
\begin{equation}
 \Gamma_{\eta \to e^+ e^-}/ \Gamma_{total}= (5\pm1)\,\times \,10^{-9} .
\end{equation}
One can see again the large theoretical uncertainty. The authors of Ref. \cite{GomezDumm:1998gw} calculated this branching ratio as follows:
\begin{equation}
 \Gamma_{\eta \to e^+ e^-}/ \Gamma_{total}= (5.8\pm0.2)\,\times \,10^{-9} .
\end{equation}
In Ref. \cite{Knecht:1999gb} there are values given for both decays $\eta \to e^+ e^-$ and $\eta\to \mu^+ \mu^-$:
\begin{eqnarray}
 \Gamma_{\eta \to e^+ e^-}/ \Gamma_{\eta\to\gamma\gamma} &=& (1.15\pm0.05)\,\times \,10^{-8} \nn \\
\Gamma_{\eta \to \mu^+ \mu^-}/ \Gamma_{\eta\to\gamma\gamma} &=& (1.4\pm0.2)\,\times \,10^{-5} .
\end{eqnarray}
The values of \cite{Knecht:1999gb} are again consistent with our calculations.

\subsection*{$\eta' \to l^+ l^-$}

The values of the $\eta'$-decays are given in Table
\ref{tab:eta'll}. Experimental data do not exist.

\begin{table}[!hbt]
\begin{center}
\renewcommand{\arraystretch}{1.3}
\begin{tabular}{lc|c|c|c|c|c}
 && un. bound & hidden gauge & modified VMD& \cite{Dorokhov:2007bd} &\cite{Dorokhov:2009xs}  \\ \hline\hline
$\Gamma_{\eta'\to e^+ e^-}/ \Gamma_{tot}$
&\!\!\!\!\!\!\!{\scriptsize$\left(10^{-10}\right)$}& $\ge 0.36$ & $1.154\pm0.058$&
$1.147\pm0.063$ & $1.178\pm0.014$ & 1.86\\ \hline
$\Gamma_{\eta'\to e^+ e^-}/\Gamma_{\eta'\to\gamma\gamma}$  &\!\!\!\!\!\!\!{\scriptsize$\left(10^{-9}\right)$}&
$\ge 1.72$ & $54.85\pm0.29$& $54.19\pm0.42$&&\\ \hline \hline
$\Gamma_{\eta'\to \mu^+ \mu^-}/\Gamma_{tot}$
&\!\!\!\!\!\!\!{\scriptsize$\left(10^{-7}\right)$}& $\ge 1.35$ & $1.17\pm0.07$& $1.14\pm0.13$&$1.364\pm0.010$&1.30\\ \hline
$\Gamma_{\eta'\to \mu^+ \mu^-}/\Gamma_{\eta'\to\gamma\gamma}$  &\!\!\!\!\!\!\!{\scriptsize$\left(10^{-6}\right)$}&
$\ge 6.38$ & $5.45\pm0.15$& $5.34\pm0.11$&&
\end{tabular}
  \caption{Unitary bound and branching ratios for the decays $\eta'\to e^+
    e^-$ and $\eta' \to \mu^+\mu^-$ calculated with different VMD models.\label{tab:eta'll}}
\end{center}
\end{table}

It is remarkable that the branching ratio for the decay $\eta' \to \mu^+ \mu^-$
is lower than the lower limit given by the unitary bound as can be seen in
our calculations and the ones of \cite{Dorokhov:2009xs}. The reason is the
additional imaginary part which occurs because one of the intermediate vector mesons can go
on-shell. For the $\eta'$ decays our calculated values are lower than both of
the other theoretical values.\\
In the $\eta'$ sector there are only ChPT calculations done by \cite{GomezDumm:1998gw}:
\begin{eqnarray}
 \Gamma_{\eta' \to e^+ e^-}/ \Gamma_{total}&=& (1.5\pm0.1)\,\times \,10^{-10} \nn \\
 \Gamma_{\eta' \to \mu^+ \mu^-}/ \Gamma_{total}&=& (2.1\pm0.3)\,\times \,10^{-7} .
\end{eqnarray}
The branching ratio for the decay $\eta' \to e^+ e^-$ is very close to other theoretical values, while the one of $\eta' \to \mu^+ \mu^-$ is higher.
\subsection*{Summary}
In the $\pi^0$-sector our calculated values are a little closer to the
experimental data, but still three standard deviations lower. Our values differ slightly from the ones of \cite{Dorokhov:2007bd} and \cite{Dorokhov:2009xs} in
the $\eta$- and $\eta'$ sector, but are still in their range. A preference of
any model can not be given.

\section{$P\to \pi^+ \pi^- \gamma$  and  $P\to \pi^+ \pi^- e^+ e^-$}
In this Section we will present the final results for the branching ratios of the
decays $P\to \pi^+ \pi^- \gamma$  and  $P\to \pi^+ \pi^- e^+ e^-$ for the
hidden gauge and the modified VMD model. There
exist several theoretical calculations (\cite{Nissler:2007zz},
\cite{Picciotto:1991ae}, \cite{Picciotto:1993aa}) and experimental
data (\cite{Amsler:2008zzb}, \cite{Ambrosino:2008cp}, \cite{Berlowski:2008zz}), especially in the
$\eta$ sector, to compare with. We will also
give predictions for the CP violating terms and the electric terms. Finally an upper limit for the
model coefficient $G$ will be given.\\

\subsection*{$\eta\to \pi^+ \pi^- \gamma$  and  $\eta\to \pi^+ \pi^- e^+ e^-$ }

\begin{figure}[!hbt]
  \begin{minipage}[b]{8.0 cm}
    \psfrag{x}[tc]{\tiny$\sqrt{s_{\pi\pi}}$ \hspace{0.2cm} [MeV]}
    \psfrag{hidden gauge}[l]{\!\!\!\!\! \tiny  hidden gauge}
    \psfrag{modified VMD}[l]{\!\!\!\!\! \tiny modified VMD}
    \psfrag{E-terms G=10}[c]{ \,\,\,\,\,\,\,\,\,\,\,\,\,\, \tiny E-terms G=10}
    \psfrag{y}[c]{\, \tiny$\partial \Gamma_{M / E}/\partial{\sqrt{s_{\pi\pi}}}$}
    \includegraphics[width=8.0cm]{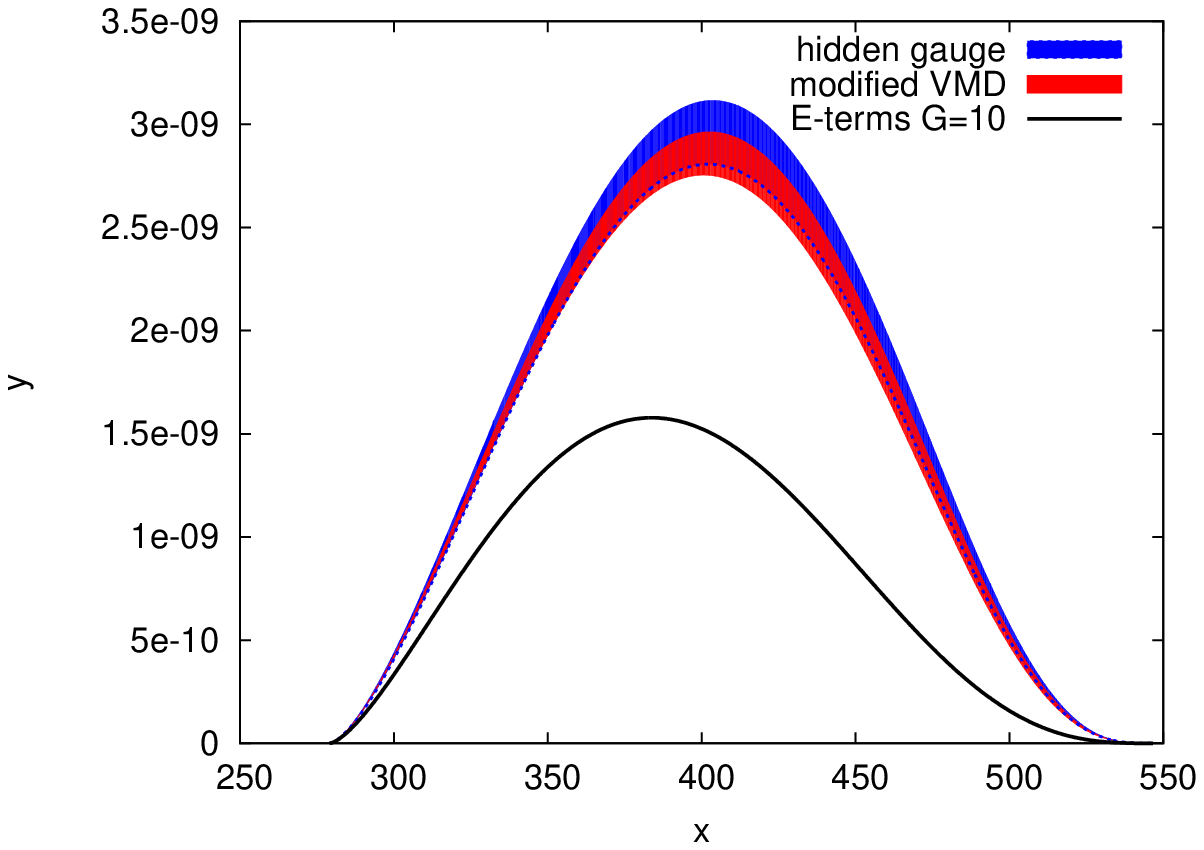}
  \end{minipage}
  \begin{minipage}[b]{8.0 cm}
    \psfrag{x}[tc]{ \tiny$\sqrt{s_{\pi\pi}}$\hspace{0.2cm} [MeV]}
    \psfrag{hidden gauge}[l]{\!\!\!\tiny hidden gauge}
    \psfrag{modified VMD}[l]{\!\!\!\!\! \tiny modified VMD}
    \psfrag{y}[c]{\tiny$-\partial \Gamma_{Asy}/\partial{\sqrt{s_{\pi\pi}}}/G$}
    \includegraphics[width=8.0cm]{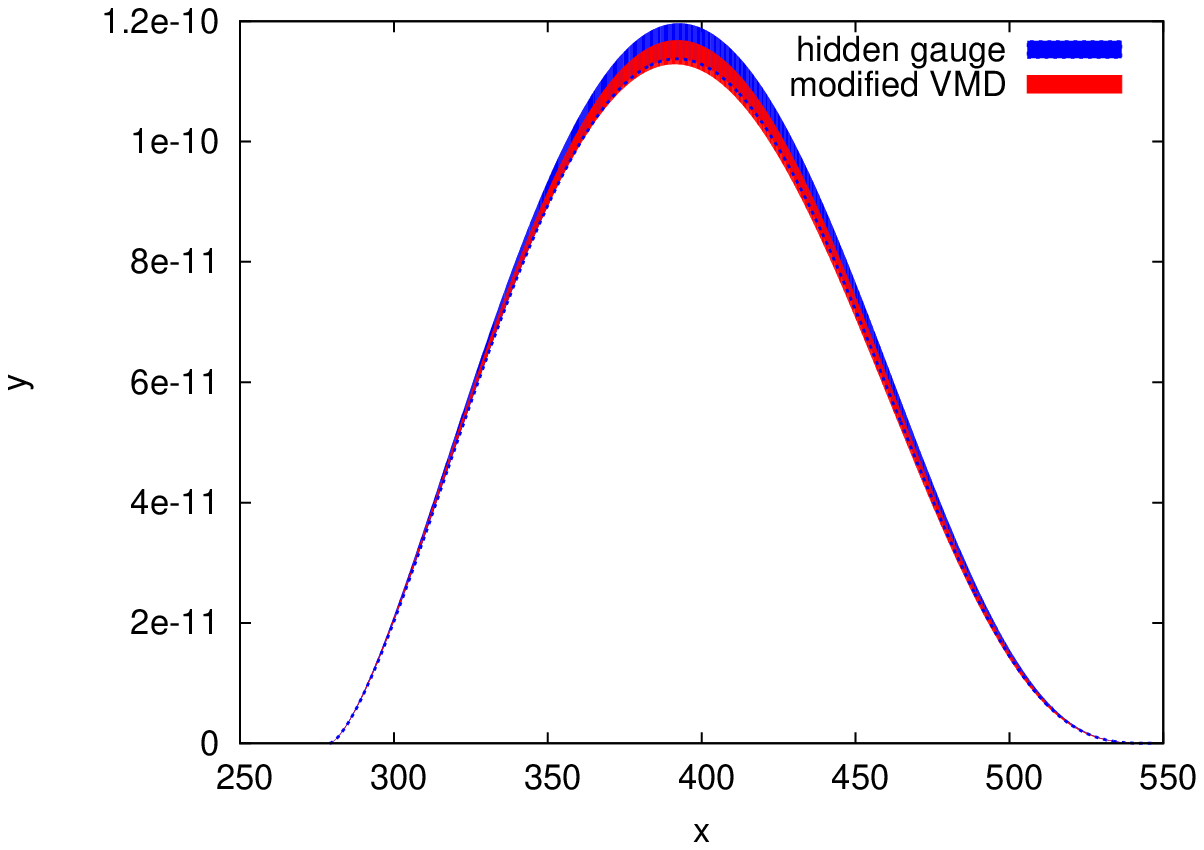}
  \end{minipage}
\caption{Dependency of the differential decay rate of the decay $\eta\to\pi^+\pi^-e^+e^-$ on
the change of the invariant mass of the pions.\label{fig:etapipieespi}}

\end{figure}

\begin{figure}[!hbt]
  \begin{minipage}[b]{8.0 cm}
    \psfrag{x}[tc]{\tiny$\sqrt{s_{ee}}$ \hspace{0.2cm} [MeV]}
    \psfrag{hidden gauge}[l]{\!\!\!\!\! \tiny hidden gauge}
    \psfrag{modified VMD}[l]{\!\!\!\tiny modified VMD}
    \psfrag{E-terms G=10}[l]{\!\!\!\!\!\! \tiny E-terms G=10}
    \psfrag{y}[c]{\tiny$\partial \Gamma_{M / E}/\partial{\sqrt{s_{ee}}}$}
    \includegraphics[width=8.0cm]{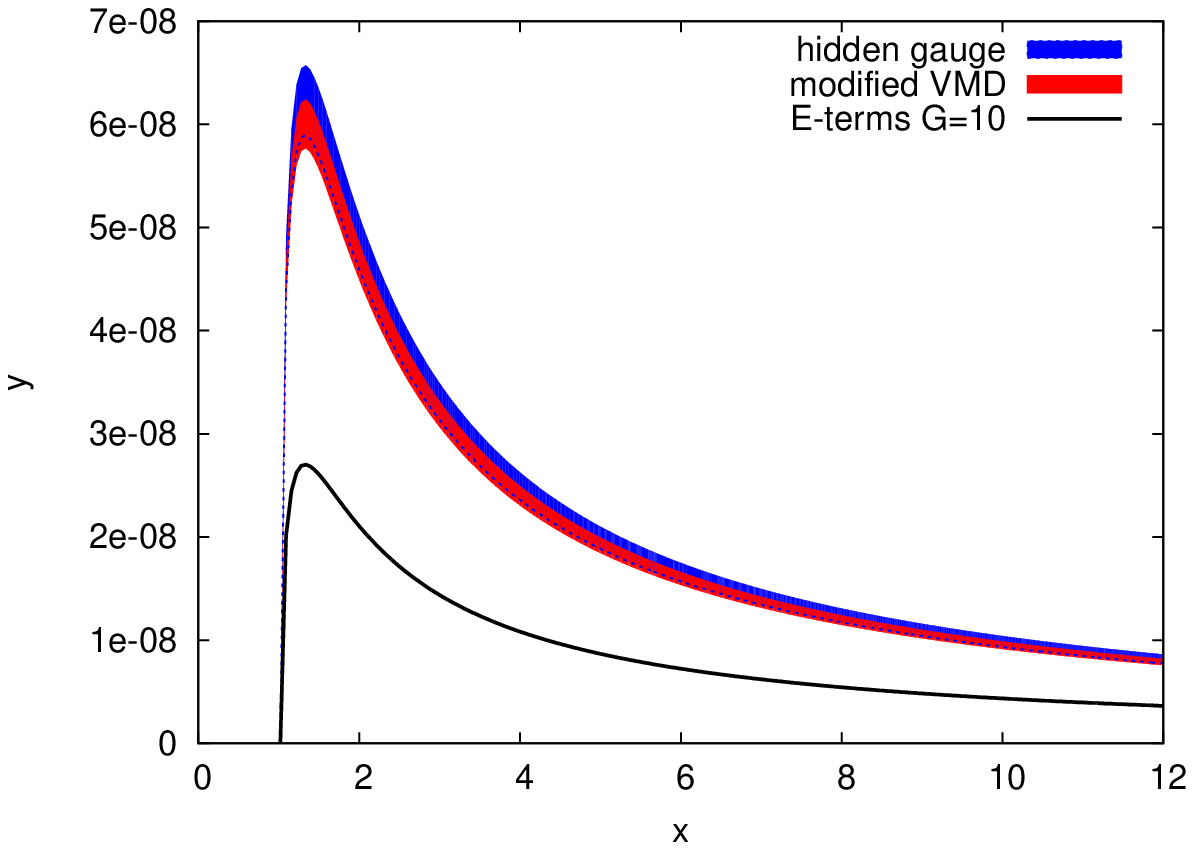}
  \end{minipage}
  \begin{minipage}[b]{8.0 cm}
    \psfrag{x}[tc]{\tiny$\sqrt{s_{ee}}$\hspace{0.2cm} [MeV]}
    \psfrag{hidden gauge}[l]{\!\!\!\!\! \tiny hidden gauge}
    \psfrag{modified VMD}[l]{\!\!\!\tiny modified VMD}
    \psfrag{y}[c]{\tiny-$\partial \Gamma_{Asy}/\partial{\sqrt{s_{ee}}}/G$}
    \includegraphics[width=8.0cm]{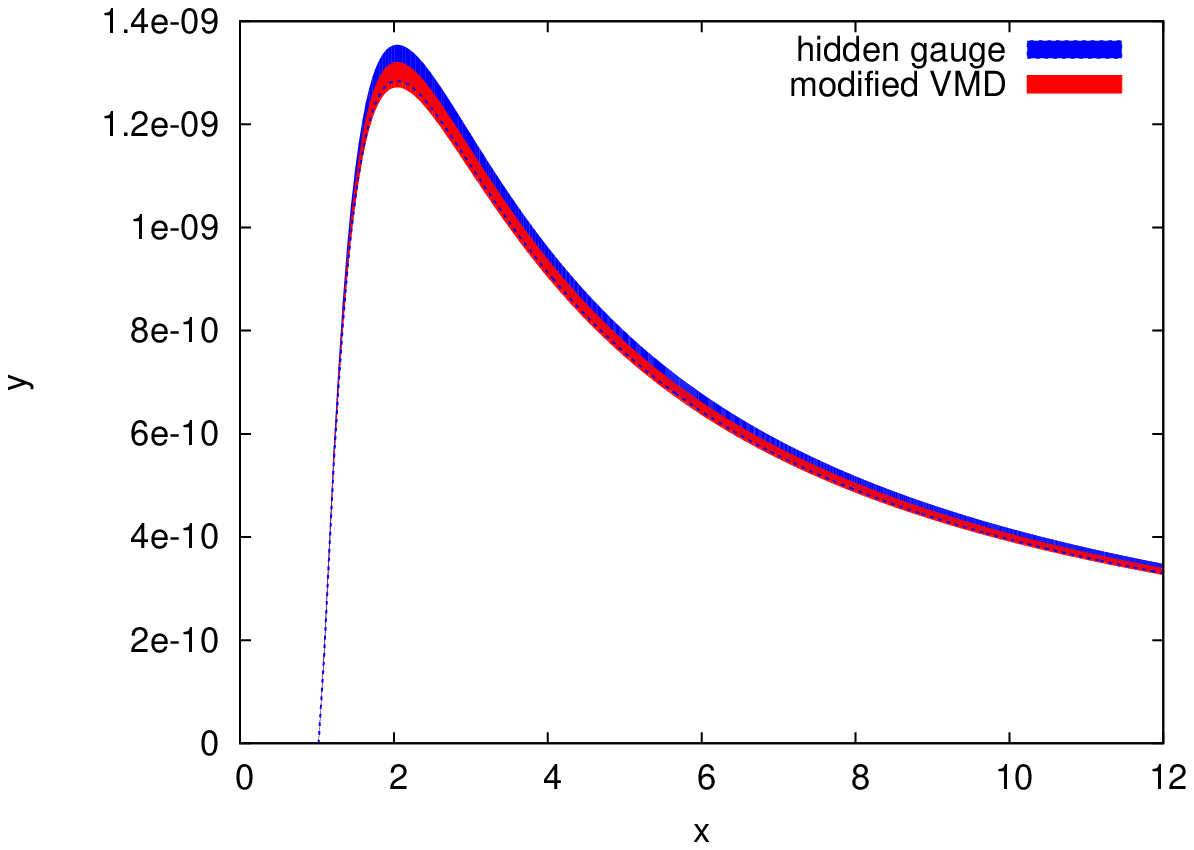}
  \end{minipage}
\caption{Dependency of the differential decay rate of the decay $\eta\to\pi^+\pi^-e^+e^-$ on
the change of the invariant mass of the electrons.\label{fig:etapipieese}}

\end{figure}

We start with the dependence of the rate of the decay $\eta \to
\pi^+\pi^- e^+ e^-$ on the invariant mass of the
decaying particle. We plot the leading (magnetic) term calculated, with the hidden gauge
model and the modified VMD model, and the electric term which is
scaled up to make it comparable (see Figures
\ref{fig:etapipieespi} and \ref{fig:etapipieese}).

\begin{figure}[!hbt]
  \begin{minipage}[b]{8.0 cm}
    \psfrag{x}[tc]{\tiny$\sqrt{s_{\pi\pi}}$ \hspace{0.2cm} [MeV]}
    \psfrag{y}[c]{\tiny$\partial \Gamma_{M / E}/\partial{\sqrt{s_{\pi\pi}}}$}
    \psfrag{hidden gauge}[c]{\,\,\,\,\,\,\,\,\,\,\,\tiny hidden gauge}
    \psfrag{modified VMD}[c]{\,\,\,\,\,\,\,\,\,\,\,\,\,\tiny modified VMD}
    \psfrag{E-terms G=5^1/2}[c]{\,\,\,\,\,\,\,\,\,\,\,\,\,\,\,\,\,\,\,\,\,\,\,\tiny E-terms $G=\sqrt{5}$}
    \includegraphics[width=8.0cm]{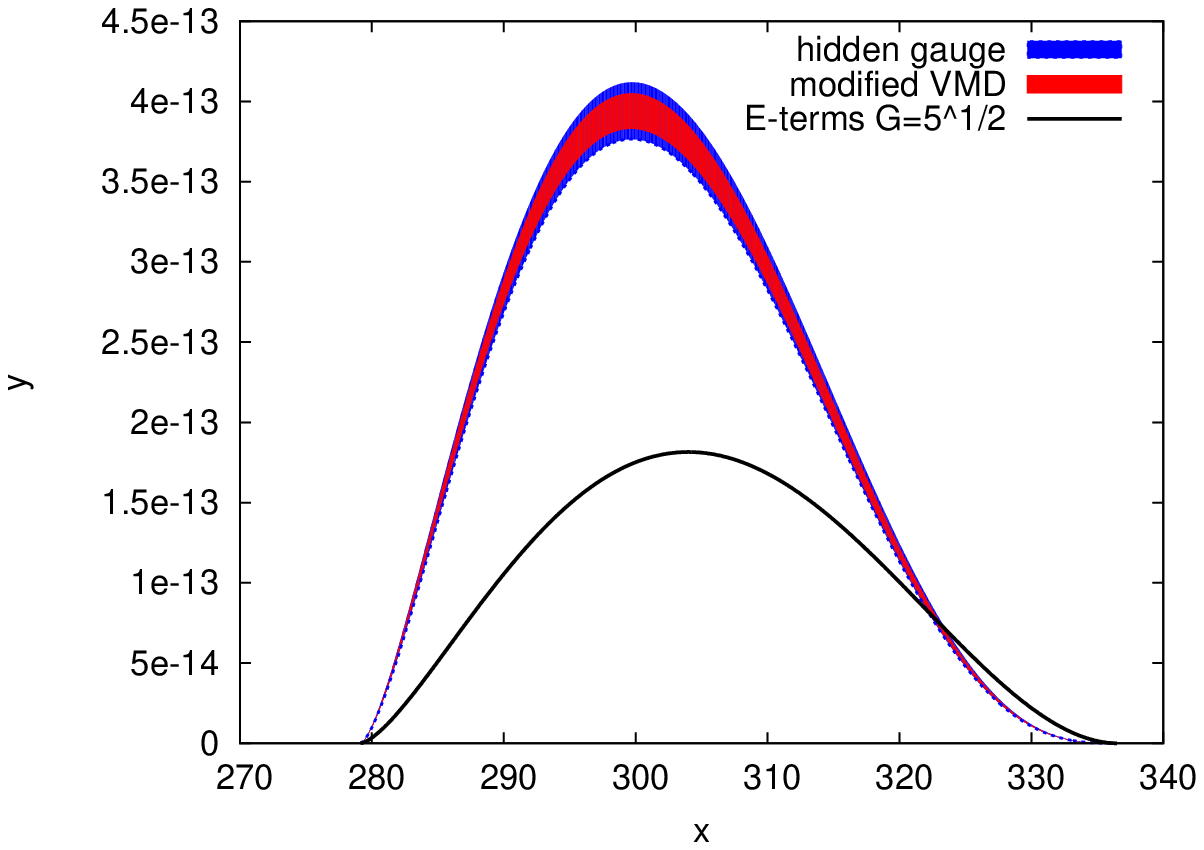}
  \end{minipage}
  \begin{minipage}[b]{8.0 cm}
    \psfrag{x}[tc]{\tiny$\sqrt{s_{\pi\pi}}$\hspace{0.2cm} [MeV]}
    \psfrag{y}[c]{\tiny-$\partial \Gamma_{Asy}/\partial{\sqrt{s_{\pi\pi}}}/G$}
    \psfrag{hidden gauge}[c]{\,\,\,\,\,\,\,\,\,\,\,\tiny hidden gauge}
    \psfrag{modified VMD}[c]{\,\,\,\,\,\,\,\,\,\,\,\,\,\tiny modified VMD}
    \includegraphics[width=8.0cm]{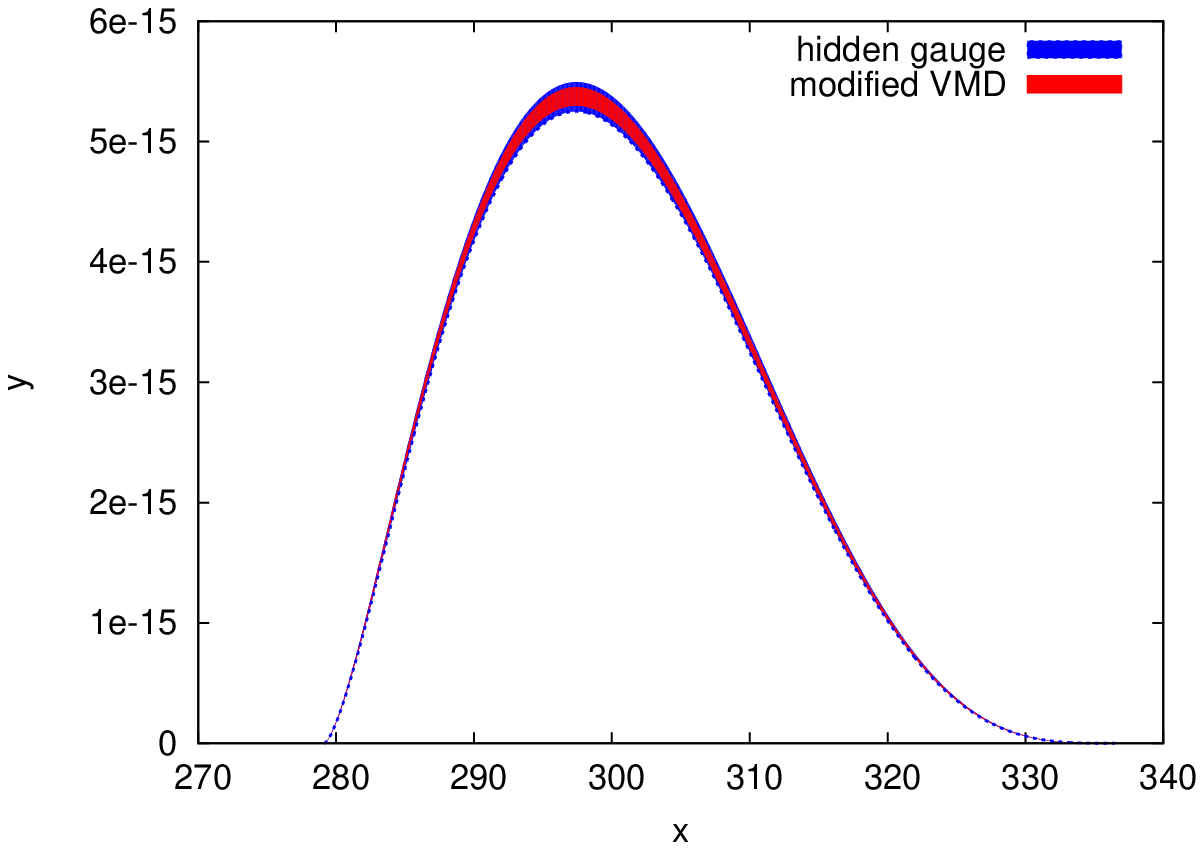}
  \end{minipage}
\caption{Dependency of the differential decay rate of the decay $\eta\to\pi^+\pi^-\mu^+\mu^-$ on
the change of the invariant mass of the pions.\label{fig:etapipimumuspi}}

\end{figure}

\begin{figure}[!hbt]
  \begin{minipage}[b]{8.0 cm}
    \psfrag{x}[tc]{\tiny$\sqrt{s_{\mu\mu}}$ \hspace{0.2cm} [MeV]}
    \psfrag{y}[c]{\tiny$\partial \Gamma_{M / E}/\partial{\sqrt{s_{\mu\mu}}}$}
    \psfrag{hidden gauge}[c]{\,\,\,\,\,\,\,\,\,\,\,\tiny hidden gauge}
    \psfrag{modified VMD}[c]{\,\,\,\,\,\,\,\,\,\,\,\,\,\tiny modified VMD}
    \psfrag{E-terms G=5^1/2}[c]{\,\,\,\,\,\,\,\,\,\,\,\,\,\,\,\,\,\,\,\,\,\,\,\tiny E-terms $G=\sqrt{5}$}
    \includegraphics[width=8.0cm]{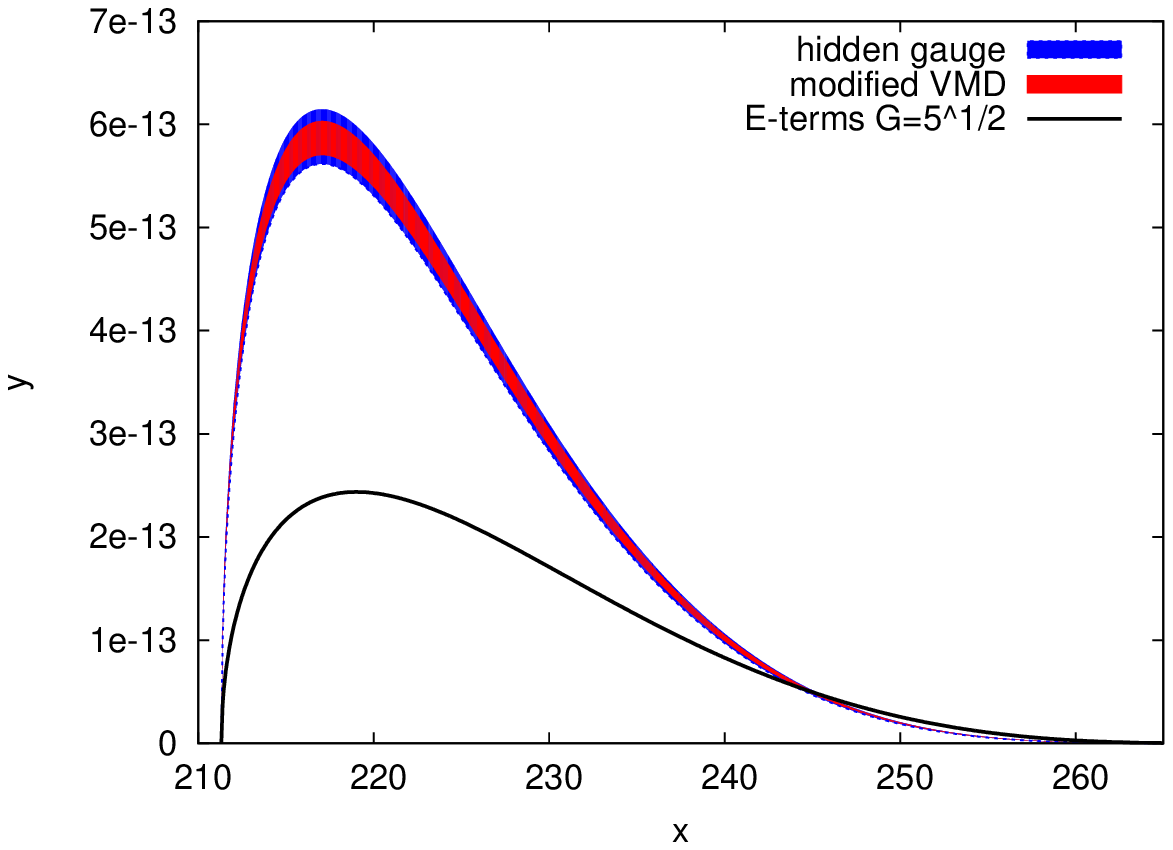}
  \end{minipage}
  \begin{minipage}[b]{8.0 cm}
    \psfrag{x}[tc]{\tiny$\sqrt{s_{\mu\mu}}$\hspace{0.2cm} [MeV]}
    \psfrag{y}[c]{\tiny-$\partial \Gamma_{Asy}/\partial{\sqrt{s_{\mu\mu}}}/G$}
    \psfrag{hidden gauge}[c]{\,\,\,\,\,\,\,\,\,\,\,\tiny hidden gauge}
    \psfrag{modified VMD}[c]{\,\,\,\,\,\,\,\,\,\,\,\,\,\tiny modified VMD}
    \includegraphics[width=8.0cm]{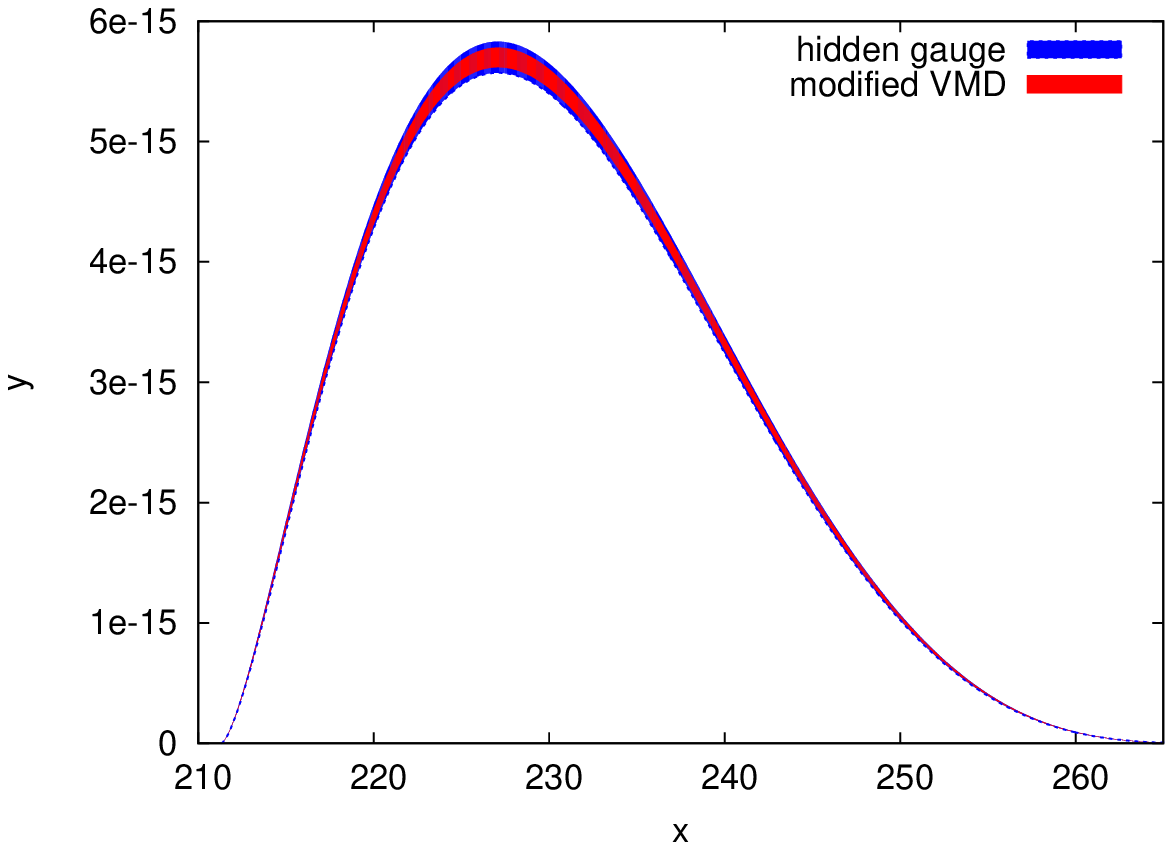}
  \end{minipage}
\caption{Dependency of the differential decay rate of the decay $\eta\to\pi^+\pi^-\mu^+\mu^-$ on
the change of the invariant mass of the muons.\label{fig:etapipimumusmu}}

\end{figure}

The behavior is quite similar to the graphs we showed before. With respect to
the dependence on
the invariant mass of the electrons, there is a large peak at low energies. The
behavior of the invariant mass dependence of the pions is much broader
with a smaller peak.\\
For the decay $\eta\to \pi^+\pi^-\mu^+\mu^-$, the graphs are shown in Figures
\ref{fig:etapipimumuspi} and \ref{fig:etapipimumusmu}.

It is interesting that in these decays the electric terms seem to be much bigger
than in the previous case. But this is actually not the case if one compares the calculated values of the decay rates and branching ratios.\\
These values for the $\eta$ sector are listed in Table
\ref{tab:etappee}. Here the branching ratios relative to the total $\eta$
decay width and to the width of the $\eta \to \pi^+ \pi^- \gamma$-decay are listed. The
asymmetry term $A_\phi$ is defined according to Eq. (2) of \cite{Gao:2002gq} with
the mixed term normalized to the total $\eta \to \pi^+ \pi^- \mu^+ \mu^-$
width, see Eq. \equa{ACP} of Chapter 3. The electric terms are independent of the various VMD models and
normalized to the total width.
\begin{table}[!hbt]
\begin{center}
\renewcommand{\arraystretch}{1.3}
\begin{tabular}{lc|c|c}
 && hidden gauge & modified VMD  \\ \hline\hline
\multicolumn{4}{l}{$\eta\to \pi^+ \pi^- \gamma$}\\\hline
$\Gamma_{\pi^+\pi^- \gamma} / \Gamma_{total}$  &{\footnotesize$\left(10^{-2}\right)$}& \:\;\;$4.97 \pm 0.27$ & \:\;\;$4.79 \pm 0.19$ \\ \hline\hline
\multicolumn{4}{l}{$\eta\to \pi^+ \pi^- e^+e^-$}\\\hline
$\Gamma_{\pi^+\pi^-e^+e^-} / \Gamma_{total} $ &{\footnotesize$\left(10^{-4}\right)$}  & \:\;\;$3.14\pm0.17$ & \:\;\;$3.02 \pm 0.12$ \\ \hline
$\Gamma_{\pi^+\pi^-e^+e^-} / \Gamma_{\pi^+\pi^- \gamma} $ &{\footnotesize$\left(10^{-3}\right)$} & $\:\;\;6.32\pm0.01$ &$\:\;\; 6.33 \pm 0.01$ \\ \hline
$A_\phi/G$          &{\footnotesize$\left(10^{-2}\right)$}    & $-3.88\pm 0.11$ & $-3.95 \pm 0.08$ \\ \hline
${\rm E-terms}$      &{\footnotesize$\left(10^{-6}\right)$}     &
\multicolumn{2}{c}{  1.6 $G^2$ independent on VMD}\\ \hline\hline
\multicolumn{4}{l}{$\eta\to \pi^+ \pi^- \mu^+\mu^-$}\\\hline
$\Gamma_{\pi^+\pi^-\mu^+\mu^-} / \Gamma_{total} $ &{\footnotesize$\left(10^{-9}\right)$}  & \:\;\;$8.65\pm 0.39$ & \:\;\;$8.64 \pm 0.25$ \\ \hline
$\Gamma_{\pi^+\pi^-\mu^+\mu^-} / \Gamma_{\pi^+\pi^- \gamma} $&{\footnotesize$\left(10^{-7}\right)$}  &
\:\;\;$1.74\pm 0.02$ & \:\;\;$1.81 \pm 0.02$  \\ \hline
$A_\phi/G  $    &{\footnotesize$\left(10^{-2}\right)$}        & $-1.28\pm 0.03$ & $-1.28 \pm0.02$ \\ \hline
${\rm E-terms}$     &{\footnotesize$\left(10^{-9}\right)$ }     &
\multicolumn{2}{c}{  0.889 $G^2$ independent on VMD}\\
\end{tabular}
  \caption{Branching ratios for the decay $\eta\to \pi^+ \pi^- \gamma$,
    $\eta\to \pi^+ \pi^- e^+ e^-$ and $\eta\to \pi^+ \pi^- \mu^+\mu^-$
    calculated with different VMD models. \label{tab:etappee}}
\end{center}
  
\end{table}

One can see that the branching ratio of $\eta \to \pi^+ \pi^- \mu^+ \mu^-$ is
much smaller than the other branching ratios. This is due to the smaller phase space. The
electric terms are about two orders of magnitude smaller for the $\eta \to \pi^+ \pi^- e^+
e^-$ decay and one order of magnitude smaller for the $\eta \to \pi^+ \pi^- \mu^+
\mu^-$ decay, even if we set $G$ to one. So it
is obvious that the magnetic terms still determines the leading
contribution. The difference in  predictions of the various VMD models are
highly visible for the case of $\eta \to \pi^+ \pi^- \gamma$ and the $\eta \to \pi^+ \pi^- e^+
e^-$-decay. It is interesting that the CP-violating asymmetry term has a
negative sign. This is due to the negative sign of the VMD term relative to the
contact term. In the case of the magnetic term this term
appears squared, so it does not affect the sign there.\\
We can now compare our values with other data, see Table \ref{tab:etappee} and \ref{tab:etappeeA}. Note that the authors of
\cite{Picciotto:1991ae} and \cite{Picciotto:1993aa} used the hidden gauge
model as well, while \cite{Nissler:2007zz} calculated in the framework of unitary chiral
perturbation theory. So it will be more interesting to compare our values with
\cite{Nissler:2007zz}, because an agreement would be a verification of both models. The discrepancies between our values calculated via the hidden gauge model, and the ones given in
\cite{Picciotto:1991ae} for the $\eta \to \pi^+ \pi^- \gamma$ and in
\cite{Picciotto:1993aa} for the $\eta \to \pi^+ \pi^- e^+ e^-$ decay can be
explained by the modern values of the constants $f_\pi$, $f_0$ and $f_8$ as well as
the $\eta/\eta'$ mixing angle. The data of \cite{Ambrosino:2008cp} are the
most recent
published data and also the most interesting.
\begin{table}[!hbt]
\begin{center}
\renewcommand{\arraystretch}{1.3}
\begin{tabular}{lc|c|c||c|c}
 && \cite{Nissler:2007zz} & \cite{Picciotto:1991ae}, \cite{Picciotto:1993aa}&
 PDG \cite{Amsler:2008zzb}& KLOE \cite{Ambrosino:2008cp} \\ \hline\hline
\multicolumn{6}{l}{$\eta\to \pi^+ \pi^- \gamma$}\\\hline
$\Gamma / \Gamma_{total}$  &{\scriptsize$\left(10^{-2}\right)$}& $4.68^{+0.09}_{-0.09}$ & 5.22&$4.6\pm0.16$& \\ \hline\hline
\multicolumn{6}{l}{$\eta\to \pi^+ \pi^- e^+e^-$}\\\hline
$\Gamma / \Gamma_{total} $
&{\scriptsize$\left(10^{-4}\right)$}  & $2.99^{+0.08}_{-0.11}$ & $3.2\pm
0.3$ &$4.2\pm1.2$ & $2.68\pm0.09\pm0.07$\\ \hline
$\Gamma / \Gamma_{\pi^+\pi^- \gamma} $ &{\scriptsize$\left(10^{-3}\right)$} & $6.39^{+0.08}_{-0.11}$ &&$9.2\pm2.5$& \\ \hline
$A_\phi$          &{\scriptsize$\left(10^{-2}\right)$} & \multicolumn{1}{c}{}& &&$-0.6\pm2.5\pm1.8$ \\ \hline
\multicolumn{6}{l}{$\eta\to \pi^+ \pi^- \mu^+\mu^-$}\\\hline \hline
$\Gamma / \Gamma_{total} $
&{\scriptsize$\left(10^{-9}\right)$}  & $7.5^{+4.5}_{-2.7}$ & &$<3.6\cdot 10^{5}$& \\ \hline
$\Gamma / \Gamma_{\pi^+\pi^- \gamma} $&{\scriptsize$\left(10^{-7}\right)$}  &$1.61^{+0.95}_{-0.55}$ & &\multicolumn{2}{c}{}
\end{tabular}
  \caption[tab:etappee]{Theoretical values ( \cite{Nissler:2007zz},
    \cite{Picciotto:1991ae} and \cite{Picciotto:1993aa}) and experimental data
    (\cite{Amsler:2008zzb} and \cite{Ambrosino:2008cp}) of the branching ratios of the decay $\eta\to \pi^+ \pi^- \gamma$, $\eta\to \pi^+ \pi^- e^+ e^-$ and $\eta\to \pi^+ \pi^- \mu^+\mu^-$. \label{tab:etappeeA}}
\end{center}
  
\end{table}

For the $\eta \to \pi^+ \pi^- \gamma$-decay all theoretical values represent
the data very well. The modified VMD result is closer to
\cite{Nissler:2007zz} than the hidden gauge result.\\
The decay $\eta \to \pi^+ \pi^- e^+ e^-$ is more interesting. One
can see by comparing Table \ref{tab:etappee} and \ref{tab:etappeeA} that the hidden gauge model does not agree with \cite{Nissler:2007zz},
but that the modified model gives nearly the same value. The experimental situation
is even more interesting. The values of the hidden gauge model actually
represent the data of \cite{Amsler:2008zzb} better than the modified model,
although both are consistent. The recent data of the KLOE measurement
\cite{Ambrosino:2008cp} are much smaller than the other experimental data and
also have extremely small errors. Neither other theoretical calculations \cite{Nissler:2007zz} nor our values have an overlap. It would be very
interesting to see whether the measurement of the WASA@COSY experiment is compatible with
the one of KLOE, when their new data appear. (The most recent published data
of WASA@CELSIUS are already included in the PDG value and can be found in Ref.
\cite{Berlowski:2008zz}.) If the data were confirmed a new theoretical view on this decay
channel would be needed.\\
The ratio $\Gamma_{\pi^+\pi^-e^+e^-} / \Gamma_{\pi^+\pi^- \gamma} $ is
relatively constant within the various VMD models, which makes sense, since
the VMD factors should approximately cancel. The errors are smaller because some prefactors cancel also. Note that our values are in agreement with
the ones of \cite{Nissler:2007zz}.\\
The comparison of this ratio to the recent KLOE data is again of special interest (see Table \ref{tab:fracetappee}).\\

\begin{table}[!hbt]
\begin{center}
\renewcommand{\arraystretch}{1.7}
\begin{tabular}{lc|c|c|c}
&&$\frac{KLOE}{CLEO}$&$\frac{KLOE}{PDG}$&PDG\\ \hline\hline
 $\frac{\Gamma_{\pi^+\pi^-e^+e^--}}{\Gamma_{\pi^+\pi^- \gamma}}$&{\scriptsize$\left(10^{-3}\right)$} & $6.77\pm0.44$&$5.83\pm0.31$&$9.2\pm2.5$\\
\end{tabular}
  \caption[tab:fracetappee]{The branching ratio $\Gamma_{\pi^+\pi^-e^+e^--} / \Gamma_{\pi^+\pi^- \gamma}$ calculated with different data of the KLOE and CLEO experiment (\cite{Ambrosino:2008cp} and \cite{Lopez:2007ppa}) and the PDG value (\cite{Amsler:2008zzb}). \label{tab:fracetappee}}
\end{center}
\end{table}
Normalized to the PDG value the fraction is smaller than the theoretical predictions and has no overlap with these values. 
The normalization to the recent CLEO data (\cite{Lopez:2007ppa}) gives a larger value, because the data of the CLEO experiment is smaller than the PDG value ($BR(\eta \to \pi^+ \pi^- \gamma)=(3.96\pm0.14\pm0.14)\times 10^{-2}$). 
This ratio has a small overlap with the UChPT value (\cite{Nissler:2007zz}), as already pointed out in Ref. \cite{Ambrosino:2008cp}, but barely overlaps with our results. 
Note that the PDG value does not include the CLEO data.
The PDG value is larger than all the other values. It has no overlap with any theoretical value. A measurement of this branching ratio in one experiment would be usefull.\\
We can now take a look at the CP violating asymmetry term
$A_\phi$, see Eq. \equa{ACP} of Chapter 3. Theoretically it was estimated by Ref. \cite{Gao:2002gq} as $A_\phi=2.0
\cdot 10^{-2} G$ which is about half of our calculated value. $A_\phi$ was
measured for the first time via the forward-backward asymmetry by Ref.
\cite{Ambrosino:2008cp}. According to these data we can give constraints for
the model coefficient $G$. To make sure that our calculated values are
consistent with the KLOE data is has to be
\begin{equation}
-0.9<G<1.2.
\end{equation}
This agrees with the assumption that $G \le O(1)$ is natural.\\
For the $\eta \to \pi^+ \pi^- \mu^+ \mu^-$-decay, the choice of the VMD model does not have a
serious effect on the branching ratio, which is remarkable. Our values are
consistent with the ones of \cite{Nissler:2007zz} and meet the upper
experimental bound easily.

\subsection*{$\eta'\to \pi^+ \pi^- \gamma$  and  $\eta'\to \pi^+ \pi^- e^+ e^-$}
\begin{figure}[!hbt]
  \begin{minipage}[b]{8.0 cm}
    \psfrag{x}[tc]{\tiny$\sqrt{s_{\pi\pi}}$ \hspace{0.2cm} [MeV]}
    \psfrag{y}[c]{\tiny$\partial \Gamma_{M / E}/\partial{\sqrt{s_{\pi\pi}}}$}
    \psfrag{hidden gauge}[c]{\,\,\,\,\,\,\,\,\,\,\,\,\,\,\,\,\,\,\,\tiny hidden gauge}
    \psfrag{modified VMD}[c]{\,\,\,\,\,\,\,\,\,\,\,\,\,\,\,\,\,\tiny modified VMD}
    \psfrag{E-terms hidde gauge G=100}[l]{ \!\!\!\!\!\!\!\!\!\tiny E-terms G=100 hidden gauge}
    \includegraphics[width=8.0cm]{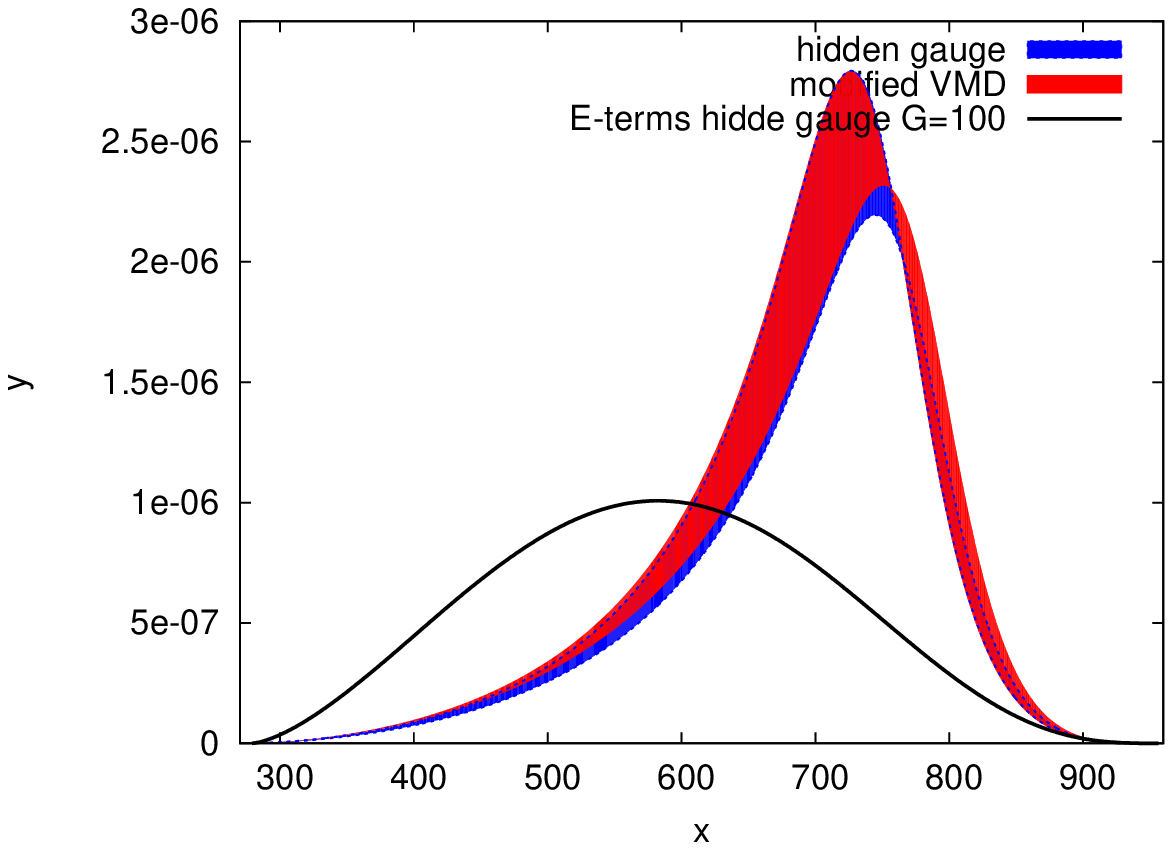}
  \end{minipage}
  \begin{minipage}[b]{8.0 cm}
    \psfrag{x}[tc]{\tiny$\sqrt{s_{\pi\pi}}$\hspace{0.2cm} [MeV]}
    \psfrag{y}[c]{\tiny$\partial \Gamma_{Asy}/\partial{\sqrt{s_{\pi\pi}}}/G$}
    \psfrag{hidden gauge}[c]{\,\,\,\,\,\,\,\,\,\,\,\,\,\,\,\,\,\tiny hidden gauge}
    \psfrag{modified VMD}[c]{\,\,\,\,\,\,\,\,\,\,\,\,\,\,\,\,\,\tiny modified VMD}
    \includegraphics[width=8.0cm]{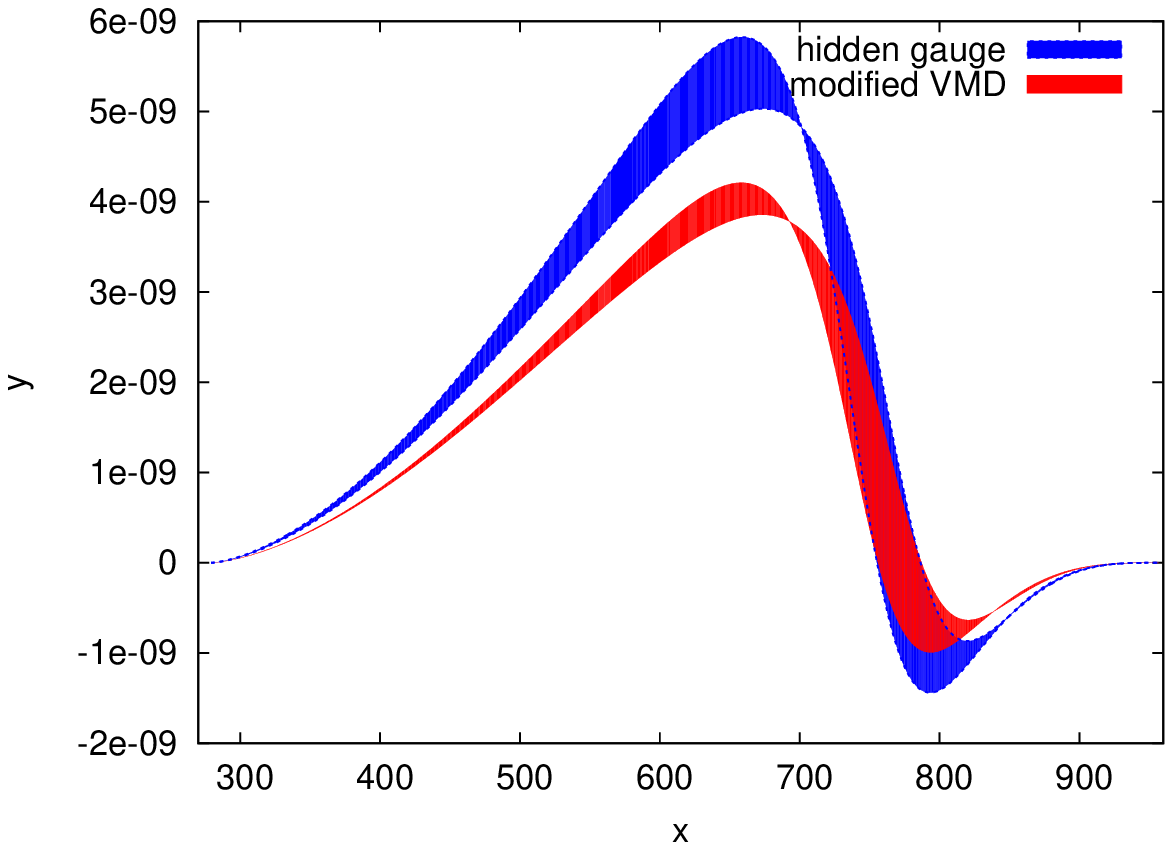}
  \end{minipage}
\caption{Dependency of the differential decay rate of the decay $\eta'\to\pi^+\pi^-e^+e^-$ on
the change of the invariant mass of the electrons.\label{fig:etappipieepi}}

\end{figure}

In the curves of the differential decay rate of the decay $\eta' \to
\pi^+\pi^- e^+ e^-$ there is a very strong contribution of the width as one
can see in Figure \ref{fig:etappipieepi}. For the leading magnetic
contribution the curve is shifted to the right for larger vector meson masses. The dependence of the asymmetry term
is very interesting.

For the branching ratios the situation is somewhat different
as we will discuss using the calculated values of the branching rations, see Table
\ref{tab:eta'ppee}.
\begin{table}[!hbt]
\begin{center}
\renewcommand{\arraystretch}{1.3}
\begin{tabular}{lc|c|c||c|c}
 && hidden gauge & modified VMD & th. values \cite{Nissler:2007zz}& exp. data \cite{Amsler:2008zzb}\\ \hline\hline
\multicolumn{6}{l}{$\eta'\to \pi^+ \pi^- \gamma$}\\\hline
$\Gamma_{\pi^+\pi^- \gamma} / \Gamma_{total}$ && $0.294 \pm 0.027$ & $0.308 \pm 0.016$ &  $0.294^{+0.027}_{-0.043}$&\\ \hline\hline
\multicolumn{6}{l}{$\eta'\to \pi^+ \pi^- e^+e^-$}\\\hline
$\Gamma_{\pi^+\pi^-e^+e^-} / \Gamma_{total} $  &{\footnotesize$\left(10^{-3}\right)$} & $2.17\pm0.21$ & $2.27 \pm 0.13$ &  $2.13^{+0.19}_{-0.32}$&$2.5^{+1.3}_{-1.0}$\\ \hline
$\Gamma_{\pi^+\pi^-e^+e^-} / \Gamma_{\pi^+\pi^- \gamma} $ &{\footnotesize$\left(10^{-3}\right)$} & $7.368\pm0.043$ & $7.370 \pm 0.059$ & $7.24^{+0.09}_{-0.15}$& \\ \hline
$A_\phi/G$            &{\footnotesize$\left(10^{-3}\right)$}  & $2.83\pm 0.29$ & $2.02 \pm 0.18$&\multicolumn{2}{l}{} \\ \hline
${\rm E-terms}$    &{\footnotesize$\left(10^{-7}\right)$}       &    $ 1.67 G^2$&$0.85 G^2$& \multicolumn{2}{l}{}\\ \hline\hline
\multicolumn{6}{l}{$\eta'\to \pi^+ \pi^- \mu^+\mu^-$}\\\hline
$\Gamma_{\pi^+\pi^-\mu^+\mu^-} / \Gamma_{total} $&{\footnotesize$\left(10^{-5}\right)$}   & $2.20\pm 0.30$ & $2.41 \pm 0.25$ &  $1.57^{+0.96}_{-0.75}$&$<23$\\ \hline
$\Gamma_{\pi^+\pi^-\mu^+\mu^-} / \Gamma_{\pi^+\pi^- \gamma} $ &{\footnotesize$\left(10^{-5}\right)$} &
$7.46\pm 0.32$ & $7.82 \pm 0.42$ & $5.4^{+3.6}_{-2.6}$& \\ \hline
$A_\phi/G  $    &{\footnotesize$\left(10^{-3}\right)$}        & $4.41\pm 0.33$ & $3.03 \pm0.19$& \multicolumn{2}{l}{}\\ \hline
${\rm E-terms}$      &{\footnotesize$\left(10^{-8}\right)$}     &    $ 1.28G^2$&$0.65G^2$& \multicolumn{2}{l}{}\\
\end{tabular}
  \caption{Branching ratios for the decay $\eta'\to \pi^+ \pi^- \gamma$,
    $\eta'\to \pi^+ \pi^- e^+ e^-$ and $\eta'\to \pi^+ \pi^- \mu^+\mu^-$
    calculated with
    different VMD models, other theoretical values and experimental data. \label{tab:eta'ppee}}
\end{center}
  
\end{table}

The calculation of the modified  model now gives larger
values than the calculations of the hidden gauge model. We can see that the
CP violating terms are one order of magnitude lower than we the ones in the
$\eta$-decays and are of positive sign. The electric terms
are visibly smaller than in the $\eta$-decays. As we pointed out before we can assume $G \le O(1)$
and so the electric terms do not have any effect.\\
In contrast to the $\eta$-decay, the agreement of our values and the ones
of \cite{Nissler:2007zz} does not hold any longer. For the $\eta' \to
\pi^+\pi^- \gamma$  and the $\eta'\to
\pi^+ \pi^- e^+ e^-$ channel the discrepancies are very small. However, for the
decay $\eta'\to \pi^+ \pi^- \mu^+\mu^-$ there is only a small overlap, although the
errors of \cite{Nissler:2007zz} are very large.\\
Both of our VMD models as well as the values of \cite{Nissler:2007zz} are
consistent with the data \cite{Amsler:2008zzb}, though our values are less constraint by
the data. More precise measurements, especially for the
$\eta'\to \pi^+ \pi^- \mu^+\mu^-$ decay, are needed to see which model
represents the data the best.\\

\subsection*{Summary}
For the $\eta$ decays the values calculated with the modified VMD model are
closer to the experimental data. One can see that the modified model is indeed
useful in the box anomaly sector. The lesser agreement for the $\eta'$-decays is
due to the scarce data. This could also be the reason why the values of
\cite{Nissler:2007zz} are different especially for the decay $\eta' \to \pi^+
\pi^- \mu^+ \mu^-$. The asymmetry term was only measured for the decay
$\eta\to \pi^+\pi^- e^+ e^-$. Although the errors given for the total
branching ratios are extremely small, the ones for the asymmetry term are
still much higher than the value itself. For the other decays we made predictions for the
asymmetry term, which will be more accurate if the parameter $G$ can be given
more precisely.

\chapter{Summary and outlook}
In this work we studied anomalous decays of pseudoscalar mesons. We calculated
explicit expressions for the decay rates and branching ratios and discussed
the relevance of the form factors of the various vector
meson dominance models. Thereafter, we presented our results of the branching
ratios for the different vector meson dominance models and compared them to
other theoretical values and experimental data.
\medskip

For the decay $P\to l^+l^-\gamma$ our results represented the experimental
data very well. Also the theoretical values calculated by other groups totally
agreed with ours. The difference between the values calculated with and without
VMD models was very pronounced for the decays $\eta / \eta' \to \mu^+\mu^-
\gamma$. We were able to conclude that the VMD factor is needed here to represent the data, but we had
no preference on one of the models. For the $\eta' \to l^+ l^- \gamma$ we also showed the contribution of the width.
\smallskip

The situation in the $P\to l^+l^-l^+l^-$-channel was slightly different. Our values and the other presented theoretical values of the decay $\pi^0
\to e^+e^-e^+e^-$ could represent the experimental data equally well. For the interference
term we found differences to the results of the older calculations of \cite{Miyazaki:1974qi} which had
also been found by \cite{Barker:2002ib}. We calculated very different values for
the decays $\eta \to l^+l^-l^+l^-$ in comparison to the other theoretical groups. We can
refer again to \cite{Barker:2002ib} where an error in the calculations of
\cite{Miyazaki:1974qi} was found. Unfortunately experimental data are very scarce and only give upper bounds
that all theoretical predictions could meet, so we could not constrain our results
there. Theoretical calculations and experimental data for the respective
$\eta'$-decays do not exist so that we could only make predictions. In the $\eta'$ sector we also found contributions of the width in the decays $\eta'\to e^+e^-e^+e^-$ and $\eta' \to \mu^+ \mu^- e^+ e^-$.
\smallskip

Our improvements in the case of $P \to l^+ l^-$ were very limited. We basically
followed the calculations of \cite{Dorokhov:2007bd} and
\cite{Dorokhov:2009xs}. Our calculated values for the decay $\pi^0 \to e^+
e^-$ are very close to the ones they
calculated and still far away from the experimental data. For the $\eta$ decays
our results fall between the values of \cite{Dorokhov:2007bd} and
\cite{Dorokhov:2009xs} and  even have an overlap with the experimental
data for the $\mu^+\mu^-$
decay channel. The values for the $\eta'$ decays are again very close to the ones of \cite{Dorokhov:2007bd} and
\cite{Dorokhov:2009xs}, but there exist no data to compare with.
\smallskip

In general, we found that the differences between the VMD models in the triangle
anomaly sector are insignificant. This result is welcome, because the
hidden gauge model could already describe the data very well.
\medskip

In the box anomaly sector, we found that the modified VMD
model led to improvements. The calculated values for the branching ratios of the decay $\eta \to \pi^+ \pi^- \gamma$
represented the data very well for both VMD models, but the improved VMD model
achieved a better
agreement with the theoretical values calculated via unitary chiral
perturbation theory (\cite{Nissler:2007zz}). This was also valid for the decay
$\eta \to \pi^+ \pi^- e^+ e^-$. On the other hand the experimental
situation is different, because the recent measurements done by KLOE supplied
distinctly smaller values. Here the modified VMD model was closer to the
experimental data and could almost reach an overlap. The calculated values for
the decay $\eta \to \pi^+ \pi^- \mu^+ \mu^-$ matched with the ones of other
theoretical calculations and met the upper experimental bound.
\smallskip

In the decay $\eta \to \pi^+ \pi^- e^+ e^-$ we also analyzed the CP-violating
asymmetry term and compared our values to recent experimental data. We
calculated an upper bound for the model specific factor $G$. This factor
scales a CP-violating, flavor-conserving local four-quark operator which is
sensitive to the $s \bar s$ content of
$\eta$ and $\eta'$ and determines the strength of the  additional electric form
factor $E$. We verified the claim that it could be of natural size. We also
gave a prediction for the asymmetry term of the decay $\eta \to \pi^+ \pi^-
\mu^+ \mu^-$.
\smallskip

For the $\eta'$ decays the situation was again different. The only existing
data were the ones of the decay $\eta' \to \pi^+ \pi^- e^+ e^-$. Here all models
represented the data very well. For the decay $\eta' \to \pi^+ \pi^- \mu^+ \mu^-$ our
values were very different from the one calculated by
\cite{Nissler:2007zz}, but the experimental situation is too scarce that a
preference of any work can be justified. We found again very interesting contributions of the width in the $\eta'$ sector. Especially for the asymmetry term we had a change in the algebraic sign in the region of the vector meson mass.\\

\bigskip

We completed the calculations concerning the anomalous decays of the $\eta$
and the $\pi^0$ in the framework of vector meson dominance. The only decays
where the experimental data could not be described by theoretical calculations
were the ones into two leptons $P\to l^+ l^-$ and the decay $\eta \to \pi^+
\pi^- e^+ e^-$. If the experimental data are
confirmed, the probability that new physics is needed to describe the data is
high indeed. The same holds for the
decay $\eta \to \pi^+ \pi^- e^+ e^-$, although the theoretical values and data
are much closer than in the decay into two leptons.\smallskip

The anomalous $\eta'$-decay $\eta'\to \pi^+ \pi^- \pi^+ \pi^-$ should be investigated. The kinematics should be very similar to the ones of the
decays $P \to l^+ l^- l^+ l^-$, especially an interference term will contribute, but the form factor would be of special interest.

\setcounter{secnumdepth}{1}

\appendix
\chapter{Kinematics}

\section{The {\em parallel} boosts}
Throughout we assume that all boosts are performed parallel (or anti-parallel) to 
the virtual- or real-photon axis which should point parallel to the $\hatz$-axis.
Therefore only the $0^{\rm th}$ and $3^{\rm rd}$ components of the 4-vectors
will be affected by the boosts ({\it e.g.}
$q^0$ and $q^z$), whereas the $1^{\rm st}$ and $2^{\rm nd}$
components $q^x$ and $q^y$ will remain untouched.
This is the reason for the notation $\mathbf{q}_{\|}$ and $\mathbf{q}_{\perp}$ 
introduced in \equa{qpara} and \equa{qperp}. Without loss of generality, we can therefore always
mangage to rewrite  \equa{pmometa} as 
\begin{eqnarray}
  \vthree{P^0}{\nullbf_\perp}{\Pbf_\para} 
  &=&   \vthree{p^0}{\nullbf_\perp}{\pbf_\para} 
    + \vthree{k^0}{\nullbf_\perp}{\kbf_\para} \nn \\
      &=&   \vthree{p_+^0}{\pplusbf_\perp}{\pplusbf_\para}   
    +  \vthree{p_-^0}{-\pplusbf_\perp}{\pminbf_\para}   
     +  \vthree{k^0}{\nullbf_\perp}{\kbf_\para}   \,.
 \label{pgenthree}\\
  &=&   \vthree{p_+^0}{\pplusbf_\perp}{\pplusbf_\para}   
    +  \vthree{p_-^0}{-\pplusbf_\perp}{\pminbf_\para}   
     +  \vthree{k_-^0}{\kminbf_\perp}{\kminbf_\para}   
    +  \vthree{k_+^0}{-\kminbf_\perp}{\kplusbf_\para} 
    \label{pgentwo}
\end{eqnarray}
Note that the frames and boosts have been chosen in such a way that 
$\Pbf_\perp = \pbf_\perp = \kbf_\perp \equiv  \nullbf_\perp$. Therefore, we always have
\begin{eqnarray}
  \pminbf_\perp \equiv -\pplusbf_\perp\,,
 \\
  \kplusbf_\perp \equiv -\kminbf_\perp \end{eqnarray}
as it was applied in \equa{pgentwo}.


\section{The relevant frames}
The coordinates and four-momenta 
in the $P$ rest frame are denoted here by a tilde~($\widetilde{\mbox{\ }}$), 
the ones
in the $p_+ p_-$ rest frame by an asterix\,($^\star$), and the ones in the
$k_+ k_-$ rest frame by a
a diamond\,($^\diamond$).\\
Thus,  the relation\,\equa{pgenthree} is given in the $P$ rest frame by
\begin{equation}
 \vthree{ { \widetilde P}^0  } { { \nullbf}_\perp}{ { \widetilde \Pbf }_{\|}}
 \equiv \vthree{ m_P } { {  \nullbf }_\perp}{ {  \nullbf }_{\|}}
 = \underbrace{\vthree{ { \widetilde p_+}^0  } { { \pplusbf }_\perp}{ { \widetilde \pplusbf }_{\|}}
+ \vthree{ { \widetilde p_-}^0  } {- { \pplusbf }_\perp}{ { \widetilde \pminbf }_{\|}}}_{
\vthree{\widetilde p^0}{\nullbf_\perp}{-\widetilde\kbf_\para}}
+ \vthree{ { \widetilde k}^0  } { { \nullbf }_\perp}{ { \widetilde \kbf }_{\|}} \,,
\end{equation}
whereas in the $p_+p_-$ rest frame it reads as
\begin{eqnarray}
 \vthree{ {P^\star}^0  } { { \nullbf}_\perp}{ {\Pbf^{\!\!\star} }_{\|}}
  &=& \vthree{ { p^\star }^0  } { { \nullbf }_\perp}{  \nullbf_{\|}}
+ \vthree{ {k^\star}^0  } { { \nullbf }_\perp}{ {\kbf^{\star} }_{\|}} 
 \equiv \vthree{ \sqrt{s_{pp} }} { { \nullbf }_\perp}{  \nullbf_{\|}}
+ \vthree{ {k^\star}^0  } { { \nullbf }_\perp}{ {\kbf^{\star} }_{\|}}  \nn
\\
 &=& \vthree{ { p_+^\star }^0  } { { \pplusbf }_\perp}{ { \pplusbf^{\!\!\star} }_{\|}}
+ \vthree{ { p_+^\star }^0  } {- { \pplusbf }_\perp}{ { -\pplusbf^{\!\!\star} }_{\|}}
+ \vthree{ {k^\star}^0  } { { \nullbf }_\perp}{ {\kbf^{\star} }_{\|}} \nn 
\\
&=& \vthree{ \half \sqrt{s_{pp}} } { { \pplusbf }_\perp}{ { \pplusbf^{\!\!\star} }_{\|}}
+ \vthree{  \half\sqrt{s_{pp}} } {- { \pplusbf }_\perp}{ { -\pplusbf^{\!\!\star} }_{\|}}
+ \vthree{ {k^\star}^0  } { { \nullbf }_\perp}{ {\kbf^{\star} }_{\|}} 
=  \vthree{ {\peta^\star}^0  } { { \nullbf}_\perp}{ \kbf^\star_\para}
 \label{petapipilast}
\,.
\end{eqnarray}
Switching the asterix\,($^\star$) with a diamond\,($^\diamond$) and $p$ with
$k$ this relation
holds in the $k_+ k_-$ rest frame. 

\section{Comparison with other kinematics}

Note that $\theta_{p_+}^\star$ is exactly the angle $\theta$ defined below Eq.\,(6)
of Ref.\,\cite{Geng:2002ua}.
Moreover, since the three-momentum $\Pbf^{\star}$
of the $P$ meson in the $p_+p_-$ rest frame is identical to the three-momentum
$\kbf^\star$  in that frame (see \equa{petapipilast}), the angle 
 $\theta_{p_+}^\star$ is identical to the angle $\theta_\pi$ defined below Eq.\,(1) 
 of Ref.\,\cite{Gao:2002gq}, {\it i.e.} identical to the angle between the $p_+$ three-momentum
 and the $P$ three-momentum in the $p_+ p_-$ rest frame:
\begin{equation}
\theta_{p_+}^\star \equiv\  \theta \ \mbox{of Ref.\,\cite{Geng:2002ua}}\ \  \equiv \ \theta_\pi\ \mbox{of Ref.\,\cite{Gao:2002gq}.}
 \label{thess}
\end{equation}
 Finally note
 \begin{equation}
\left( \pplusbf_{\!\perp}\right)^2 = \left(\widetilde\pbf _+\right)^2 \sin^2 \widetilde\theta_{p_+}
= \left(\pbf _+^\star \right)^2 \sin^2 \theta_{p_+}^\star 
\equiv \left(\pbf _+^\star \right)^2 \sin^2 \theta_{p}\,, 
\label{pess}
 \end{equation} 
 and in total analogy 
 \begin{equation}
\left( {\kbf_-}_{\!\perp}\right)^2 = \left(\widetilde\kbf _-\right)^2 \sin^2 \widetilde\theta_{k^-}
= \left(\kbf _-^\diamond \right)^2 \sin^2 \theta_{k^-}^\diamond 
\equiv \left(\kbf _-^\diamond \right)^2 \sin^2 \theta_{k} 
 \,.
 \label{kess}
 \end{equation}

See Ref.\cite{Gao:2002gq} for the definition of $\theta_k$:  {\it i.e.} $\theta_k$ is
 the angle between the $k_-$ three-momentum and the $P$ three-momentum in the $k_-k_+$
 rest frame.  Note that  the $P$ three-momentum in the $k_-k_+$
 rest frame points into the negative $z$-direction,  therefore $\theta_k$ should be replaced
 by $\pi-\theta_k$ if a common coordinate system (which involves $\theta_p$) is used.
  Equations \equa{thess},  \equa{pess} and \equa{kess} are the essential
  formulae.
\\

The kinematic of the decay $P(P) \to p_1^+(p_+)\,p_2^-(p_-)\, p_3^+(k_+)\, p_4^-(k_-)$ is
very close to the kinematics of the $K_{l4}$ decay,  the decay  process of one kaon into two
pions, one anti-lepton and one (to the lepton corresponding) neutrino, {\it e.g.}:$
K^+(p_K)\to \pi^+(p_+)\, \pi^-(p_-) \,l^+(k_l)\, \nu_l(k_\nu)\,.
$ The corresponding kinematics were introduced and described in Cabibbo and 
Maksymovicz\,\cite{Cabibbo:1965zz}, 
Pais and Treiman\,\cite{Pais:1968zz} and 
Bijnens et al.\,\cite{Bijnens:1994me}. 
There is, however, one subtlety: the mass of the neutrino $\nu_l$ is of course 
assumed to be zero, whereas here both leptons have the non-vanishing mass $m_e$.
This induces changes in {\it e.g.} \equa{kppkmkpmkm}, \equa{kpmkmkpmkm}
\equa{ppppmkpmkm}, \equa{pmmpmkpmkm}, \equa{epsrel} see below.

\label{sec:Cabibbo:1965zz}

Refs.\,\cite{Cabibbo:1965zz}, \cite{Pais:1968zz} and \cite{Bijnens:1994me} utilize the follwing five variables (see Fig.\,1 of
Ref.\,\cite{Cabibbo:1965zz} or Fig.\,5.1 of Ref.\,\cite{Bijnens:1994me}) 
transcribed to the decay $P \to p_1 p_2 p_3 p_4$:
\begin{figure}[hbt]
  \includegraphics{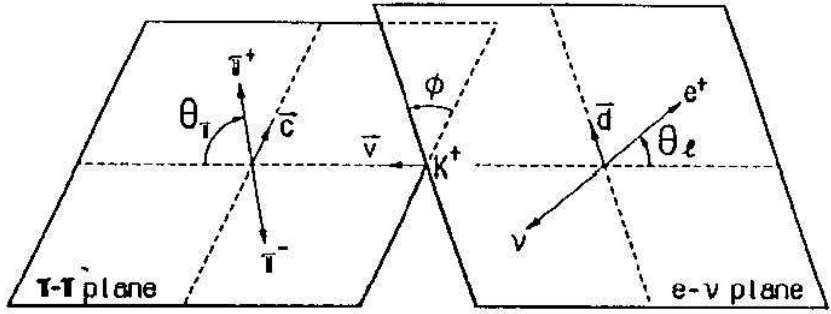}
\caption{ \label{fig:1}\footnotesize{Kinematics of the $K_{l4}$ decay \cite{Bijnens:1994me}}}
\end{figure}
\begin{enumerate}
\item 
$s_{pp}\equiv (p_+ + p_-)^2$ (corresponding to $R^2$  in \cite{Cabibbo:1965zz}
and $s_\pi$ in \cite{Pais:1968zz,Bijnens:1994me}), the effective mass squared
of the $p_1 p_2$ system,
\item 
$s_{kk}\equiv (k_+ +k_-)^2$ (corresponding to $K^2$ in \cite{Cabibbo:1965zz} and $s_l$ in
\cite{Pais:1968zz,Bijnens:1994me}), the effective mass squared of the $p_3 p_4$ system,
\item 
$\theta_{p}$ (corresponding to $\theta$ in \cite{Cabibbo:1965zz} and $\theta_\pi$ in
\cite{Pais:1968zz,Bijnens:1994me}), 
the polar angle ($0\!\leq\! \theta_{p}\!\leq\!\pi$) of the $p_1(p^+)$ in the $p_1p_2$ 
rest frame with respect to the 
direction of flight of the $p_1p_2$ in the $P$ rest frame,
\item 
$\theta_{k}$ (corresponding to $\zeta$ in \cite{Cabibbo:1965zz} and $\theta_l$ in Refs.\,\cite{Pais:1968zz,Bijnens:1994me}), 
the polar angle ($0\!\leq\! \theta_{k}\!\leq\!\pi$)
of the $p_4(k^-)$ in the $p_3 p_4$ rest frame with respect
to the direction of flight of the $p_3 p_4$ in the $P$ rest frame, and
\item 
$\varphi$ ($\phi$ in \cite{Cabibbo:1965zz,Bijnens:1994me} and $\varphi$ in \cite{Pais:1968zz}), the azimuthal 
angle ($0\!\leq\! \varphi\! \leq \! 2\pi$) between the plane formed by $p_1p_2$ in the $P$ rest frame  and the
corresponding plane formed by the $p_3 p_4$.
\end{enumerate}
Ref.\,\cite{Pais:1968zz} uses a convention for the metric opposite to the standard Bjorken-Drell metric.
The latter is applied here and in Refs.\cite{Cabibbo:1965zz,Bijnens:1994me}. 
Thus\\

\begin{center}
$\theta_{p} =  \theta \,\mbox{of Ref.\,\cite{Geng:2002ua}}\   =
  \ \theta_\pi\,\mbox{of Ref.\,\cite{Gao:2002gq}}\  = \pi-\theta\,\mbox{of
    Ref.\,\cite{Cabibbo:1965zz}}\  $\\
$=\pi -\theta_\pi\,\mbox{of 
Ref.\,\cite{Pais:1968zz,Bijnens:1994me}}=\pi -\theta_{12}\,\mbox{of 
Ref.\,\cite{Barker:2002ib}} $ 
\end{center}

and\\

\begin{center}
$\theta_{k}   \ \ = \theta_e\ \mbox{of Ref.\,\cite{Gao:2002gq}}\  \ = \zeta\ \mbox{of Ref.\,\cite{Cabibbo:1965zz}}\  \ =\theta_l\ \mbox{of 
Ref.\,\cite{Pais:1968zz,Bijnens:1994me}.} =\theta_{34}\ \mbox{of 
Ref.\,\cite{Pais:1968zz,Bijnens:1994me}.} $ 
\end{center}

In Ref.\,\cite{Cabibbo:1965zz,Bijnens:1994me} the following explicit construction can be found:
Let $\textbf{v}$ be a unit vector along the direction of flight of the $\pi^+\pi^-$ in the $P$ rest frame
($\Sigma_P$), {\it i.e.} $\textbf{v} =-\hat \kbf$. Furthermore, let  $\textbf{c}$ ($\textbf{d}$) a unit vector along the projection of $\pbf_+$ ($\kbf_+$) 
perpendicular to $\textbf{v}$ ($-\textbf{v}$),
\begin{eqnarray}
   \textbf{c} &=& \frac{\pbf_+ - \textbf{v} (\textbf{v}\cdot \pbf_+)}{\sqrt{\pbf_+^2 +(\textbf{v}\cdot \pbf_+)^2|}}  \equiv   \!\!\!\!\hat{\ \ \pbf_+}_{\!\perp} \, \\
 \textbf{d} &=& \frac{\kbf_+ - \textbf{v} (\textbf{v}\cdot \kbf_+)}{\sqrt{\kbf_+^2 +(\textbf{v}\cdot \kbf_+)^2|}}
 \ \equiv\  \!\!\!\!\hat{\ \ \kbf_+}_{\!\perp} = - \!\!\!\!\hat{\ \ \kbf_-}_{\!\perp}
 \,.
 \end{eqnarray}
Then one has
\begin{eqnarray}
   \cos \theta_p &=& 
    -\textbf{v}\cdot \hat\pbf_{+} \ =  \hat\kbf \cdot \hat \pbf_+ = \cos\theta_{p}^\star\,, 
 \label{khatpp} \\
     \cos\theta_k &=& -\textbf{v}\cdot\hat\kbf_{+} 
      = \hat\kbf\cdot \hat\kbf_{+}= -\hat\kbf\cdot\hat\kbf_- =\cos\theta_{k}^\diamond\,, 
  \label{khatkm}\\
       \cos\varphi &=& \textbf{c}\cdot \textbf{d} = \!\!\!\!\hat{\ \ \pbf_+}_\perp \cdot \!\!\!\!\hat{\ \ \kbf_+}_\perp=
      -\!\!\!\!\hat{\ \ \pbf_+}_{\!\perp} \cdot \!\!\!\!\hat{\ \ \kbf_-}_{\!\perp} =-\cos\phi\,,  
 \label{pppkmp}
\\
    \sin\varphi &=& (\textbf{c}\wedge\textbf{v})\cdot \textbf{d} = -\textbf{v} \cdot (\textbf{c} \wedge\textbf{d})\nn
   \\
   &=& 
    \hat\kbf\cdot\left(\!\!\!\!\hat{\ \ \pbf_+}_{\!\perp}\wedge (-\!\!\!\!\hat{\ \ \kbf_{-}}_{\!\perp})\right) 
    = -\sin\phi\,.
    \label{cross}
\end{eqnarray}
Thus $\varphi = \pi +\phi$.

\section{Invariant expressions of the decay momenta}

The interesting relations read:
\begin{eqnarray}
 p^2=(p_+ + p_-)^2 &=& 2 m_p^2 +2 p_+ \cdot p_- = s_{pp}\,, 
 \\
 (p_+ + p_-)\cdot (p_+ - p_-) & =& m_p^2-m_p^2 =0\,
\\
  (p_+ -  p_-)^2 &=& 2 m_p^2 - 2 p_+\cdot p_- = 4 m_p^2 - s_{pp} = -s_{pp}\beta_p^2\,, 
\\
  k^2=(k_+ + k_-)^2 &=&  2 m_k^2 + 2 k_+ \cdot k_- = s_{kk} \,,
\\
  (k_+ + k_-)\cdot(k_+-k_-) &=& m_k^2 - m_k^2 = 0\,,
   \label{kppkmkpmkm}
\\
  (k_+ -  k_-)^2 &=&  2 m_k^2  - 2 k_+\cdot k_- = 4m_k^2 - s_{kk}=-s_{kk}\beta_k^2\,, 
 \label{kpmkmkpmkm}
\\
(p_+ + p_- + k_+ +k_ -)^2 &=& (p+k)^2=P^2=m_P^2\, 
\\
  (p_+ + p_-)\cdot (k_+ + k_-) &=& P \cdot (k_+ \!+\! k_-) -s_{kk} = (p_+\!+\!p_-)\cdot P-s_{pp}\nn \\ 
  &=&\half \left(m_P^2 - s_{pp} -s_{kk}\right) \,
   \label{ppppmkppkm}
\\
  (p_+ - p_-)\cdot(k_+ + k_-) &=& (p_+ -p_-)\cdot P \nn \\
&=&     - \beta_p\, \half \lambda^{1/2}(s_{pp},m_P^2, s_{kk})\cos\theta_p \,,
  \label{ppmpmkppkm}    
 \\
  (p_+ + p_-)\cdot (k_+ - k_-) &=&  P \cdot(k_+-k_-) \nn \\
&=&   +\beta_k \half \lambda^{1/2}(s_{kk},m_P^2,s_{pp} ) 
  \cos\theta_k \,,
 \label{ppppmkpmkm}
 \end{eqnarray}
 \begin{eqnarray}
  (p_+ \!-\! p_-)\cdot(k_+ \!-\! k_-) &=&-\half\beta_{p}\beta_k (m_{P}^2-s_{pp}-s_{kk})\cos\theta_{p} \cos\theta_k \nn \\
&&\mbox{}+\sqrt{s_{pp}}\beta_{p}\sqrt{s_{kk}}\beta_{k}\sin\theta_{p}  \sin\theta_{k} \cos\phi\nn\\
  \label{pmmpmkpmkm}
  \\
  \varepsilon_{\mu\nu\alpha\beta}(p_+\! -\! p_-)^\mu p^\nu (k_+\! -\! k_-)^\alpha 
  k^\beta
   &=& -m_P \varepsilon^{ijk} \,\widetilde k^i (\widetilde p_+\!-\!\widetilde p_-)^j 
   (\widetilde k_+\! -\!  \widetilde k_-)^k\nn \\
   &=&  {\textstyle\frac{1}{8}} {\lambda^{1/2}(m_P^2, s_{pp},s_{kk})} 
                          \sqrt{s_{pp}} \beta_p \sqrt{s_{kk}} \beta_k\nn \\
             &&    \quad \mbox{}\times \sin\theta_p\,\sin\theta_k\, \sin\phi \nn
\end{eqnarray}
\begin{equation}
    =       \frac{\left[\lambda(m_P^2, s_{pp},s_{kk}) 
                    \lambda( s_{pp},m_p^2,m_p^2) \lambda( s_{kk},m_k^2,m_k^2)\right]^{1/2}}{8 
                    \sqrt{ s_{pp} s_{kk}} }
                  \sin\theta_p\,\sin\theta_k\, \sin\phi. 
  \label{epsrel}
\end{equation}
Some of the relations given above deserve some additional remarks: 
Equation \equa{ppppmkppkm}  can be rewritten as
\begin{eqnarray}
 k \cdot \underbrace{(p_+ + p_-)}_p &=& \half (m_P^2 -s_{pp}- s_{kk})\,.\label{kppm}
 \end{eqnarray}
 It can be combined with correspondingly rewritten equation \equa{ppmpmkppkm}
 \begin{eqnarray}
 k \cdot( p_+ - p_-) &=& - \kbf^\star\cdot \left (\pbf_+^\star - \pbf_-^\star \right) 
 =      -2|\kbf^\star |\, |\pbf_+^\star| \,\cos\!\theta_p     \nn \\
 &=& -2\,\frac{\lambda^{1/2}(s_{pp},m_P^2,s_{kk})}{2\, \sqrt{s_{pp} } } \ 
         \half \sqrt{ s_{pp}}\,\beta_p\, \cos\!\theta_p\nn \\
 &=&  -\half \beta_p \lambda^{1/2}(s_{pp},m_P^2,s_{kk}) \, \cos\!\theta_p\,,       
 \label{kpmm}
\end{eqnarray}
such that the following relations hold:
\begin{eqnarray}
  k \cdot p_+ &=& \fourth\left(m_P^2-s_{pp}-s_{kk}\right)
  - \fourth \beta_p\,\lambda^{1/2}(s_{pp},m_P^2,s_{kk})\,\cos\!\theta_p\,,
\label{kpp}\\
    k \cdot p_- &=& \fourth\left(m_P^2-s_{pp}-s_{kk}\right)
  +\fourth\beta_p\,\lambda^{1/2}(s_{pp},m_P^2,s_{kk})\,\cos\!\theta_p\,,
\label{kpm}\\
 p \cdot k_+ &=& \fourth\left(m_P^2-s_{pp}-s_{kk}\right)
  + \fourth \beta_k\,\lambda^{1/2}(s_{pp},m_P^2,s_{kk})\,\cos\!\theta_k\,,
\label{pkp}\\
    p \cdot k_- &=& \fourth\left(m_P^2-s_{pp}-s_{kk}\right)
  -\fourth\beta_k\,\lambda^{1/2}(s_{pp},m_P^2,s_{kk})\,\cos\!\theta_k\,.
\label{pkm}
 \end{eqnarray}
 Note that these expressions deviate from Eqs.(10) and (11) of Ref.\,\cite{Gao:2002gq}, where
there is a factor $\mp 1/2$ instead of $\mp 1/4$ in front of $\beta_p$.
In fact, the normalization of Eq.\,\equa{kpmm} exactly agrees with the normalization
of the first relation of Eq.\,(3) from Ref.\,\cite{Pais:1968zz}, when Eqs.\,(2') and (3') of that reference 
are inserted into this relation.\\
Using  \equa{ppppmkpmkm} and \equa{pmmpmkpmkm} we will get the following expressions:\\
\begin{eqnarray}
  (k_+-k_-) \cdot p_+ &=& \fourth\biggl[\lambda^{1/2}(s_{pp}, m_pp^2,s_{kk}) - 
\left(m_P^2-s_{pp}-s_{kk}\right) \beta_p\,\,\cos\!\theta_p\biggr] \cdot \beta_k \cos\theta_k\nn \\
&&\mbox{}+ \half \sqrt{s_{pp}}\beta_{p}\sqrt{s_{kk}}\beta_{k}\sin\theta_p  \sin\theta_k \cos\phi\nn\,, \\
(k_+-k_-) \cdot p_- &=& \fourth\biggl[\lambda^{1/2}(s_{pp}, m_P^2,s_{kk}) + 
\left(m_P^2-s_{pp}-s_{kk}\right) \beta_p\,\,\cos\!\theta_p\biggr] \cdot \beta_k \cos\theta_k\nn \\
&&\mbox{}- \half \sqrt{s_{pp}}\beta_{p}\sqrt{s_{kk}}\beta_{k}\sin\theta_p  \sin\theta_k \cos\phi \,, \nn \\
\label{kmkppm}
  (p_+-p_-) \cdot k_+ &=& - \fourth\biggl[\lambda^{1/2}(s_{pp}, m_P^2,s_{kk}) + 
\left(m_P^2-s_{pp}-s_{kk}\right) \beta_k\,\,\cos\!\theta_k\biggr] \cdot \beta_p \cos\theta_p\nn \\
&&\mbox{}+ \half \sqrt{s_{pp}}\beta_{p}\sqrt{s_{kk}}\beta_{k}\sin\theta_p  \sin\theta_k \cos\phi\nn\,, \\
(p_+-p_-) \cdot k_- &=& - \fourth\biggl[\lambda^{1/2}(s_{pp}, m_P^2,s_{kk}) - 
\left(m_P^2-s_{pp}-s_{kk}\right) \beta_k\,\,\cos\!\theta_k\biggr] \cdot \beta_p \cos\theta_p\nn \\
&&\mbox{}- \half \sqrt{s_{pp}}\beta_{p}\sqrt{s_{kk}}\beta_{k}\sin\theta_p  \sin\theta_k \cos\phi \,. \nn \\
 \end{eqnarray}
In order to derive the relation \equa{pmmpmkpmkm} we used the relation
\begin{equation}
|\kbf_-^\star|_\para=\frac{1}{4\sqrt{s_{pp}}}\biggl( \lambda^{1/2}(s_{kk},m_P^2,s_{pp})-(m_{P}^2-s_{pp}-s_{kk})\beta_k \cos\theta_k \biggr)\,.
\end{equation}
This expression can be calculated via
\[ 
  |\kbf_-^\star|_\para^2=|\kbf_-^\star|^2-|\kbf_-|_\perp^2\quad\mbox{with}\quad
 |\kbf_-^\star|=\sqrt{(k_-^{\star0})^2-m_k^2}\,.
\]
The zero component $k_-^{\star0}$ can be expressed via \equa{ppppmkppkm} and
\equa{ppppmkpmkm} to\\
\begin{equation}
k_{\pm}^{\star0}=\frac{1}{4\sqrt{s_{\pi\pi}}}\biggl(
(m_{P}^2-s_{pp}-s_{kk})\pm\beta_k \lambda^{1/2}(s_{kk},m_P^2,s_{pp} ) 
  \cos\theta_k \biggr)\,.
\end{equation}
With this relations we can calculate the expression\\
\begin{eqnarray}
&&- \underbrace{(\pbf_+^\star\!-\!\pbf_-^\star)_\para}_{2{\pbf_+^\star}_{\!\para}}
  \cdot \underbrace{(\kbf_+^\star\!-\!\kbf_-^\star)_\para}_{\kbf_\para^\star-2 {\kbf_-^\star}_{\!\para}} 
  - (\pbf_+\!-\!\pbf_-)_{\!\perp}\cdot (\kbf_+\!-\!\kbf_-)_{\!\perp}\nn \\
  &&=  -2 \underbrace{( \pbf_+^\star \cdot \hat \kbf )}_{Eq.\,\equa{khatpp}}  ( \kbf^\star \cdot \hat \kbf )
 +4 ( \pbf_+^\star \cdot \hat\kbf )( \kbf_-^\star\cdot\hat\kbf )
 + 4 \underbrace{{\pbf_+}_{\!\perp} \cdot {\kbf_-}_{\!\perp} }_{Eq.\,\equa{pppkmp}}\nn \\
 &&= -2|\pbf_+^\star| \cos\theta_p |\kbf^\star|
                         +4 |\pbf_+^\star|\cos\theta_p |\kbf_-^\star|_\para
                         \nn\\
   &&            \quad          \mbox{}
  +4 |\pbf_+^\star| |\kbf_-^\diamond| \sin\theta_p  \sin\theta_k \cos\phi
\end{eqnarray}
which is essential for the structure of \equa{pmmpmkpmkm}.\\
Finally we list the scalar products:\\
\begin{eqnarray}
  k_+ \cdot p_- &=& \frac{1}{8} \lambda^{1/2}(s_{pp}, m_P^2,s_{kk}) \left[\beta_k \cos\theta_k + 
    \beta_p\,\,\cos\!\theta_p\right]\nn \\
&&\mbox{} + \frac{1}{8} \left(m_P^2-s_{pp}-s_{kk}\right) \left[1+\beta_k \beta_p \cos\theta_k \cos\!\theta_p\right] \nn \\
&&\mbox{}  - \fourth \sqrt{s_{pp}}\beta_{p}\sqrt{s_{kk}}\beta_{k}\sin\theta_p
\sin\theta_k \cos\phi\nn\,, \\
  k_- \cdot p_+ &=& - \frac{1}{8} \lambda^{1/2}(s_{pp}, m_P^2,s_{kk}) \left[\beta_k \cos\theta_k + 
    \beta_p\,\,\cos\!\theta_p\right]\nn \\
&&\mbox{} + \frac{1}{8} \left(m_P^2-s_{pp}-s_{kk}\right) \left[1+\beta_k \beta_p \cos\theta_k \cos\!\theta_p\right] \nn \\
&&\mbox{}  - \fourth
\sqrt{s_{pp}}\beta_{p}\sqrt{s_{kk}}\beta_{k}\sin\theta_p
\sin\theta_k \cos\phi\nn\,, \\
 k_+ \cdot p_+ &=&  \frac{1}{8} \lambda^{1/2}(s_{pp}, m_P^2,s_{kk}) \left[\beta_k \cos\theta_k - 
    \beta_p\,\,\cos\!\theta_p\right]\nn \\
&&\mbox{} + \frac{1}{8} \left(m_P^2-s_{pp}-s_{kk}\right) \left[1 - \beta_k \beta_p \cos\theta_k \cos\!\theta_p\right] \nn \\
&&\mbox{}  + \fourth
\sqrt{s_{pp}}\beta_{p}\sqrt{s_{kk}}\beta_{k}\sin\theta_p
\sin\theta_k \cos\phi\nn\,, \\
 k_- \cdot p_- &=& -  \frac{1}{8} \lambda^{1/2}(s_{pp}, m_P^2,s_{kk}) \left[\beta_k \cos\theta_k - 
    \beta_p\,\,\cos\!\theta_p\right]\nn \\
&&\mbox{} + \frac{1}{8} \left(m_P^2-s_{pp}-s_{kk}\right) \left[1 - \beta_k \beta_p \cos\theta_k \cos\!\theta_p\right] \nn \\
&&\mbox{}  + \fourth \sqrt{s_{pp}}\beta_{p}\sqrt{s_{kk}}\beta_{k}\sin\theta_p  \sin\theta_k \cos\phi\nn\,. \\
 \end{eqnarray}

\section{Projection tensor}
The projection tensor, which we needed in Chapter 3, can be constructed via the
summation over the final spins $s_-$ and $s_+$ of the current $j_\mu(k_-,s_-;k_+,s_+)$
\begin{eqnarray}
&& \!\!\!\! {\cal O}_{\mu \mu'}(k_-,k_+) \equiv 
  \!\!\sum_{s_-=-1/2}^{1/2}  \sum_{s_+=-1/2}^{1/2} 
\!\!  j_\mu(k_-,s_-;k_+,s_+) \, j^\dagger _{\mu'}(k_-,s_-;k_+,s_+)  \nn 
 \\
  &=&  e^2 \sum_{s_-=-1/2}^{1/2} \sum_{s_+=-1/2}^{1/2} 
{\rm Tr} \left[ u(k_-,s_-) \bar u(k_-,s_-) \gamma_\mu v(k_+,s_+) \bar v(k_+,s_+) \gamma_{\mu'} \right]
\nn \\
&=& e^2 {\rm Tr}\left[ (k_-\!\!\!\!\!\!/ \,\,+ m) \gamma_\mu  (k_+\!\!\!\!\!\!/ \,\,- m) \gamma_{\mu'}\right] 
  \nn
\\
&=& 4 e^2 \left[ (k_-)_\mu (k_+)_{\mu'} +  (k_+)_\mu (k_-)_{\mu'}-g_{\mu \mu'} (k_- \cdot k_+ + m^2)
 \right]  \nn \\
 &=& -2 e^2 \left[  g_{\mu \mu'} (\underbrace{k_+ \!+\!k_-}_{=k})^2 
  - (k_+ \!+\! k_-)_\mu (k_+ \!+\! k_-)_{\mu'} 
  +  (k_+ \!-\! k_-)_\mu (k_+ \!-\! k_-)_{\mu'} \right] \nn
\\
  &=& e^2 k^2 \times 2  \left[  -\left( g_{\mu \mu'}  - \frac{k_\mu k_{\mu'}}{k^2} \right) -
  \frac{(k^+ \!-\! k^-)_\mu (k^+ \!-\! k^-)_{\mu'}}{k^2} \right] .
\label{projtensor}
\end{eqnarray}
\section{Decay rate}
We will now discuss the decay rate for a decay of a particle $P$ with momentum
$p_P$ into four particles with momenta $p_+,p_-,k_+$ and $k_-$. The
notation of the particles follows the ones of the momenta. The decay rates into two and three particles are also
given, \equa{2dr} and \equa{3dr}.\\
The decay rate is defined as follows:\\
\begin{equation}
{\rm d}\Gamma=\frac{(2\pi)^4}{2m_{\eta}}|A|^2 {\rm d} \Phi_4 (p_P;p_+,p_-,k_+,k_-)\\
\end{equation}
with the four body phase space:
\begin{eqnarray}
&& {\rm d}\Phi_4(p_P;p_+,p_-,k_+,k_-)\nn \\
&& =
\delta^4\!\left( p_P\!-\!(p_+  \!+\!p_-  \!+\!k_+   \!+\!k_-  )  \right)
\frac{{\rm d}^3p_+}{ (2\pi)^3 2 E_{p_+}}\frac{{\rm d}^3p_-}{(2\pi)^3 2E_{p_-}}\frac{{\rm d}^3k_+}{(2\pi)^3 2 E_{k_+}}\frac{{\rm d}^3k_-}{(2\pi)^3 2E_{k_-}}.\nn\\
\end{eqnarray}
$|A|^2$ is the matrix element squared.\\
We investigate the case where $P$ decays into three particles with momenta $p_+,p_-$ and $k$, where the third particle $k$, in our decay it is always a (off-shell) photon, decays into the remaining two particles $k_+, k_-$ via a two-body decay:
\begin{equation}
  {\rm d}\Phi_4(p_P;p_+,p_-,k_+,k_-)=(2\pi)^3 {\rm d}\Phi_2(k;p_+,p_-)\cdot {\rm d}\Phi_{3}(p_P;p_+,p_-,k)\,{\rm d}k^2 \\
\end{equation}
with the following expressions for the two body and three body decays \cite{Amsler:2008zzb}:
\begin{equation}
{\rm d}\Phi_2(k;p_+,p_-)=\frac{1}{16\pi^2}\frac{1}{(2\pi)^4}\frac{|\kbf_+^\diamond|}{\sqrt{s_{k_+k_-}}}
{\rm d}\Omega_{k_+}^\diamond ,
\label{2dr}
\end{equation}
\begin{equation}
{\rm d}\Phi_{3}(p_P;p_+,p_-,k)=\frac{1}{(2\pi)^9}\frac{1}{8m_P}|\pbf_+^*| |\widetilde \kbf|
{\rm d}m_{p_+p_-} {\rm d}\Omega_{p_+}^\ast {\rm d} \widetilde \Omega_k.
\label{3dr}
\end{equation}
The decay rate becomes
\begin{equation}
  {\rm d}\Gamma=|A|^2\,\frac{1}{m_P^2\sqrt{s_{k_+k_-}}}\, 
  \frac{1}{2^{14}}\, \frac{1}{\pi^{8}}\,| \kbf_+^\diamond |\, |\pbf_+^*| \,|\widetilde k |\, {\rm d}m_{p_+p_-} {\rm d}\Omega _{k_+}^\diamond {\rm d}\Omega_{p_+}^\ast {\rm d}\widetilde\Omega_k \, {\rm d}k^2 .
\end{equation}
The angles are defined as ${\rm d}\Omega={\rm d}\phi\, {\rm
  d}\!\cos\theta$; ${\rm d}\Omega _{k_+}^\diamond$ is the solid angle of the
$k^+$ in the $k^+k^-$ rest frame, ${\rm d}\Omega_{p_+}^\ast$ is the angle of
$p^+$ in the $p^+p^-$ rest frame; and ${\rm d}\widetilde \Omega_k$ is the angle of $k$ in the $P$ rest frame.\\
We use the relation 
${\rm d}\sqrt{s_{p^+p^-}}=\frac{1}{2\sqrt{s_{p^+p^-}}}{\rm d}s_{p^+p^-}$ in the integration. The momenta are the same as defined in the previous Section.\\
Because the squared matrix element only contains the angles
$\theta_k^\diamond$, $\theta_p^\ast$ and $\phi$ we can integrate over the
remaining angles. Therefore, we combine the azimuthal angles of $\Omega
_{k_+}^\diamond$ and $\Omega_{p_+}^\ast$ to the corresponding
angle $\varphi$ as the difference between the planes and integrate over the
remaining independent angle in Fig. \ref{fig:1}. The angles $\Omega_k$ stay untouched and can be
integrated out.
The decay rate becomes\\
\begin{equation}
{\rm d}\Gamma=|A|^2\frac{1}{m_P^2\sqrt{s_{k_+k_-}}}\,
\frac{1}{2^{12}} \,\frac{1}{\pi^{6}}\, |\kbf_+^\diamond|\, |\pbf_+^\ast|\, |\widetilde \kbf|\, \frac{1}{\sqrt{s_{\pi \pi}}}{\rm d}s_{p^+p^-}\, {\rm d}\!\cos\theta_k^\diamond \,{\rm d}\!\cos\theta_{p}^\ast 
\,{\rm d}\varphi\, {\rm d}k^2\,.
\label{4dr}
\end{equation}

\chapter{Translation formulae for the mixed and interference term of the decay $P \to l^+ l^- l^+ l^-$}
Most of the expressions are invariant and independent of the different rest frames and
therefore easy to be calculated in terms of the equations given in Appendix A:
\begin{eqnarray}
&& s_{14} = \frac{1}{4} \biggl[m_P^2 - \beta_{12}^2 s_{12} - \beta_{34}^2
  s_{34} + \lambda^{1/2}(m_{P}^2,s_{14},s_{23})\bigl[ \beta_{12} \cos
    \theta_{12}+ \beta_{34} \cos\theta_{34} \bigr] \nonumber\\
&& \qquad\qquad + (m_\eta^2 - s_{12} -  s_{34}) \beta_{12}\beta_{34} \cos\theta_{12}\cos\theta_{34} \nonumber\\
&& \qquad\qquad - 2 \sqrt{s_{12}}  \sqrt{s_{34}}\beta_{12}\beta_{34} \sin\theta_{12} \sin\theta_{34} \cos\phi
  \biggr]   \label{llllid1},\\
&& s_{23} = \frac{1}{4} \biggl[m_P^2 - \beta_{12}^2 s_{12} - \beta_{34}^2
  s_{34} - \lambda^{1/2}(m_{P}^2,s_{14},s_{23})\bigl[ \beta_{12} \cos
    \theta_{12}+ \beta_{34} \cos\theta_{34} \bigr]  \nonumber\\
&& \qquad\quad + (m_\eta^2 - s_{12} -
  s_{34}) \beta_{12}\beta_{34} \cos\theta_{12}\cos\theta_{34}  \nonumber\\
&& \qquad\qquad - 2 \sqrt{s_{12}}
  \sqrt{s_{34}}\beta_{12}\beta_{34} \sin\theta_{12} \sin\theta_{34} \cos\phi
  \biggr] .
 \label{llllid2}
\end{eqnarray}
Using the above expressions, one can easily
calculate $\lambda(m_{P}^2,s_{14},s_{23})$ as well as $\beta_{14}$ and $\beta_{23}$.\\
The expressions for the angles $\theta_{14}$,
$\theta_{23}$ and $\widetilde\phi$ are more complicated, but can also be given
in terms of the invariants:
\begin{eqnarray}
&& \cos^2\theta_{14} = \biggl(\frac{s_{12} - s_{34} - \frac{1}{2}
  \lambda^{1/2}(m_{P}^2,s_{12},s_{34})
  [ \beta_{12} \cos \theta_{12} - \beta_{34} \cos \theta_{34} ] }{ \beta_{14}
  \lambda^{1/2}(m_{P}^2,s_{12},s_{34})}\biggr)^2 ,
 \\
&& \cos^2\theta_{23} = \biggl(\frac{s_{12} - s_{34} + \frac{1}{2}
  \lambda^{1/2}(m_{P}^2,s_{12},s_{34})
  [ \beta_{12} \cos \theta_{12} - \beta_{34} \cos \theta_{34} ] }{ \beta_{23}
  \lambda^{1/2}(m_{P}^2,s_{12},s_{34})}\biggr)^2,
 \\
&& \sin \theta_{14} \sin \theta_{23} \cos \widetilde\phi = \frac{1}{2
  \sqrt{s_{14}} \sqrt{s_{23}} \beta_{14} \beta_{23}}  \nn \\
&& \qquad\qquad\qquad\qquad \times \biggl[ s_{12} + s_{34} -
  \frac{1}{2} (m_P^2 - s_{12} - s_{34}) \left[1 - \beta_{12} \beta_{34} \cos
    \theta_{12} \cos \theta_{23} \right]  \nn \\
&&\qquad\qquad\qquad\qquad\quad - \sqrt{s_{12}} \beta_{12} \sqrt{s_{34}} \beta_{34} \sin \theta_{12} \sin
 \theta_{34} \cos \phi  \nn \\
&& \qquad\qquad\qquad\qquad\quad + \biggl( m_P^2 - \frac{1}{2} (m_P^2 - s_{12} -s_{34}) \left[1 + \beta_{12}
   \beta_{34} \cos \theta_{12} \cos \theta_{34} \right] \nn \\
&& \qquad\qquad\qquad\qquad\qquad + \sqrt{s_{12}} \beta_{12} \sqrt{s_{34}} \beta_{34} \sin \theta_{12} \sin
 \theta_{34} \cos \phi \biggr) \beta_{12} \beta_{34} \cos \theta_{12} \cos
 \theta_{34} \biggr] . \nn \\
\label{idangles}
\end{eqnarray}\\

\chapter{Further calculations for the decay $P\to l^+l^-$}
\section{Numerator of  the $P\to l^+l^-$ amplitude Eq. \equa{All}}
To calculate the numerator of the Integral in $A$ we first deal with $Tr \left[ v(p',s_+) \bar u(p,s_-)  \right]$
in \equa{All}. This form is rather unusual, because normally the completeness
relations have a form with the same particle operator and the same momenta, e.g. $u(p,s) \bar u(p,s)$. The derivation of such terms can be found in
Appendix A of \cite{Martin:1970ai}. Therefore we calculate structures with
different momenta $\left( u(k,s) \bar u(p,s') \right)$ first and insert the
different operators via the expressions $v(k,s) = \gamma_5 u(k,-s)$,
respectively, $\bar v(k,-s) = - \bar u(k,s)\gamma_5$. A general form for all
structures is given in equation (A6) of that paper. Using this expressions general forms
of projection operators on the singlet and triplet state of an outgoing lepton
pair can be calculated. Because the total angular momentum of the lepton pair is $J=0$
it can only be
either in a singlet $^1S_0$ or in a triplet $^3P_0$ state. The decaying particle is a pseudoscalar meson with $C=+1$ and $P=-1$, such that $CP=-1$. Since $CP$ of the lepton pair is given by $CP =
(-1)^{L+S}(-1)^{s+1} = (-1)^{s+1}$ the outgoing lepton pair has
to be in a singlet state to keep $CP$-invariance. We can now replace the term
in \equa{All} by the projector on the singlet state given in equation (A16) of \cite{Martin:1970ai}. This
reads,
\begin{eqnarray}
P^{(0)}(p,k) &=& \frac{1}{\sqrt{2}}\left[ v(p,+) \bar u(k,-) + v(p,-) \bar
  u(p,+) \right]\nn \\
&&= \frac{1}{2\left(m_pm_k + p \cdot k\right)^{\half}}\left( -A_\mu \gamma^\mu \gamma_5 +
  T_{\mu\nu}\sigma^{\mu\nu} +P \gamma_5  \right)\nn  \\
&&= \frac{1}{2\sqrt{2t}} \left[ -2m(p+k)_\mu \gamma^\mu \gamma^5 + \half
  \varepsilon_{\mu\nu\rho\sigma}(k^\rho p^\sigma - p^\rho k^\sigma) + t \gamma_5)  \right],
\end{eqnarray}
where 
\begin{eqnarray}
A_\mu &=& m k_\mu + m p_\mu ,\nn \\
T_{\mu\nu}&=&\frac{1}{4}\varepsilon_{\mu\nu\rho\sigma}(k^\rho p^\sigma - p^\rho
k^\sigma) ,\nn \\
{\rm and} \,\,\, t= 2P &=& (p+k)^2 = \half (m^2 + p \cdot k)
\end{eqnarray}
were inserted. With this the amplitude changes to
\begin{equation}
A = \frac{e^4}{f_\pi} \frac{i}{\pi^2} \int \frac{dk}{(2 \pi)^4}
\frac{ \varepsilon_{\mu\nu\sigma\tau}k^\sigma q^\tau L^{\mu\nu} }{\left[ (p-k)^2 -m_l^2  \right] k^2 (q-k)^2 } \times VMD(k^2,(q-k)^2) .
\end{equation}
$L_{\mu\nu}$ is the singlet projection of the final lepton pair, which can be
given by calculating the trace
\begin{equation}
L_{\mu\nu} = Tr\left[P^{(0)}(p,p') \gamma_\mu \left( {\slashed p} -{\slashed p'}
  +m_l \right) \gamma_\nu  \right] = -2 i
\sqrt{2}\frac{m_l}{m_P}\varepsilon_{\mu\nu\sigma\tau} p'^\sigma q^\tau .
\end{equation}
Combining the remaining momenta and dealing with the two total antisymmetric
tensors, we can derive
\begin{equation}
\varepsilon_{\mu\nu\sigma\tau}k^\sigma q^\tau L^{\mu\nu} = -\frac{4i}{\sqrt{2}}\frac{m_l}{m_P} \varepsilon^{\mu\nu\sigma\tau}
\varepsilon_{\mu\nu\rho\lambda}k_\sigma q_\tau k^\rho q^\lambda =
\frac{8im_l}{\sqrt{2}m_P}\left( m_P^2 k^2 - (k \cdot q)^2  \right),
\end{equation}
where we used the well known expression
\begin{equation}
\varepsilon^{\mu\nu\sigma\tau} \varepsilon_{\mu\nu\rho\lambda} = -2\left(
  \delta^\sigma_\rho\delta^\tau_\lambda - \delta^\sigma_\lambda\delta^\tau_\rho \right).
\end{equation}

\section{Calculation of the imaginary part of the reduced $P \to l^+ l^-$ amplitude ${\cal A}$}
To calculate the imaginary part of the integral ${\cal A}$ we use the rules of Cutkosky
(\cite{Cutkosky:1960sp}) where the off-shell
propagators are replaced by $\delta$-functions:
\begin{eqnarray}
\frac{1}{k^2} &\rightarrow& -2\pi i \delta (k^2) \nn \\
\frac{1}{(q-k)^2} &\rightarrow&  -2\pi i \delta ((q-k)^2) \nn
\end{eqnarray}
Moreover we switch to the rest frame of the decaying particle $P$ to rewrite its momentum as $q=\begin{pmatrix} m_P \\ \vec 0 \end{pmatrix}$. The VMD part of
course is equal to unity for on-shell photons. This leads to
\begin{eqnarray}
{\rm Im} {\cal A} = \frac{i}{\pi^2} \int \frac{d^4k}{m_P}\frac{m_P^2 k^2
  -m_P^2k_0^2}{p^2-2p\cdot k + k^2 - m_l^2} \times (-2\pi i)^2 \delta (k^2)
\delta ((q-k)^2).
\end{eqnarray}
We can now insert $k^2-k_0^2=\vec k ^2$ and rewrite the $\delta$-functions
\begin{eqnarray}
\delta (k^2) &=& \Theta (k_0) \delta (k^2) = \frac{1}{2k_0}\delta (k_0-|\vec k|^2) , \nn \\
\delta ((q-k)^2)&=& \delta (m_P^2-2m_Pk_0) . \nn
\end{eqnarray}
After integrating over spherical coordinates this reads
\begin{equation}
{\rm Im} {\cal A} = -\frac{\pi}{4} \frac{m_P}{|\vec p|^2} {\rm ln} \left(\frac{p_o-|\vec
  p|}{p_o+|\vec p|}\right), \nn
\end{equation}
which is exactly the desired result
\begin{equation}
{\rm Im} {\cal A} = \frac{\pi}{2\beta}  {\rm ln} \left(\frac{1-\beta}{1+\beta}\right).
\end{equation}

\section{Derivation of the reduced amplitude ${\cal A}(0)$}
We will give the derivation of the subtraction constant ${\cal A}(0)$ following the
work of \cite{Dorokhov:2008cd}.
To evaluate the amplitude ${\cal A}(0)$ we first transform the integral from Minkowski to Euclidean space by $k_0\rightarrow i k_4$. Now we will rewrite the VMD form factor
in terms of a double Mellin transformation. According to {\it e.g.} \cite{Sneddon:1972ev} a Mellin
transformation is defined as
\begin{equation}
A(t)\equiv M[\tilde A (s);t]= \frac{1}{2\pi i}\int_0^\infty \tilde A (s) t^{-s}dt
\end{equation}
with the inverse Mellin transformation given as
\begin{equation}
\tilde A (s) = \int_0^\infty dt A (t) t^{s-1}dt.
\end{equation}
So, the double-mellin-transformed form factor $VMD$ is given by
\begin{equation}
VMD(k^2,(q-k)^2)=\frac{1}{(2\pi i)^2}\int_{\sigma + i R^2}dz
\Phi(z_1,z_2) \left(\frac{\Lambda^2}{k^2}\right)^{z_1} \left(\frac{\Lambda^2}{(k-q)^2}\right)^{z_2} .
\end{equation}
Here $\Lambda$ is the characteristic scale for the form factor, in our case
the vector meson mass $m_V$, $dz=dz_1dz_2$, the vector $\sigma=(\sigma_1,\sigma_2)\in
\mathbb{R}^2$, and $\Phi(z_1,z_2)$ is the inverse Mellin transform of the VMD form factor
given as
\begin{equation}
\Phi(z_1,z_2)=\int_0^\infty dt_1 \int_0^\infty dt_2 t_1^{z_1-1} t_2^{z_2-1} VMD(t_1,t_2) .
\end{equation}
The integral \equa{integralR} can be rewritten using Feynman parameters. Therefore the
denominator reads (see e.g. equation (6.42) of \cite{Peskin:1995ev}):
\begin{eqnarray}
  &&\frac{1}{(k^2)^{z_1+1}[(k-q)^2]^{z_2+1}[(p_--k)^2+m^2]}\nn \\
&=&\frac{\Gamma(3+z_1+z_2)}{\Gamma(z_1+1)\Gamma(1)\Gamma(z_2+1)}\int
\prod \limits_{i=1}^3 d\alpha_i \, \delta\left( 1-\sum_{i=1}^3 \alpha_i \right)\frac{\alpha_1^{z_1}\alpha_2^{z_2}(\alpha_3^{1-1})}{[k^2+D]^{3+z_1+z_2}}
\end{eqnarray}
with $D=\alpha_3^2m^2-\alpha_1\alpha_2 q^2$. Applying this we can calculate the loop integral:
\begin{eqnarray}
&&\frac{2}{q^2}\int\frac{d^4 k}{\pi^2}\frac{(q\cdot k)^2
  -q^2k^2}{[k^2+D]^{3+z_1+z_2}}\nn \\
&=&\frac{\Gamma(z_1+z_2)}{\Gamma(3+z_1+z_2)D^{z_1+z_2}}\left[ -3+2\frac{\alpha_3^2}{D}\left(
  m^2-\frac{1}{4}q^2 \right) (z_1+z_2) \right] .
\label{loop}
\end{eqnarray}
Using the Mellin transformed form factor and the loop integral \equa{loop}
this leads to \cite{Dorokhov:2008cd}
\begin{eqnarray}
{\cal A}(q^2)&=&\frac{1}{(2\pi i)^2}\int_{\sigma + i R^2} dz\frac{\Phi(z_1,z_2)(\Lambda^2)^{z_1+z_2}\Gamma(z_1+z_2)}{\Gamma(z_1+1)\Gamma(z_2+1)}\nn\\
&& \times \int \Pi_{i=1}^3 d\alpha_i \delta\left( 1-\sum_{i=1}^3 \alpha_i
\right)\frac{\alpha_1^{z_1}\alpha_2^{z_2}}{(\alpha_3^2m^2-\alpha_1\alpha_2
  q^2)^{z_1+z_2}}\nn \\
&& \times \left[ -3+2\frac{\alpha_3^2\left(
  m^2-\frac{1}{4}q^2 \right)}{\alpha_3^2m^2-\alpha_1\alpha_2 q^2} (z_1+z_2) \right] .
\end{eqnarray}
Because we are only interested in ${\cal A}(0)$ we set $q^2=0$. Using again equation
(6.42) of \cite{Peskin:1995ev} to integrate over the Feynman parameters
and setting $\xi^2=m^2/\Lambda^2$ and $z_{12}=z_1+z_2$ this reads
\begin{eqnarray}
{\cal A}(0)&=&\frac{1}{(2\pi i)^2}\int_{\sigma + i R^2} dz
\left(\xi^2\right)^{-(z_1+z_2)}\frac{\Gamma(z_1)\Gamma(z_2)\Gamma(z_{12})\Gamma(1-2z_{12})}{\Gamma(3-z_{12})}\nn
\\
&&\mbox{}\qquad\qquad\qquad\times \left[ \frac{(-3+2z_{12})\Phi(z_1,z_2)}{\Gamma(z_1)\Gamma(z_1)} \right] .
\end{eqnarray}
The integral can be solved, as in \cite{Dorokhov:2008cd},with the help of the
residues:
\begin{equation}
{\cal A}(0)=\sum_{z_r \in \Pi_\Delta} {\rm res}_{z_r} [{\rm Integrand}R(0)] .
\end{equation}
The authors of \cite{Dorokhov:2008cd} followed the work of
\cite{Passare:1996db} and found two contributions to the integral according to
the residues
\begin{eqnarray}
\mbox{} a) \qquad\qquad z_2+\epsilon &=& -\alpha ,\nn\\
z_1&=&-\beta ;
\end{eqnarray}
\begin{eqnarray}
\mbox{} b) \quad\qquad z_2+\epsilon &=& -\alpha ,\nn\\
z_1&=&\alpha-\beta .
\end{eqnarray}
They give the following contributions to the integral \cite{Dorokhov:2008cd}:
\begin{eqnarray}
{\cal A}(0)&=&{\cal A}_a(0)+{\cal A}_b(0) ,\\
{\cal A}_a(0)&=& \sum_{a,b=0}^\infty
\frac{(-1)^{\alpha+\beta}}{\alpha!\beta!}\left(\xi^2\right)^{\alpha+\beta+\epsilon}\frac{\Gamma(-\alpha-\beta-\epsilon)\Gamma(1+2(\alpha+\beta+\epsilon))}{\Gamma(3+\alpha+\epsilon)} ,\nn
\\
&&\mbox{}\qquad\times(-3-2(\alpha+\beta+\epsilon)) \left[
  \frac{\Phi(-\alpha,-\beta)}{\Gamma(-\alpha)\Gamma(-\beta)} \right]\label{CResidueA}\\
{\cal A}_b(0)&=&\sum_{a,b=0}^\infty\frac{(-1)^{\alpha+\beta}}{\alpha!\beta!}\left(\xi^2\right)^{\beta}\frac{\Gamma(1+2\beta)}{\Gamma(3+\beta)}(3-2\beta)\left[\frac{\Phi(\alpha-\beta+\epsilon,-\alpha)}{\Gamma(-\alpha)}\right] .
\label{CResidueB}
\end{eqnarray}
We need to give expressions for the Mellin transformed form factor. This can
be done using Mellin transforms of derivatives as mentioned in Chapter 4.2 of \cite{Sneddon:1972ev}. It can be proven that the following relation holds:
\begin{equation}
M\left[ x^n F^{(n)}(x);s \right]=(-1)^n \frac{\Gamma(s+n)}{\Gamma(s)}\tilde F (s) .
\end{equation}
Here $F^{(n)}(x)$ denotes the derivatives of order of $n$ with respect to the corresponding arguments of the form factor. 
Thus the following relation obviously holds:
\begin{equation}
\int_0^\infty dt t^{-\alpha-1} F(t)=\int_0^\infty dt t^{-\alpha-1} t^n
F^{(n)}(t)\left( (-1)^{-n} \frac{\Gamma(-\alpha)}{\Gamma(-\alpha+n)} \right) .
\end{equation}
Setting $n=\alpha+1$ we can easily integrate over $t$:
\begin{eqnarray}
\int_0^\infty dt t^{-\alpha-1} F(t)&=& (-1)^{\alpha+1} \Gamma(-\alpha)  \int_0^\infty dt F^{(\alpha+1)}(t)\nn \\
&=& (-1)^{\alpha} \Gamma(-\alpha) F^\alpha(0) .
\end{eqnarray}
In the last step we used $F^\alpha(\infty)=0$, which is obviously the case for our VMD form factor. If we extend this relations to two arguments of
$F$, we can now give the following expressions for the Mellin-transformed VMD
form factors \cite{Dorokhov:2008cd}.
\begin{eqnarray}
\frac{\Phi(-\alpha,-\beta)}{\Gamma(-\alpha)\Gamma(-\beta)}&=&(-1)^{\alpha+\beta}VMD^{(\alpha,\beta)}(0,0),\nn\\
\frac{\Phi(z,-\beta)}{\Gamma(-\alpha)}&=&(-1)^{\alpha}\int_0^\infty dt t^{z-1}
VMD^{(0,\alpha)}(t,0).
\label{CMellinB}
\end{eqnarray}
Inserting
\equa{CMellinB} in \equa{CResidueA} and \equa{CResidueB} one gets the following
result:
\begin{eqnarray}
{\cal A}_a(0)&=&-\sum_{n=0}^\infty
\frac{VMD^n(0)}{n!}\left(\xi^2\right)^{n+\epsilon}\frac{\Gamma(-n-\epsilon)\Gamma(1+2(n+\epsilon))}{\Gamma(3+n+\epsilon)}(3+2(n+\epsilon)) , \nn
\\
{\cal A}_b(0)&=&\sum_{n=0}^\infty\frac{(-1)^{n}}{n!}\left(\xi^2\right)^{n}\frac{\Gamma(1+2n)}{\Gamma(3+n)\Gamma(1-\epsilon+n)}(3+2n)
\int_0^\infty dt t^\epsilon VMD^{(n+1)}(t) . \nn \\
\end{eqnarray}
This can be expanded in $\epsilon$, and in the limit $\epsilon \rightarrow
0$ we get
\begin{eqnarray}
{\cal A}(0)&=&\sum_{n=0}^\infty
\left(-\xi^2\right)^{n}\frac{\Gamma(1+2n)}{\Gamma(1+n)\Gamma(3+n)}\Biggl\{
VMD^2(0) \left[ 2+(3+2n)\left( \ln 4\xi^2 -\Psi(n+3)
  \right)\right] \nn\\
&& \mbox{} \qquad\qquad\qquad\qquad\qquad\qquad\quad +(3+2n)\int_0^\infty dtVMD^{n+1}(t) \ln t \Biggr\} .
\end{eqnarray}
To the lowest order of $\xi^2$ one gets the desired result \cite{Dorokhov:2008cd}
\begin{equation}
{\cal A}^{(0)}(0)=\half\left[ 3\ln \xi^2 -\frac{5}{2} +3\int_0^\infty dt VMD^{(1)}(t) \ln t\right] .
\end{equation}
To the next order one finds:
\begin{eqnarray}
{\cal A}^{(1)}(0)=-\xi^2 \frac{1}{3}\left[ VMD^{(1)}(0) \left( 5 \ln \xi^2 +\frac{13}{6}
\right) +5\int_0^{\infty} dt VMD^{(2)} (t) \ln t \right] .
\end{eqnarray}

\bibliography{Diplom}

\begin{thebibliography}{10}

\bibitem{Samios:1961zz}
N.~P. Samios,
\newblock Phys. Rev. {\bf 121}, 275 (1961).

\bibitem{Budagov:1960zz}
Y.~A. Budagov, S.~Viktor, V.~P. Dzhelepov, and P.~F. Ermolov,
\newblock Sov. Phys. JETP {\bf 11}, 755 (1960).

\bibitem{Joseph:1960zz}
D.~W. Joseph,
\newblock Nuovo Cim. {\bf 16}, 997 (1960).

\bibitem{Beddall:2008zza}
A.~Beddall and A.~Beddall,
\newblock Eur. Phys. J. {\bf C54}, 365 (2008).

\bibitem{Landsberg:1986fd}
L.~G. Landsberg,
\newblock Phys. Rept. {\bf 128}, 301 (1985).

\bibitem{Kroll:1955zu}
N.~M. Kroll and W.~Wada,
\newblock Phys. Rev. {\bf 98}, 1355 (1955).

\bibitem{Miyazaki:1974qi}
T.~Miyazaki and E.~Takasugi,
\newblock Phys. Rev. {\bf D8}, 2051 (1973).

\bibitem{Barker:2002ib}
A.~R. Barker, H.~Huang, P.~A. Toale, and J.~Engle,
\newblock Phys. Rev. {\bf D67}, 033008 (2003), hep-ph/0210174.

\bibitem{Lih:2009np}
C.-C. Lih,
\newblock (2009), 0912.2147.

\bibitem{Deshpande:1993zn}
A.~Deshpande {\em et~al.},
\newblock Phys. Rev. Lett. {\bf 71}, 27 (1993).

\bibitem{McFarland:1993wv}
K.~S. McFarland {\em et~al.},
\newblock Phys. Rev. Lett. {\bf 71}, 31 (1993).

\bibitem{Dorokhov:2007bd}
A.~E. Dorokhov and M.~A. Ivanov,
\newblock Phys. Rev. {\bf D75}, 114007 (2007), 0704.3498.

\bibitem{Dorokhov:2008cd}
A.~E. Dorokhov and M.~A. Ivanov,
\newblock JETP Lett. {\bf 87}, 531 (2008), 0803.4493.

\bibitem{Dorokhov:2009xs}
A.~E. Dorokhov, M.~A. Ivanov, and S.~G. Kovalenko,
\newblock Phys. Lett. {\bf B677}, 145 (2009), 0903.4249.

\bibitem{Abouzaid:2008cd}
KTeV, E.~Abouzaid {\em et~al.},
\newblock Phys. Rev. Lett. {\bf 100}, 182001 (2008), 0802.2064.

\bibitem{Berlowski:2008zz}
M.~Berlowski {\em et~al.},
\newblock Phys. Rev. {\bf D77}, 032004 (2008).

\bibitem{Ambrosino:2008cp}
KLOE, F.~Ambrosino {\em et~al.},
\newblock (2008), 0812.4830.

\bibitem{Akhmetshin:2000bw}
CMD-2, R.~R. Akhmetshin {\em et~al.},
\newblock Phys. Lett. {\bf B501}, 191 (2001), hep-ex/0012039.

\bibitem{Picciotto:1991ae}
C.~Picciotto,
\newblock Phys. Rev. {\bf D45}, 1569 (1992).

\bibitem{Geng:2000fs}
C.~Q. Geng, C.~C. Lih, and W.-M. Zhang,
\newblock Phys. Rev. {\bf D62}, 074017 (2000), hep-ph/0007252.

\bibitem{Picciotto:1993aa}
C.~Picciotto and S.~Richardson,
\newblock Phys. Rev. {\bf D48}, 3395 (1993).

\bibitem{Nissler:2007zz}
R.~Nissler and B.~Borasoy,
\newblock AIP Conf. Proc. {\bf 950}, 188 (2007).

\bibitem{Gao:2002gq}
D.-N. Gao,
\newblock Mod. Phys. Lett. {\bf A17}, 1583 (2002), hep-ph/0202002.

\bibitem{Wess:1971yu}
J.~Wess and B.~Zumino,
\newblock Phys. Lett. {\bf B37}, 95 (1971).

\bibitem{Witten:1983tw}
E.~Witten,
\newblock Nucl. Phys. {\bf B223}, 422 (1983).

\bibitem{Bando:1984pw}
M.~Bando, T.~Kugo, and K.~Yamawaki,
\newblock Prog. Theor. Phys. {\bf 73}, 1541 (1985).

\bibitem{Fujiwara:1984mp}
T.~Fujiwara, T.~Kugo, H.~Terao, S.~Uehara, and K.~Yamawaki,
\newblock Prog. Theor. Phys. {\bf 73}, 926 (1985).

\bibitem{Meissner:1986ka}
U.-G. Meißner, N.~Kaiser, A.~Wirzba, and W.~Weise,
\newblock Phys. Rev. Lett. {\bf 57}, 1676 (1986).

\bibitem{Benayoun:2007cu}
M.~Benayoun, P.~David, L.~DelBuono, O.~Leitner, and H.~B. O'Connell,
\newblock Eur. Phys. J. {\bf C55}, 199 (2008), 0711.4482.

\bibitem{Benayoun:2009im}
M.~Benayoun, P.~David, L.~DelBuono, and O.~Leitner,
\newblock Eur. Phys. J. {\bf C65}, 211 (2010), 0907.4047.

\bibitem{Benayoun:2009fz}
M.~Benayoun, P.~David, L.~DelBuono, and O.~Leitner,
\newblock (2009), 0907.5603.

\bibitem{Holstein:2001bt}
B.~R. Holstein,
\newblock Phys. Scripta {\bf T99}, 55 (2002), hep-ph/0112150.

\bibitem{Yost:1988ke}
Particle Data Group, G.~P. Yost {\em et~al.},
\newblock Phys. Lett. {\bf B204}, 1 (1988).

\bibitem{Amsler:2008zzb}
Particle Data Group, C.~Amsler {\em et~al.},
\newblock Phys. Lett. {\bf B667}, 1 (2008).

\bibitem{Adler:1969gk}
S.~L. Adler,
\newblock Phys. Rev. {\bf 177}, 2426 (1969).

\bibitem{Bell:1969ts}
J.~S. Bell and R.~Jackiw,
\newblock Nuovo Cim. {\bf A60}, 47 (1969).

\bibitem{Fujikawa:1979ay}
K.~Fujikawa,
\newblock Phys. Rev. Lett. {\bf 42}, 1195 (1979).

\bibitem{Peskin:1995ev}
M.~E. Peskin and D.~V. Schroeder,
\newblock Reading, USA: Addison-Wesley (1995) 842 p.

\bibitem{Bjorken:1979dk}
J.~D. Bjorken and S.~D. Drell,
\newblock Bibliograph.Inst./Mannheim 1967, 409 P.(B.I.-
  Hochschultaschenbücher, Band 101).

\bibitem{Kaymakcalan:1983qq}
O.~Kaymakcalan, S.~Rajeev, and J.~Schechter,
\newblock Phys. Rev. {\bf D30}, 594 (1984).

\bibitem{Gomm:1984at}
H.~Gomm, O.~Kaymakcalan, and J.~Schechter,
\newblock Phys. Rev. {\bf D30}, 2345 (1984).

\bibitem{Kaymakcalan:1984bz}
O.~Kaymakcalan and J.~Schechter,
\newblock Phys. Rev. {\bf D31}, 1109 (1985).

\bibitem{Jain:1988se}
P.~Jain, N.~W. Park, J.~Schechter, R.~Johnson, and U.-G. Meißner,
\newblock In *Storrs 1988, Proceedings, 4th Meeting of the Division of
  Particles and Fields of the APS* 587-589.

\bibitem{Bando:1985rf}
M.~Bando, T.~Kugo, and K.~Yamawaki,
\newblock Nucl. Phys. {\bf B259}, 493 (1985).

\bibitem{Meissner:1987ge}
U.-G. Meißner,
\newblock Phys. Rept. {\bf 161}, 213 (1988).

\bibitem{Harada:2003jx}
M.~Harada and K.~Yamawaki,
\newblock Phys. Rept. {\bf 381}, 1 (2003), hep-ph/0302103.

\bibitem{Kawarabayashi:1966kd}
K.~Kawarabayashi and M.~Suzuki,
\newblock Phys. Rev. Lett. {\bf 16}, 255 (1966).

\bibitem{Riazuddin:1966sw}
Riazuddin and Fayyazuddin,
\newblock Phys. Rev. {\bf 147}, 1071 (1966).

\bibitem{Sakurai:1960ju}
J.~J. Sakurai,
\newblock Annals Phys. {\bf 11}, 1 (1960).

\bibitem{Bando:1987br}
M.~Bando, T.~Kugo, and K.~Yamawaki,
\newblock Phys. Rept. {\bf 164}, 217 (1988).

\bibitem{Furui:1986ep}
S.-Y. Furui, R.~Kobayashi, and K.~Ujiie,
\newblock Prog. Theor. Phys. {\bf 76}, 963 (1986).

\bibitem{Jain:1987sz}
P.~Jain, R.~Johnson, U.-G. Meißner, N.~W. Park, and J.~Schechter,
\newblock Phys. Rev. {\bf D37}, 3252 (1988).

\bibitem{Bijnens:1988kx}
J.~Bijnens, A.~Bramon, and F.~Cornet,
\newblock Phys. Rev. Lett. {\bf 61}, 1453 (1988).

\bibitem{Ametller:2001nq}
L.~Ametller,
\newblock Phys. Scripta {\bf T99}, 45 (2002), hep-ph/0111278.

\bibitem{PhysRevD.66.076014}
J.~L. Goity, A.~M. Bernstein, and B.~R. Holstein,
\newblock Phys. Rev. D {\bf 66}, 076014 (2002).

\bibitem{PhysRevD.51.4939}
B.~Moussallam,
\newblock Phys. Rev. D {\bf 51}, 4939 (1995).

\bibitem{Ioffe:2007eg}
B.~L. Ioffe and A.~G. Oganesian,
\newblock Phys. Lett. {\bf B647}, 389 (2007), hep-ph/0701077.

\bibitem{PrimEx:2006}
Prim Ex, M.~Kubantsev, I.~Larin, and A.~Gasparyan,
\newblock (2006), physics/0609201v1.

\bibitem{Ericson:1988gk}
T.~E.~O. Ericson and W.~Weise,
\newblock Oxford, UK: Clarendon (1988) 479 P. (The International Series of
  Monographs on Physics, 74).

\bibitem{Itzykson:1980rh}
C.~Itzykson and J.~B. Zuber,
\newblock New York, USA: McGraw-Hill (1980) 705 P.(International Series In Pure
  and Applied Physics).

\bibitem{Vermaseren:2000nd}
J.~A.~M. Vermaseren,
\newblock (2000), math-ph/0010025.

\bibitem{Young:1967zz}
B.-Y. Young,
\newblock Phys. Rev. {\bf 161}, 1620 (1967).

\bibitem{Pratap:1972tb}
M.~Pratap and J.~Smith,
\newblock Phys. Rev. {\bf D5}, 2020 (1972).

\bibitem{Martin:1970ai}
B.~R. Martin, E.~De~Rafael, and J.~Smith,
\newblock Phys. Rev. {\bf D2}, 179 (1970).

\bibitem{Bergstrom:1982zq}
L.~Bergstrom,
\newblock Zeit. Phys. {\bf C14}, 129 (1982).

\bibitem{Bergstrom:1983ay}
L.~Bergstrom, E.~Masso, L.~Ametller, and A.~Bramon,
\newblock Phys. Lett. {\bf B126}, 117 (1983).

\bibitem{Silagadze:2006rt}
Z.~K. Silagadze,
\newblock Phys. Rev. {\bf D74}, 054003 (2006), hep-ph/0606284.

\bibitem{Dorokhov:2008uk}
A.~E. Dorokhov,
\newblock Nucl. Phys. Proc. Suppl. {\bf 181-182}, 37 (2008), 0805.0994.

\bibitem{Cutkosky:1960sp}
R.~E. Cutkosky,
\newblock J. Math. Phys. {\bf 1}, 429 (1960).

\bibitem{Mandelstam:1959bc}
S.~Mandelstam,
\newblock Phys. Rev. {\bf 115}, 1741 (1959).

\bibitem{Bijnens:1994me}
J.~Bijnens, G.~Colangelo, G.~Ecker, and J.~Gasser,
\newblock (1994), hep-ph/9411311.

\bibitem{Geng:2002ua}
C.~Q. Geng, J.~N. Ng, and T.~H. Wu,
\newblock Mod. Phys. Lett. {\bf A17}, 1489 (2002), hep-ph/0201191.

\bibitem{Gorchtein:2008pe}
M.~Gorchtein,
\newblock (2008), 0803.2906.

\bibitem{Geng:1997ws}
C.~Q. Geng, C.~C. Lih, and W.-M. Zhang,
\newblock Phys. Rev. {\bf D57}, 5697 (1998), hep-ph/9710323.

\bibitem{Poblaguev:1990tv}
A.~A. Poblaguev,
\newblock Phys. Lett. {\bf B238}, 108 (1990).

\bibitem{CERNLIB}
CERNLIB,
\newblock http://cernlib.web.cern.ch/cernlib/ .

\bibitem{Bijnens:1999jp}
J.~Bijnens and F.~Perrsson,
\newblock (1999), hep-ph/0106130.

\bibitem{Abouzaid:2006kk}
KTeV, E.~Abouzaid {\em et~al.},
\newblock Phys. Rev. {\bf D75}, 012004 (2007), hep-ex/0610072.

\bibitem{Babu:1982yz}
K.~S. Babu and E.~Ma,
\newblock Phys. Lett. {\bf B119}, 449 (1982).

\bibitem{Ametller:1993we}
L.~Ametller, A.~Bramon, and E.~Masso,
\newblock Phys. Rev. {\bf D48}, 3388 (1993), hep-ph/9302304.

\bibitem{Savage:1992ac}
M.~J. Savage, M.~E. Luke, and M.~B. Wise,
\newblock Phys. Lett. {\bf B291}, 481 (1992), hep-ph/9207233.

\bibitem{GomezDumm:1998gw}
D.~Gomez~Dumm and A.~Pich,
\newblock Phys. Rev. Lett. {\bf 80}, 4633 (1998), hep-ph/9801298.

\bibitem{Knecht:1999gb}
M.~Knecht, S.~Peris, M.~Perrottet, and E.~de~Rafael,
\newblock Phys. Rev. Lett. {\bf 83}, 5230 (1999), hep-ph/9908283.

\bibitem{Abegg:1994wx}
R.~Abegg {\em et~al.},
\newblock Phys. Rev. {\bf D50}, 92 (1994).

\bibitem{Lopez:2007ppa}
CLEO, A.~Lopez {\em et~al.},
\newblock Phys. Rev. Lett. {\bf 99}, 122001 (2007), 0707.1601.

\bibitem{Cabibbo:1965zz}
N.~Cabibbo and A.~Maksymowicz,
\newblock Phys. Rev. {\bf 137}, B438 (1965).

\bibitem{Pais:1968zz}
A.~Pais and S.~B. Treiman,
\newblock Phys. Rev. {\bf 168}, 1858 (1968).

\bibitem{Sneddon:1972ev}
I.~H. Sneddon,
\newblock McGraw-Hill, Inc (1972).

\bibitem{Passare:1996db}
M.~Passare, A.~K. Tsikh, and A.~A. Cheshel,
\newblock Theor. Math. Phys. {\bf 109}, 1544 (1997), hep-th/9609215.

\end{thebibliography}
\bibliographystyle{h-physrev3}

%
%
%

\end{document}